\pdfoutput=1

\documentclass[12pt,nohyper,twosided]{JHEP3}
\usepackage{epsfig}
\usepackage{amsmath}
\usepackage{color}
\usepackage{array}

\usepackage{wick}
\usepackage{cite}

\renewcommand{\bar}{\overline}

\renewcommand{\tilde}{\widetilde}

\definecolor{cr_blue}{rgb}{0.2, 0, 0.8}
\definecolor{cr_red}{rgb}{0.7, 0.126, 0.259}
\definecolor{cr_side1}{rgb}{0,0,0}
\definecolor{cr_side2}{rgb}{0,0,0}
\definecolor{paper_blue}{rgb}{0.3,0.2,0.75}
\definecolor{paper_blue}{rgb}{0.25,0.15,0.75}
\definecolor{paper_red}{rgb}{0.65,0.1,0.15}
\definecolor{paper_green}{rgb}{0.05,0.35,0.125}
\definecolor{paper_grey}{gray}{0.375}

\newcommand{\eq}[1]{\begin{equation}#1\end{equation}}
\newcommand{\eqs}[1]{\begin{equation}\begin{split}#1\end{split}\end{equation}}
\newcommand{\ab}[1]{\langle #1\rangle}
\newcommand{\smallminus}{{\rm\rule[2.4pt]{6pt}{0.65pt}}}
\newcommand{\smallplus}{\hspace{0.5pt}\text{{\small+}}\hspace{-0.5pt}}
\newcommand{\mi}{\smallminus}
\newcommand{\pl}{\smallplus}
\newcommand{\newcap}{{\small\mathrm{\raisebox{0.95pt}{{$\,\bigcap\,$}}}}}

\renewcommand{\tilde}{\widetilde}
\newcommand{\la}{\langle}
\newcommand{\ra}{\rangle}
\newcommand{\figBox}[4]{\mbox{\hspace{#1 cm}\raisebox{#2 cm}{\includegraphics[scale=#3]{#4}}}}

\newcommand{\be}{\begin{eqnarray}}
\newcommand{\ee}{\end{eqnarray}}

\newcommand{\Li}{\text{Li}_2}
\newcommand{\deltatilde}{\widetilde{\Delta}_4}

\title{\mbox{\hspace{-0.35cm}{\LARGE Local Integrals for Planar Scattering Amplitudes}}}
\author{N. Arkani-Hamed$^a$, J. Bourjaily$^{a,b}$, F. Cachazo$^{c}$, and J. Trnka$^{a,b}$\\
{\it $^{a}$ School of Natural Sciences, Institute for Advanced Study, Princeton, NJ 08540}\\
{\it $^{b}$ Department of Physics, Princeton University, Princeton, NJ 08544}\\
{\it $^{c}$ Perimeter Institute for Theoretical Physics, Waterloo, Ontario N2J W29, CA}}
\preprint{2010}

\abstract{Recently, an explicit, recursive formula for the all-loop integrand of
planar scattering amplitudes in ${\cal N}$=4 SYM has been described,
generalizing the BCFW formula for tree amplitudes, and making manifest the
Yangian symmetry of the theory. This has made it possible to easily
study the structure of multi-loop amplitudes in the theory. In this
paper we describe a remarkable fact revealed by these investigations:
the integrand can be expressed in an amazingly simple and manifestly
local form when represented in momentum-twistor space using a set of
chiral integrals with unit leading singularities. As examples, we
present very-concise expressions for all 2- and 3-loop MHV
integrands, as well as all 2-loop NMHV integrands.  We also describe a natural set of manifestly IR-finite
integrals that can be used to express IR-safe objects such as the
ratio function. Along the way we give a pedagogical introduction to the foundations of
the subject. The new local forms of the integrand are closely
connected to leading singularities --- matching only a small subset of all leading singularities remarkably suffices to determine the full integrand. These results
strongly suggest the existence of a theory for the integrand directly
yielding these local expressions, allowing for a more direct
understanding of the emergence of local spacetime physics.
}

\begin{document}


\newpage
\section{Invitation to Local Loop Integrals and Integrands}

The {\it integrand} for scattering amplitudes in planar theories is a
well-defined, rational function of external- and loop-momenta at all orders of perturbation theory \cite{ArkaniHamed:2010kv}. Recently, an explicit
recursion for the integrand of planar scattering amplitudes
in ${\cal N}=4$ SYM was presented \cite{ArkaniHamed:2010kv}, generalizing the BCFW recursion
for tree amplitudes \cite{Britto:2004ap,Britto:2005fq}. The integrand is most naturally
presented in momentum-twistor space. All the objects appearing in the
recursion relation have simple interpretations in terms of canonical
operations on Yangian-invariants derived from the Grassmannian
integral \cite{ArkaniHamed:2009dn}, making the Yangian invariance of the theory (up to total
derivatives) manifest at the level of the integrand. It has also been
recently realized that the integrand has a beautiful dual
interpretation as a natural supersymmetric Wilson loop, resolving a
long-standing open problem \cite{CaronHuot:2010ek, Mason:2010yk}. This proposal has been checked to satisfy
the all-loop recursion relation at the level of the integrand \cite{CaronHuot:2010ek},
providing a proof of the duality between scattering amplitudes and
Wilson-loops \cite{Alday:2008yw}.

The recursion relation gives a complete definition for the integrand,
making no explicit reference to spacetime notions either in the usual
or dual spacetimes. The words ``spacetime",``Lagrangian", ``path
integral" and ``gauge symmetry" make no appearance. A reflection of
this fact is that, as familiar from the BCFW computation of tree
amplitudes, individual terms in the integrand are riddled with non-local poles
that cancel in the sum. But also familiar from BCFW at tree-level, the
recursion relation is a very powerful calculational tool, and has allowed
us to gather a huge amount of  ``data" about the properties of
multi-loop amplitudes.

In this paper we report on a remarkable property of the loop integrand
revealed by examining this ``data", amplifying a theme already
stressed in \cite{ArkaniHamed:2010kv}. Loop integrands take an amazingly simple form
{\it when expressed in a manifestly local way}. This is surprising,
since the enormous complexity of Feynman diagrams is inexorably tied
to locality, while by contrast, the great simplicity of BCFW recursion
is inexorably tied to the presence of non-local poles. What we are
finding is a {\it new} local form of the integrand---certainly not
following from Feynman diagrams!---which is even simpler than the forms
obtained from BCFW recursion.

This great simplicity is apparent only when the integrand is written
in momentum-twistor space, using a special set of objects that are almost completely
{\it chiral},  and have {\it unit leading singularities}. For instance, all
2-loop MHV amplitudes are given as a sum over a single type of object,\vspace{-0.4cm}
\vspace{-0.0cm}\eq{\mathcal{A}_{\mathrm{MHV}}^{2\mathrm{-loop}}=\displaystyle\underset{\substack{i<j<k<l<i}}{\frac{1}{2}\!\!\text{{\Huge$\sum$}}\phantom{\frac{1}{2}\!\!}}\hspace{-0.2cm}\raisebox{-1.25cm}{\includegraphics[scale=0.425]{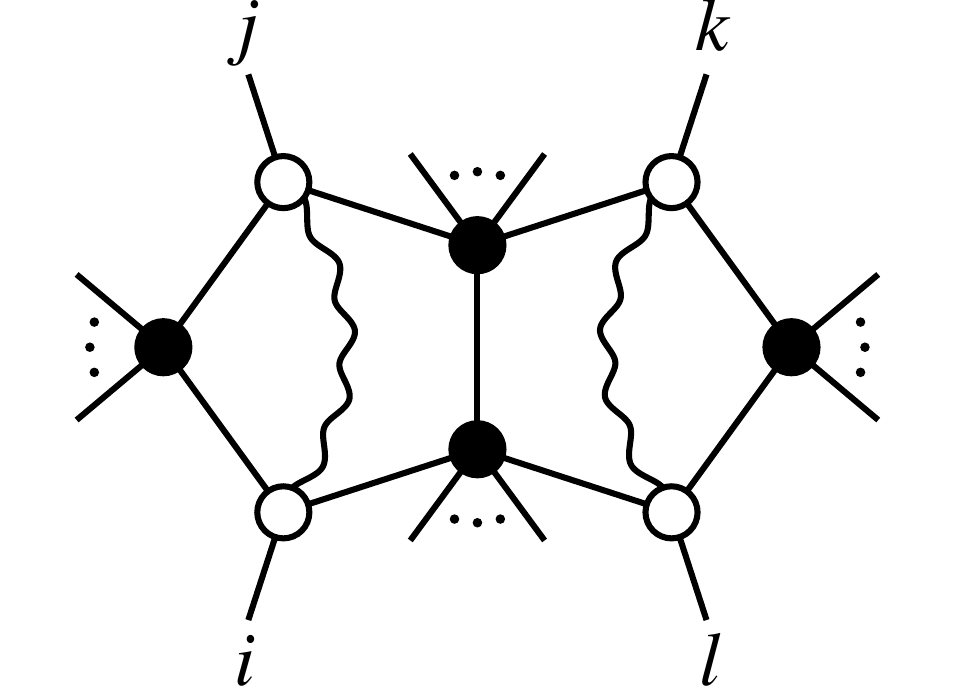}}}
This result was already presented (albeit in a slightly more clumsy
form) in \cite{ArkaniHamed:2010kv}. We will describe these objects in much more detail
in the body of the paper; here, it suffices to say that these are simple
double-pentagon integrals with a special tensor-numerator structure which is indicated
by the wavy lines, and that the notation `$i\!<\!j\!<\!\cdots\!<\!k\!<\!i$' in the summand should be understood as the sum of all cyclically-ordered sets of labels $i,j,\ldots,k$ for each $i\in\{1,\ldots,n\}$.

All 2-loop NMHV amplitudes are also associated with similar  integrands; indeed, the $n$-point NMHV scattering amplitude's integrand is simply given by,
\vspace{-1.45cm}\eq{\hspace{-0.45cm}\mathcal{A}_{\mathrm{NMHV}}^{2\mathrm{-loop}}=\!\!\!\displaystyle\underset{\substack{i<j<l<m\leq k< i\\i<j<k<l<m\leq i\\i\leq l<m\leq j<k<i}}{\!\!\phantom{\frac{1}{2}}\!\text{{\Huge$\sum$}}\phantom{\!\!\frac{1}{2}\!}}\hspace{-0.2cm}\begin{array}{c}~\\[-0.1cm]\figBox{0}{-1.5}{0.5}{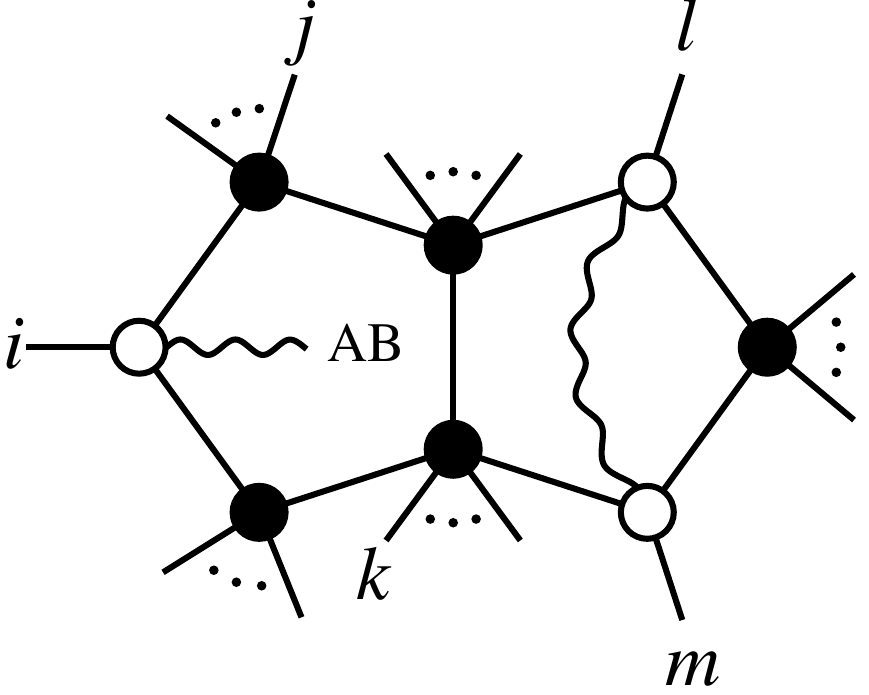}\\\times\left[i,j,j+1,k,k+1\right]\phantom{\times}\end{array}\hspace{0.3cm}\hspace{-0.3cm}+\underset{\substack{i<j<k<l<i}}{\!\!\frac{1}{2}\!\text{{\Huge$\sum$}}\phantom{\frac{1}{2}\!\!\!}}\hspace{-0.6cm}\begin{array}{c}~\\[0.78cm]\figBox{0}{-1.65}{0.5}{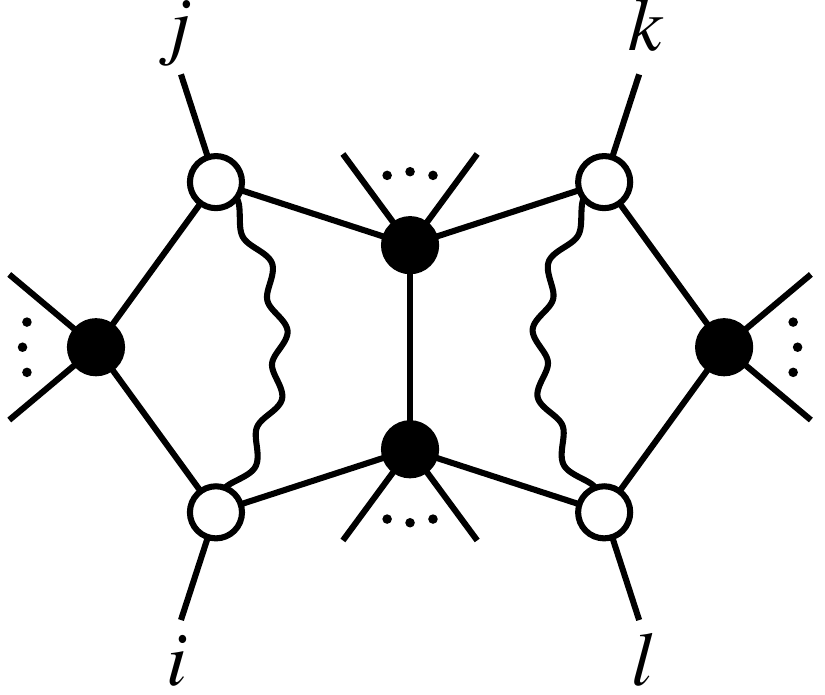}\\\times\left\{{\footnotesize\begin{array}{c}\mathcal{A}_{\mathrm{NMHV}}^{\mathrm{tree}}(j,\ldots,k;\;l,\ldots,i)\\+\mathcal{A}_{\mathrm{NMHV}}^{\mathrm{tree}}(i,\ldots,j)\\+\mathcal{A}_{\mathrm{NMHV}}^{\mathrm{tree}}(k,\ldots,l)\end{array}}\right\}\phantom{\times}\end{array}\hspace{-0.75cm}\label{NMHV2intro}}
Here, $\left[i\,j\,k\,l\,m\right]$ denotes the familiar dual-superconformal invariant of five particles, \eq{\hspace{-0.3cm}\left[i\,j\,k\,l\,m\right]\equiv\frac{\delta^{0|4}\left(\ab{j\,k\,l\,m}\eta_i+\ab{k\,l\,m\,i}\eta_j+\ab{l\,m\,i\,j}\eta_k+\ab{m\,i\,j\,k}\eta_l+\ab{i\,j\,k\,l}\eta_m\right)}{\ab{i\,j\,k\,l}\ab{j\,k\,l\,m}\ab{k\,l\,m\,i}\ab{l\,m\,i\,j}\ab{m\,i\,j\,k}}.} This result dramatically
simplifies the way this result was presented in \cite{ArkaniHamed:2010kv} for the 6- and 7-particle 2-loop NMHV integrands.

Finally, all 3-loop MHV amplitude integrands are given by a sum over the same types of \nolinebreak objects,\vspace{-0.2cm}
\vspace{-0.2cm}\eq{\nonumber\hspace{-1cm}\mathcal{A}_{\mathrm{MHV}}^{3\mathrm{-loop}}=\!\!\!\!\displaystyle\underset{\substack{i_1\leq i_2<j_1\leq\\\leq j_2<k_1\leq k_2<i_1}}{\frac{1}{3}\!\!\text{{\Huge$\sum$}}\phantom{\frac{1}{2}\!\!}}\!\!\!\figBox{0}{-2.15}{0.55}{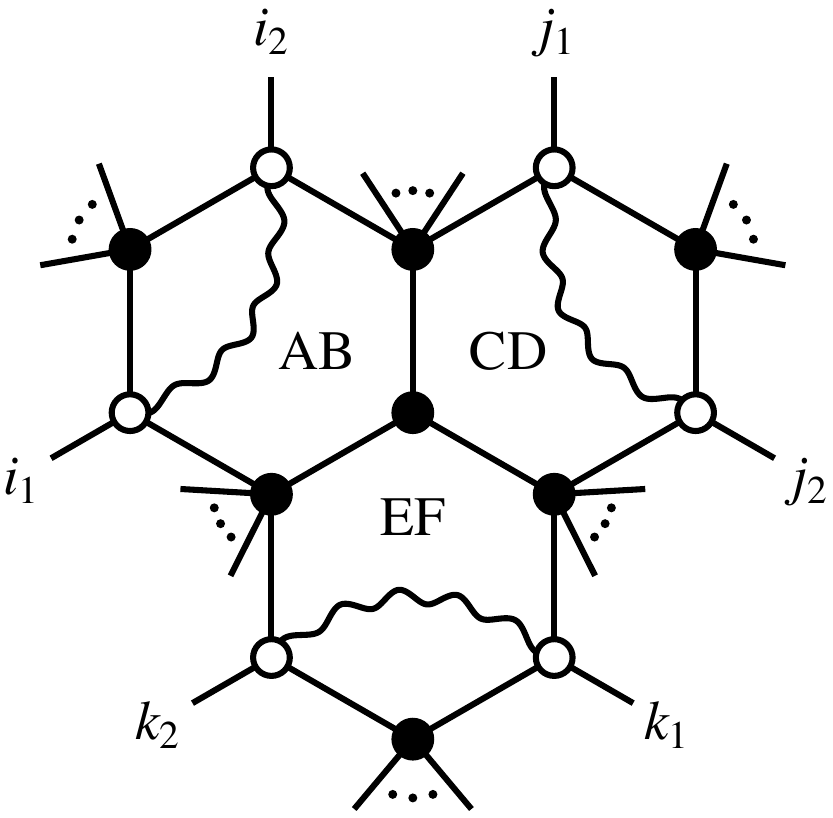}+\!\!\displaystyle\underset{\substack{i_1\leq j_1< k_1<\\< k_2\leq j_2< i_2<i_1}}{\frac{1}{2}\!\text{{\Huge$\sum$}}\phantom{\frac{1}{4}\!\!}}\!\!\!\!\figBox{0}{-1.8}{0.55}{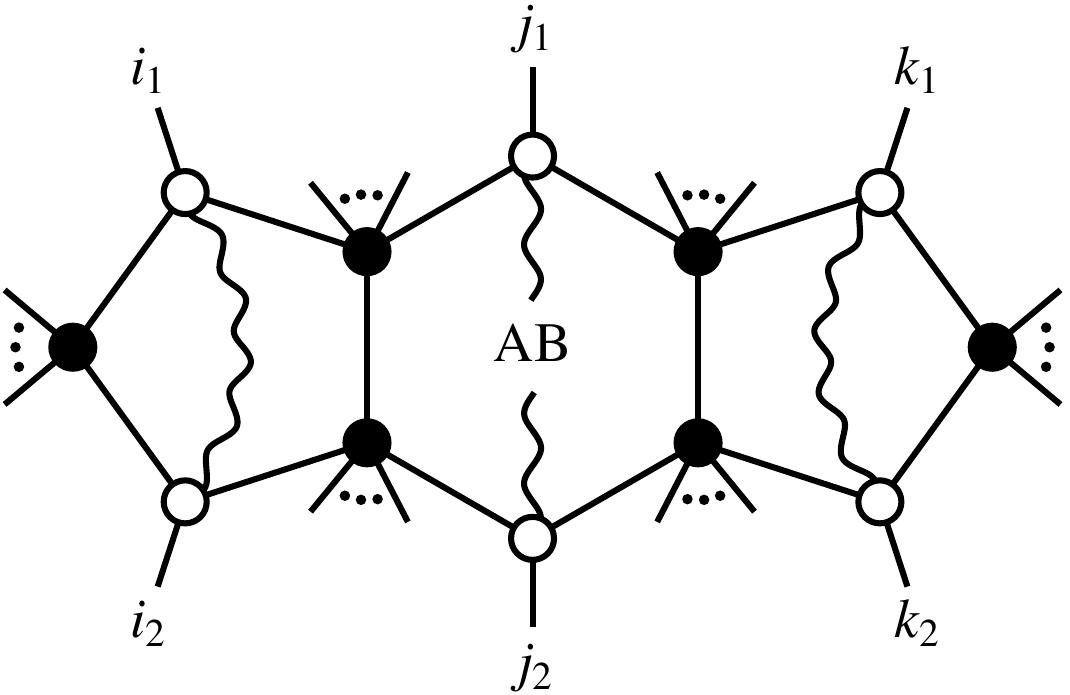}\vspace{-0.4cm}}

These explicitly-local, manifestly cyclic results for all 2-loop NMHV and 3-loop MHV
amplitudes are new, and stunningly-simple---even simpler than
the form produced by the loop-level recursion formula.

As we will see, these extremely simple expressions are very closely
related to the leading singularity structure of the theory. The reason for the dramatic simplicity of these results relative to the ones
presented in \cite{ArkaniHamed:2010kv} is that in \cite{ArkaniHamed:2010kv}, each integrand was straightforwardly expanded in terms of a fixed basis of chiral integrals with unit
leading singularities, while here we are tailoring the objects that
appear directly to the amplitude.
The structures are motivated by matching a particularly simple set of
leading singularities of the theory; this is made possible only by using
chiral integrands with unit leading singularities, which is why these
objects play such a crucial role in the story. What is remarkable is
that matching only a small subset of leading singularities in this way
suffices to determine the full result. Of course, we confirm this not
by laboriously matching all leading singularities, but rather by
directly checking the conjectured local forms against what we obtain from
the all-loop recursion relation.

We do not yet have a satisfactory understanding for the origin of this
amazing simplicity. Certainly, these expressions differ from the BCFW
form in that they are not term-by-term Yangian invariant. This
suggests the existence of a deeper theory for the integrand that will
directly produce these new local forms, allowing a more direct
understanding of the emergence of local spacetime physics. We strongly
suspect that it is {\it this} formulation that will also help explain
the amazing simplicity \cite{polytopePaper} seen in the {\it integrals} yielding
the physical amplitudes, and also form the point of contact with the
remarkable integrable structures of $\mathcal{N}=4$ SYM---Y-systems and Yang-Yang
equations---seen at strong coupling and also in some collinear limits
\cite{Alday:2010vh, Alday:2010ku,Gaiotto:2010fk}.

In \cite{polytopePaper}, a geometric picture for scattering amplitudes is advanced,
building on a beautiful paper of Hodges \cite{Hodges:2009hk}, which may shed some
light on the origin of these new local expressions. Hodges interpreted
NMHV tree amplitudes as the volume of certain polytopes in
momentum-twistor space, and showed that a natural class of
triangulations of this polytope correspond to different BCFW
representations of the amplitude. In \cite{polytopePaper}, it is shown that at an
even simpler triangulation of the same polytope is possible, yielding
a new, manifestly-local formula for NMHV tree-amplitudes. Also in \cite{polytopePaper}, a
completely analogous `polytope' formulation is presented for all 1-loop MHV amplitudes. Again, one natural set of
triangulations leads to the BCFW form of the integrand, while even
simpler triangulations directly lead to a number of new, manifestly
local forms for the integrand. While this polytope picture has not yet
been generalized beyond these most elementary cases of NMHV tree and
MHV 1-loop amplitudes, the extremely simple local forms for higher
loop amplitudes we present in this paper strongly encourages the thought that an
appropriate extension of this idea must be possible.

We should stress that when we say our results for the integrand are ``manifestly local", we mean that the  poles 
involving the loop integration variables are local. Of course the integrand should be ``ultralocal", that is, the poles involving both the loop integration 
variables as well as the external momenta must be local. The MHV integrands we present trivially have this property, but for NMHV amplitudes, our expressions involve 
the standard $R$-invariants which have spurious poles as function of the external particle momenta. Given the beautiful, local form of the 
NMHV tree amplitude obtained from the polytope picture \cite{polytopePaper}, it is quite likely that there is an even nicer representation of loop amplitudes which are not only local but ultralocal. This fascinating possibility certainly merits further exploration, but is beyond the scope of the present paper. 

We close this invitation with an outline for the rest of the
paper. We begin with a pedagogical introduction to some of the
foundations of the subject in \mbox{section \ref{foundations_section}} starting with a review of
momentum-twistors and some of the associated projective geometry in
$\mathbb{CP}^3$. We also discuss how planar loop integrals are written
in momentum-twistor space; while our focus in the paper is on ${\cal
N} = 4$ SYM, we expect that the momentum-twistor representation of
loop amplitudes will be extremely useful for {\it any} planar theory.
We discuss the way that momentum-twistors make integral reduction
trivial, and illustrate this by showing how the 1-loop integrand can
be reduced to a sum over pentagon integrals. Finally we discuss
leading singularities at 1-loop and beyond in momentum-twistor
language. The standard exercise of determining quadruple-cuts in
momentum space is mapped in momentum-twistor language to a simple,
beautiful and classic problem in enumerative geometry first posed by
Schubert in the 1870's, and we discuss the solution of these
``Schubert problems" in detail.   

In section \ref{pure_loop_integrals_section} we introduce chiral integrals with unit leading
singularities which play a central role in our story. We illustrate
how they work starting with the simplest case of 1-loop MHV
amplitudes.

In section \ref{finite_integrals_section}, we discuss another feature of chiral integrals with
unit leading singularities---generic integrals of this form are
manifestly infrared finite, and can be used to express finite objects
related to scattering amplitudes, such as the ratio function \cite{Drummond:2008vq}.

In section \ref{one_loop_integrands_and_basis}, we construct a basis for all 1-loop integrands,
whose building blocks are not the familiar boxes or even pentagons,
but a natural set of chiral octagons with unit leading singularities.
We also compute the finite 1-loop integrals explicitly, and use
these results to give a simple formula for the NMHV ratio-function at
1-loop, for any number of particles.

In section \ref{multiloop_amplitudes_section}, we discuss multi-loop amplitudes. We
describe our heuristic strategy for using leading singularities to
tailor momentum-twistor integrals to the amplitude, and show how this
works for the 1-loop MHV amplitude, reproducing one of the local forms
first derived using the polytope picture of \cite{polytopePaper}. We also discuss
the 1-loop NMHV amplitudes in the same way. We then extend these
methods to two loops and beyond, and show how to ``glue" the 1-loop
expressions together to produce natural conjectures for all 2- and 3-loop MHV amplitudes, as well all 2-loop NMHV amplitudes. These
conjectures are verified by comparing with the integrand derived
from the all-loop recursion relation.

A number of appendices discuss various technical points
needed in the body of the paper, including a detailed discussion of the 2-loop NMHV and 3-loop MHV integrands.

\section{Foundations}\label{foundations_section}

In theories with massless particles, a well-known and convenient way of trivializing the constraint $p_a^2=0$ for each particle is to introduce a pair of spinors $\lambda^{(a)}$ and $\tilde\lambda^{(a)}$, replacing $p_a^{\mu}\mapsto(p_a)_{\underline{\alpha}\,\underline{\dot\alpha}}\equiv p_a^{\mu}(\sigma_{\mu})_{\underline{\alpha}\underline{\dot{\alpha}}}\equiv \lambda^{(a)}_{\underline{\alpha}}\tilde\lambda_{\underline{\dot\alpha}}^{(a)}$. Of course, this map is not invertible, as any rescaling $\{ \lambda,\tilde\lambda\}\to \{ t\lambda,t^{-1}\tilde\lambda\}$ leaves $p$ invariant. This reflects that these variables come with a new source of redundancy; in the case of particles with spin, this redundancy is quite welcomed as it allows the construction of functions that transform with fixed projective weights as S-matrix elements under Lorentz transformations. This is all well-known under the name of the {\it spinor-helicity formalism} \nolinebreak\cite{Berends:1981rb,DeCausmaecker:1981bg,Kleiss:1985yh,Gunion:1985vca,Xu:1986xb}.

Amplitudes are supported on momenta that satisfy momentum conservation. Clearly, it would be convenient to find variables where this constraint, $\sum_a p_a=0$, is trivial. In planar theories, where color ordering is available, there is a natural way to achieve this, by choosing instead to express the external momenta in terms of what are known as dual-space coordinates, writing $p_a\equiv x_{a}-x_{a-1}$, \cite{Drummond:2006rz}.

To see the role played by planarity, consider the standard decomposition of scattering amplitudes according to the overall color structure, keeping only the leading color part: 
\be
A_n = {\rm Tr}(T^{a_1}T^{a_2}\ldots T^{a_n}) \mathcal{A}_n(1,2,\ldots ,n) + {\rm permutations};
\ee
here, each {\it partial amplitude} $\mathcal{A}_n(1,2,\ldots , n)$ can be expanded in perturbation theory, and we denote the $L$-loop contribution by $\mathcal{A}^{L\mathrm{-loop}}_n$. Partial amplitudes are computed by summing over Feynman diagrams with a given color-ordering structure. 


In this paper we only consider the planar sector of the theory, and therefore $\mathcal{A}^{L\mathrm{-loop}}_n$ will always refer to the leading-color, partial amplitude in the planar limit.

Restricted to a particular partial amplitude, say, $\mathcal{A}_n(1,2,\ldots, n)$, each momenta can be expressed as the difference of two ``spacetime" points. More precisely, we make the identification \mbox{$p_a \equiv x_a -x_{a-1}$}, with $p_1=x_1-x_n$. It is clear that momenta obtained in this way automatically satisfy $\sum_a p_a = 0$---and the redundancy introduced in this case is a translation $x_a\to x_a+y$ by any fixed vector $y$.

Now, the only poles that can occur in $\mathcal{A}_n(1,2,\ldots , n)$ are of the form $\sum_{m=a}^b p_m$, {\it i.e.}, only the sum over consecutive momenta can appear. In the dual variables these become $\sum_{m=a+1}^b p_m = x_a-x_b$. The same kind of simplifications happen in planar Feynman diagrams to all orders in perturbation theory as we will describe.

Now we have the variables $\{\lambda,\tilde\lambda\}$ which make the null condition trivial while ignoring momentum conservation, while the dual-space variables do the opposite. It is perfectly natural to wonder if there exists any way to combine these two constructions which makes both the null-condition and momentum conservation trivial. It turns out that such a set of variables does exist: they are known as {\it momentum-twistors} and were introduced by Hodges in \cite{Hodges:2009hk}.

The standard twistor construction developed in the 1960's \cite{Penrose:1967wn} starts by making a connection between points in an auxiliary space---twistor-space---and null rays in spacetime.
Likewise, a complex line in twistor space is related to a point in spacetime. The key formula is called the {\it incidence relation}, according to which a point $x$ in spacetime corresponds to set of twistors $Z=(\lambda, \mu)$ which satisfy
\be
\mu_{\underline{\dot\alpha}} = x_{\underline{\alpha}\,\underline{\dot\alpha}}\lambda^{\underline{\alpha}}.
\ee
Twistors satisfying this relation form a projective line in $\mathbb{CP}^3$. Even though $Z$ has the components of a point in $\mathbb{C}^4$, the incidence relation cannot distinguish $Z$ from $ t Z$, and therefore the space is projectivized.

In order to specify a line in twistor space---and therefore a point in spacetime---all that is needed is a pair of twistors, say $Z_A$ and $Z_B$, that belong to the line. Given the twistors, the line or spacetime point is found by solving the four equations coming from imposing the incidence relation for $Z_A$ and $Z_B$ with $x$. It is easy to check that the solution is,
\be
x_{\underline{\alpha}\,\underline{\dot\alpha}}  =\frac{\lambda_{A,\underline{\alpha}}\mu_{B,\underline{\dot\alpha}}}{\langle \lambda_A\,\lambda_B\rangle}+\frac{\lambda_{B,\underline{\alpha}}\mu_{A,\underline{\dot\alpha}}}{\langle \lambda_B\,\lambda_A\rangle}.
\ee
(Here, we have made use of the familiar Lorentz-invariant contraction of two spinors $\ab{\lambda_A\,\lambda_B}\equiv\epsilon_{\underline{\alpha}\,\underline{\beta}}\lambda_A^{\underline{\alpha}}\lambda_{B}^{\underline{\beta}}$).

\begin{figure}[t!]\begin{center}\caption{Defining the connections between momentum-twistors, dual-coordinates, and cyclically-ordered external four-momenta\label{momentum_twistor_geometry}}\figBox{0}{-1}{0.85}{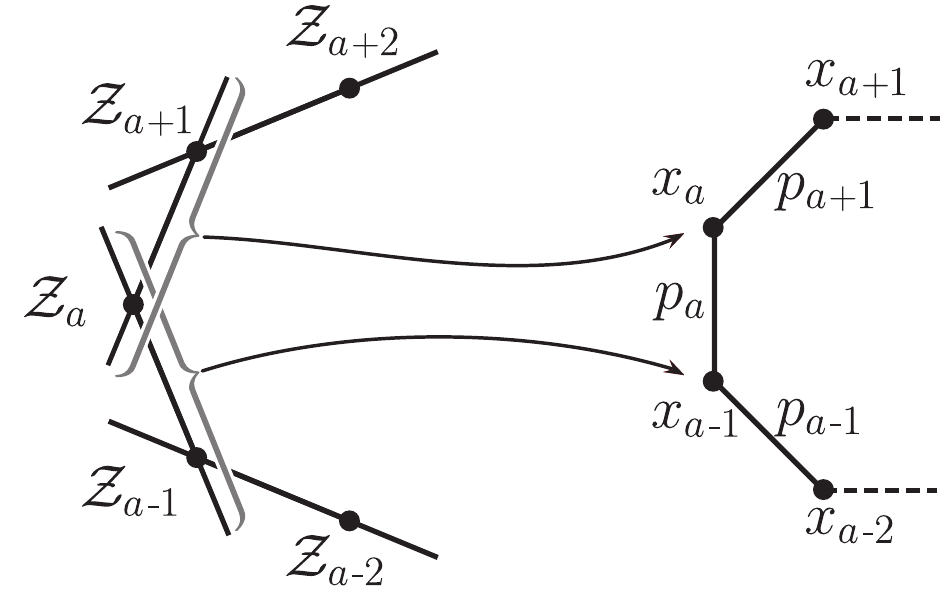}\end{center}\vspace{-1cm}\end{figure}

Hodges' construction starts with any set of $n$ twistors $\{ Z_1,\ldots ,Z_n\}$. Using the association $x_a \leftrightarrow (Z_a,Z_{a+1})$, $n$ spacetime points are defined. Quite nicely, it is trivial that $p_a^2 = (x_a-x_{a-1})^2=0$ because the corresponding lines, or ($\mathbb{CP}^1$s), intersect. This is illustrated in \mbox{Figure \ref{momentum_twistor_geometry}}.

Given the importance of this latter fact, it is worth giving it a slightly more detailed discussion than we have so far. If two lines in twistor-space intersect, {\it i.e.}\ share a twistor $Z_{\rm int}$, then the corresponding spacetime points, say $x$ and $y$, associated with the lines are null-separated. To see this, take the difference of the incidence relations for $Z_{\rm int}$,
$$\mu_{\underline{\dot\alpha}}^{\rm int} = x_{\underline{\alpha}\,\underline{\dot\alpha}}\lambda^{\underline{\alpha}}_{\rm int}, \quad  \mu_{\underline{\dot\alpha}}^{\rm int} = y_{\underline{\alpha}\,\underline{\dot\alpha}}\lambda^{\underline{\alpha}}_{\rm int},$$
to get
$$(x-y)_{\underline{\alpha}\,\underline{\dot\alpha}}\lambda^{\underline{\alpha}}_{\rm int}=0;$$
which means that the $2\times2$-matrix $(x-y)$ has a non-vanishing null eigenvector, {\it i.e.}\ $\lambda^{\underline{\alpha}}_{\rm int}$, and therefore the determinant of $(x-y)$ vanishes. But the determinant is proportional to $(x-y)^2$ when $x$ and $y$ are taken as vectors; and therefore $x$ and $y$ are null separated.

As useful background for the rest of the paper let us discuss the null-separation condition, which is a conformally invariant statement, in twistor space. Consider again two generic spacetime points $x$ and $y$ and choose two representatives of the lines associated to them in twistor space, say, $(Z_A, Z_B)$ and $(Z_C, Z_D)$. Treating each twistor as a vector in $\mathbb{C}^4$ there is a natural $SL(4)$ (conformal) invariant that can be constructed. This is done by contracting all four twistors with the completely antisymmetric tensor $\epsilon_{IJKL}$ to produce
\be
\label{dett}
\langle Z_A Z_B Z_C Z_D\rangle = \epsilon_{IJKL}Z^I_A Z^J_BZ^K_C Z^L_D .
\ee
Clearly, this conformally-invariant quantity must encode information about how $x$ and $y$ are causally related. The Lorentz invariant separation $(x-y)^2$ is not conformally-invariant because it is not a cross ratio. However, the way to relate the two quantities is simple
\be
\label{locon}
(x-y)^2 = \frac{\langle Z_A Z_B Z_C Z_D\rangle}{\langle \lambda_A ~\lambda_B\rangle \langle \lambda_C~\lambda_D\rangle}.
\ee
%
This relation is consistent with our earlier finding that if the points $x$ and $y$ are null-separated, then the twistors $Z_A$, $Z_B$, $Z_C$ and $Z_D$, are coplanar as points in $\mathbb{CP}^3$. In other words, the two complex lines intersect.

When twistors are used to produce a configuration of points in spacetime which are pairwise null separated and then used to build momenta, the corresponding twistor space is called {\it momentum-twistor space} \cite{Hodges:2009hk}.

This twistor construction is in fact slightly more involved when one is interested in {\it real} slices of spacetime. In our discussion so far, we have been assuming that momenta are complex and hence the dual spacetime is complexified. This is useful for {\it e.g.}\ defining the usual unitarity cuts of loop amplitudes. In this paper, the complex version suffices and we refer the interested reader to \cite{Hodges:2009hk, Hodges:2010kq}.

A related construction is called dual momentum twistor space. Here `dual' refers to the usual geometric---`Poincar\'{e}'---dual of a space. In other words, the dual space is the space of planes in $\mathbb{CP}^3$. Points in the new space which is also a $\mathbb{CP}^3$ are denoted by $W_I$. The construction maps points to planes and lines to lines. In Hodges' construction \cite{Hodges:2009hk}, there is a natural definition of dual points associated to the planes defined by consecutive lines of the polygon in momentum twistor space of \mbox{Figure \ref{momentum_twistor_geometry}}.

The construction defines a dual polygon by introducing dual momentum twistors $W_a$ defined by
\be
(W_a)_I =\frac{\epsilon_{IJKL}Z_{a-1}^JZ_a^KZ_{a+1}^L}{\langle \lambda_{a\mi1}\,\lambda_a\rangle\langle \lambda_a\,\lambda_{a\pl1}\rangle}.
\ee
This definition is made so that $W_a$ contains $\tilde\lambda_a$ as two of its components.

\subsection{Loop Integrals}

The focus of this paper is loop integrands and integrals. Here too, it is well known that in planar theories, loop integrals are very naturally expressed in terms of dual spacetime coordinates. Consider a very simple 1-loop integral, known as a zero mass integral,
\be
\figBox{0}{-1.5}{0.45}{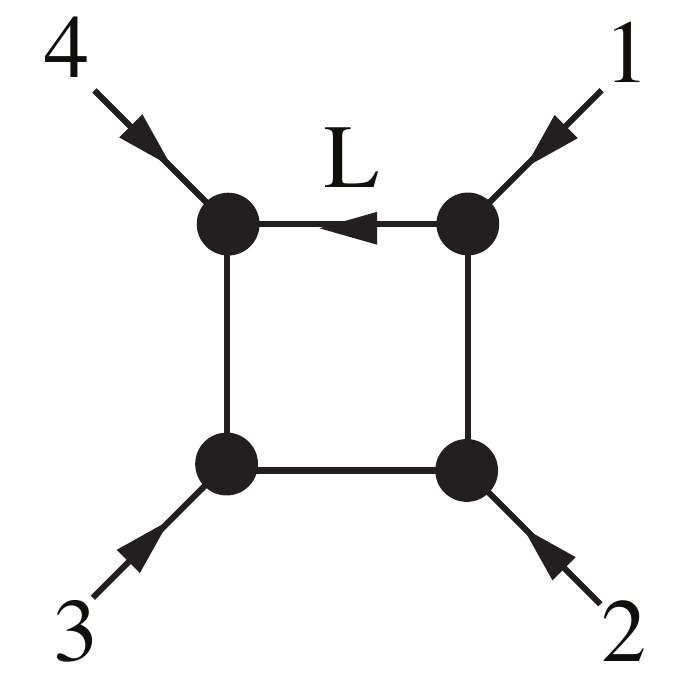}=\int d^4L\frac{N}{L^2(L-p_1)^2(L-p_1-p_2)^2(L-p_1-p_2-p_3)^2}
\ee
where the external momentum at each of the four vertices is null (hence the name) and $N=(p_1+p_2)^2(p_2+p_3)^2$ is a convenient normalization factor. Momentum conservation gives $p_4 = -p_1-p_2-p_3$; and introducing the dual-coordinates \mbox{$p_a= x_a-x_{a-1}$}, it is easy to see that the unique choice of $L$ that makes translation invariance (in $x$-space) manifest is $L=x-x_4$. The integral becomes \cite{Drummond:2006rz}
\be
\label{easy}
\figBox{0}{-1.5}{0.45}{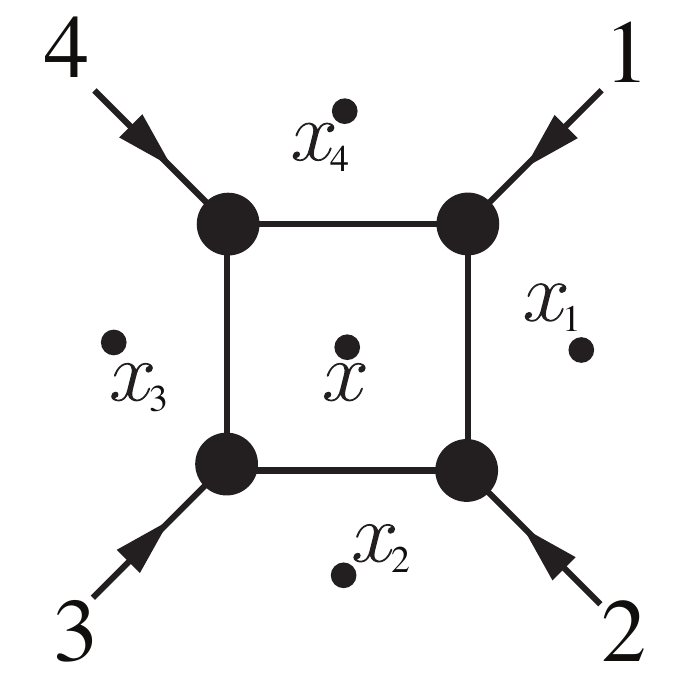}=\int d^4x\frac{N}{(x-x_1)^2(x-x_2)^2(x-x_3)^2(x-x_4)^2},
\ee
where $N=(x_1-x_3)^2(x_2-x_4)^2$. Imposing translation-invariance gives rise to the same integral in $x$-space regardless of the original definition of $L$ in the loop diagram. In other words, a different propagator could have been chosen to be $L$ and the form (\ref{easy}) would still be the same. This uniqueness plays a crucial role in the definition of the integrand of the theory.

Integrating over all points $x$ in spacetime is the same as integrating over all $\mathbb{CP}^1$'s in $\mathbb{CP}^3$. As before, each line in twistor space can be represented by a pair of twistors $x \leftrightarrow (Z_A,Z_B)$. Clearly, any $GL(2,\mathbb{C})$ transformation on the $A,B$ ``indices" leaves the line invariant. Therefore the integral over spacetime is the same as the integral over the pairs $(Z_A,Z_B)$ modulo $GL(2)$. This is nothing but the Grassmannian $G(2,4)$ which can be parameterized by a $2\times 4$ matrix
\be
\label{STgras}
\left(\begin{array}{cccc} Z_A^1 & Z_A^2 & Z_A^3 & Z_A^4 \\  Z_B^1 & Z_B^2 & Z_B^3 & Z_B^4 \end{array} \right) = \left(\begin{array}{cccc} \lambda_A^1 & \lambda_A^2 & \mu_A^{\dot 1} & \mu_A^{\dot 2} \\ \lambda_B^1 & \lambda_B^2 & \mu_B^{\dot 1} & \mu_B^{\dot 2} \end{array} \right).
\ee
We can immediately write a measure which is $GL(2)$-invariant by integrating over all $Z_A$'s and $Z_B$'s together with a combination of $2\times 2$ minors of the matrix (\ref{STgras}) with total weight $-4$. It turns out that the precise measure that corresponds to a $d^4x$ integration is
\be
\int d^4x \Leftrightarrow\int \frac{d^4Z_Ad^4Z_B}{{\rm Vol}GL(2)\times \langle \lambda_A~\lambda_B\rangle^4},
\ee
where $\langle \lambda_A~\lambda_B\rangle$ is the $(1~2)$ minor of (\ref{STgras})---the determinant of the first two columns of the $2\times4$ matrix (\ref{STgras}). In the twistor literature this is written as $\langle \lambda_A~\lambda_B\rangle = \langle Z_AZ_B\,I_{\infty}\rangle$ where $(I_{\infty})^{KL}$ is the infinity twistor which is block diagonal with the only nonzero diagonal element equal to $\epsilon_{ab}$. $I_{\infty}$ is called the infinity twistor because it corresponds to a choice of the point at infinity in spacetime and therefore a line in twistor space. Its presence therefore breaks conformal invariance. This is not surprising as the measure $d^4x$ `knows about' the metric in spacetime.

Since the integration over lines will appear in many different contexts in this paper we introduce a special notation for it. Let's define
\be
\label{defInt}
\int\limits_{AB} \Leftrightarrow \int \frac{d^4Z_Ad^4Z_B}{{\rm Vol}GL(2)}.
\ee
The reason we have not included the factor $\langle \lambda_A~\lambda_B\rangle^4$ in the definition is that in this paper we mostly deal with ${\cal N}=4$ SYM and in its integrand factors with infinity twistors cancel.

Going back to the loop integral in $x$-space (\ref{easy}), one can introduce the four momentum twistors in Hodges' construction $\{Z_1,Z_2,Z_3,Z_4\}$ to describe the external particles. Using the relation between the Lorentz invariant separations and momentum twistor invariants in (\ref{locon}), the integral (\ref{easy}) becomes
\be
\label{zeromass}
\int\limits_{AB}\frac{\langle 1234\rangle^2}{\langle AB\,12\rangle\langle AB\,23\rangle \langle AB\,34\rangle\langle AB\,41\rangle}.
\ee
where $\langle ijkl\rangle$ stands for the determinant of the $4\times 4$ matrix with columns given by four twistors $Z_i,Z_j,Z_j,Z_k$ defined in (\ref{dett}).

One of the remarkable facts about (\ref{zeromass}) is that all factors involving the infinity twistor have disappeared. This means that the integral is formally conformal invariant under the conformal group that acts on the dual spacetime. This is why it is said to be {\it dual conformally invariant} (DCI). 

Clearly, if we had started with a triangle integral then the factor $\langle Z_1 I Z_2\rangle = \langle \lambda_1~\lambda_2\rangle$ would not have canceled and would have remained with power one in the denominator as if it were a propagator. Indeed, this viewpoint trivializes the surprising connections made in the past between the explicit form of triangle and box integrals. In other words, one can think of a triangle integral as a box where one of the points is at infinity.

Once again, a careful definition of the contour which should correspond to only points in a real slice of complexified spacetime is not needed in this paper. It suffices to say that on the physical contour, the integrals can have infrared divergences (IR). This is the reason why we said that the integral was `formally' DCI. We postpone a more detailed discussion of IR divergences to section \ref{finite_integrals_section}.

The purpose of this section is to show how momentum twistors are the most natural set of variables to work with loop amplitudes in planar theories. In order to do this we will first show how many familiar results can be translated into momentum twistors. Not infrequently, momentum twistors will completely clarify physics points which have been misunderstood in the literature.


\newpage
\subsubsection*{Integral Reduction at One-Loop Level}\label{foundations_integral_reduction}

In a general theory, 1-loop integral reduction techniques allow scattering amplitudes to be expressed as linear combinations of a basic set of scalar integrals\footnote{This is true in theories with no rational terms or in general theories for what is known as the cut-constructible part of them. See \cite{Dixon:1996wi} for more details. In ${\cal N}=4$ SYM rational terms are absent. This is why we do not elaborate more on this point.}. The integrals have the topology of bubbles, triangles or boxes.

Let us start this section by translating each of the integrals in the standard basis into momentum twistor language. Their corresponding form in momentum twistor space is
\begin{align}
\label{BTB}
\nonumber\hspace{-1.505cm}I_{\rm Box} =\figBox{0}{-1.3}{0.4}{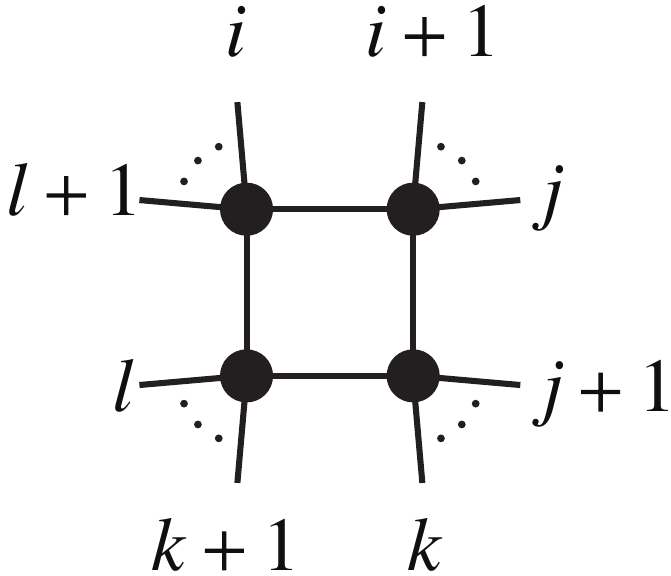} = & \int\limits_{AB} \frac{\langle i~i\pl1 \rangle\langle j~j\pl1 \rangle\langle k~k\pl1 \rangle\langle l~l\pl1 \rangle}{\langle AB~i~i\pl1\rangle\langle AB~j~j\pl1\rangle\langle AB~k~k\pl1\rangle\langle AB~l~l\pl1\rangle};\\[-0.05cm]
\hspace{-1.25cm}I_{\rm Triangle}   =\hspace{-0.4cm}\figBox{0}{-1.5}{0.4}{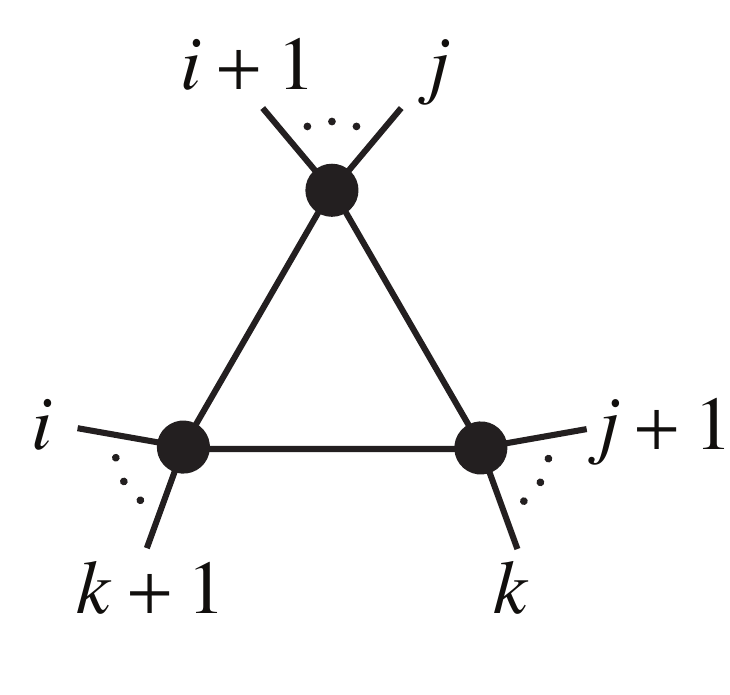}\hspace{-0.3cm}= & \int\limits_{AB}\frac{1}{\langle AB\rangle}\frac{\langle i~i\pl1 \rangle\langle j~j\pl1 \rangle\langle k~k\pl1 \rangle}{\langle AB~i~i\pl1\rangle\langle AB~j~j\pl1\rangle\langle AB~k~k\pl1\rangle};\\[-0.95cm]
\nonumber\hspace{-1.25cm}I_{\rm Bubble} =\hspace{-0.2cm}\figBox{0}{-1.3}{0.4}{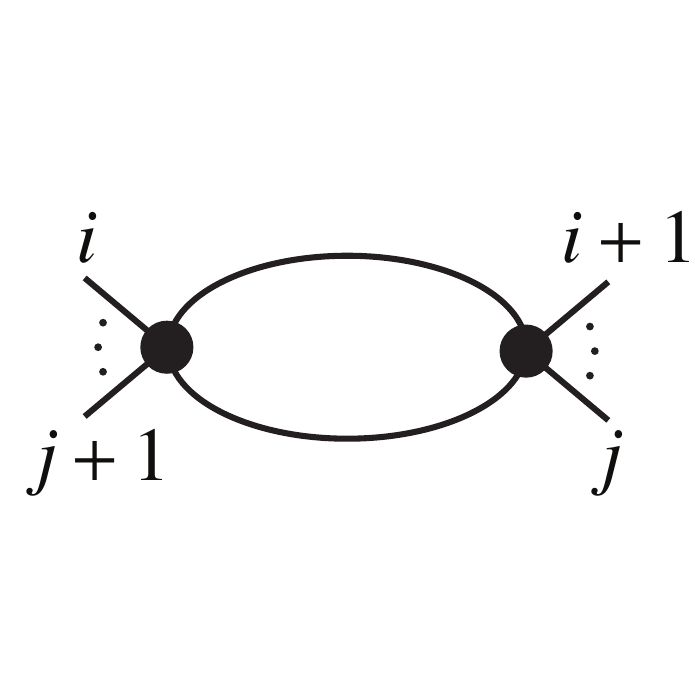}\hspace{-0.2cm}= & \int\limits_{AB}\frac{1}{\langle AB\rangle^2}\frac{\langle i~i\pl1 \rangle\langle j~j\pl1 \rangle}{\langle AB~i~i\pl1\rangle\langle AB~j~j\pl1\rangle}.
\end{align}
Note that here we have translated the plain scalar integrals without any normalization factors. Once again, only boxes are dual conformal invariant except for an overall factor which only depends on the external data. This factor involving 2-brackets and hence the infinity twistor can always be removed by a proper normalization as done in the zero-mass example (\ref{zeromass}). Scalar boxes in momentum twistor space have also been recently studied in \cite{Hodges:2010kq, Mason:2010pg}.

A well known fact about ${\cal N}=4$ SYM is that at 1-loop level, bubbles and triangles are absent and all one needs are scalar box integrals. However, as we will see, this point of view is not the most natural one and actually turns out to be misleading.

In order to understand this point, one needs to review the reduction procedures used to reach this conclusion. Before doing that let us mention some useful facts about momentum twistors.

In loop integrals, combinations of momentum twistors of the form $Z_A^{[I}Z_B^{J]}$ make an appearance in every expression (where the brackets mean that the indices are anti-symmetrized), reflecting the fact that it is the {\it line} $(AB)$ that is being integrated-over, and not the individual twistors $Z_A$ and $Z_B$.

These two-index objects are a class of more general ones called bitwistors. A generic bitwistor is a rank-two antisymmetric tensor $Y^{IJ}$. Given two bitwistors, $Y$ and $\widetilde Y$, the conformally-invariant inner-product is given by $\langle Y~\widetilde Y\rangle = \epsilon_{IJKL}Y^{IJ}\widetilde Y^{KL}$. A bitwistor which can be written in terms of two twistors as $Z_A^{[I}Z_B^{J]}$ is called {\it simple}. It is easy to show that a bitwistor is simple if and only if $Y^2=0$ with the product defined as above.

The reason for discussing bitwistors is that they provide a very natural integral reduction procedure. The procedure can be applied to integrals at any loop order but in this section we concentrate on only 1-loop integrals. The procedure we are about to present is in part the momentum twistor analog of the one introduced by van-Neerven and Vermaseren in \cite{vanNeerven:1983vr}.

At one loop one starts with general Feynman integrals of the form
\be
T_{\mu_1\ldots\mu_m}\int d^4L \frac{L^{\mu_1}\ldots L^{\mu_m}}{\prod_{i=1}^n (L-P_i)^2}
\ee
where the tensor $T$ is made out of polarization vectors, momenta of external particles and the spacetime metric.

By Lorentz invariance, it is clear that one can decompose integrals of this type as linear combinations of momentum twistor tensor integrals of the form
\be
\label{redu}
\int\limits_{AB} \frac{1}{\langle AB I_{\infty}\rangle^{4-(n-m)}}\frac{\langle AB\,Y_1\rangle\langle AB\,Y_2\rangle\ldots \langle AB\,Y_{m}\rangle }{\langle AB\,12\rangle\langle AB\,23\rangle\ldots \langle AB\,n\mi1~n\rangle\langle AB\,n\,1\rangle}
\ee
where $Y_a$ are generic bitwistors.

The reduction procedure relies on the fact that a generic bitwistor has six degrees of freedom and can therefore be expanded in a basis of any six independent bitwistors. To reduce the integrals in (\ref{redu}) simply choose any six of the bitwistors that appear in the denominator, say, $Z_1Z_2$, $Z_2Z_3$, $\ldots$, $Z_6Z_7$ and expand any of the bitwistors in the numerator as
\be
(Y_j)^{IJ} = \alpha_1 Z_1^IZ_2^J + \alpha_2 Z_2^IZ_3^J +\ldots +\alpha_6 Z_6^IZ^J_7.
\ee
The coefficients can be found by contracting with enough bitwistors two get six independent equations. More explicitly, one can consider equations of the form
$$\langle Z_2Z_3Y_j\rangle= \alpha_4\langle 2345\rangle + \alpha_5\langle 2356\rangle+\alpha_6\langle 2367\rangle.$$
and solve for the $\alpha's$. Once this is done, the factor $\langle AB\,Y_j\rangle$ becomes a linear-combination of factors in the denominator, thus reducing the degree of the denominator and numerator by one.

The integral in (\ref{redu}) is for a general quantum field theory with a planar sector. One can continue with the integral procedure in this case but it will take us too far away from the main line of the paper. Therefore we concentrate directly on ${\cal N}=4$ SYM. In ${\cal N}=4$ SYM it has been known since the 1990's \cite{Bern:1994zx} that all integrals satisfy $n-m=4$. In modern language, this means that the integrals are dual conformally-invariant as discussed in the simple example of the all massless box integral (\ref{zeromass}).

Iterating the reduction procedure, we can write the any amplitude as a sum over pentagons and boxes. But as far as we have seen, the reduction procedure we have described so far does not reduce the pentagons any further. Notice that the pentagons we have described here are not {\it scalar} pentagons, but {\it tensor} pentagons---and they are {\it manifestly} DCI. However, one is always free to choose a basis of bitwistors including $Y=I_{\infty}$ to obtain scalar pentagons, but only at the cost of manifest dual conformal invariance.

But doesn't the reduction procedure of van-Neerven and Vermaseren, when applied to ${\cal N}=4$ SYM, allow for a reduction all the way down to only scalar boxes? One might wonder why our analysis so far does not generate this familiar `box-expansion'. The answer is that the reduction to box-integrals is {\it not valid at the level of the integrand}---only the reduction to boxes {\it and pentagons} (scalar or otherwise) is valid at the level of the integrand. In order to obtain the all-too familiar box-expansion, it is necessary to parity-symmetrize the integrand---a step that is only justified when integrated on a parity-invariant contour, and one which does violence to the highly {\it chiral} loop-integrands of a quantum field theory such as $\mathcal{N}=4$ SYM.

Here, we should briefly clarify a point which has been unnecessarily confused in the literature on $\mathcal{N}=4$. Because {\it integrand}-level reduction must terminate with boxes {\it and} pentagons, and box-integrals are both manifestly parity-even and DCI while {\it scalar} pentagons---which have a factor of $\ab{AB\,I_{\infty}}$ in the numerator---are not DCI, the corrections to the box-expansion needed to match the full integrand of $\mathcal{N}=4$ were first expressed in terms of parity-odd combinations of scalar pentagons. This led some researchers to suppose that there was some connection between DCI and parity. There is of course no such connection: as evidenced by the extension of BCFW to all-loop orders, the full $\mathcal{N}=4$ loop-integrand is DCI.

Especially for theories such as $\mathcal{N}=4$ which are DCI, one should strictly avoid parity-symmetrization at one-loop or higher. Although scalar pentagon integrals are quite familiar, {\it chiral} pentagons are slightly novel---although they have already played an important role in the literature (see e.g.\ \cite{ArkaniHamed:2010kv, Drummond:2010mb}). The first appearance of pentagon integrals occurs for five particles, and there are essentially two possibilities that arise:
\vspace{-0.2cm}\eq{\figBox{0}{-1.05}{0.4}{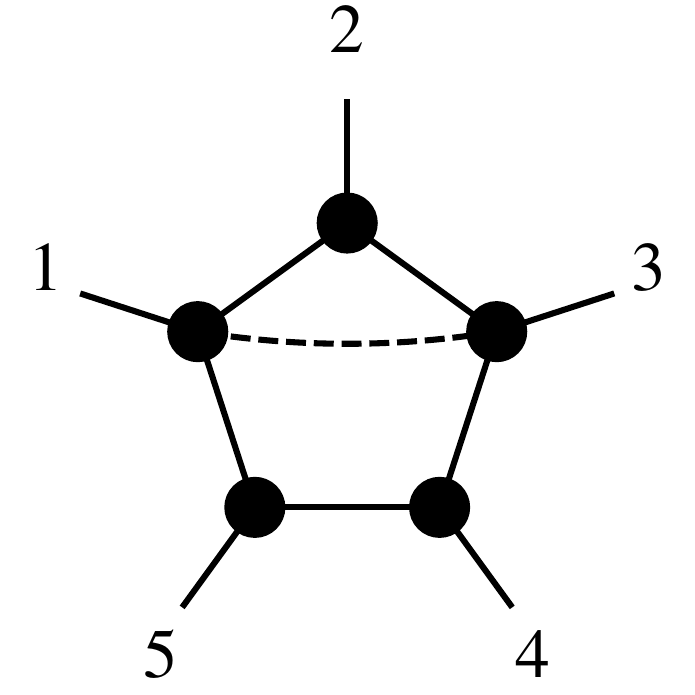}\Longleftrightarrow\int\limits_{AB}
\frac{\ab{AB\,Y}\times\ab{2\,3\,4\,5}\ab{4\,5\,1\,2}}{\ab{AB\,12}\ab{AB\,23}\ab{AB\,34}\ab{AB\,45}\ab{AB\,51}},\label{baby_pentagon_example1}\vspace{-0.2cm}}
where $\ab{2\,3\,4\,5}\ab{4\,5\,1\,2}$ in the numerator is for normalization\footnote{We will see that this normalization follows from the requirement that the integral have {\it unit leading-singularities}, and its sign is fixed by parity relative to the `wavy-line' pentagon drawn below it. In fact, as we will describe in \mbox{section \ref{pure_loop_integrals_section}}, the dashed-line in the figure dictates both the bitwistor $Y\equiv Z_1Z_3$ and the normalization of the integral.} and the bitwistor $Y$ is simply $Z_1Z_3$ (this is indicated by the dashed-line in the associated figure); and,
\eq{\figBox{0}{-1.05}{0.4}{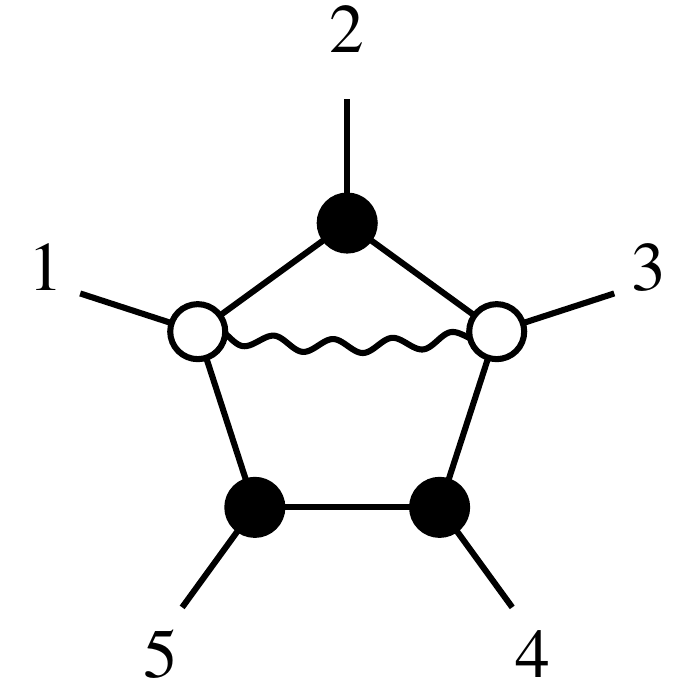} \Longleftrightarrow\int\limits_{AB}\frac{\ab{AB\,\widetilde{Y}}\times\ab{3\,4\,5\,1}}{\ab{AB\,12}\ab{AB\,23}\ab{AB\,34}\ab{AB\,45}\ab{AB\,51}},\label{baby_pentagon_example2}\vspace{-0.2cm}}
where the factor $\ab{3\,4\,5\,1}$ in the numerator is for normalization, and the bitwistor $\widetilde{Y}\equiv`(512)\newcap(234)$' is the line in twistor-space which lies along the intersection of the planes spanned by twistors $(Z_5,Z_1,Z_2)$ and $(Z_2,Z_3,Z_4)$---which is indicated in the figure by the `wavy-line'. As the first of many such examples, it is useful to write-out $\widetilde{Y}$ explicitly: \eqs{\widetilde{Y}\equiv({\color{paper_red}512})\newcap({\color{paper_blue}234})&=Z_{{\color{paper_red}5}}Z_{{\color{paper_red}1}}\ab{{\color{paper_red}2}\,{\color{paper_blue}2\,3\,4}}+Z_{{\color{paper_red}1}}Z_{{\color{paper_red}2}}\ab{{\color{paper_red}5}\,{\color{paper_blue}2\,3\,4}}+Z_{{\color{paper_red}2}}Z_{{\color{paper_red}5}}\ab{{\color{paper_red}1}\,{\color{paper_blue}2\,3\,4}},\\&=\qquad\,\,\,0\,\,\,\qquad+Z_1Z_2\ab{5\,2\,3\,4}+Z_2Z_5\ab{1\,2\,3\,4},} where we have used the fact that $\ab{2\,2\,3\,4}=0$. (The translation between statements such as `the line along the intersection of two planes' and explicit representative formulae such as the above will be explained in more detail below; here, we merely quote the result in a way from which we hope it will easy to guess the general case.)

These two integrals are examples of a very important class of integrals that we call {\it chiral integrals with unit leading singularities}, or {\it pure} integrals. In each case, the bitwistor appearing in the numerator (together with the integrand's normalization) is completely specified by the dashed- or wavy-line in the corresponding figure. We will explain many of the important features of these integrals together with the way their graphical representations in more detail in \mbox{section \ref{pure_loop_integrals_section}}. It is worth noting in passing, however, that the two integrals are parity conjugates of one another, and special bitwistors $Y$ and $\widetilde{Y}$ represent the two lines in twistor-space which simultaneously intersect the four lines $(51),(12),(23),$ and $(34)$; this means that $\ab{Y\,51}=\ab{Y\,12}=\ab{Y\,23}=\ab{Y\,34}=0$, and similarly for $\widetilde{Y}$. Because of this, they represent the two isolated points in $AB$-space for which these four propagators go on-shell.

Before moving-on to discuss loop integrands, we should emphasize that because the primary focus of this paper is the loop {\it integrand}---the sum of all the Feynman diagrams, as a rational function---there is nothing to say about the regulation of IR-divergent integrals such as the zero-mass box integral and the pentagons integrals given above. The only {\it integrals} we will evaluate explicitly are all {\it manifestly} finite (in a precise sense which will be described in \mbox{section \ref{finite_integrals_section}}), and hence are well-defined without any regulator. However, it is important to mention that IR divergent integrals can also easily be regulated and evaluated. In fact, the most natural way to add a regulator is also a very physical
one, given by moving out on the Coulomb branch \cite{Alday:2009zm} of the theory.

\subsection{The Loop {\it Integrand}}

A simple but far-reaching consequence of writing each Feynman integral in a loop amplitude using the dual variables  is that one can meaningfully combine all integrals appearing in a particular amplitude under the same integral sign. This leads to the concept of {\it the} loop {\it integrand} \cite{ArkaniHamed:2010kv}. We stress again that planarity and the use of dual variables plays a crucial role in making this possible--for a general theory, there is no natural origin of loop momentum space and therefore no canonical way of combining all Feynman diagrams under a common loop integral. 

It is easy to characterize the structure of the $n$ particle 1-loop integrand for ${\cal N}=4$ SYM using momentum-twistor space integrals. All the terms in the integrand can be combined defining a universal denominator containing all $n$ physical propagators of the form  $\langle AB\,a ~ a\pl1\rangle$. If a particular Feynman diagram has fewer propagators, then the numerator is chosen so as to cancel the extra propagators. The loop amplitude is given as an 
an integral over a single rational function,
\be
\label{integrand}
\mathcal{A}_n = \int\limits_{AB}\frac{\sum_{i}c_i\langle AB\,Y^i_1\rangle\langle AB\,Y^i_2\rangle\ldots \langle AB\,Y^i_{n-4}\rangle }{\langle AB\,12\rangle\langle AB\,23\rangle\ldots \langle AB\,n\mi1~n\rangle\langle AB\,n\,1\rangle}
\ee
where $\mathcal{A}_n$ is the full 1-loop partial amplitude. This formula is already written using the simplifications that arise in ${\cal N}=4$ SYM, in other words, it is manifestly DCI. However, the integrand exists in any planar theory: for a theory which is not DCI, (\ref{integrand}) would necessarily contain also terms with powers of $\ab{AB\,I_{\infty}}$.

At higher loops, say $L$ loops, scattering amplitudes are given as linear combination of integrals of the form
\be
\int \prod_{i=1}^L d^4\ell_i \frac{\prod_{j=1}^L N(\ell_j)}{\prod_{k=1}^L P(\ell_k)}\times \frac{1}{R(\ell_1,\ldots \ell_L)},
\ee
where $N,P,$ and $R$ are products of Lorentz invariants constructed out of Feynman propagators and which depend on the variables shown and on the external momenta. Written in this form, there is clearly a large amount of redundancy in the definitions of the internal loop momenta.

Since we are dealing with only planar integrals, for each Feynman diagram there exists a dual diagram (the standard dual graph of a planar graph). Consider for example the following four-point two-loop integral: \eq{\figBox{0}{-1.675}{0.5}{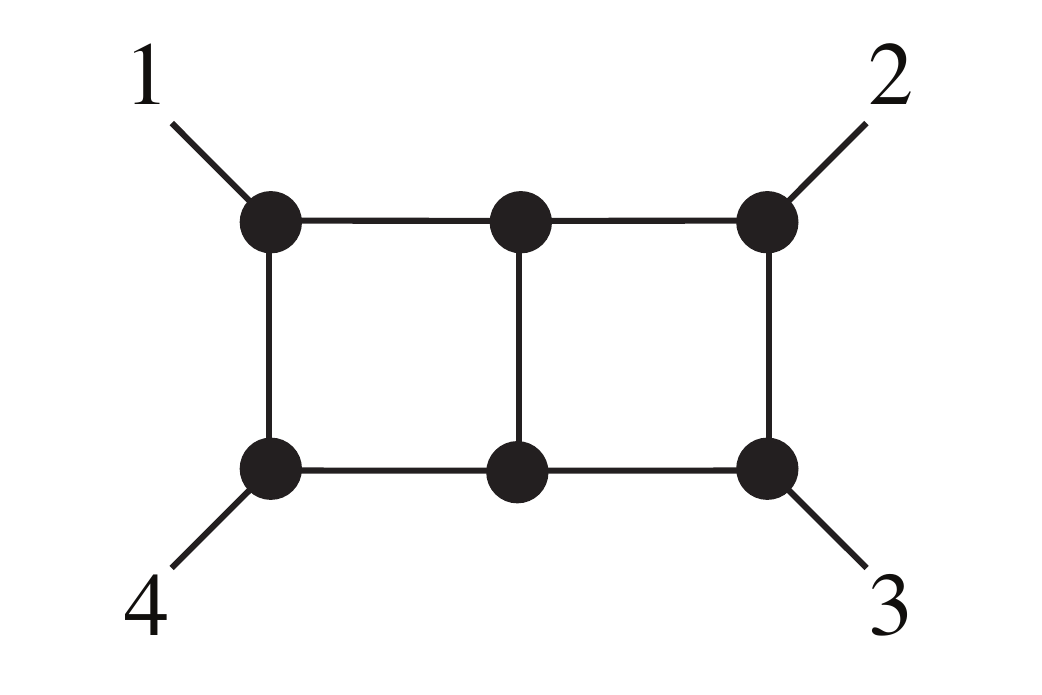}\Longleftrightarrow\figBox{0}{-1.675}{0.5}{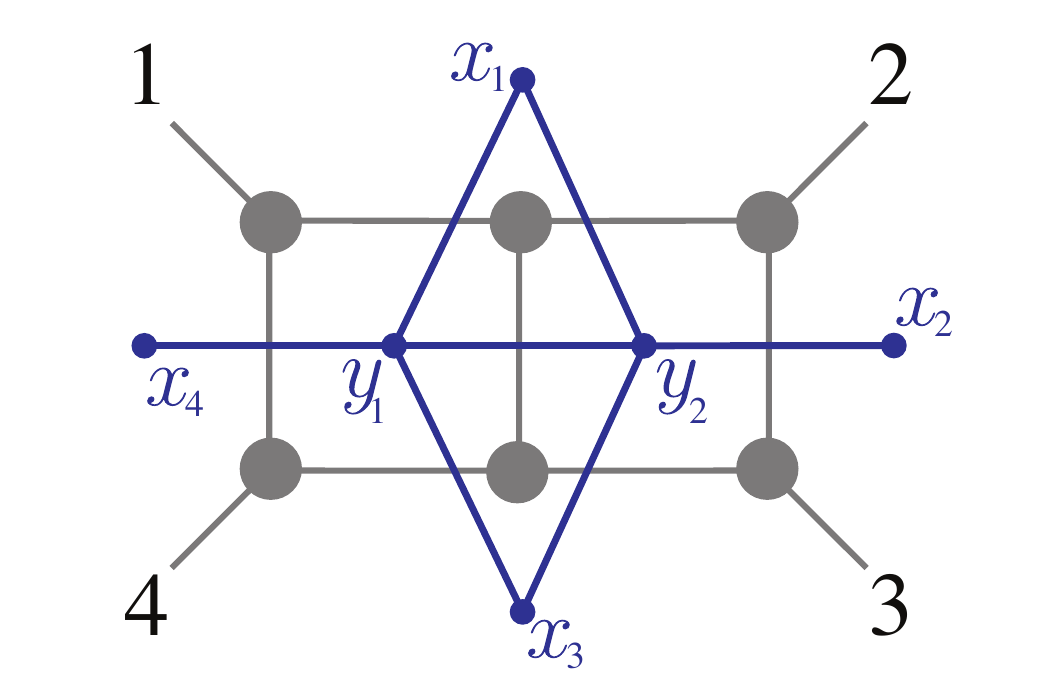}\label{four_point_two_loop_graphs}}
Using $x_i$ to denote the dual coordinates of the external momenta and $y_i$ to denote the internal points, one can write any planar $L$-loop integral in dual  coordinates. There is, however, one slight subtlety in using such a prescription to uniquely define `the' integrand: while the definition of the external points $x_i$ is unique, the labeling of the internal points is not (when $L>1$). But the solution to this problem is very simple: we are always free to completely symmetrize the integrand with respect to all $L!$ permutations of the internal loop-variable labels. Although we will often write multi-loop integrands in some particular representative choice of the labels for internal propagators, complete-symmetrization over all permutations of indices is always implied (including a factor of $1/L!$ from this symmetrization).

Consider for example the simplest two-loop integral, given above in (\ref{four_point_two_loop_graphs}). Written in dual-coordinates, the integral would be given by
\be
\hspace{-1.15cm}\int\!\!\!\left(\frac{d^4y_1d^4y_2}{2}\right)\frac{((x_1-x_3)^2)^2(x_2-x_4)^2}{(y_1-x_3)^2(y_1-x_4)^2(y_1-x_1)^2(y_2-x_1)^2(y_2-x_2)^2(y_2-x_3)^2(y_1-y_2)^2}+(y_1\leftrightarrow y_2),
\nonumber
\ee
---where the numerator was chosen in order to make the integral dual-conformally invariant, and the factor of $1/2$ in the measure reflects the complete-symmetrization.

Of course, as we will see repeatedly throughout this paper, (multi-)loop integrands are much more naturally expressed in terms of momentum-twistor variables. To translate the integral (\ref{four_point_two_loop_graphs}) in momentum-twistor variables, we need to associate a pair of twistors to each of the two loop variables. This we can do by making the association \eq{\label{notaloops} y_1\leftrightarrow (Z_A,Z_B)\quad\mathrm{and}\quad y_2\leftrightarrow(Z_C,Z_D).}

Using this notation and the translation of propagators in terms of momentum twistors given in (\ref{locon}) one finds
\be
\hspace{-0.05cm}\int\limits_{(AB,CD)}\!\!\!\!\!\!\!\frac{\langle 1234 \rangle^2\ab{2341}}{\langle AB\,41\rangle\langle AB\,12\rangle\langle AB\,23\rangle\langle CD\,23\rangle\langle CD\,34\rangle\langle CD\,41\rangle\langle AB\,CD\rangle},\nonumber
\ee
where `$(AB,CD)$' implies that the integration measure carries with it a factor of $1/2$ from the symmetrization of $(AB)\leftrightarrow(CD)$. We should mention here that for 3-loops, we will use $(Z_E,Z_F)$ to denote the line corresponding to $y_3$---but of course, a convention such as that of associating ($Z_{A_m},Z_{B_m}$) with $y_m$ would be increasingly preferable at high-loop order.

Before we leave the topic of {\it the} loop-integrand in general, we should mention that the form of the integrand obtained via BCFW as described in \cite{ArkaniHamed:2010kv} makes it completely manifest that the loop-integrands in $\mathcal{N}=4$ enjoy the full Yangian symmetry of the theory. (Of course, the choice of an integration contour which introduces IR divergences, such as the physical contour, breaks this symmetry.) 

However, just as with the BCFW recursion relations at tree level, the formulae obtained from the recursion do not enjoy manifest locality or manifest cylcic invariance. The restriction that we impose throughout this work, however, is that loop-integrand be expanded in a way which makes use of only planar, local propagators. As we have stressed a number of times, we will find amazingly simple, manifestly cyclically symmetric and local expressions for multi-loop amplitudes, that are significantly simpler and more beautiful than their BCFW counterparts! Taken together with the parallel results presented in \cite{polytopePaper}, this strongly suggests the existence of a formulation for scattering amplitudes directly yielding these remarkable local forms.

The local formulae presented in this paper are very closely related to and influenced by the concept of the {\it leading singularities} of scattering amplitudes, 
which we proceed to presently describe.

\subsection{Leading Singularities}

\subsubsection*{Definition}

The concept of leading singularities was introduced in the 1960's in the context of massive scalar theories \cite{ELOP}. More recently, in 2004, the same concept was modified to accommodate massless particles and this was exploited for Yang-Mills in \cite{Britto:2004nc}. The original definition of `leading-singularity' refers to a discontinuity of a scattering amplitude across a singularity of the highest possible co-dimension. At 1-loop, for example, leading singularity discontinuities are computed using a generalization of a unitarity cut, but where four propagators are cut instead of two. Using ${\cal A}_i$ for $i=1,\ldots,4$ to denote the four partial amplitudes, each with their associated momentum-conserving $\delta$-function, one has what can be called leading-singularity discontinuity,\vspace{-0.3cm}
\begin{align}\phantom{=}&\figBox{4.48}{-2.05}{0.5}{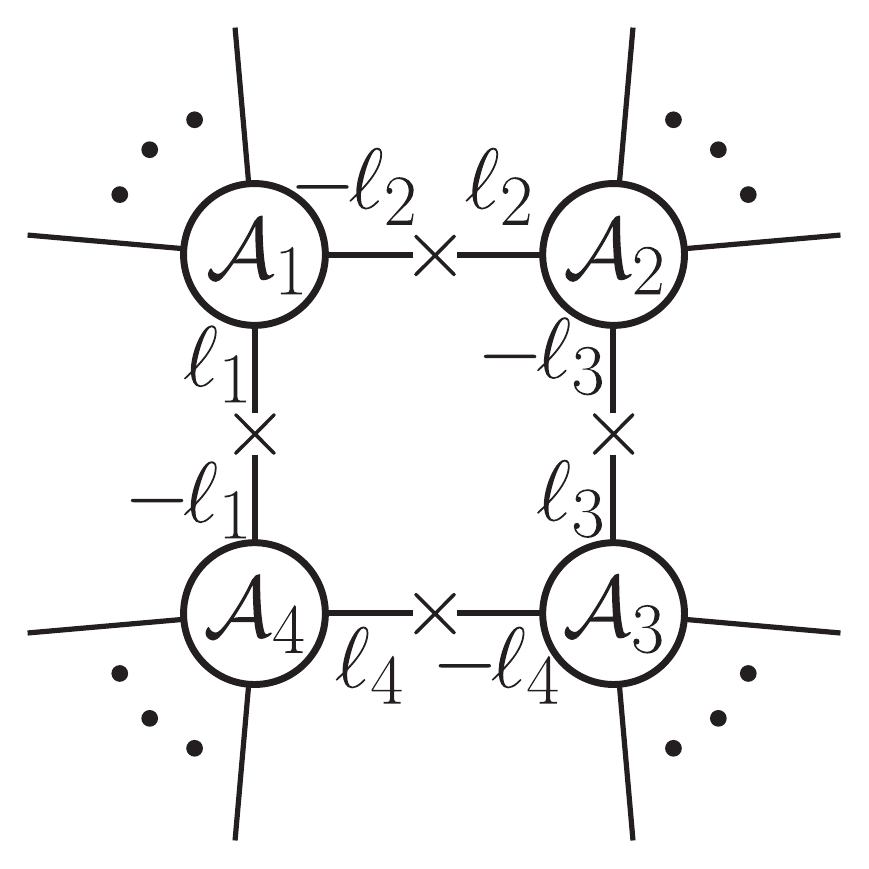}\label{oldLS}\\=&\int\!\!\prod_{r=1}^4d^4\tilde\eta_r d^4\ell_r\delta (\ell^2_r)\,\,{\cal A}_1(\{\ell_1,\tilde\eta_1\},\{-\ell_2,\tilde\eta_2\},\ldots )\times {\cal A}_2(\{\ell_2,\tilde\eta_2\},\{-\ell_3,\tilde\eta_3\},\ldots ) \phantom{\,.}\nonumber \\ &\hspace{2.95cm} \times{\cal A}_3(\{\ell_3,\tilde\eta_3\}, \{-\ell_4,\tilde\eta_4\},\ldots )\times
{\cal A}_4(\{\ell_4,\tilde\eta_4\},\{-\ell_1,\tilde\eta_1\},\ldots)\,.\nonumber\end{align}
Here, the integrations over the internal loop momenta are there only to remind us that we are to sum-over all solutions to the conditions imposed by the $\delta$-functions, and the integral over the Grassmann coordinate $\tilde\eta_i$ of each internal particle $\ell_i$ is there to remind us that we are to sum-over the exchange of all possible internal particles---which in the case of $\mathcal{N}=4$ means the full super-multiplet.\footnote{Here, we are using an on-shell superspace formalism which allows us to talk about all particles in the same super-multiplet as a single 1-particle state. We assume familiarity with this concept, but for careful definitions, more references and applications see \cite{ArkaniHamed:2008gz}.}

This point of view of leading-singularities has been very useful and allows a complete determination of 1-loop amplitudes in ${\cal N}=4$  and in ${\cal N}=8$ supergravity amplitudes when thought of as linear combinations of scalar box integrals with rational coefficients. The rational coefficients can be computed using the notion of generalized unitarity. Clearly, the notion of discontinuities is not related to the existence of an integrand and this is the reason it works in ${\cal N}=8$, supergravity where an analog of `the integrand'---which requires a way to combine integrals with different cyclic orderings---has not yet been found.

As mentioned in our discussion of reduction procedures in ${\cal N}=4$ SYM, the expansion in terms of boxes cannot give the physical integrand. The physical integrand is defined as that which coincides with the one from Feynman diagrams, prior any to reduction techniques, as rational functions---and, as we will see, the Feynman diagrams of $\mathcal{N}=4$ in a given $R$-charge sector are {\it chiral}.

Once we think about the integrand as being the object we are after, we can try to model it by using some appropriate basis of functions, dictated by a general reduction procedure. Clearly, the set of all DCI tensor pentagons and boxes should be enough. Nevertheless, we will find that such a basis would still possess many of the unattractive features of the box-expansion, and so we will introduce much more refined choice in section \ref{one_loop_integrands_and_basis}.

The importance of dealing with a specific rational function is that we can integrate it on {\it any} choice of contour we'd like---not just the real-contour which defines the Feynman integral. This allows us to define a more refined notion of a leading-singularity---the previous notion, motivated by generalized unitarity, is much coarser version of the one we will use now. In \cite{Cachazo:2008vp}, this more refined notion was introduced, and it was used to match the full $\mathcal{N}=4$ integrand for several 1-loop and 2-loop examples. However, in \cite{Cachazo:2008vp} the deep reason for why the idea was working, {\it i.e.}, the existence of the integrand, was not appreciated.

Whether written in ordinary momentum space, using dual-coordinates, or using momentum-twistors, loop integrals can be thought of as complex contour integrals on $\mathbb{C}^4$ with the choice of contour corresponding to $\mathbb{R}^4$---the real-slice. However, this choice of contour is known to break many of the symmetries of the theory, and is littered with IR-divergences, etc. that can be the source of confusion. From various viewpoints, the most natural contours would instead be those which compute the {\it residues} of the integrand. These are always finite, are often vanishing, and make manifest the full Yangian symmetry of the theory. We refer the reader to \cite{Griffiths:1978a} for a mathematical definition of residues in several complex variables; here we hope the reader will find the definitions a natural generalization of the one-dimensional residues with which everyone is familiar.
\begin{figure}[b]\vspace{-0.5cm}\centering\figBox{0}{-4.05}{0.55}{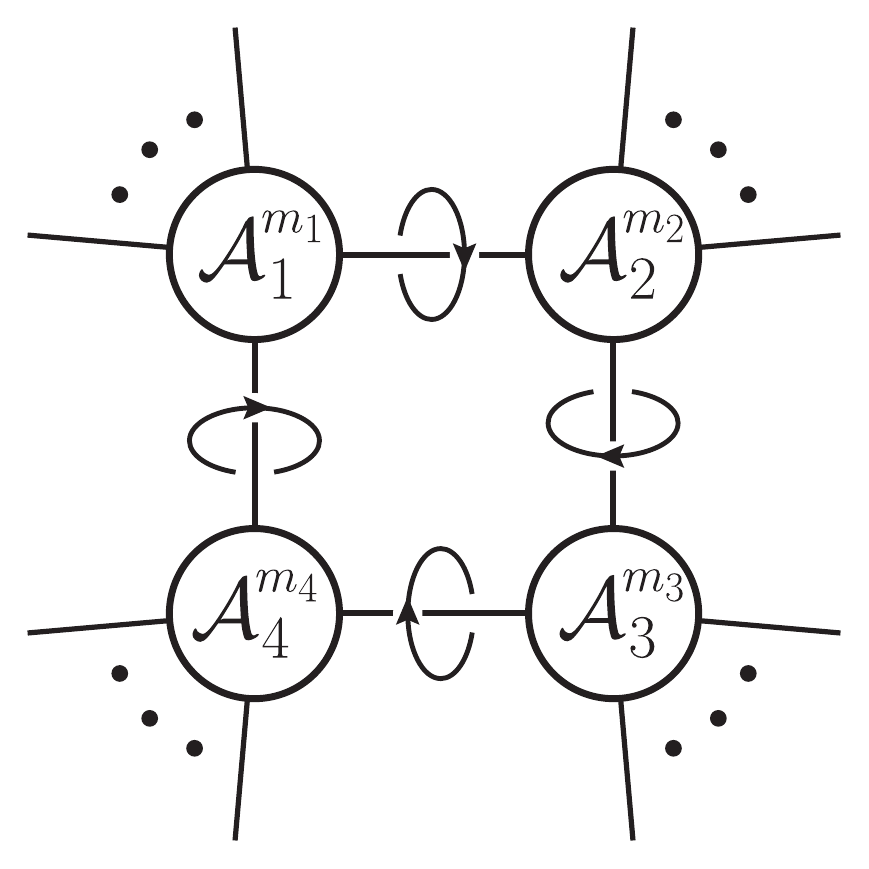}\caption{A `{\it quad-cut}' one-loop leading-singularity viewed as a $T^4$ contour-integral which `encircles' the point in $\mathbb{C}^4$ where four-propagators are made on-shell.\label{one_loop_leading_singularities_contour}}\end{figure}

Let us present the definition using $x$ variables first. Consider a contour of integrations with the topology of a $T^4 = (S^1)^4$. In order to compute a particular residue one has to choose four propagators $(x-x_{a_i})^2$, with $i=1,\ldots,4$ and integrate over the $T^4$, defined by $|(x-x_{a_i})|=\epsilon_i$ where $\epsilon_i$ are small positive real numbers near one of the solutions. The circles, $S^1$ are parametrized by the phases and are given a particular orientation.

The definition of a multidimensional residue is very natural if one defines variables $u_i = (x-x_{a_i})^2$. Performing the change of variables the integral becomes
\be
\int \prod_{i=1}^4 \frac{du_i}{u_i}\times \frac{1}{J}\times \{{\rm The~rest~of~the~integrand}\}
\ee
where now the contour becomes small circles around $u_i=0$. $J$ is the Jacobian of the change of variables. The residue is then the Jacobian times the rest of the integrand evaluated at $u_i=0$. The Jacobian
\be
J = {\rm det}\left( \frac{\partial (u_1,u_2,u_3,u_4)}{\partial (x_1,x_2,x_3,x_4)}\right),
\ee
is clearly antisymmetric in the order of the columns. Different orderings can differ by a sign and this is related to the orientation of the contour. These signs are important when discussing the generalization of residue theorems to the multidimensional case, which will play an important role momentarily.

From now on we call each individual residue a {\it leading-singularity}. As before, these are given by the product of four on-shell tree amplitudes as shown in \mbox{Figure \ref{one_loop_leading_singularities_contour}}. The reason for the appearance of the tree amplitudes is that the residue of the poles is computed where the four propagators vanish and therefore internal particles can be taken on-shell.

Leading singularities at higher loop-level can also be defined as residues of a complex, multidimensional integral over $\mathbb{C}^{4L}$ where $L$ is the loop order. This means that in order to define a residue one has to define a $T^{4L}$ torus as a contour of integration. Na\"ively, residues can only be defined for integrals with at least $4L$ propagators. However, noticing that propagators are quadratic in the loop-momentum, one can define {\it composite leading singularities} which involve less than $4L$ propagators as done in \cite{Buchbinder:2005wp, Cachazo:2008vp, ArkaniHamed:2009dn}, using the self-intersection of curves defined by the on-shell condition to define isolated points in $\mathbb{C}^4$ about which the $T^{4L}$ contour should `encircle.'

We will not discuss composite leading singularities in detail here simply because we will present evidence that when a special set of integrals, we call {\it chiral integrals with unit leading-singularities}, are used, matching non-composite leading-singularities appears to suffice to fix the entire amplitude. Moreover, we will see that only a very small subset of non-composite leading-singularities need to be considered to accomplish this.

\subsubsection*{Chirality of Leading Singularities}

It turns out that for nonsingular external momenta, there are exactly two solutions to the equations $(x-x_{a_i})^2=0$, with $i=1,\ldots,4$, and therefore two residues of each choice of four propagators. (This has a beautiful geometric interpretation in momentum twistors as we will see shortly.) This means that for an $n$-particle amplitude, there are $2{\small \left( \begin{array}{c} n\\4\end{array}\right)}$ (non-composite) one-loop leading-singularities.

Consider any box integral, say, an integral with two massless legs and two massive, known as the `two-mass-easy' integral: \vspace{-0.5cm}
\be
\hspace{-1.4cm}\figBox{0}{-1.75}{0.6}{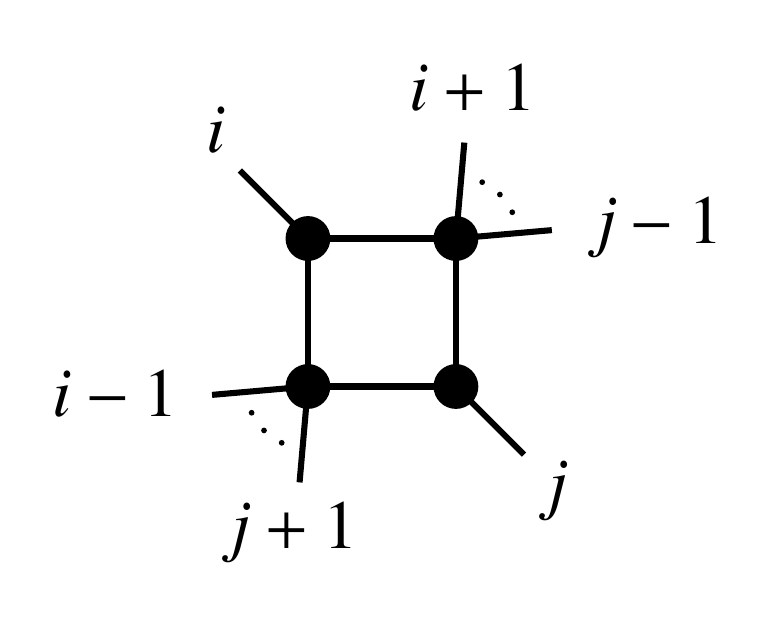}\hspace{-0.5cm}\Leftrightarrow\mathcal{I}_{2\mathrm{me}}= \int\!\!\!d^4x \frac{N}{(x-x_{i-1})^2(x-x_{i})^2(x-x_{j-1})^2(x-x_{j})^2}\,,\vspace{-0.2cm}
\ee
where $N$ is just some normalization that need not concern us presently. The equations $$(x-x_{i-1})^2=(x-x_{i})^2=(x-x_{j-1})^2=(x-x_{j})^2=0$$ have two solutions, and therefore a residue can be computed for each such point separately. We'll soon see that these two solutions are easily found and differentiated when written with momentum-twistor variables; but for now, let us suppose the two solutions have been found, and denote the corresponding contours $T^4_1$ and $T^4_2$.

A very important tool that will make an appearance many times is multidimensional analogue of Cauchy's theorem, called the {\it Global Residue Theorem} (GRT). The GRT states that---given a suitable condition at infinity---the sum over all the residues of a given rational function vanishes (see chapter 6 of \cite{Griffiths:1978a}). This means, in the present case, that
\be
{\rm res}_{T^4_1}({\cal I}_{\rm 2me}) + {\rm res}_{T^4_2}({\cal I}_{\rm 2me})  = 0
\ee
Moreover, we can choose the normalization $N$ is such that, say ${\rm res}_{T^4_1}({\cal I}_{\rm 2me})=1$. Such a choice is possible for all box integrals, following from the simple fact that all box-integrals---having only four propagators---must have residues which are proportional equal and opposite. We refer to this fact by saying that scalar box integrals are {\it not chiral}. The use of the word chiral is justified by the fact that the locations of the leading singularities, as points in $\mathbb{C}^4$, are mapped into each other by parity---which is just complex conjugation. And so the corresponding contours are mapped into each other up to orientation by parity. If use $(T^4_1)^*$ to denote the parity conjugate contour of $T^4_1$, then ${\rm res}_{(T^4_1)^*} = -{\rm res}_{T^4_2}$ and the GRT implies that
\be
{\rm res}_{T^4_1}({\cal I}_{\rm 2me}) = {\rm res}_{(T^4_1)^*}({\cal I}_{\rm 2me}).
\ee

Let us now consider the leading-singularities of the one-loop integrands of ${\cal N}=4$ Yang-Mills. We'll see that, as scattering amplitudes of $\mathcal{N}=4$ in a given $R$-charge sector are chiral, so are the one-loop leading-singularities of field theory! In other words, the two residues associated with the two solutions of cutting four-propagators are {\it not} the same. Let us see this in an example. The simplest possible example is the five-particle MHV amplitude\footnote{The only DCI object for four-particles is the zero-mass box integral. This is why both leading singularities are equal to the tree amplitude.}. Let us consider taking the leading singularities of the field-theory integrand which encircles the point in $\mathbb{C}^4$ where the following four propagators go on-shell: \eq{\figBox{-0.5}{-1.95}{0.65}{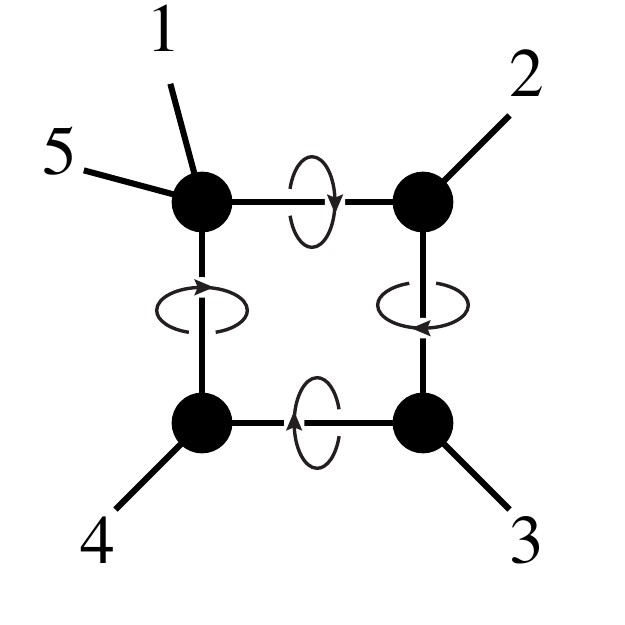}\hspace{-0.4cm}\Longleftrightarrow(x-x_1)^2=(x-x_2)^2=(x-x_3)^2=(x-x_4)^2=0.\vspace{-0.3cm}\label{five_point_quad_cut}} It was noticed already in \cite{Britto:2004nc} that on one solution ${\cal N}=4$ SYM gives the tree amplitude, $A^{\rm tree}_5$, while it vanishes on the second.

The vanishing of leading singularities can be understood from pure supersymmetry. Consider an amplitude in the $R$-charge sector $m$. Recall the N$^{m-2}$MHV classification of amplitudes in $\mathcal{N}=4$: under a rescaling of all $\tilde\eta_a$ variables by $t\tilde\eta_a$, an N$^{m-2}$MHV amplitude picks up a factor of $t^{4m}$. From the definition of leading singularities as the product of tree amplitudes connected by internal on-shell states we see that every internal line contributes $(-1)$ to the $R$-charge counting coming from the integration over $\tilde\eta$ variables. At 1-loop, we have four tree-amplitudes and four propagators. If the $R$-charge of each tree-amplitude is $m_i$ (see \mbox{Figure \ref{one_loop_leading_singularities_contour}}), then the $R$-charge of the leading singularity is $m_1+m_2+m_3+m_4-4$.

Returning to the five-particle example, because we are interested in a one-loop MHV amplitude, all its leading-singularities must have $m=2$. The four-particle vertex (in the upper-left of the figure above) can only have $m_1=2$ and therefore the three-particle vertices have to satisfy $m_2+m_3+m_4 = 4$. Since the possible values for $m$ for a three-particle amplitude are $1$ and $2$, two vertices must have $m=1$ and one must have $m=2$. This leaves only the possibilities shown below: \vspace{-0.2cm}\eq{\nonumber\figBox{0}{-1.5}{0.5}{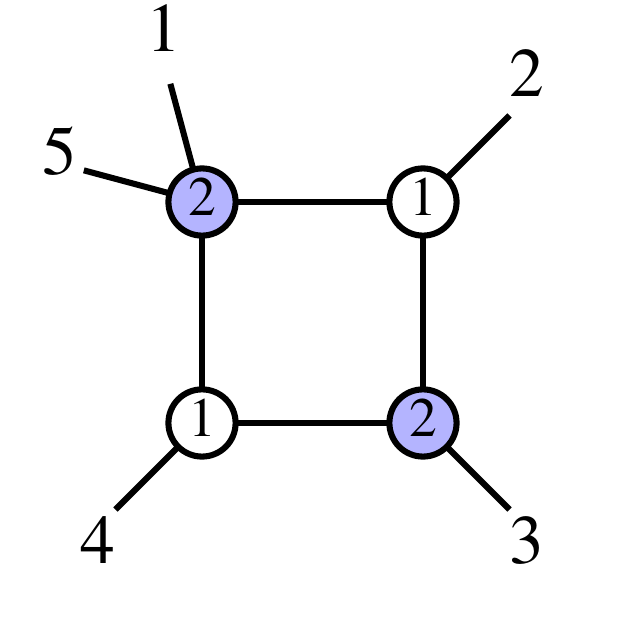}\qquad\figBox{0}{-1.5}{0.5}{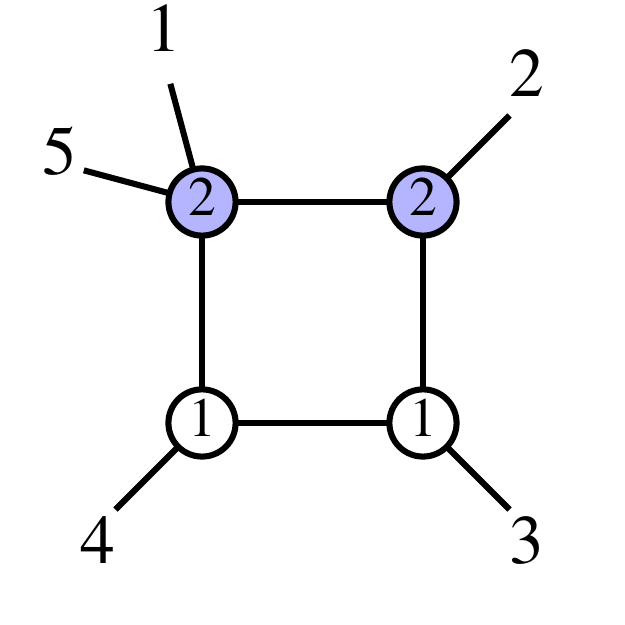}\qquad\figBox{0}{-1.5}{0.5}{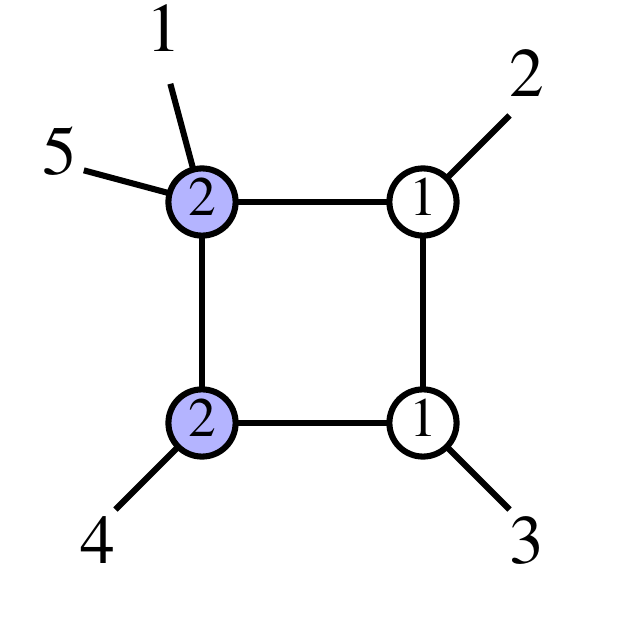}\vspace{-0.3cm}}
Of these three possible leading-singularities of field theory, it turns out that the first one is equal to the five-point MHV tree-amplitude, and the latter two vanish for generic external momenta. In fact, whenever one is considering a leading singularity which involves 3-particle vertices, some very simple and powerful rules prove very useful: 1. any leading singularity involving adjacent three-particle vertices with the same $R$-charge will vanish for generic external momenta (momentum conservation in this case, requires that the external particles attached to these vertices must be collinear); and 2. leading singularities involving three-particle amplitudes are almost always {\it chiral}---the only exception being the four-particle amplitude.

In the case of the five particle example under consideration, we see that the residue from the contour encircling one of the two solutions to the quad-cut equations in (\ref{five_point_quad_cut}) is equal to $\mathcal{A}_{5,\mathrm{MHV}}^{\mathrm{tree}}$, while the conjugate contour integral vanishes. We will explore this in more detail once we introduce the geometric point of view.

\subsubsection*{Dual Formulation of Leading Singularities}

In the rest of the paper we will make much use of the fact that leading-singularities satisfy many relations. These relations can be seen as resulting from residue theorems of the integrals which compute them. As a final comment before exploring the connection between leading singularities and the classic enumerative problems in the projective geometry of momentum twistor space let us briefly introduce the Grassmannian formulation.

In \cite{ArkaniHamed:2009dn}, leading singularities were proposed as completely IR-finite quantities that were likely to contain all the information needed to compute the S-Matrix of ${\cal N}=4$ SYM. Moreover, it was conjectured that all leading singularities of the theory, which can be obtained to arbitrarily higher loop order, are computed by a contour integral over a Grassmannian manifold\footnote{The Grassmannian $G(m,n)$, a natural generalization of ordinary projective space, is the space of $m$-dimensional planes in $n$-dimensions. Each point in $G(m,n)$ can be represented by the $m$ $n$-vectors which span the plane, modulo a $GL(m)$ redundancy.} $G(m,n)$  called ${\cal L}_{m,n}$. Here $m$ determines the $R$-charge sector of the theory under consideration.

The integral was first presented in twistor space
\be
{\cal L}_{m,n}({\cal W}_a) = \int \frac{d^{nm}C_{\alpha a}}{{\rm vol}(GL(m))}\frac{\prod_{\alpha=1}^m\delta^{4|4}\left(\sum_{a=1}^nC_{\alpha a}{\cal W}_a\right)}{(1\,2\,\cdots\,m)(2\,3\,\cdots\,m\pl1)\cdots (n\,1\,\cdots\,m\mi1)}.
\ee
In this presentation, residues of this integral are manifestly superconformal invariant (that is, superconformally-invariant in ordinary spacetime). Here we have introduced the concept of dual super twistor space $W = (\tilde\lambda,\mu,\tilde\eta)$. This particular space will not play a significant role in this work, so we refer the interested reader to \cite{ArkaniHamed:2009dn, ArkaniHamed:2009si} for more details.

This formula can be transformed to momentum-space and then to momentum-twistor space. Very remarkably, the formula in momentum-twistor space also turns out to be an integral over a Grassmannian, with the MHV-tree-amplitude arising as the Jacobian from the change of variables. Specifically,
\be
{\cal L}_{m,n}|_{\rm momentum-space}(\lambda, \tilde\lambda, \tilde\eta ) = {\cal L}_{2,n}\times {\cal R}_{k,n},
\ee
where $k=m-2$ and
\be
\label{rkn}
{\cal R}_{k,n}({\cal Z}_a) = \int \frac{d^{nk}D_{\alpha a}}{{\rm vol}(GL(k))}\frac{\prod_{\alpha=1}^k\delta^{4|4}\left(\sum_{a=1}^nD_{\alpha a}{\cal Z}_a\right)}{(1\,2\,\cdots\,k)(2\,3\,\cdots\,k\pl1)\cdots (n\,1\,\cdots\,k\mi1)}.
\ee
This representation in momentum twistor space makes {\it dual} superconformal invariance manifest \cite{ArkaniHamed:2009vw, Mason:2009qx}. With some more effort one can prove that residues of this formula are also invariant under level one generators of the Yangian of the dual superconformal algebra and hence invariant under the whole Yangian \cite{Drummond:2010uq}. The level one generators are nothing but the superconformal generators when passed through ${\cal L}_{2,n}$.

It has now been proven that all leading singularities are Yangian invariant and that all Yangian invariants are residues of the integral (\ref{rkn}). From the physical point of view the problem has been solved. It might also be interesting to go further and prove that all residues of (\ref{rkn}) correspond to some leading singularity but we will not discuss this issue any further.

\subsubsection*{Momentum Twistors and Schubert Problems}

Statements like the number of solutions to setting four propagators to zero is two are non-obvious from the dual space $x$ point of view. In terms of momentum twistors, this statement turns out be a simple, classic problem of the enumerative geometry of ${\mathbb{CP}^3}$, solved by Schubert in the 1870's \cite{Schubert:1879,schubert:Bio}.

Recall that an $n$-particle 1-loop amplitude can be written as 
\be
\mathcal{A}_n = \int\limits_{AB} \frac{\sum_{i}c_i\langle AB\,Y^i_1\rangle\langle AB\,Y^i_2\rangle\cdots \langle AB\,Y^i_{n-4}\rangle }{\langle AB\,12\rangle\langle AB\,23\rangle\cdots \langle AB\,n\mi 1~n\rangle\langle AB\,n\,1\rangle}.
\ee
Each one-loop leading-singularity is associated with a {\it point} in the space of loop-momenta for which some choice of four propagators simultaneously become on-shell, \eq{\nonumber\figBox{-1.05}{-1.35}{0.5}{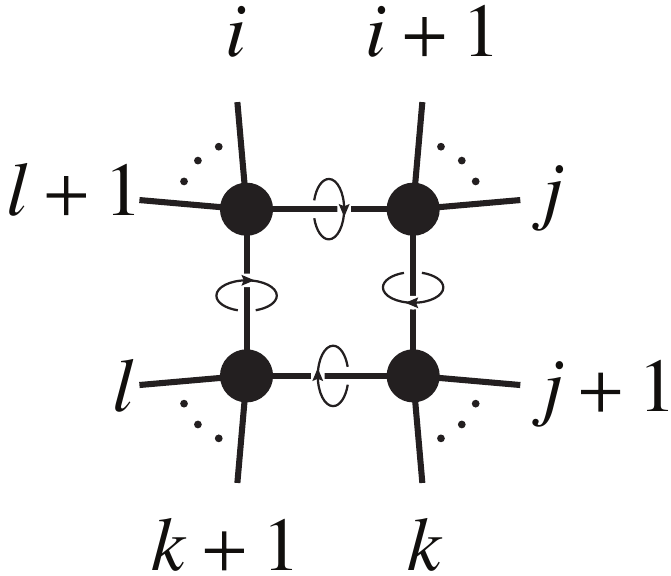}\Longleftrightarrow\langle AB\,i~i\pl1\rangle=\langle AB\,j~j\pl1\rangle=\langle AB\,k~k\pl1\rangle=\langle AB\,l~l\pl1\rangle=0;}
Because the loop momentum is represented in momentum-twistors as the {\it line} (AB), the solution to these four equations should correspond to a particular configuration for the line $(AB)$. We will see that for all leading-singularities which involve a three-particle vertex (a `massless leg'), the two solutions to four equations above are cleanly distinguished geometrically, allowing for a richly-chiral description of the integrand.

Before describing the full problem of putting four propagators on-shell, let us briefly consider the geometric significance of having a single factor, say $\ab{AB\,i\,i\pl1}$, vanish. Recall that the four-bracket $\ab{\cdot\,\cdot\,\cdot\,\,\cdot}$ is nothing but the determinant of the $4\times 4$ matrix of components of its four momentum-twistor arguments (viewed as elements of $\mathbb{C}^4$). As such, $\ab{AB\,i\,i\pl1}=0$ if and only if the vectors $Z_A,Z_B,Z_i,Z_{i+1}$ are {\it not} linearly independent, implying the existence of some linear relation among the four twistors of the form $\alpha_AZ_A+\alpha_BZ_B+\alpha_iZ_i+\alpha_{i+1}Z_{i+1}=0$. Trivially rearranging we see that \eq{\alpha_AZ_A+\alpha_BZ_B=-(\alpha_iZ_i+\alpha_{i+1}Z_{i+1}),} which we may read as saying there is a point on the line spanned by $Z_A,Z_B$---namely $(\alpha_AZ_A+\alpha_BZ_B)$---which lies along the line spanned by $Z_i,Z_{i+1}$. Which is to say, the lines $(AB)$ and $(Z_iZ_{i+1})$ {\it intersect}; and because two intersecting lines describe a plane, we say that the four points $Z_A,Z_B,Z_i,Z_{i+1}$ are {\it coplanar}.

Therefore, the problem of finding the particular lines $(AB)$ for which four propagators simultaneously vanish is equivalent to finding the set of lines in $\mathbb{CP}^3$ which simultaneously intersect four given lines (which are presumed fixed by the external data). The number of solutions to this problem is one of the classic examples of the enumerative geometry developed by Schubert in the 1870's. For this reason we call these problems {\it Schubert problems}.

The answer to the number of lines which intersect a given four turns out to be remarkably robust: provided the four lines are sufficiently generic, there are always 2 solutions, and an infinite number otherwise.\footnote{To be precise, we must count solutions with multiplicity; however, for a generic set of lines in the problem, the 2 solutions will always be distinct.} (An example of a {\it non-generic} configuration would be one for which three or more of the lines were coplanar; these are never found for generic external momenta.)

Schubert derived the number of such solutions with an argument that is deceptively simple. The idea is to consider a particular configuration where it is easy to count the number of solutions. Schubert intuited that the answers to such enumerative questions should be topological in nature, and therefore should not depend on the particular configuration in question. Therefore, one can analyze the most convenient possible configuration (for which the number of solutions is not infinite) and the answer found for that case, should be the answer in general. Said another way, the number of solutions to a given Schubert problem should not change when a particular special configuration is smoothly moved into a more general position.

Perhaps the easiest configuration for which we can count the number of solutions to the Schubert problem of finding the lines $(AB)$ that intersect four given lines in $\mathbb{CP}^3$ is the {\it zero-mass} configuration; it is so-called because it is the configuration which corresponds to the box integral with {\it zero} of its four corners massive,
\vspace{-0.2cm}\eq{\nonumber\figBox{0}{-1.5}{0.45}{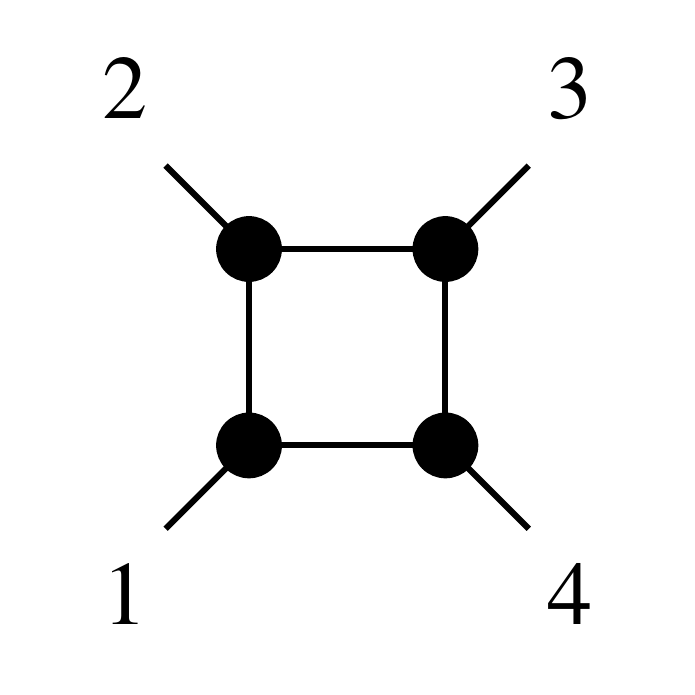}\hspace{-0.5cm}\Longleftrightarrow\;\;\;\int\limits_{AB}\frac{\ab{1234}\ab{2341}}{\ab{AB\,12}\ab{AB\,23}\ab{AB\,34}\ab{AB\,41}},\vspace{-0.3cm}\label{zero-mass-integral}} which is an integral we have seen before. Explicitly, we would like to find all the lines $(AB)$ which intersect all the four lines $(12),(23),(34),$ and $(41)$. This problem is indeed easy to solve, and the two solutions are drawn below.
\eq{\hspace{-0.95cm}\nonumber\begin{array}{c}\figBox{0}{-1.65}{0.5}{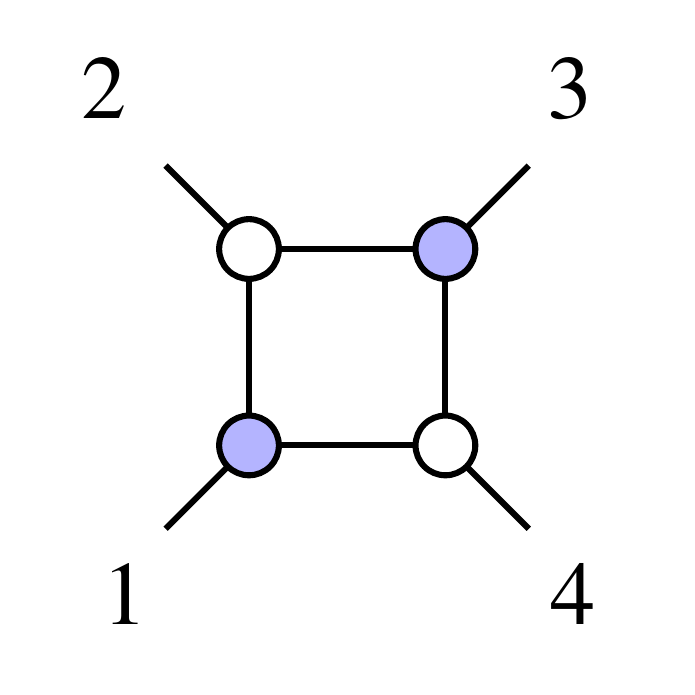}\hspace{-0.8cm}\Leftrightarrow\figBox{0}{-1.5}{0.55}{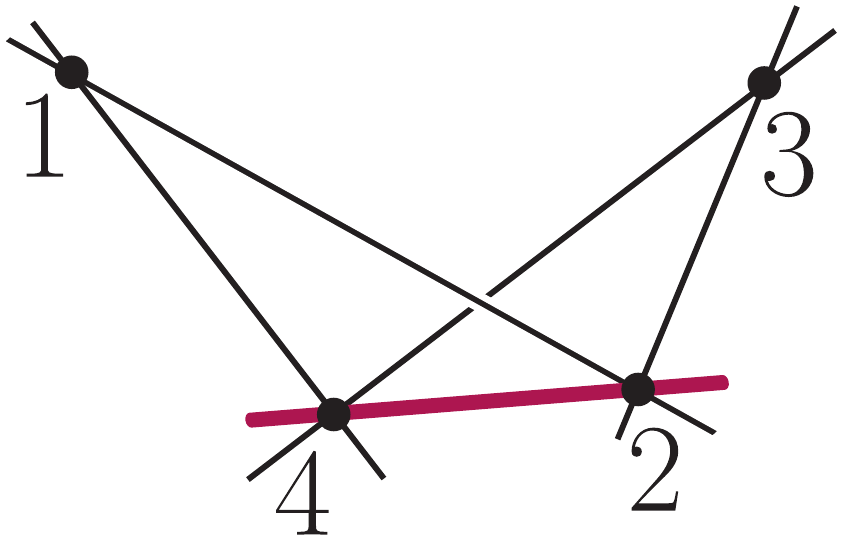}\\(AB)=(24)\end{array}\quad\begin{array}{c}\figBox{0}{-1.5}{0.55}{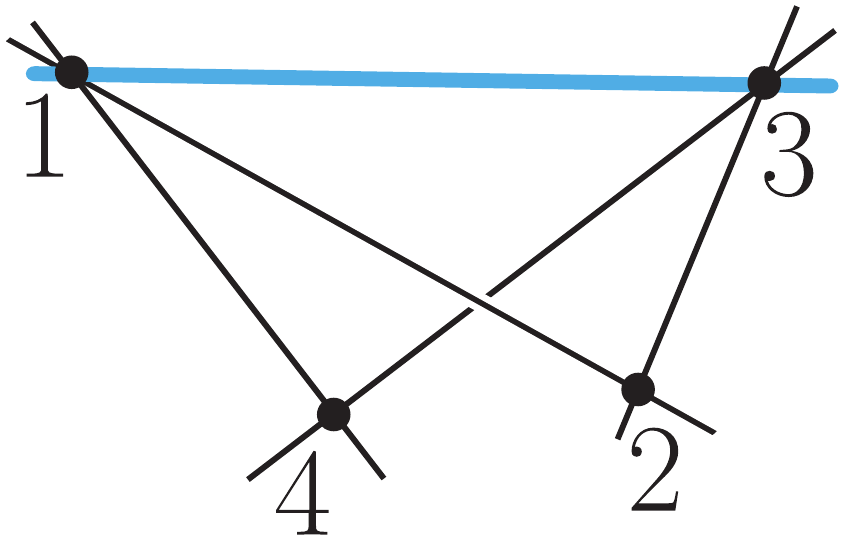}\Leftrightarrow\hspace{-0.8cm}\figBox{0}{-1.65}{0.5}{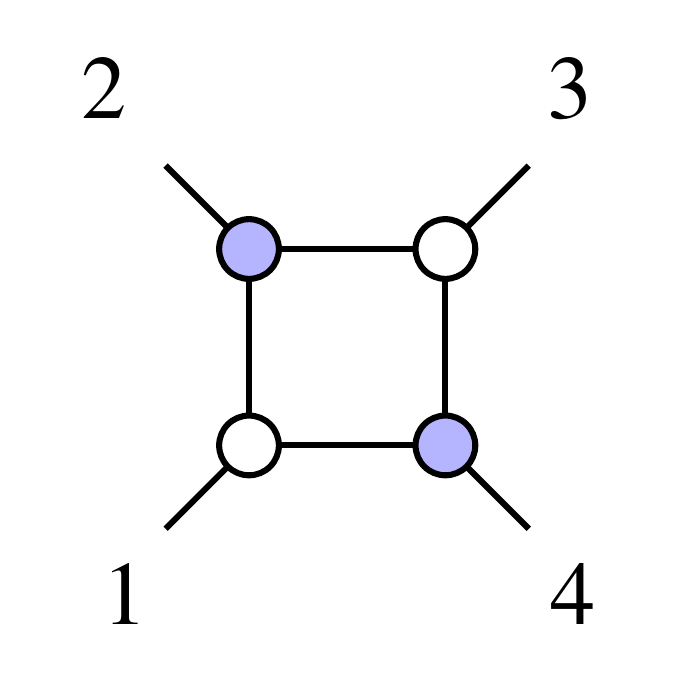}\\(AB)=(13)\end{array}}
Clearly, because $(12)\newcap(23)\supset Z_2$ and $(34)\newcap(41)\supset Z_4$, the line $(AB)=(24)$ intersects all four lines, as desired; this is drawn in red above. The same argument also applies to the second solution, the line $(AB)=(13)$, drawn in blue above. Also in this figure, we have indicated which leading-singularities have non-vanishing support on the corresponding (complex) point in the space of loop-momenta which corresponds to the particular line $(AB)$. As explained above, each three-particle MHV \mbox{($m=2$)}---colored blue in the figure above---or $\overline{\mathrm{MHV}}$ ($m=1$)---colored white---vertex of a leading singularity vanishes for every leading-singularity, and so which of the 2 three-particle amplitudes is non-vanishing for this value of the loop-momentum determines the chirality of the contour.

As a convenient way to gain some intuition about momentum-twistor geometry that will prove useful in the rest of this paper and to establish some of the notation that will be ubiquitous throughout, we will study each of the 1-loop Schubert problems in turn.\\

\noindent{\it One-Mass Schubert Problem:}

A `one-mass' 1-loop leading singularity is one for which three of the four legs are massless, and is associated with the following archetypical box-integral:
\vspace{-0.2cm}\eq{\figBox{0}{-1.5}{0.45}{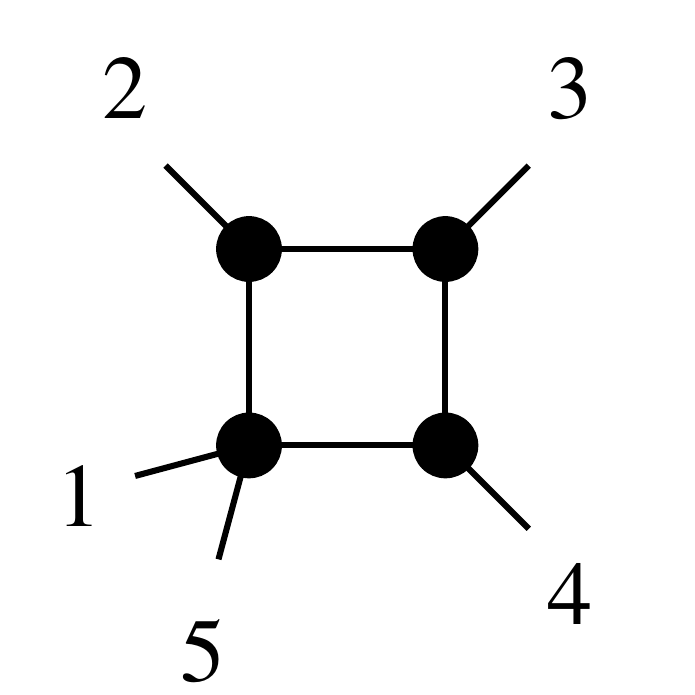}\hspace{-0.5cm}\Longleftrightarrow\;\;\;\int\limits_{AB}\frac{\ab{12\,34}\ab{23\,45}}{\ab{AB\,12}\ab{AB\,23}\ab{AB\,34}\ab{AB\,45}}\;.\vspace{-0.3cm}\label{one-mass-integral}}
In momentum-twistor space, the leading-singularities of this integral are associated with the lines $(AB)$ which intersect the four lines $(12),(23),(34),$ and $(45)$.
Considering the configuration of lines, it is not hard to find the two configurations which solve this Schubert problem:
\eq{\hspace{-0.0cm}\nonumber\begin{array}{c}\figBox{0}{-1.65}{0.5}{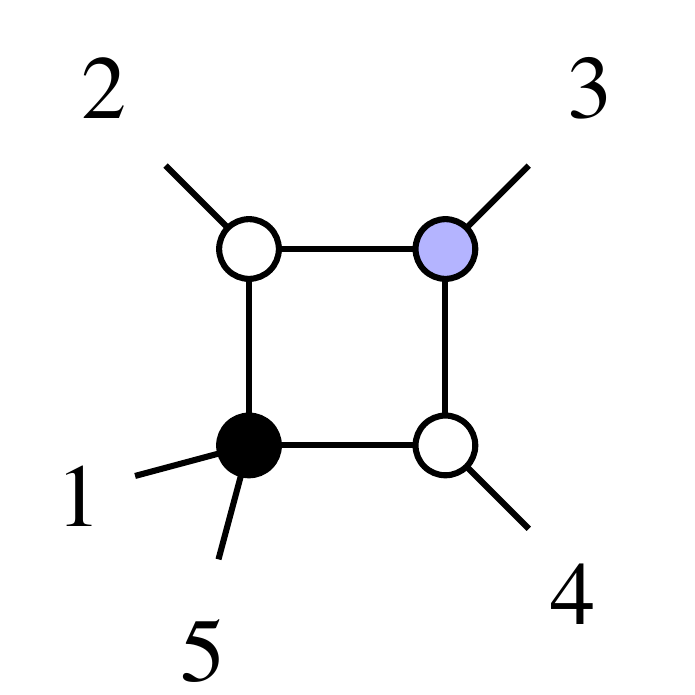}\hspace{-0.4cm}\Leftrightarrow\figBox{0}{-1.75}{0.5}{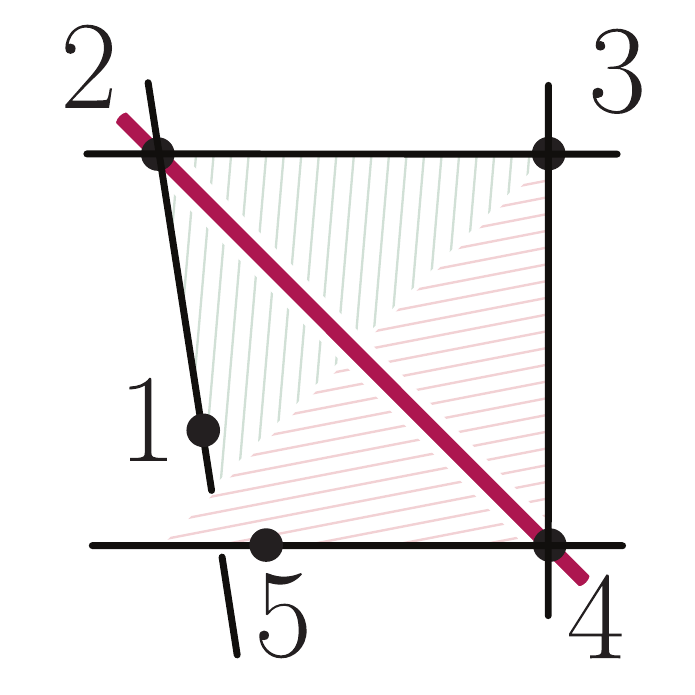}\\(AB)=(24)\end{array}\qquad\begin{array}{c}\figBox{0}{-1.75}{0.5}{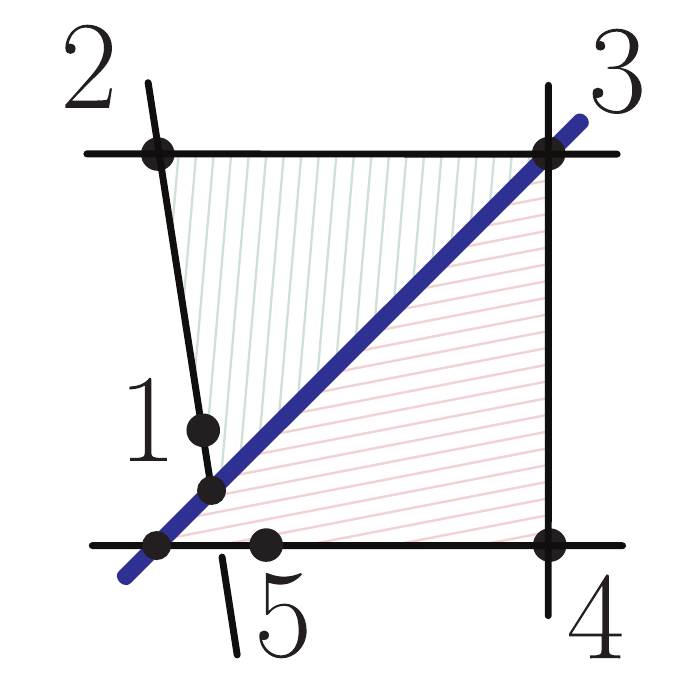}\Leftrightarrow\hspace{-0.4cm}\figBox{0}{-1.65}{0.5}{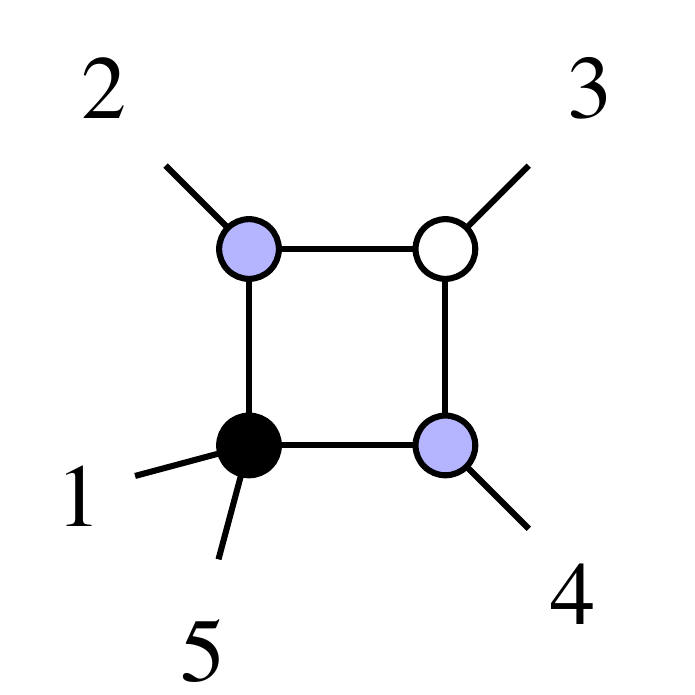}\\(AB)=(123)\newcap(345)\end{array}}
As before, because $(12)\newcap(23)\supset Z_{2}$ and $(34)\newcap(45)\supset Z_4$, the line $(AB)=(24)$ intersects all four lines. The second solution, however, is new. This solution is drawn in blue in the figure above, and represents the line of the intersection of the planes spanned by $(Z_1,Z_2,Z_3)\equiv(123)$ and $(Z_3,Z_4,Z_5)\equiv(345)$. Although geometrically clear, it is worthwhile to recall that any generic line in the plane $(123)$ will intersect the lines $(12),(23)$, and $(31)$, and any generic line in the plane $(345)$ will intersect the lines $(34),(45),$ and $(53)$. Therefore, the line $(AB)=(123)\newcap(345)$ will intersect all four lines, as required.

Similar to the case discussed in the context of the pentagon with a `wavy-line' numerator (\ref{baby_pentagon_example2}), the line $(123)\newcap(345)$ can easily be expanded in terms of ordinary bitwistors as: $(23)\ab{1\,345}+(31)\ab{2\,345}$. This follows from a more general rule which review presently.

~\\
\noindent{\it On the Intersection of Planes in Twistor-Space}

In general, the intersection of the planes $(abc)\newcap(def)$ is can be canonically expanded in either of the following ways:
\eqs{({\color{paper_red}abc})\newcap({\color{paper_blue}def})&=Z_{{\color{paper_red}a}}Z_{{\color{paper_red}b}}\ab{{\color{paper_red}c}\,{\color{paper_blue}d\,e\,f}}+Z_{{\color{paper_red}b}}Z_{{\color{paper_red}c}}\ab{{\color{paper_red}a}\,{\color{paper_blue}d\,e\,f}}+Z_{{\color{paper_red}c}}Z_{{\color{paper_red}a}}\ab{{\color{paper_red}b}\,{\color{paper_blue}d\,e\,f}};\\&=\ab{{\color{paper_red}a\,b\,c}\,{\color{paper_blue}d}}Z_{{\color{paper_blue}e}}Z_{{\color{paper_blue}f}}+\ab{{\color{paper_red}a\,b\,c}\,{\color{paper_blue}f}}Z_{{\color{paper_blue}d}}Z_{{\color{paper_blue}e}}+\ab{{\color{paper_red}a\,b\,c}\,{\color{paper_blue}e}}Z_{{\color{paper_blue}f}}Z_{{\color{paper_blue}d}}.}
Alternatively, when expanding a four-bracket of the form $\ab{xy\,(abc)\newcap(def)}$, the manifest dependence on the two planes can be preserved at the cost of breaking the manifest dependence on the line $(xy)$, as follows:
\eq{\ab{{\color{paper_green}xy}\,({\color{paper_red}abc})\newcap({\color{paper_blue}def})}=\ab{{\color{paper_green}x}\,{\color{paper_red}abc}}\ab{{\color{paper_green}y}\,{\color{paper_blue}def}}-\ab{{\color{paper_green}y}\,{\color{paper_red}abc}}\ab{{\color{paper_green}x}\,{\color{paper_blue}def}}.}

~\\
\noindent{\it Two-Mass-Easy Schubert Problem}

The two-mass-easy Schubert problem is associated with the following one-loop archetypical box-integral,
\vspace{-0.2cm}\eq{\figBox{0}{-1.5}{0.45}{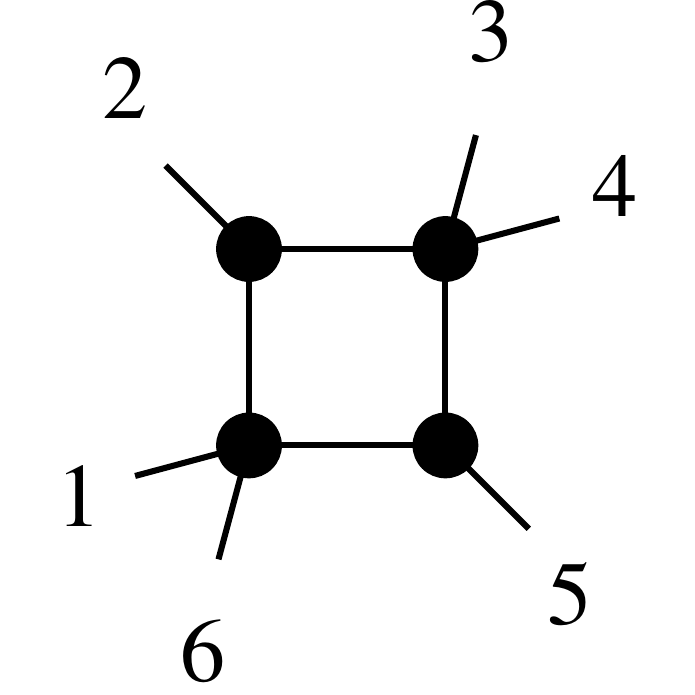}\hspace{-0.5cm}\Longleftrightarrow\;\;\;\int\limits_{AB}\frac{\ab{123\,5}\ab{2\,345}}{\ab{AB\,12}\ab{AB\,23}\ab{AB\,45}\ab{AB\,56}}\;,\vspace{-0.3cm}\label{one-mass-integral}}
which has leading singularities supported on the configuration $(AB)$ which intersect all four of the lines $(12),(23),(45),$ and $(56)$. The two solutions are essentially the same as for the one-mass Schubert problem, and are illustrated in the Figure below:
\eq{\hspace{-0.0cm}\nonumber\begin{array}{c}\figBox{0}{-1.65}{0.5}{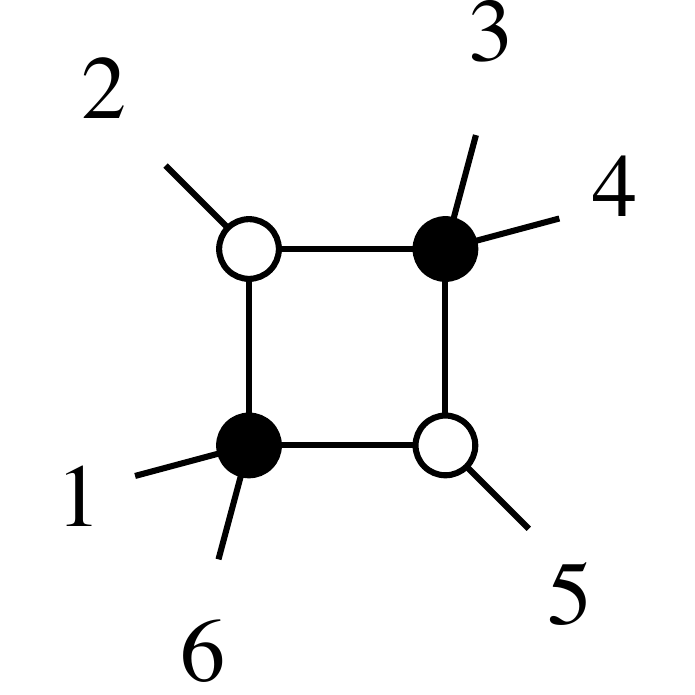}\hspace{-0.4cm}\Leftrightarrow\figBox{0}{-1.75}{0.5}{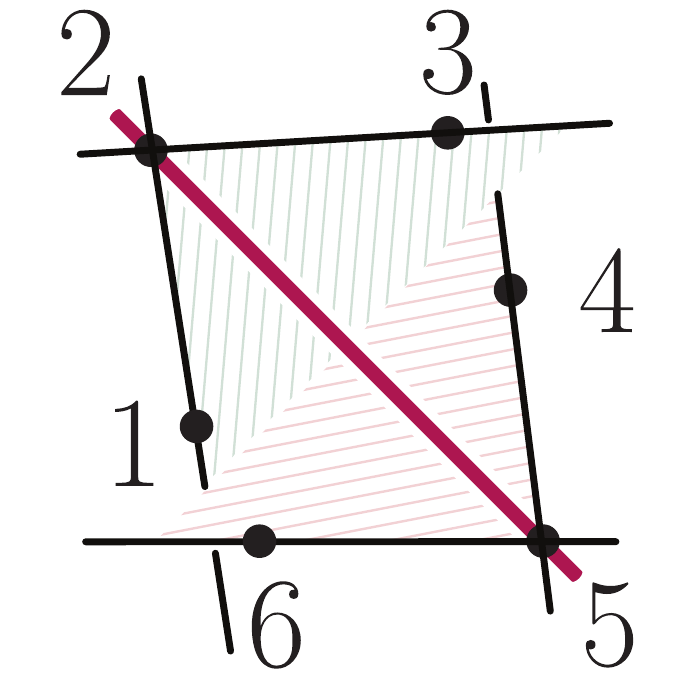}\\(AB)=(25)\end{array}\qquad\begin{array}{c}\figBox{0}{-1.75}{0.5}{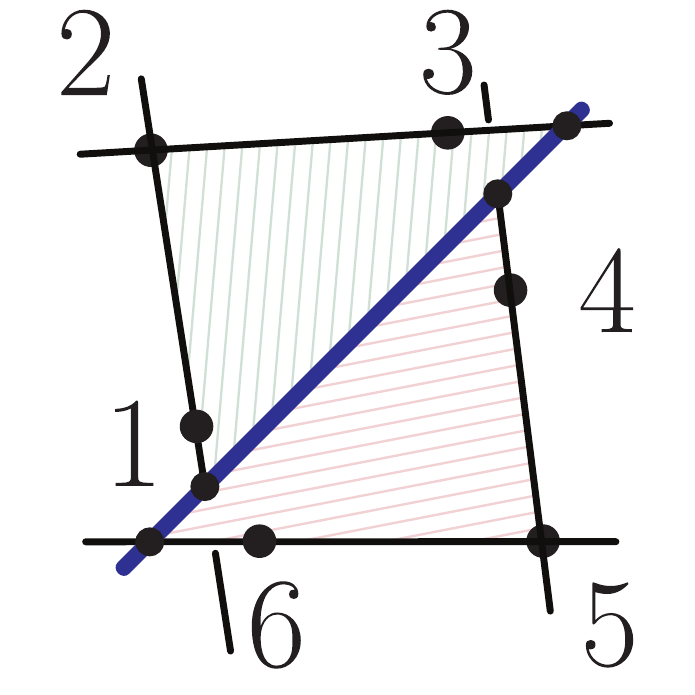}\Leftrightarrow\hspace{-0.4cm}\figBox{0}{-1.65}{0.5}{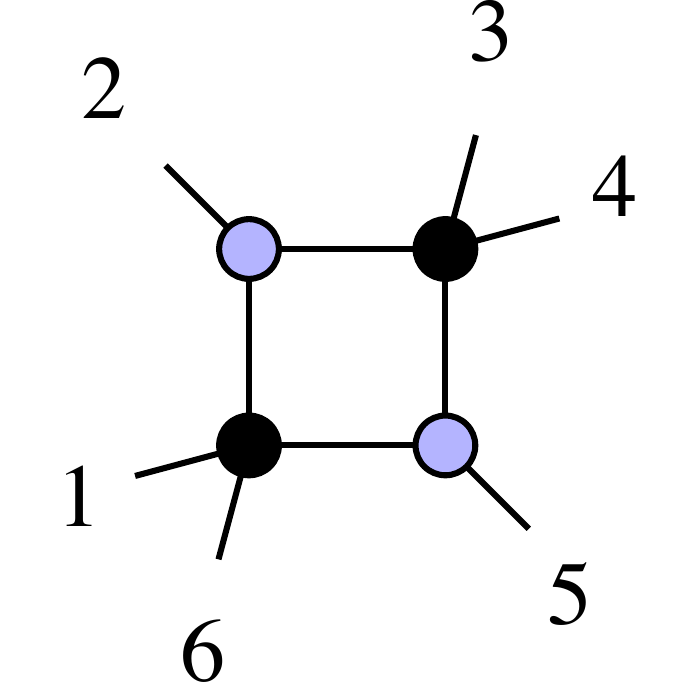}\\(AB)=(123)\newcap(456)\end{array}}
Once again, there is a very easy solution, in this case the line $(AB)=(25)$ which obviously intersects the four lines. And using the same reasoning as int the one-mass Schubert problem, it is easy to see that the second solution is simply the intersection of the planes $(123)\newcap(456)$.

~\\
\noindent{\it Two-Mass-Hard Schubert Problem}

The two-mass-hard Schubert problem differs from the two-mass easy problem in that the two massless corners are adjacent---making the Schubert problem slightly less `easy' (which at least partially justifies the name). It is associated with the following archetypical one-loop integral,
\vspace{-0.2cm}\eq{\figBox{0}{-1.5}{0.45}{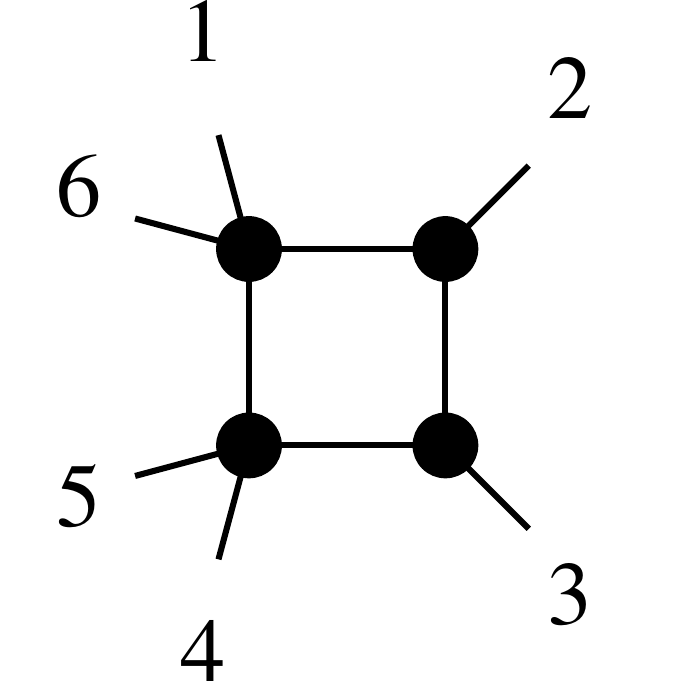}\hspace{-0.5cm}\Longleftrightarrow\;\;\;\int\limits_{AB}\frac{\ab{12\,34}\ab{23\,56}}{\ab{AB\,12}\ab{AB\,23}\ab{AB\,34}\ab{AB\,56}}\;,\vspace{-0.3cm}\label{one-mass-integral}}
and has leading singularities supported where the line $(AB)$ intersects the four lines $(12),(23),(34),$ and $(56)$. The two solutions are shown in the Figure below:
\eq{\hspace{-0.0cm}\nonumber\begin{array}{c}\figBox{0}{-1.65}{0.5}{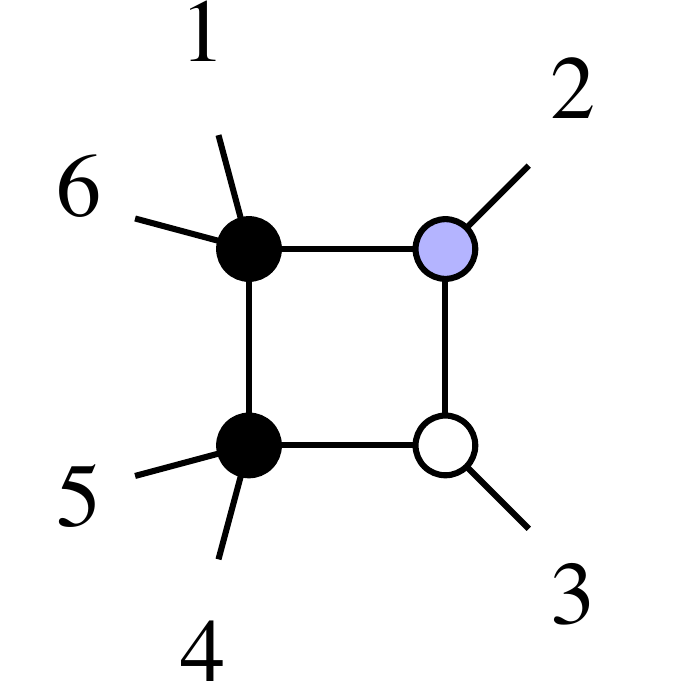}\hspace{-0.4cm}\Leftrightarrow\figBox{0}{-1.75}{0.5}{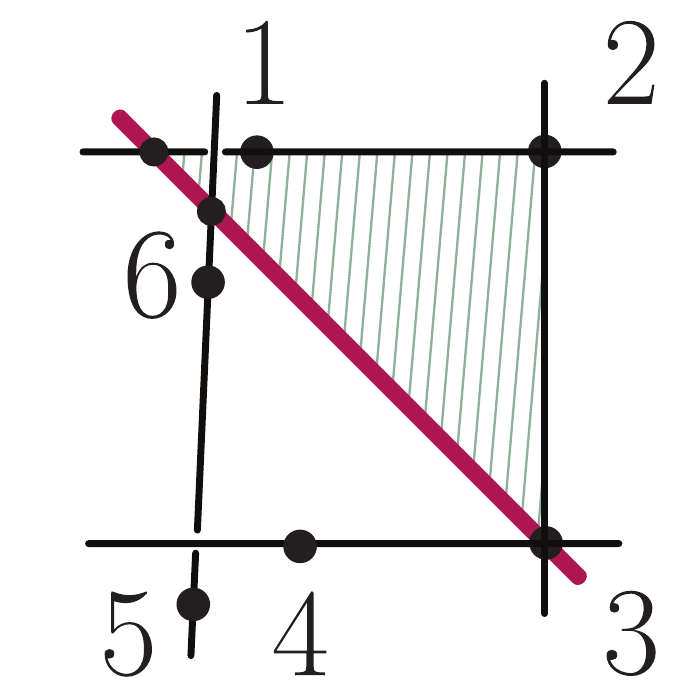}\\(AB)=(123)\newcap(356)\end{array}\qquad\begin{array}{c}\figBox{0}{-1.75}{0.5}{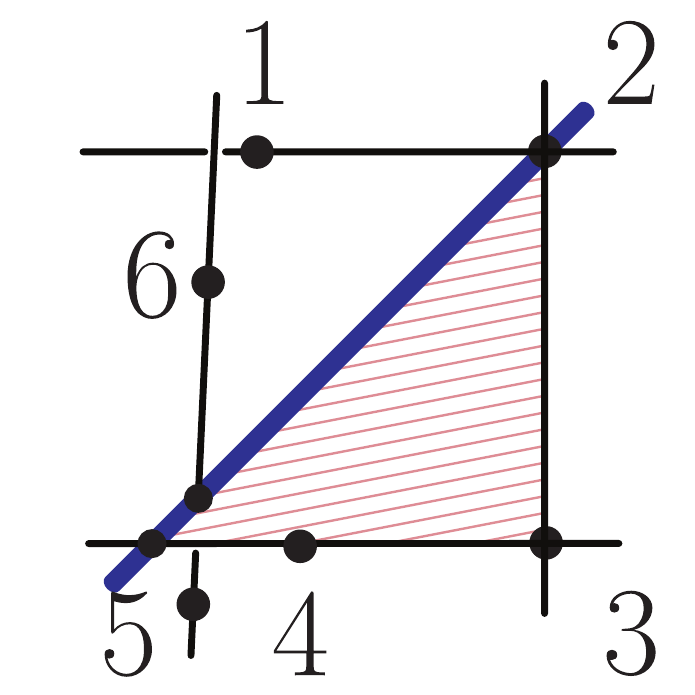}\Leftrightarrow\hspace{-0.4cm}\figBox{0}{-1.65}{0.5}{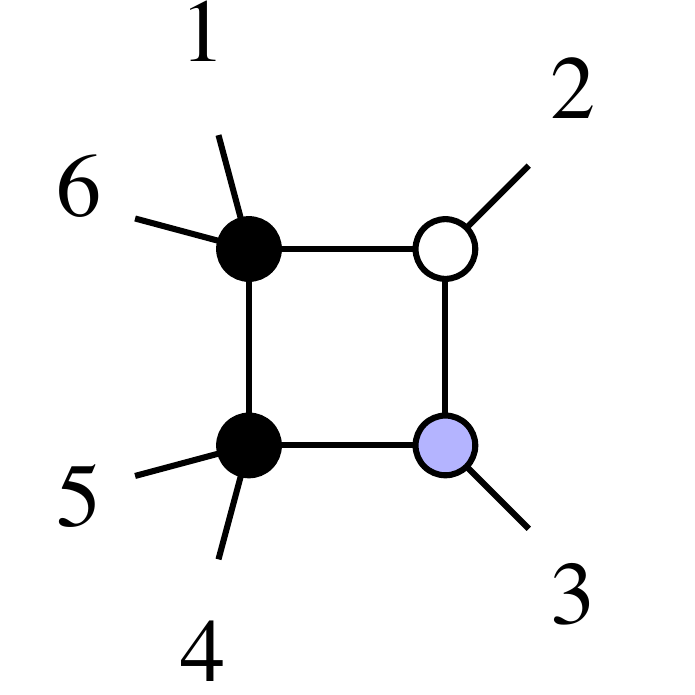}\\(AB)=(562)\newcap(234)\end{array}}
Let us briefly discuss the first of the two solutions. Here, the line $(AB)=(123)\newcap(356)$ intersects the lines $(23),(34)$ trivially because $Z_3\subset(123)\newcap(356)$, and it intersects the lines $(12)$ and $(56)$ because any generic line in the plane $(123)$ intersects $(12)$, and any generic line in the plane $(356)$ intersects $(56)$.

~\\
\noindent{\it Three-Mass Schubert Problem}

The last Schubert problem that involves a massless corner is known as the `three-mass' problem, and is associated with the following archetypical one-loop integral:
\vspace{-0.2cm}\eq{\figBox{0}{-1.5}{0.45}{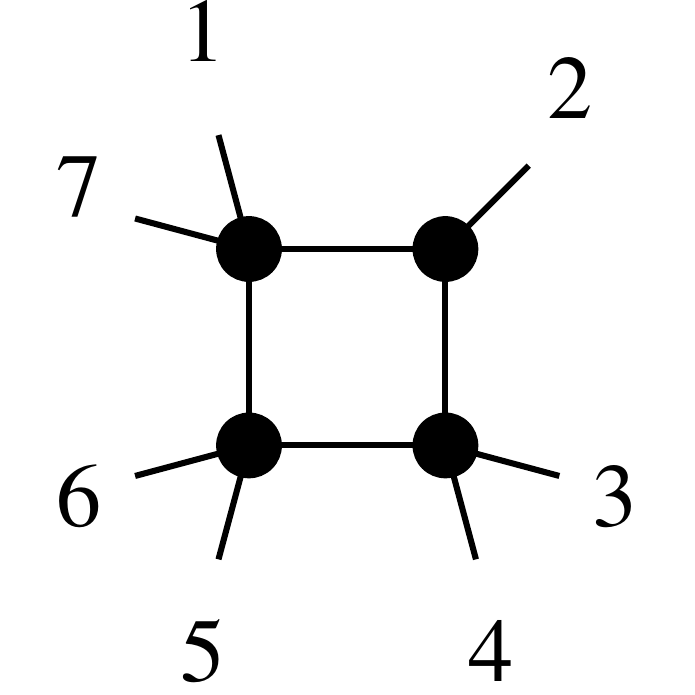}\hspace{-0.5cm}\Longleftrightarrow\;\;\;\int\limits_{AB}\frac{\ab{1\,(245)\newcap(672)\,3}}{\ab{AB\,12}\ab{AB\,23}\ab{AB\,34}\ab{AB\,45}}\;.\vspace{-0.3cm}\label{one-mass-integral}}
This integral is the most general one which involves a massless corner, and supports leadings singularities where the line $(AB)$ intersects the four lines $(12),(23),(45),$ and $(67)$. The two solutions are indicated in the Figure below.
\eq{\hspace{-0.0cm}\nonumber\begin{array}{c}\figBox{0}{-1.65}{0.5}{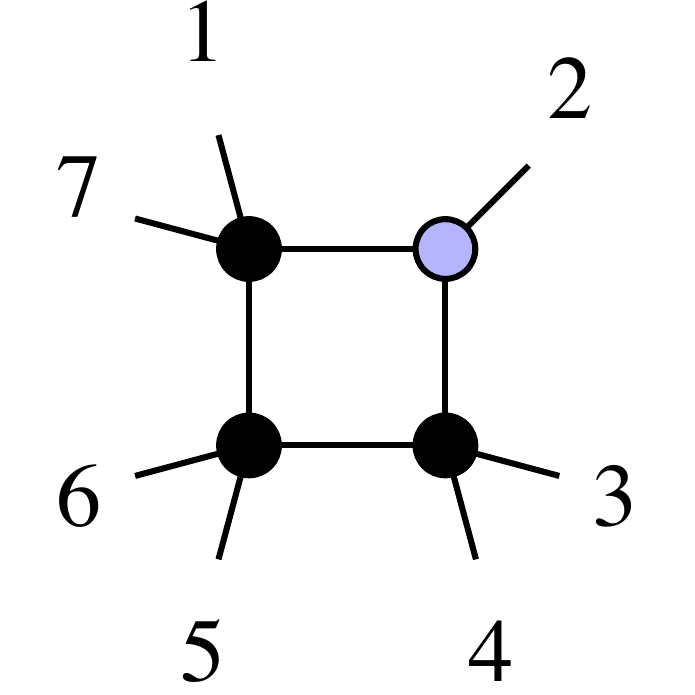}\hspace{-0.4cm}\Leftrightarrow\figBox{0}{-1.75}{0.5}{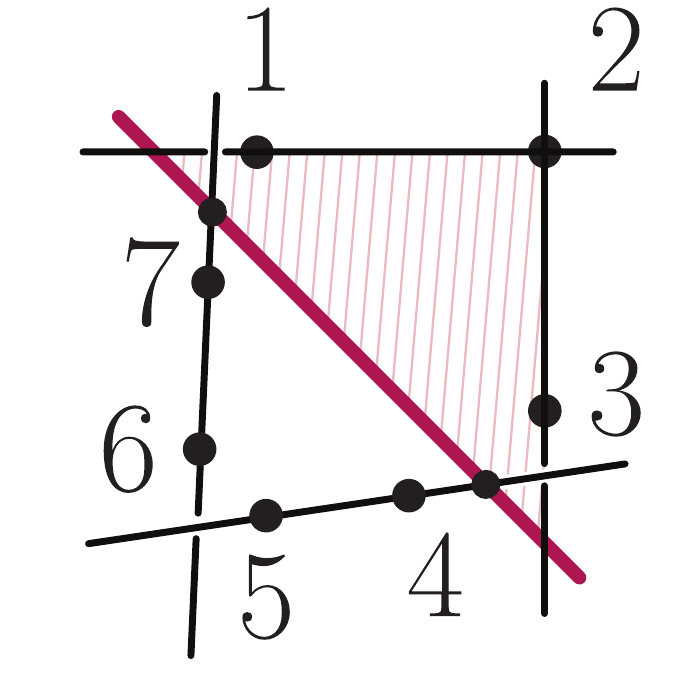}\\(AB)=\Big((123)\newcap(45),(67)\newcap(123)\Big)\end{array}\qquad\begin{array}{c}\figBox{0}{-1.75}{0.5}{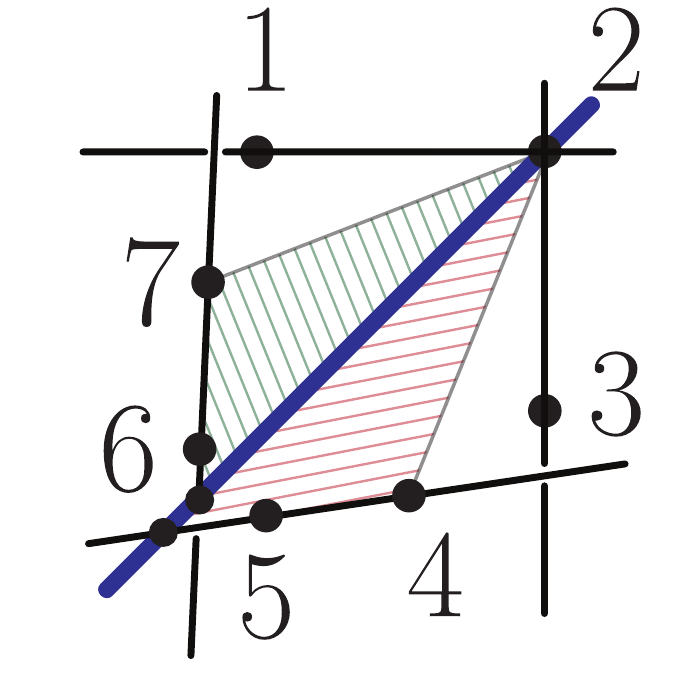}\Leftrightarrow\hspace{-0.4cm}\figBox{0}{-1.65}{0.5}{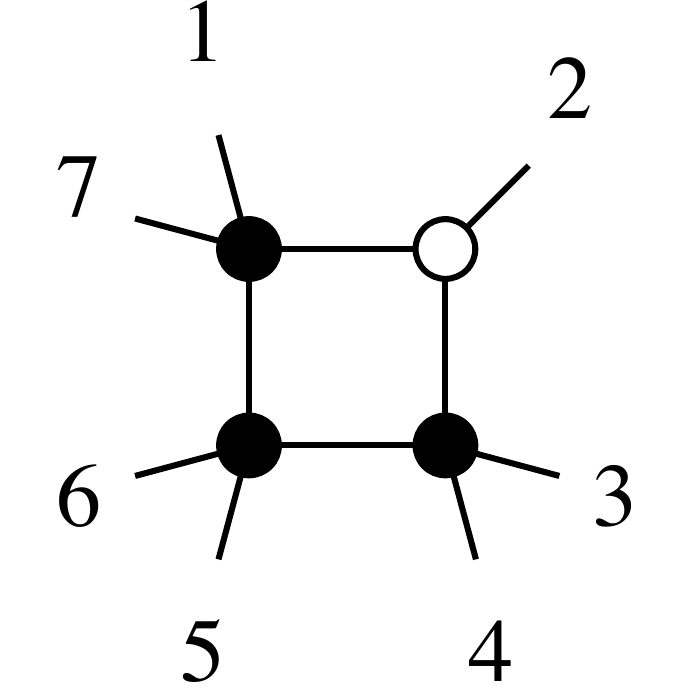}\\(AB)=(245)\newcap(672)\end{array}}
Here, the notation `$(ab)\newcap(cde)$' has been used to indicate the {\it point} in twistor-space where the line $(ab)$ intersects the plane $(cde)$. We will discuss the expansion of such geometrically-defined objects more generally at the end of this subsection; for now, let us merely quote the result: \eq{\nonumber({\color{paper_red}ab})\newcap({\color{paper_blue}cde})\equiv Z_{{\color{paper_red}a}}\ab{{\color{paper_red}b}\,{\color{paper_blue}c\,d\,e}}+Z_{{\color{paper_red}b}}\ab{{\color{paper_blue}c\,d\,e}\,{\color{paper_red}a}}=-\Big(Z_{{\color{paper_blue}c}}\ab{{\color{paper_blue}d\,e}\,{\color{paper_red}a\,b}}+Z_{{\color{paper_blue}d}}\ab{{\color{paper_blue}e}\,{\color{paper_red}a\,b}\,{\color{paper_blue}c}}+Z_{\color{paper_blue}e}\ab{{\color{paper_red}a\,b}\,{\color{paper_blue}c\,d}}\Big);} and similarly,
\eq{\nonumber({\color{paper_blue}cde})\newcap({\color{paper_red}ab})\equiv Z_{{\color{paper_blue}c}}\ab{{\color{paper_blue}d\,e}\,{\color{paper_red}a\,b}}+Z_{{\color{paper_blue}d}}\ab{{\color{paper_blue}e}\,{\color{paper_red}a\,b}\,{\color{paper_blue}c}}+Z_{\color{paper_blue}e}\ab{{\color{paper_red}a\,b}\,{\color{paper_blue}c\,d}}=-\Big( Z_{{\color{paper_red}a}}\ab{{\color{paper_red}b}\,{\color{paper_blue}c\,d\,e}}+Z_{{\color{paper_red}b}}\ab{{\color{paper_blue}c\,d\,e}\,{\color{paper_red}a}}\Big);}
so that $(ab)\newcap(cde)=-(cde)\newcap(ab)$.

~\\
\noindent{\it On Schouten-Identities and Projective Geometry}

Perhaps the single most useful identity for momentum-twistor geometry is known as `{\it the} five-term identity:' any arbitrary set of five twistors $\{Z_a,Z_b,Z_c,Z_d,Z_e\}$ will satisfy the following identity, \eq{Z_a\ab{b\,c\,d\,e}+Z_b\ab{c\,d\,e\,a}+Z_c\ab{d\,e\,a\,b}+Z_d\ab{e\,a\,b\,c}+Z_e\ab{a\,b\,c\,d}=0.\label{five_term_id}}
This identity merely reflects the general solution to a homogeneous, linear system of equations in four-variables, and as such, has analogues in any number of dimensions.  For example, in two dimensions, we have that for any $\{\lambda_a,\lambda_b,\lambda_c\}\subset\mathbb{C}^2$, there is an identity \eq{\lambda_a\ab{b\,c}+\lambda_b\ab{c\,a}+\lambda_c\ab{a\,b}=0,} where we have naturally extended the definition of `$\ab{\cdot\,\cdot}$' to be the determinant of the components of the corresponding two-vectors. This two-dimensional identity represents the general solution to a homogeneous, linear system of equations in 2 unknowns, and by contracting it with a fourth two-vector $\lambda_d$, we obtain the familiar `Schouten identity:' \eq{\ab{d\,a}\ab{b\,c}+\ab{d\,b}\ab{c\,a}+\ab{d\,c}\ab{a\,b}=0.}

This familiar identity of course has an analogue descending from equation (\ref{five_term_id}). By contracting equation (\ref{five_term_id}) with any arbitrary plane $(f\,g\,h)$, we find the following 5-term identity which we will therefore call `a Schouten identity:'\eq{\hspace{-0.35cm}\ab{f\,g\,h\,a}\ab{b\,c\,d\,e}+\ab{f\,g\,h\,b}\ab{c\,d\,e\,a}+\ab{f\,g\,h\,c}\ab{d\,e\,a\,b}+\ab{f\,g\,h\,d}\ab{e\,a\,b\,c}+\ab{f\,g\,h\,e}\ab{a\,b\,c\,d}=0.\nonumber}

In addition to being quite useful for simplifying formulae, equation (\ref{five_term_id}) can be trivially re-arranged to yield the solutions to some of the most often-encountered problems in momentum-twistor geometry:
\begin{enumerate}
\item the expansion of any arbitrary twistor $Z_a$ into a basis composed of any four linearly-independent twistors $\{Z_b,Z_c,Z_d,Z_e\}$: \eq{\nonumber Z_a\ab{b\,c\,d\,e}=-\Big(Z_b\ab{c\,d\,e\,a}+Z_c\ab{d\,e\,a\,b}+Z_d\ab{e\,a\,b\,c}+Z_e\ab{a\,b\,c\,d}\Big);}
\item the point along the line $(ab)$ which intersects the plane $(cde)$: \eq{\nonumber(ab)\newcap(cde)\equiv Z_a\ab{b\,c\,d\,e}+Z_b\ab{c\,d\,e\,a}=-\Big(Z_c\ab{d\,e\,a\,b}+Z_d\ab{e\,a\,b\,c}+Z_e\ab{a\,b\,c\,d}\Big);}
\item the point on the plane $(abc)$ which intersects the line $(de)$: \eq{\nonumber(abc)\newcap(de)\equiv Z_a\ab{b\,c\,d\,e}+Z_b\ab{c\,d\,e\,a}+Z_c\ab{d\,e\,a\,b}=-\Big(Z_d\ab{e\,a\,b\,c}+Z_e\ab{a\,b\,c\,d}\Big);}
\end{enumerate}
and so-on.

\subsubsection*{Matching All Leading Singularities}

We close this introductory section to momentum twistor integrals and leading singularities with a physical point. We have seen that the leading singularities of ${\cal N}=4$ SYM are chiral while those of scalar boxes are non-chiral. This means that if we want to construct the integrand of the theory it is impossible to do it using scalar boxes. Momentum twistors already give the solution to this problem. Since leading singularities are Yangian invariant and in particular dual conformal invariant (DCI), one should use the reduction procedure to go down to tensor pentagons and boxes and not any further. Even going down to scalar pentagons would be doing something brutal to the manifestly DCI structure of the amplitudes.

In the rest of the paper we will find that by using a special class of integrals known as {\it chiral unit leading singularity} integrals, the full integrand of scattering amplitudes can be reproduced yielding to stunningly simple forms.

\newpage

\section{Chiral Integrals with Unit Leading Singularities}\label{pure_loop_integrals_section}

Given the success of the recently introduced recursion relations
for the construction of the integrand to all orders in perturbation theory \cite{ArkaniHamed:2010kv},
it is clear that the physical integrand is the important object to obtain.

In the previous section we showed that the usual constructions of, say, one-loop amplitudes in ${\cal N}=4$ SYM as a
linear combination of scalar boxes cannot possibly be the physical integrand. Of course, the answer obtained from
scalar boxes gives the same integrals as the one originally defined from Feynman diagrams. However, as we will see, insisting in obtaining the physical integral leads to stunningly simple formulas for one and higher loop amplitudes. These new formulas are possible thanks to the use of a new suit of integrals with very special properties. These are {\it chiral integrals with unit leading singularities}.

\subsection{Integrals with Unit Leading Singularities, or {\it Pure} Integrals}

Let us start by given a definition of integrals with unit leading singularities. As we will see, it is appropriate to call these {\it pure integrals}.

Consider a particular DCI $L$-loop integral and compute all possible residues. If all non-vanishing residues are the same up to a sign then the integral can be normalized so that all residues are $\pm 1$ or $0$. When this is done, the integral is said to have {\it unit leading singularities} or to be a {\it pure integral}.

We already encountered examples of pure integrals in the previous section. The zero mass box (\ref{zeromass}), the general scalar box (\ref{BTB}) (properly normalized), and the pentagon integrals in (\ref{baby_pentagon_example1}) and (\ref{baby_pentagon_example2}).

Using the global residue theorem, we proved in section 2 that boxes are pure integrals. However, it is not obvious that the pentagons in (\ref{baby_pentagon_example1}) and (\ref{baby_pentagon_example2}) satisfy the requirement.

Consider first pentagons of the first class
\be
\int\limits_{AB} \frac{\langle AB\,13\rangle N}{\langle AB\,12\rangle\langle AB\,23\rangle\langle AB\,34\rangle\langle AB\,45\rangle\langle AB\,51\rangle}
\ee
where $N=\langle 12\,45\rangle\langle 23\,45\rangle$.

In order to see that all non-vanishing leading singularities are equal up to a sign let us use a global residue theorem. In section 2 we gave a very imprecise definition of the global residue theorem (GRT) which was enough for the purposes of that section. Here we have to be more precise. The GRT states that given a choice of a map $f:\mathbb{C}^4\to \mathbb{C}^4$ made from polynomial factors in the denominator, the sum over all the residues associated with the zeroes of the map vanishes.

In the present case, consider the map given by $f=(f_1,f_2,f_3,f_4)$ where
$$f_1=\langle AB\,12\rangle,~ f_2 = \langle AB\,23\rangle,~f_3 = \langle AB\,34\rangle,~f_4=\langle AB\,45\rangle\langle AB\,51\rangle.$$
It is easy to see that the map $f$ has four zeroes (see section 2 for more details). The GRT assures that the sum over the four residues vanishes. How can we prove that residues are equal if the GRT only gives relations among four residues?

The answer has to do with our choice of numerator. Consider the value of $\langle AB\,13\rangle$ on the four zeroes. Each zero is a line which is the solution to some Schubert problem\footnote{A Schubert problem was defined in section 2 as the projective geometry problem of finding lines that intersect four given lines which can be in special configurations called one-mass, two-mass-easy, two-mass-hard, and three-mass, as well as in generic positions which we call four-mass configurations.}. The four solutions are the lines $(24)$, $(123)\newcap (345)$, $(13)$ and $(512)\newcap (234)$ (see the end of the section or section 2 for the notation). It is a simple exercise to show that $\langle AB13\rangle$ vanishes on the second and third solutions and it is non zero on the first and fourth. This means that the GRT implies that two leading singularities are equal and opposite in sign. The first is one of the two solutions to $\langle AB\,12\rangle=\langle AB\,23\rangle=\langle AB\,34\rangle=\langle AB\,45\rangle=0$ while the fourth is one of the two solutions to $\langle AB\,12\rangle=\langle AB\,23\rangle=\langle AB\,34\rangle=\langle AB\,51\rangle=0$. Let us denote these non-vanishing residues by $r_{(12),(23),(34),(45)}$ and $r_{(12),(23),(34),(51)}$ respectively. Therefore the GRT states that
$$(0+r_{(12),(23),(34),(45)}) + (r_{(12),(23),(34),(51)}+0) = 0$$
which implies the equality of the residues up a sign. 

The pentagon integral as $10$ leading singularities. This means that more work is needed to show that it has unit leading singularity. Consider a GRT associated to the map
$$f_1=\langle AB\,12\rangle\langle AB\,51\rangle,~ f_2 = \langle AB\,23\rangle,~f_3 = \langle AB\,34\rangle,~f_4=\langle AB\,45\rangle.$$
Once again, there are four zeroes of this map. Two of them are shared with the map we constructed before, {\it i.e.},  $(24)$ and $(123)\newcap (345)$. The two new solutions are $(35)$ and $(234)\newcap (451)$. As before, the numerator vanishes on $(123)\newcap (345)$. Very nicely, it also vanishes on $(35)$. We can denote by $r_{(12),(23),(34),(45)}$ and $r_{(51),(23),(34),(45)}$ the corresponding non-zero residues. Therefore the GRT gives
$$(0+r_{(12),(23),(34),(45)}) + (r_{(51),(23),(34),(45)}+0) = 0$$
This means that the GRT sets equal the non vanishing leading singularity in $\langle AB\,51\rangle=\langle AB\,23\rangle=\langle AB\,34\rangle=\langle AB\,45\rangle=0$ with the ones we found before.

This procedure can be continued three more times by shifting the labels in the map by one. We leave it as an exercise for the reader to verify that in every case, the numerator vanishes on one solution implying that the GRT sets all non-zero leading singularities to be the same.

In order to compute the normalization and also to show how the GRT makes obvious statements that require computations to be verified, even in this trivial case, let us compute explicitly the two residues in the first GRT discussed above.

Consider the ones in the first step. In other words, let's evaluate the residue on the solution $(24)$ to the system $\langle AB\,12\rangle=\langle AB\,23\rangle=\langle AB\,34\rangle=\langle AB\,45\rangle=0$. The residue is given by
\be
\label{norma}N\frac{\langle 2413\rangle}{\langle 2451\rangle (\langle 1234\rangle\langle 2345\rangle)}
\ee
Here the terms in parenthesis are the Jacobian in the residue computation. A geometric way to see that the Jacobian has to contain the factors $\langle 1234\rangle$ and $\langle 2345\rangle$ is that on the special configurations where either one of them vanishes, the number of solutions to the Schubert problem becomes infinite. For example, consider the configuration where $\langle 1234\rangle=0$. In this case, any line on the plane $(123)$ which passes through $Z_4$ solves the Schubert problem. Using the scaling of each momentum twistor, the Jacobian must be what we found. It might be instructive to see the full computation of the jacobian using momentum twistors. This is carried out in detail in \mbox{appendix \ref{residue_appendix}}.

In order to have a properly normalized integral we require (\ref{norma}) to be equal to one. This means that $N =\langle 5124\rangle\langle 2345\rangle$ which is the factor first given in section 2 in (\ref{baby_pentagon_example1}).

Consider now the residue coming the second Schubert problem, $\langle AB\,12\rangle=\langle AB\,23\rangle=\langle AB\,34\rangle=\langle AB\,51\rangle=0$. The non-zero residue is associated with the solution $(512)\newcap (234)$. This is a one-mass Schubert problem and one explicit form of $Z_A$ and $Z_B$ was given in section 2. Let us use $Z_A=Z_2$ and $Z_B=-\langle 1234\rangle Z_5+\langle 5234\rangle Z_1$ and compute the residue. The Jacobian is the same as before but with labels shifted back by one. The residue is then
\be
 N\frac{\langle 1234\rangle \langle 2513\rangle}{\langle 2345\rangle \langle 5124\rangle (\langle 1234\rangle \langle 2513\rangle)}.
\ee
Using the normalization derived above this quantity equals one as expected.

In section 2 we also presented a second pentagon integral which differs from the first one only in the choice of numerator. We leave it as an exercise for the reader to repeat the analysis done here and show that with the new numerator this is a pure integral\footnote{Of course, one could simply translate the whole problem into dual momentum twistor space to find exactly the same integral as before. However, it is still an instructive exercise to do it in momentum twistor space.}. Let us rewrite the integral here with the numerator given in geometric form
\be
\int\limits_{AB} \tilde{N}\frac{\langle AB\, (512)\newcap (234)\rangle}{\langle AB\,12\rangle\langle AB\,23\rangle\langle AB\,34\rangle\langle AB\,45\rangle\langle AB\,51\rangle}.
\ee
Now it should be obvious that the comment made in section 2 is true. The special numerators are made from lines, $(13)$ and $(512)\newcap (234)$, which are the two solutions to a Schubert problem.

In section 4 we study a less trivial example; a hexagon integral where the special choice of numerator also allows the use of the GRT to show that all non-vanishing residues are equal. In the hexagon case, checking the statement that all residues are equal algebraically requires many applications of 4-bracket Schouten identities.

\subsubsection*{Basic Diagrammatic Notation}

We find it convenient to introduce a diagrammatic representation for numerators. Note that with our definition of dual variables $p_a=x_a-x_{a-1}$ and of momentum twistors $x_a\leftrightarrow (Z_a,Z_{a+1})$, there is a natural diagrammatic relation between loop integrals and momentum twistor configurations. Consider a general one-loop amplitude as a polygon with $n$-sides. Attached to each vertex there is some momentum $p_a$. In momentum twistor space, we also have an $n$-sided polygon and attached to each vertex there is a momentum twisor $Z_a$. Following the intuitive correspondence between the two diagrams we are led to denote denominators (propagators) as lines connecting points depending on their geometric configuration. These are denoted by solid lines. In order to distinguish numerators, we also introduce dashed and wavy lines.

{\it Dashed lines:} Numerators which correspond to factors of the form $\ab{AB\,e\,f}$, where $(ef)$ represents a line in momentum twistor space specified by two momentum twistors $Z_e$ and $Z_f$ is represented by a dashed line connecting points $e$ and $f$ as in
\eq{\figBox{0}{-1.4}{0.5}{five_point_one_loop_ratio_1.pdf}=\int\limits_{AB}\frac{\ab{AB\,1\,3}\ab{1245}\ab{2345}}{\ab{AB\,12}\ab{AB\,23}\ab{AB\,34}\ab{AB\,45}\ab{AB\,51}}\label{pent5pt}}

{\it Wavy lines:} We also allow points to represent dual twistors. In this case the second class of numerators constructed as intersection of planes can also be represented by a line connecting two points. In order to distinguish this from the previous case we use wavy lines. In the example where the numerator corresponds to the line $(512)\newcap (234)$ or in dual twistors terminology to the point $(13)_W$, one has
\eq{\figBox{0}{-1.4}{0.5}{five_point_one_loop_ratio_2.pdf}=\int\limits_{AB}\frac{\ab{AB\,(512)\newcap (234)}\ab{1345}}{\ab{AB\,12}\ab{AB\,23}\ab{AB\,34}\ab{AB\,45}\ab{AB\,51}}\label{pent5ptwavy}}

\subsection{Chiral Integrals}

From the discussion of the pentagons, it is clear that there is a striking difference between a pentagon with a special numerator and plain scalar box integrals. Even though both kind of objects can be made pure integrals, each Schubert problem in the case of the pentagon has a single non-vanishing residue while in the boxes both solutions give rise to a residue.

When an integral has the property that the residues associated to at least one of its Schubert problems are not the same, we say that the integral is {\it chiral}. The reason for the terminology comes from the fact that the two contours associated to a given Schubert problem are exchanged under parity (see section 2 for more details). This means that one can have chiral, pure, or chiral and pure integrals.

At one-loop, one can have an even more especial class of integrals. When an integral has a numerator where at most one of the solutions to each Schubert problem gives a non-zero residue then we say that the integral is {\it completely chiral}.

Let us give two more examples in this section. The first is the most general class of chiral pure pentagon integrals. This is an integral where only two of the five legs needs to be massless. Moreover, it is clear that in order to write a special numerator the two massless legs cannot be adjacent. The claim is that the following family of integrals is (completely) chiral and pure.
\eq{\figBox{-1.25}{-1.5}{0.5}{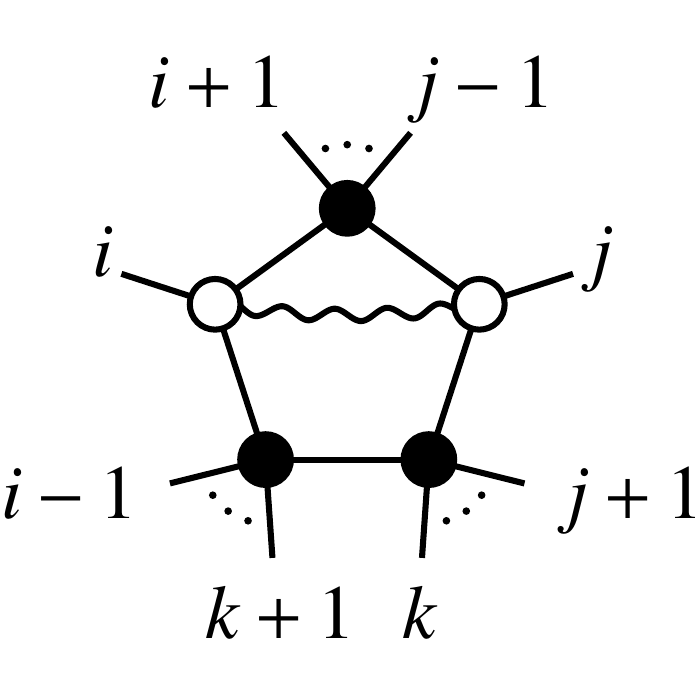}=\int\limits_{AB}
\frac{\ab{AB\,(i\mi1\,i\,i\pl1)\newcap(j\mi1\,j\,j\pl1)}\ab{i\,j\,k\,k\pl1}}{\ab{AB\,i\mi1\,i}\ab{AB\,i\,i\pl1}\ab{AB\,j\mi1\,j}\ab{AB\,j\,j\pl1}\ab{AB\,k\,k\pl1}}\label{Allpent}\hspace{-1cm}}
In this case, the GRT can also be applied to show that all residues are the same. In order to show that the normalization gives unit leading singularities, identities of the form discussed at the end of this section are needed.

Next, let us give a six-point two-loop example. Consider the following integral
\eq{\nonumber\hspace{-1cm}\figBox{0}{-1.67}{0.5}{two_loop_mhv_integrand.pdf}=\left\{\begin{array}{l} \displaystyle\frac{\ab{AB\,(i\mi1\,i\,i\pl1)\newcap(j\mi1\,j\,j\pl1)}\ab{i\,j\,k\,l}}{\ab{AB\,i\mi1\, i}\ab{AB\,i\,i\pl1}\ab{AB\,j\mi1\,j}\ab{AB\,j\,j\pl1}\ab{AB\,CD}}\\\displaystyle\times\frac{\ab{CD\,(k\mi1\,k\,k\pl1)\newcap(l\mi1\,l\,l\pl1)}}
{\ab{CD\,k\mi1k}\ab{CD\,k\,k\pl1}\ab{CD\,l\mi1\,l}\ab{CD\,l\,l\pl1}}\end{array}\right\}\label{MHVInt}}

This integral has the structure of two of the general pentagon integrals joined by the all massive edge. Consider a residue of the full integral over $\mathbb{C}^8$ which computes a residue of the pentagon on the left. The contour integral in $Z_A$ and $Z_B$ is the same as before except that the normalization is different and therefore the residue is not equal to one. The residue must then be the ration of the two normalizations, {\it i.e.}, $\ab{i\,j\,k\,l}/\ab{i\,j\,CD}$. Plugging this in the integral over $Z_C$ and $Z_D$ we now find a properly normalized integral and therefore the remaining part of residue computation gives one.

One might be tempted at this point to think that all completely chiral integrals are pure. In section 4, we describe in detail the example of a hexagon with a wavy line and a dashed line in the numerator. This integral is in fact completely chiral but it is not pure.

\subsection{Evaluation of Pure Integrals}

Evaluating integrals explicitly can be very hard and many techniques have been developed for this purpose. At one-loop, all integrals appearing in the standard reduction techniques are known analytically. At higher loops, very few examples have been evaluated analytically. Many of our chiral pure integrals turn out to be completely IR finite and therefore their evaluation can be made directly four dimensions without any regulators.  

Consider the family of pentagon integrals discussed above. The evaluation of the integrals for generic $j$ and $k$ gives
\eqs{I_5(i,j,k)=&\int\limits_{AB}\frac{\ab{AB\,(i\mi1\,i\,i\pl1)\newcap(j\mi1\,j\,j\pl1)}\ab{i\,j\,k\,k\pl1}}{\ab{AB\,i\mi1\,i}\ab{AB\,i\,i\pl1}\ab{AB\,j\mi1\,j}\ab{AB\,j\,j\pl1}\ab{AB\,k\,k\pl1}},\\=&\phantom{\,+\,}\log\left(u_{j,k,i-1,j-1}\right) \log\left(u_{k,i-1,i,j}\right)+\Li\left(1-u_{j,k,i-1,j-1}\right) +\Li\left(1-u_{k,,i-1,i,j}\right)\\& -\Li\left(1-u_{j,k,i,j-1}\right) -\Li\left(1-u_{i,j-1,k,i-1}\right) +\Li\left(1-u_{i,j-1,j,i-1}\right) \label{AllpentaEval}}
where
\eq{u_{i,j,k,l}\equiv\frac{\ab{i\,i\pl1\,j\,j\pl1}\ab{k\,k\pl1\,l\,l\pl1}}{\ab{l\,l\pl1\,j\,j\pl1}\ab{k\,k\pl1\,i\,i\pl1}}.}
For special values of $j$ and $k$ the integral becomes IR divergent and a regulator is needed. We postpone this discussion to section 4.

The reason for presenting the explicit form of the pentagon integrals is to note a general fact about pure integrals: The explicit evaluation of the integrals must be a linear combination of functions known as iterated integrals, such as polylogarithms, all with coefficient one. 

It is striking that the coefficients do not depend on kinematic invariants but this is a consequence of having unit leading singularities. This is the motivation for the terminology: pure integrals. Roughly speaking, the coefficients of the different polylogarithms are the leading singularities of the integrals. Having a pure integral ensures that no coefficient can depend on kinematical invariants.

Once again, the hexagon with a wavy and a dashed line in the numerator given in section 4 will be an example of a completely chiral and IR finite integral which is not pure and its evaluation gives products of logarithms with different coefficients that depend on kinematic invariants.

\subsection{Example: One-Loop MHV Amplitude}

Up until know we have been studying integrals individually. This is a good point to actually use them to determine the full physical integral of the simplest set of amplitudes. These are one-loop MHV amplitudes. Historically, one-loop MHV amplitudes were the very first set of amplitudes to be computed for all $n$ as a linear combination of scalar box integrals \cite{Bern:1994zx}. It was found that the answer is very simple; an overall prefactor, proportional to the tree-level amplitude, and a sum over all one-mass and two-mass-easy box integrals with coefficient one, when properly normalized. In our modern terminology, the normalization was such that only pure integrals appear. It was realized that this form of the amplitude was not equivalent to the Feynman diagram amplitude as an expansion in the dimensional regularization parameter but it differs from it only at ${\cal O}(\epsilon)$. In our language this is nothing but the fact that an expansion in terms of box integrals cannot possible reproduce the physical integrand of the theory as stressed a number of times already.

Now that we have a set of chiral pure integrals, the natural question is how much more complicated the amplitude will look like if written in a form that matches the physical integrand. It turns out that the full integrand is stunningly simple
\vspace{-0.3cm}\eq{\hspace{-1cm}\mathcal{A}_{\mathrm{MHV}}^{\mathrm{1-loop}}=
\scalebox{2}{$\displaystyle\sum_{\text{{\tiny{$i\!<\!j$}}}}$}\figBox{0}{-1.5}{0.5}{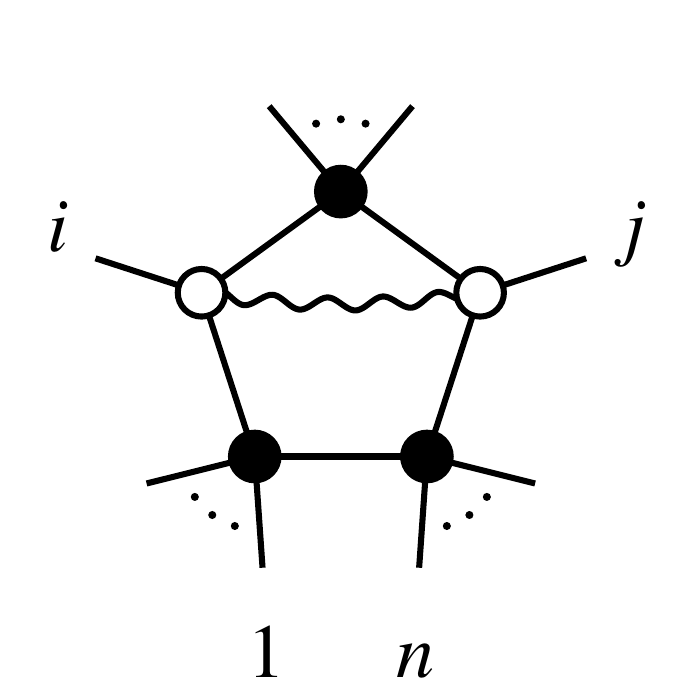}
\label{MHV1loop}\vspace{-0.1cm}}

\noindent where the propagator $\ab{AB\, n1}$ is present in all terms. Note that not all integrals in the sum are chiral pure integrals. There are boundary terms which are box integrals. Consider for example $j=i+1$. In this case the
numerator cancels one of the propagators leaving us with the box. We give no derivation for this formula here and postpone a more detailed discussion to section 6. A final comment, even though the line $(n1)$ seems especial, the amplitude is cyclic as it should be!

\section{Finite Integrals}\label{finite_integrals_section}

We have seen that the chiral integrals with unit leading
singularities, naturally written in momentum-twistor space, provide a
natural basis of objects to express the loop integrand.
In this section we will see that they have another beautiful
property---most such integrals are manifestly infrared finite.

Let us begin by illustrating with a simple example. Consider a general 1-loop integral for 6 particles, which we can write as
\be
\int\limits_{AB} \frac{\langle AB\,X \rangle \langle AB\,Y \rangle}{\langle
AB\,12 \rangle \langle AB\,23 \rangle \cdots \langle AB\,61 \rangle}.
\ee
Here $X,Y$ are generic bitwistors. Of course, like almost all generic
integrals with massless external legs, this integral is infrared
divergent.
Recall that the infrared divergences arise when the loop momentum
$l$ become collinear to a massless external momentum $p_a$, {\it i.e.}\ when
$l \cdot p_a \to 0$. The extra soft logarithmic divergence can
be thought of as an even more special case of this situation, where
the loop momentum becomes collinear to two consecutive momenta so that
$l \cdot p_a, l \cdot p_{a+1} \to 0$.
In the dual co-ordinate space, the collinear divergence arises when
the loop integration point $x$ approaches one of the  edges of the
Wilson loop, connecting
$x_a$ with $x_{a+1}$, and of course the extra soft divergence occurs
when $x$ gets close to both the lines $(x_{a-1}\, x_a)$ as well as $(x_a\,
x_{a+1})$, that is when it is close to the point $x_a$ itself. But
again the IR divergence is fundamentally a collinear one, with the
soft divergence being thought of as ``double-collinear".

We can finally describe these IR divergent regions in momentum-twistor
language. The collinear divergence associated with $l \cdot p_a \to 0$
corresponds to the region where the line $(AB)$ in momentum twistor space, associated to the loop integration point, passes through
$Z_a$ while lying the in the plane $(Z_{a-1} Z_a Z_{a+1})$. Note that
this region is quite nicely parity invariant. Recall that in
momentum-twistor variables, parity is just the poincare duality, and
exchanges the point $Z^I_a$ with the plane $W_{a I} = (Z_{a-1} Z_a
Z_{a+1})_I$ naturally paired with $Z_a$. Thus, the condition is that
the line $(AB)^{IJ}$ passes through $Z^I_a$, and also that the dual
line $(AB)_{IJ}$ passes through $W_{aI}$.

While a generic integral will indeed be IR-divergent, we see a simple
way of getting completely IR finite integrals. If the bitwistors
$X,Y$ are chosen to have a zero in all the dangerous IR divergent
configurations, then the integrals will be finite. This is very simple
to achieve. For instance, let us choose $X=(13)$ and $Y=(46)$; we can
write out the integral again as,
\be
\figBox{0}{-1.65}{0.5}{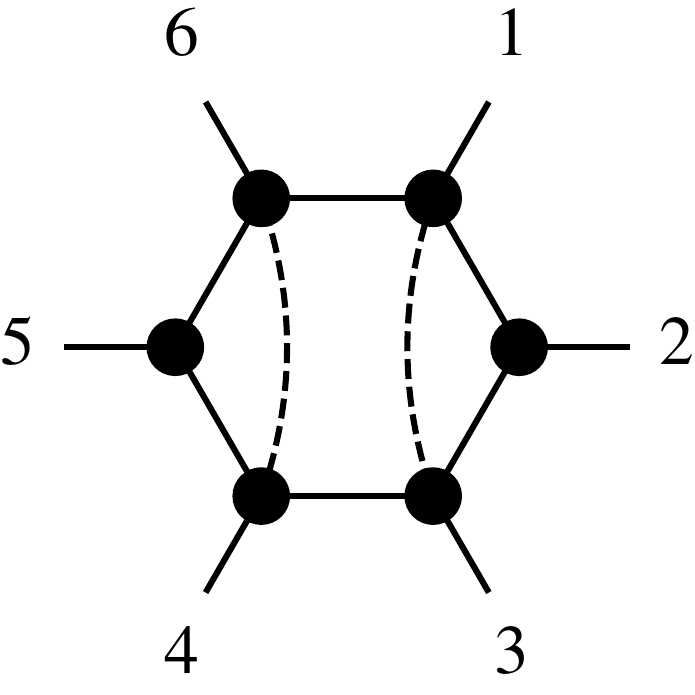}=\int\limits_{AB} \frac{\langle AB\,13 \rangle \langle AB\,46
\rangle\ab{5\,6\,1\,2}\ab{2\,3\,4\,5}}{\langle AB\,12 \rangle \langle AB\,23 \rangle\ab{AB\,34}\ab{AB\,45}\ab{AB\,56}\langle AB\,61 \rangle}
\label{example_hexagon}
\ee
Let us check that the numerator has a zero in all the IR-divergent
regions. Consider first collinearity with $p_3$. We need to see what
the numerator does when $AB$ passes through $Z_3$ while lying in the
plane $(234)$. However, the numerator factor $\langle AB\,1 3 \rangle$
vanishes simply if $AB$ passes through $1$ or $3$, regardless of
whether or not it also happens to lie in the plane $(234)$. In this way,
we can see that the collinear divergences with $1,3,4,6$ are all
killed by the numerator. Next, consider what happens when $AB$ passes
through $2$, lying in the plane $(123)$. Since $AB$ lies in $(123)$,
it necessarily intersects the line $(13)$, and therefore, $\langle AB\,
13 \rangle$ = 0, regardless of whether or not $(AB)$ also happens to
pass through $2$. A completely analogous argument holds for the
collinear divergence associated with particle $5$.

Thus we see that with this numerator, {\it all} the regions with collinear divergences
are killed by the numerator factors, and the integral is completely IR-finite! There are other choices for $X,Y$ that will do the same job;
our argument above also holds if one or both of the numerator factors $(13),(46)$ were replaced by their parity-conjugates, $(612)\newcap(234)$ and $(345)\newcap(561)$, respectively---changing one or more of the dashed-lines in (\ref{example_hexagon}) to wavy-lines.

Now, these finite integrals are clearly chiral. And when the two numerators are of the same kind, 
they have, quite nicely and
non-trivially, unit leading singularities. As usual,
verifying by direct computation requires manipulating non-trivial sequences of
4-bracket Schouten identities, but the result follows much more
transparently from an application of the global residue theorem to
this integral. Consider for instance the GRT following from choosing
$f_1 = \langle AB\,34 \rangle, f_2 = \langle AB\,45 \rangle, f_3 =
\langle AB\,56 \rangle$ and $f_4 = \langle AB\, 61 \rangle \langle AB\,12
\rangle \langle AB\,23 \rangle$. We have three different Schubert
problems to consider, with the lines $(34),(45),(56)$ combined with
$(61), (12), (23)$. Consider first the Schubert problem with the four
lines $(34),(45),(56),(61)$. This is a one-mass configuration, and it is
easy to see that the numerator kills the solution where $(AB)$ is the line $(46)$, 
only leaving the solution passing through $5$. Let us
call this non-vanishing residue $r_{(34),(45),(56),(61)}$. Similarly,
for the Schubert problem with lines $(34),(45),(56)$ and $(12)$, the
numerator kills the solution passing through $4$ while leaving the one
passing through $5$; we can call this single non-vanishing residue
$r_{(34),(45),(56),(12)}$. Finally, for the Schubert problem with
lines $(34),(45),(56),(23)$, we can see that {\it both} solutions---the
line $35$ as well the line passing through $4$---are killed by the
numerator, so both of these residues vanish.
The GRT then tells us that
\begin{eqnarray}
\left( 0 + r_{(34),(45),(56),(61)} \right) + \left(0 +
r_{(34),(45),(56),(12)}\right) + (0 + 0) = 0 \nonumber \\
\rightarrow r_{(34),(45),(56),(61)} = -  r_{(34),(45),(56),(12)}
\end{eqnarray}

It is possible to repeat this argument for other GRT's, finding a
sequence of 2-term identities relating {\it all} the non-vanishing
residues, showing that the integral is not only chiral but has {\it
unit} leading singularities. Thus, we see in this instance something that can be checked also to be true for all other
residues: the integral is completely chiral; at
most one of the two solutions to each Schubert problem are
non-vanishing, and sometimes both vanish.

Given that this integral has unit leading singularities, it is
instructive to expand it in terms of boxes, which will then also have
unit coefficients. This simple, finite momentum-twistor integral in
fact expands into the sum of nine boxes:
\eq{\nonumber\hspace{-1.25cm}\figBox{0}{-1.65}{0.5}{hexagon_integral_box_decomposition_0.pdf}=
\left\{\raisebox{-0.2cm}{$\begin{array}{lll}\phantom{\,+\,}\!\figBox{0}{-1.65}{0.5}{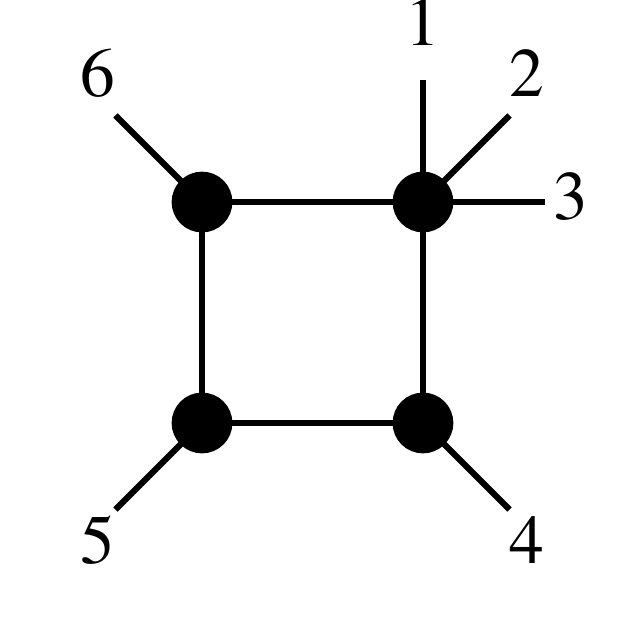}+
\figBox{0}{-1.65}{0.5}{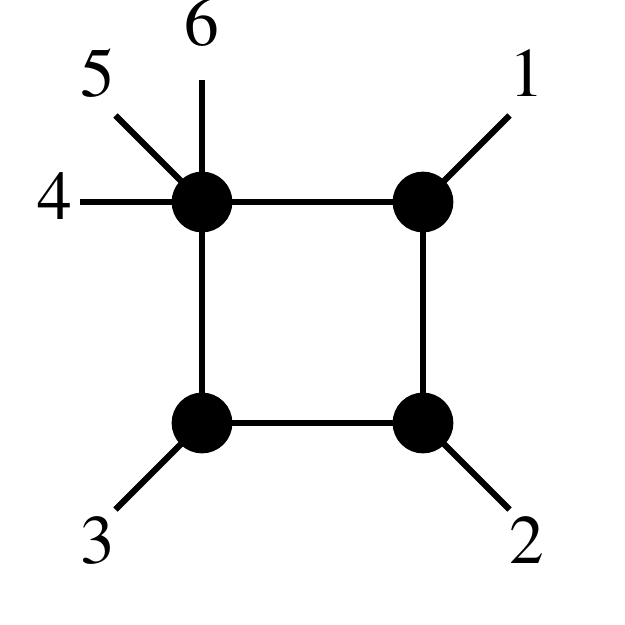}+\figBox{0}{-1.65}{0.5}{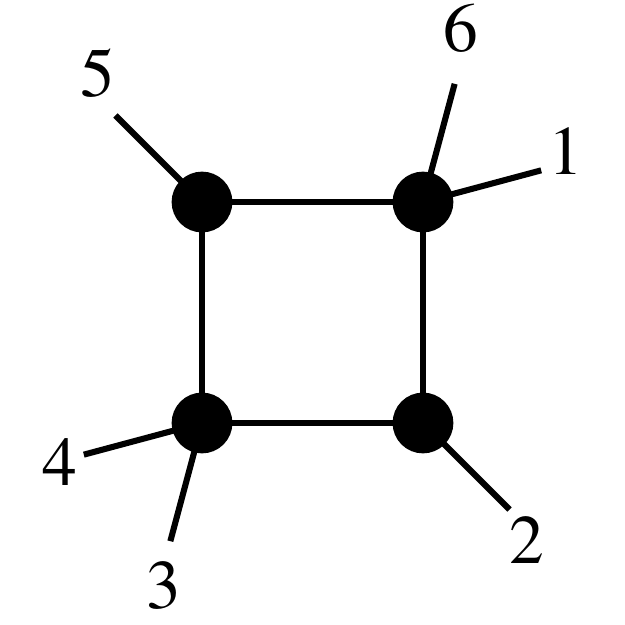}\\+
\figBox{0}{-1.65}{0.5}{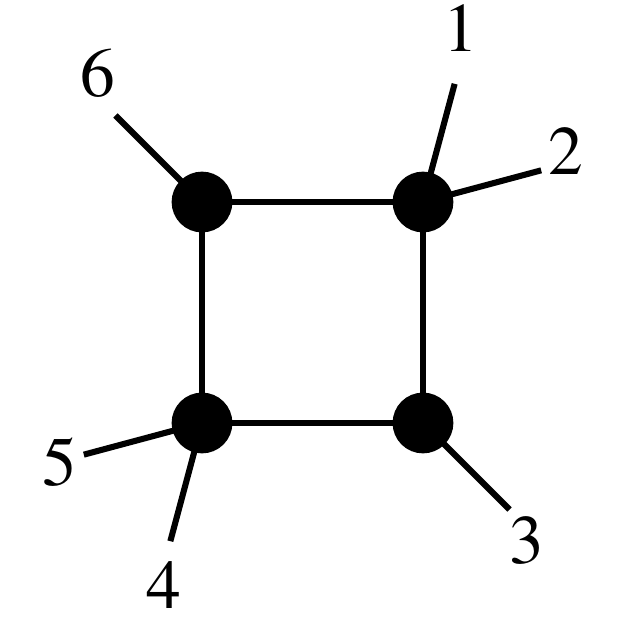}+\figBox{0}{-1.65}{0.5}{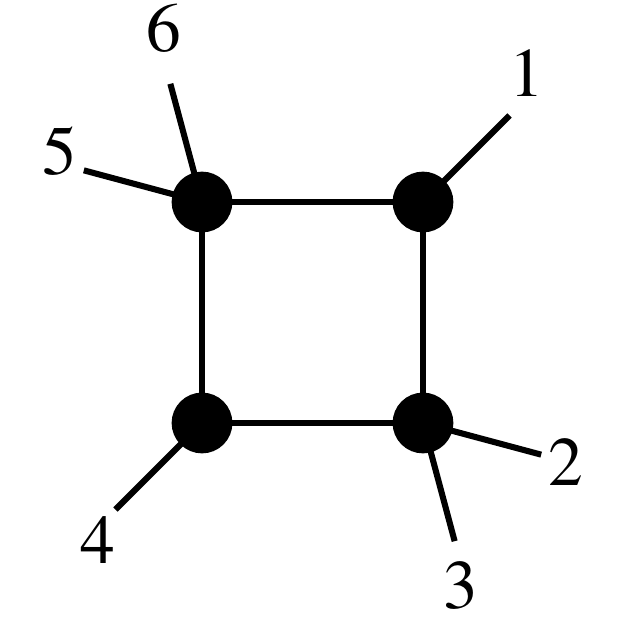}-
\figBox{0}{-1.65}{0.5}{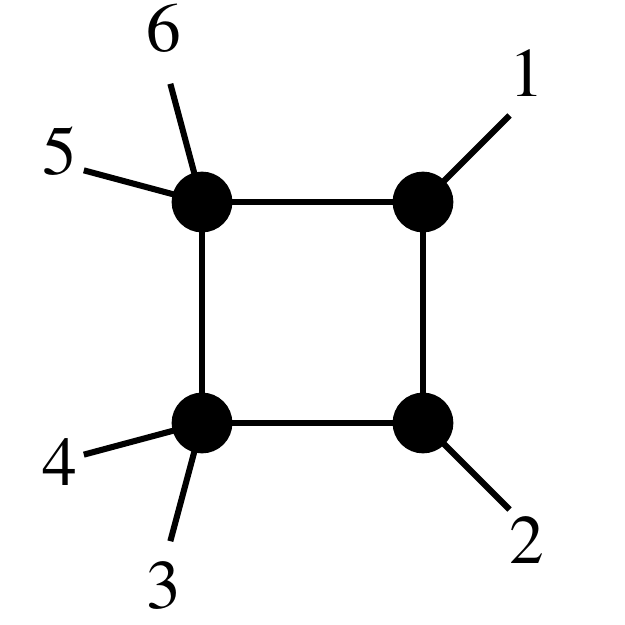}\\-\figBox{0}{-1.65}{0.5}{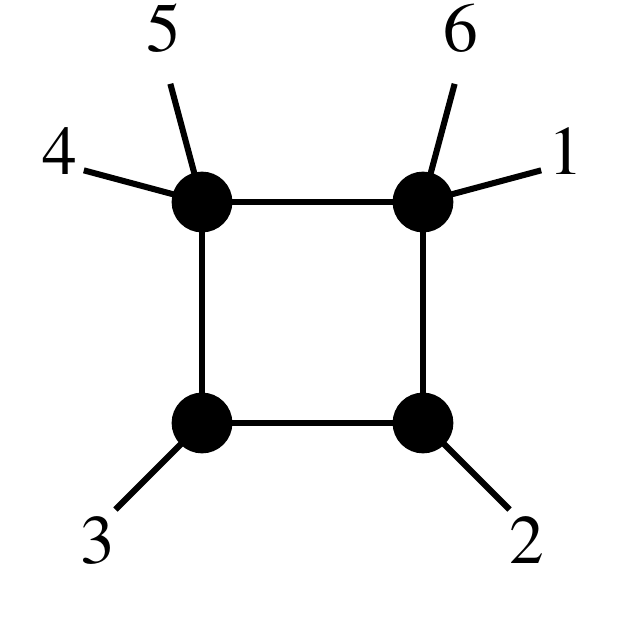}-
\figBox{0}{-1.65}{0.5}{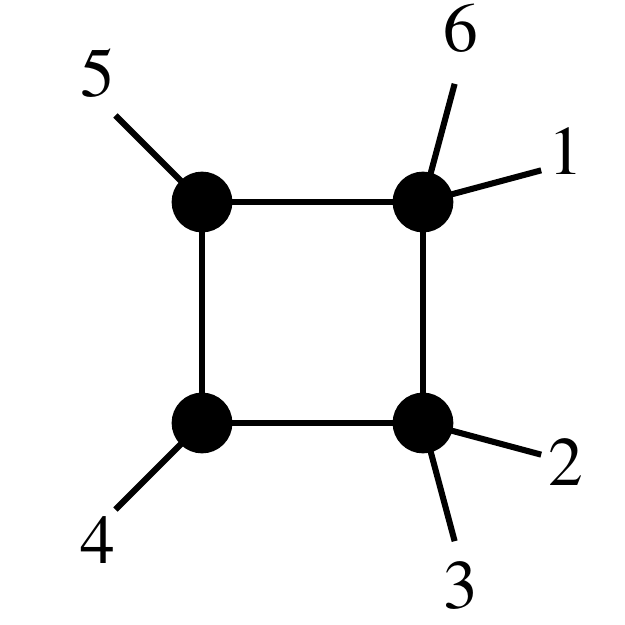}-\figBox{0}{-1.65}{0.5}{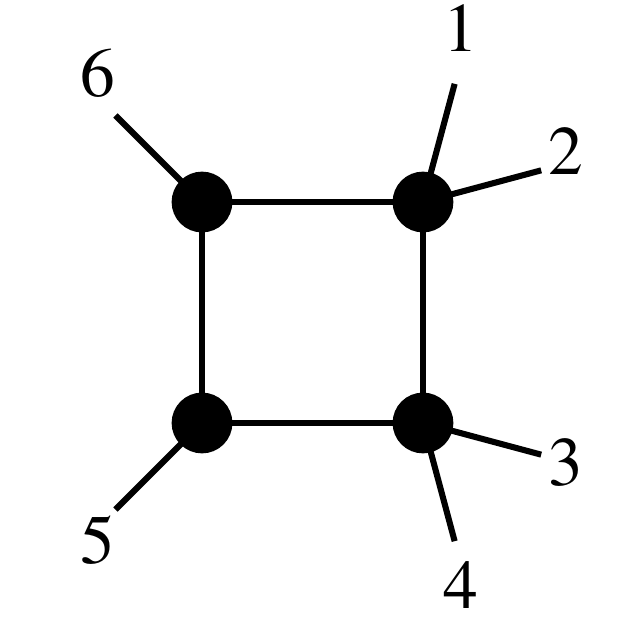}\end{array}$}\right.\hspace{-1cm}}
The seemingly complicated combinations of a large number of boxes have been
encountered before in the computation of finite 1-loop objects, such
as the NMHV ratio function \cite{Drummond:2008bq, Drummond:2008vq, Brandhuber:2009kh, Elvang:2009ya} ---the ratio function for the full
superamplitude is simply defined to be
\eq{\mathcal{R}_{n,k}^{1\mathrm{-loop}}=\mathcal{A}_{n,k}^{1\mathrm{-loop}}-\mathcal{A}_{n,k}^{\mathrm{tree}}\cdot\mathcal{A}_{n,k=2}^{1\mathrm{-loop}}.}

Note that in the box expansion, every integral is
individually IR-divergent, the IR-divergences only canceling in
the sum. Moreover, the boxes themselves are not dual conformal invariant---again, only become dual conformal invariant in the sum. But since the hexagon in which we are interested is manifestly finite and dual conformal invariant\footnote{In the literature on ratio functions, some authors have found what were claimed to be ``finite'' combinations of boxes that did {\it not} end up being dual-conformal invariant. In every case, the combinations of boxes in question were not honestly IR-finite: the divergences from different regions of the integration contour canceling between each-other. Such a cancellation is is highly regulator-dependent, and is not very meaningful.}, we can evaluate it directly---for example, using Feynman parameterization directly without any regularization. A straightforward computation shows,
\be
\hspace{-1.75cm}\figBox{0}{-1.65}{0.5}{hexagon_integral_box_decomposition_0.pdf} = {\rm Li}_2(1 - u_1) + {\rm Li}_2(1 - u_2) + {\rm Li}_2(1 - u_3) +
{\rm log}(u_3) {\rm log}(u_1) - \frac{\pi^2}{3},
\ee
where the $u_i$ are the familiar six-point cross-ratios
\eq{u_1\equiv\frac{\ab{12\,34}\ab{45\,61}}{\ab{12\,45}\ab{34\,61}},\qquad u_2\equiv\frac{\ab{23\,45}\ab{56\,12}}{\ab{23\,56}\ab{45\,12}},\quad\mathrm{and}\quad u_3\equiv\frac{\ab{34\,56}\ab{61\,23}}{\ab{34\,61}\ab{56\,23}}.}

It is easy to find examples of integrals which are finite and chiral, but which do not have unit leading singularities. For example, changing one the `dashed-line' numerator factor $\ab{AB\,13}$ in the integral above to a `wavy-line' $\ab{AB\,(612)\newcap(234)}$ will leave the integral finite and chiral, but spoil the equality of its leading singularities. Indeed, as it is also finite and dual-conformally invariant, the `mixed' hexagon integral can also be evaluated without any regularization, and one finds that,
\begin{align}\nonumber\hspace{-1.5cm}&\hspace{1.9cm}\figBox{0}{-1.65}{0.5}{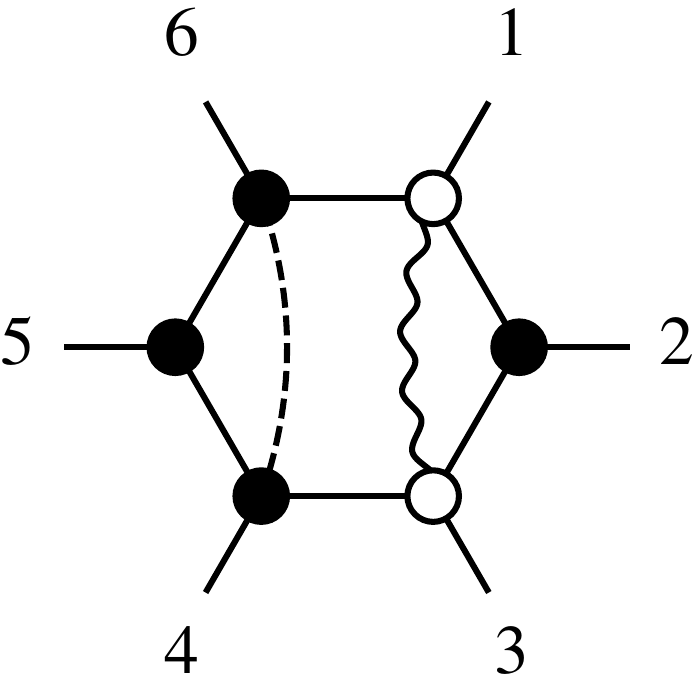}=\int\limits_{AB}\frac{\ab{AB\,(612)\newcap(234)}\ab{AB\,46}}{\ab{AB\,12}\ab{AB\,23}\ab{AB\,34}\ab{AB\,45}\ab{AB\,56}\ab{AB\,61}}\\\nonumber\hspace{-1.9cm}=&\left(\frac{\ab{1234}}{\ab{1345}\ab{1235}}\right)\log(u_1)\log(u_2)+\left(\frac{\ab{6134}}{\ab{1345}\ab{5613}}\right)\log(u_3)\log(u_1)+\left(\frac{\ab{6123}}{\ab{1235}\ab{3561}}\right)\log(u_2)\log(u_3).\hspace{1.5cm}\end{align}

In order for GRTs to yield the two-term identities necessary to guarantee that all the leading singularities are equal up-to a sign, the numerator must force vanishing residues for all but two Schubert problems. In the case of the `mixed-numerator' hexagon integral, for example, GRTs can only be used to show that the coefficients of the logarithms sum to zero:
\eq{\left(\frac{\ab{1234}}{\ab{1345}\ab{1235}}\right)+\left(\frac{\ab{6134}}{\ab{1345}\ab{5613}}\right)+\left(\frac{\ab{6123}}{\ab{1235}\ab{3561}}\right)=0.}

It is clear that these chiral momentum-twistor integrals with unit
leading singularities give us the simplest and most transparent way of
talking about finite integrals.

Just as a trivial example, the 6-point NMHV ratio function, which is typically written in terms of all 15 six-point box-integrals, with many $R$-invariants as coefficients, is given simply by
\eq{\mathcal{R}_{\mathrm{NMHV}}^{1\mathrm{-loop}}\underset{\substack{\mathrm{post-}\\\mathrm{integration}}}{=}\left(1+g+g^2\right)
\left\{\begin{array}{c}\figBox{0}{-1.65}{0.45}{hexagon_integral_box_decomposition_0.pdf}\\\times\left(\left[2\,3\,4\,5\,6\right]-
\left[3\,4\,5\,6\,1\right]+\left[4\,5\,6\,1\,2\right]\right)\phantom{\times}\end{array}\right\},\label{six_point_ratio_function}} where $g:i\mapsto i+1$ acts on both the integrand and its coefficient. Also recall the definition of the $R$-invariants given in section 1,
\eq{\hspace{-0.3cm}\left[i\,j\,k\,l\,m\right]\equiv\frac{\delta^{0|4}\left(\ab{j\,k\,l\,m}\eta_i+\ab{k\,l\,m\,i}\eta_j+\ab{l\,m\,i\,j}\eta_k+
\ab{m\,i\,j\,k}\eta_l+\ab{i\,j\,k\,l}\eta_m\right)}{\ab{i\,j\,k\,l}\ab{j\,k\,l\,m}\ab{k\,l\,m\,i}\ab{l\,m\,i\,j}\ab{m\,i\,j\,k}}.}

\newpage

\section{One-Loop Integrands, Integrals, and Amplitudes}\label{one_loop_integrands_and_basis}

As described in section \ref{foundations_integral_reduction}, one can use elementary tensor-reduction to express any 1-loop integrand in $\mathcal{N}=4$ in terms of pentagon and box integrands. These of course would form a complete basis for any 1-loop integrand in $\mathcal{N}=4$ SYM. However, such a basis would necessarily include many integrands which are {\it non}-chiral (including all boxes), and which have {\it non}-uniform leading singularities; moreover, such a basis would allow for linear combinations of IR-divergent integrals to be ultimately IR-finite and non-vanishing. But we saw in the last section that there are integrands---pentagons and hexagons with `magic' numerators---which avoid all of these shortcomings, and these integrands closely mirror the leading singularities of MHV-amplitudes, suggesting that they may be well-suited to express amplitudes more generally.

It is therefore natural to wonder if there exists a complete basis of 1-loop integrands involving only chiral, manifestly dual-conformally invariant integrands with unit leading-singularities, and for which no non-vanishing linear-combination of IR-divergent integrands is IR-finite. We'll see momentarily that the answer is affirmative, and extremely beautiful. 

Before trying to construct such a basis, however, we can gain some intuition about what to expect by assessing its size---that is, finding the dimension of the space of planar, 1-loop integrands. Recall that every $n$-point 1-loop planar integral can be written in the form \eq{\int\limits_{AB}\frac{\ab{AB\,Y_1}\cdots\ab{AB\,Y_{n-4}}}{\ab{AB\,1\,2}\ab{AB\,2\,3}\ab{AB\,3\,4}\cdots\ab{AB\,n\mi1\,n}\ab{AB\,n\,1}}.} When $n=5$, the space of 1-loop integrands is just the space of bitwistors $Y$, which is six-dimensional---which explains how the complete $5$-point 1-loop integrand could be constructed in \cite{Cachazo:2008vp} through the introduction of a single pentagon integrand.

For $n=6$, the most general integrand is a hexagon with $\ab{AB\,Y_1}\ab{AB\,Y_2}$ in the numerator. Now, each $Y_i$ is a $\mathbf{6}$-dimensional representation of $SU_4$, and of course $\mathbf{6}\otimes\mathbf{6}=\mathbf{1}\oplus\mathbf{15}\oplus\mathbf{20}$. Ordinary multiplication being commutative, the antisymmetric part, the $\mathbf{15}$-component, clearly vanishes. By expanding each $Y_i$ into a basis of six simple bitwistors, it is easy to see that the trace component, $\mathbf{1}$, also vanishes, as $\ab{Y_i\,Y_i}=0,$ when $Y_i$ is simple. Therefore, the space of $6$-point 1-loop integrands is $\mathbf{20}$-dimensional\footnote{We thank Simon Caron-Huot for helpful discussions regarding this counting.}.
\newpage

More generally, it is not hard to see that the dimension, $d$, of 1-loop integrands is the same as the dimension of the space of symmetric $n-4$-fold symmetric, traceless tensors of $\mathbf{6}$'s of $SU_4$, which is simply
\vspace{-0.2cm}\eq{d=\,\left(\begin{array}{c}n\\4\end{array}\right)+\left(\begin{array}{c}n-1\\4\end{array}\right).\label{dimension_of_one-loop_integrands}\vspace{-0.2cm}}
Recall that box-integrands form a complete basis of parity-even integrands, and that there are precisely {\footnotesize$\left(\begin{array}{c}n\\4\end{array}\right)$} boxes, all of which are independent. Therefore, we may separate $d$ in \mbox{equation (\ref{dimension_of_one-loop_integrands})}, according to $d=d_{even}+d_{odd}$ with $d_{even}=${\footnotesize $
\left(\begin{array}{c}n\\4\end{array}\right)$} and $d_{odd}=${\footnotesize $\left(\begin{array}{c}n-1\\4\end{array}\right)$}. Once we have a basis of integrands which makes parity manifest, this will allow us to count the number of relations satisfied by (parity-odd) integrands.

\subsection{The Chiral Octagon: A Basis of One-Loop Integrands}
As we can see from \mbox{equation (\ref{dimension_of_one-loop_integrands})}, the number of independent integrands grows asymptotically like $\mathcal{O}(n^4)$. In contrast, the number of chiral pentagons grows only like $\mathcal{O}(n^2)$. It is not hard to see that the simplest class of chiral integrands which number $\mathcal{O}(n^4)$ are the chiral octagons. As we will see presently, it turns out that chiral octagon-integrands indeed form an (over)-complete basis for all 1-loop integrands that satisfies all the desired criteria listed above. The most general chiral octagon integral is given by,
\vspace{-0.35cm}\begin{align}\hspace{-2cm}&\hspace{3.5cm}I_8(i,j,k,l)\equiv\figBox{0}{-1.65}{0.5}{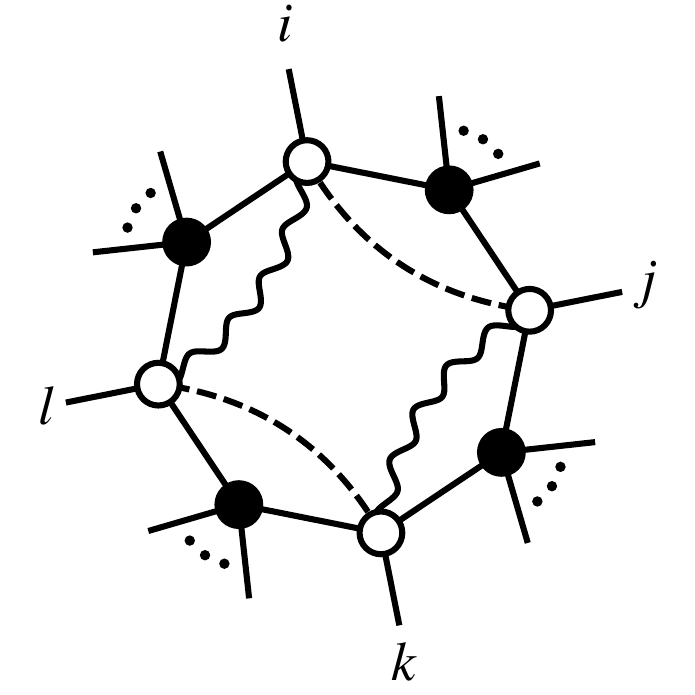}\label{genral_octagon_definition}\quad\mathrm{for}\quad i\!<\!j\!<\!k\!<\!l\!<\!i\\&\hspace{-0.8cm}=\int\limits_{AB}\!\!\frac{\ab{AB\,i\,j}\ab{AB(j\smallminus1\,j\,j\smallplus1)\newcap(k\smallminus1\,\,k\,\,k\smallplus1)}\ab{AB\,k\,\,l}\ab{AB(l\smallminus1\,\,l\,\,l\smallplus1)\newcap(i\smallminus1\,\,i\,\,i\smallplus1)}}{\ab{AB\,i\smallminus1\,i}\ab{AB\,i\,i\smallplus1}\ab{AB\,j\smallminus1\,j}\ab{AB\,j\,j\smallplus1}\ab{AB\,k\smallminus1\,k}\ab{AB\,k\,k\smallplus1}\ab{AB\,l\smallminus1\,l}\ab{AB\,l\,l\smallplus1}}.\hspace{2cm}{~}\nonumber\vspace{-0.3cm}\end{align}
Notice that parity acts according to $\mathbb{P}:I_8(i,j,k,l)\mapsto I_8(j,k,l,i)$, making it trivial to define parity-even/parity-odd sectors:
\vspace{-0.75cm}\eq{\hspace{-1.05cm}I^{\mathrm{even}/\mathrm{odd}}_8(i,j,k,l)\equiv I_8(i,j,k,l)\pm I_8(j,k,l,i)=\!\!\figBox{0}{-1.65}{0.5}{general_octagon_integral.pdf}\!\!\!\pm\!\!\!\figBox{0}{-1.65}{0.5}{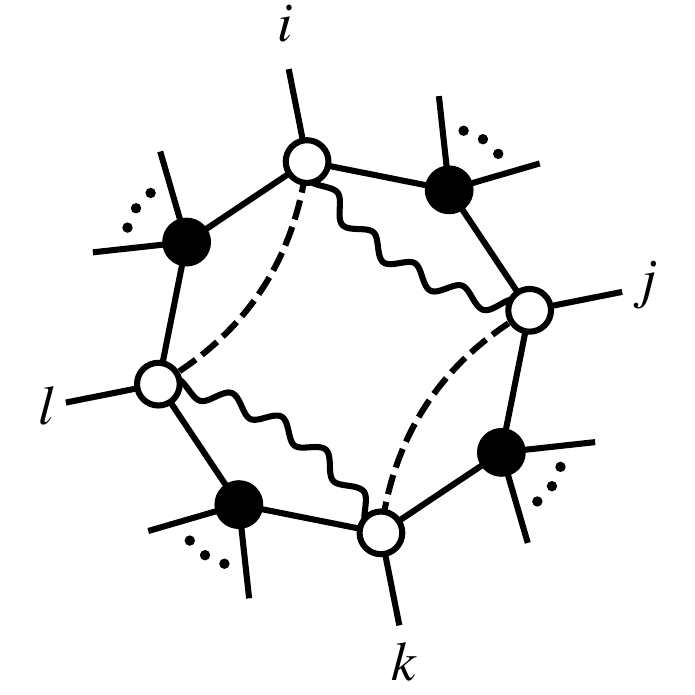}\hspace{-2cm}}

\begin{figure}[t!]\caption{The possible degenerations of the general octagon integrand.\label{degenerations_of_the_octagon}}\figBox{0}{-1}{1}{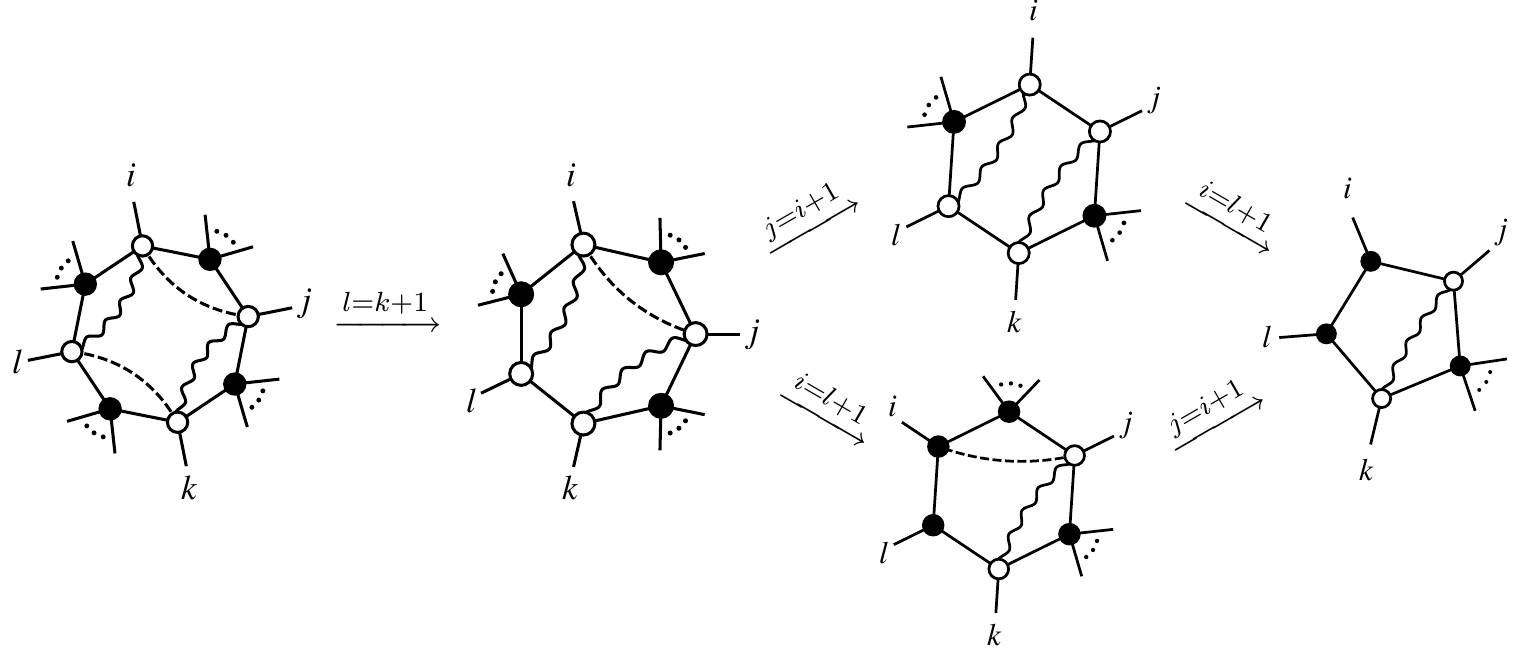}\vspace{-0.6cm}\end{figure}

Notice that these octagon integrands are well-defined for any distinct set of indices $\{i,j,k,l\}$, including those for which the `octagon' degenerates into lower-polygons. For example, when $l=k+1$, the extra (duplicated) propagator in \mbox{equation (\ref{genral_octagon_definition})}, $\ab{AB\,k\,k\pl1}$ is cancelled by the dashed-line term $\ab{AB\,k\,l}\to\ab{AB\,k\,k\pl1}$ in the numerator. A complete sampling of degenerate `octagons' is illustrated in \mbox{Figure \ref{degenerations_of_the_octagon}}.
One important advantage of this presentation is that all but the most degenerate of the octagons are {\it manifestly} finite. Indeed, only the pentagons---octagons of the from $I_8(i,i\pl1,i\pl2,i\pl3,i\pl4)$---and the lower hexagons in \mbox{Figure \ref{degenerations_of_the_octagon}}---octagons of the form $I_8(i,i\pl1,i\pl2,j)$---are IR-divergent. Specifically, we have the following separation into manifestly IR-finite and IR-divergent basis integrands.
\vspace{-0.2cm}\eqs{\text{IR-finite:}&\figBox{0}{-1.75}{0.5}{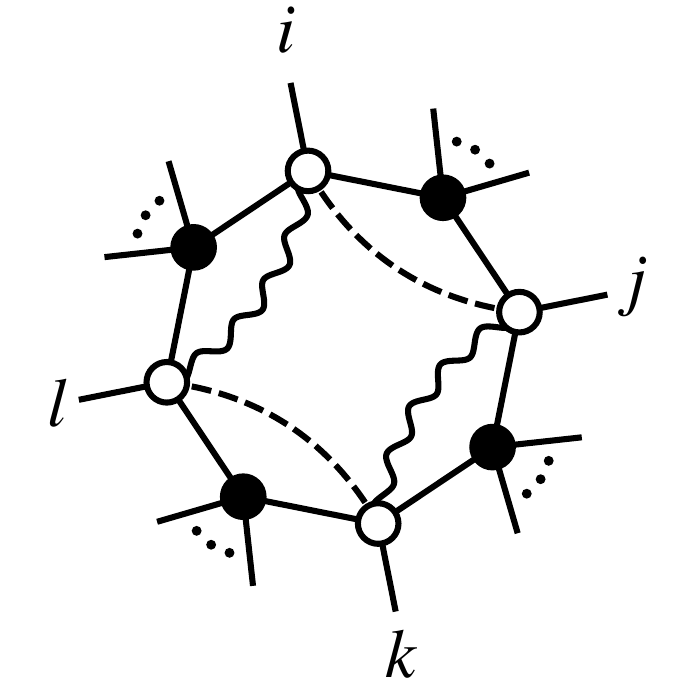}\qquad\figBox{0}{-1.75}{0.5}{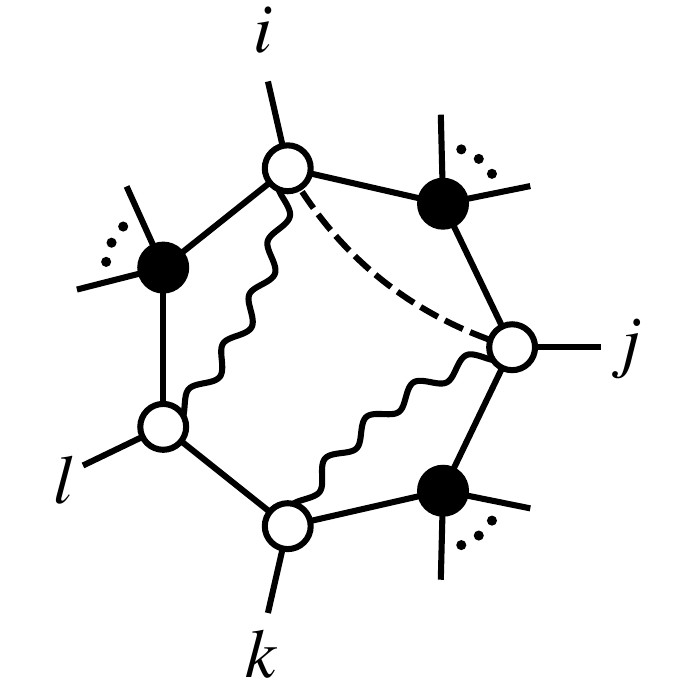}\qquad\figBox{-0.5}{-1.775}{0.45}{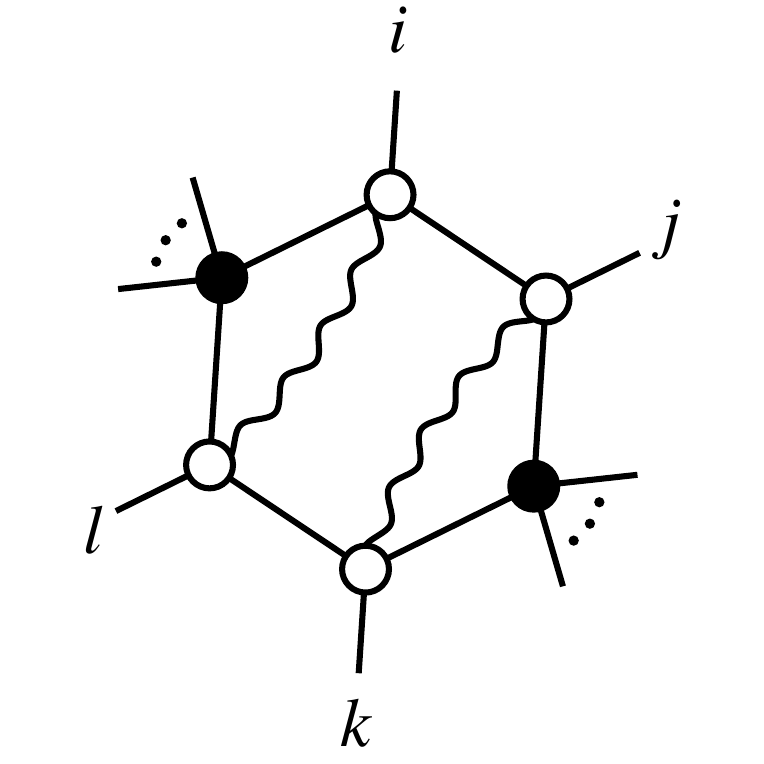}\\[-0.5cm]\text{IR-divergent:}\hspace{-2.25cm}&\hspace{2.25cm}\figBox{0}{-1.675}{0.45}{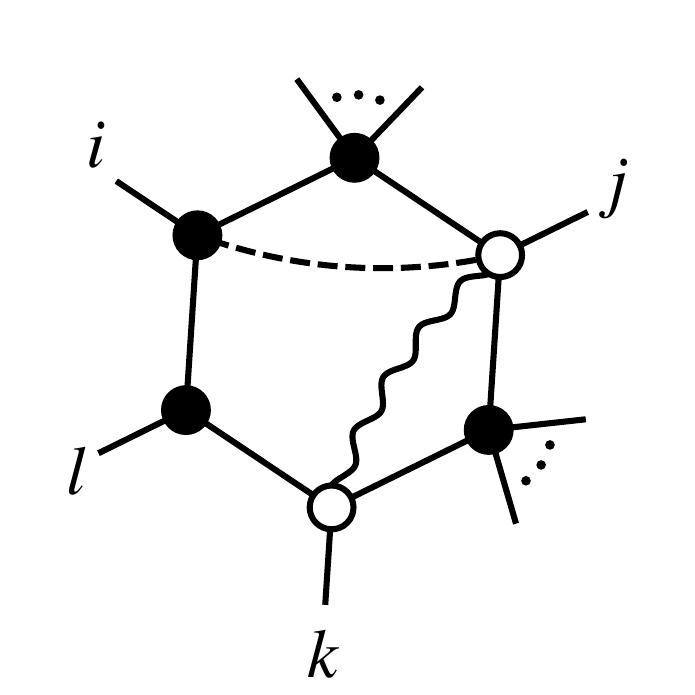}\qquad\figBox{0}{-1.675}{0.45}{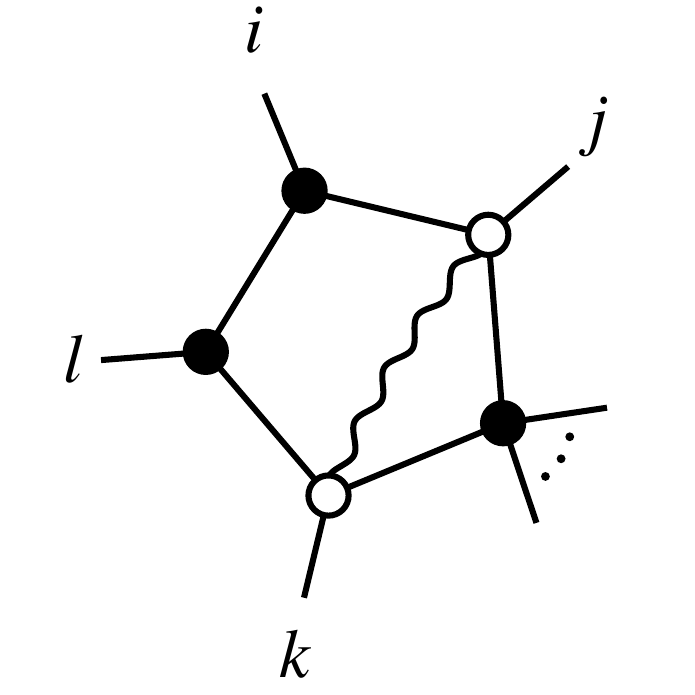}\vspace{-0.2cm}}
It is worth emphasizing that the only IR-finite combinations of IR-divergent integrals in this basis are parity-odd, which automatically vanish upon integration. Furthermore, as discussed above, because the criterion for local divergences in the region of integration is itself parity-invariant,\footnote{Recall that a local, IR-divergence develops in the region of integration when the line $(AB)$ passes through a point $Z_i$ while simultaneously lying on the plane $(Z_{i-1}\,Z_i\,Z_{i+1})$.} parity-odd combinations of integrands are in fact manifestly {\it locally finite}.

Parity-symmetrizing, and parity anti-symmetrizing, it is clear that there {\footnotesize$2\left(\begin{array}{c}n\\4\end{array}\right)$} octagon integrands, evenly split between parity-odd and parity-even. As we described above, among the parity-odd combinations of integrands only {\footnotesize$\left(\begin{array}{c}n-1\\4\end{array}\right)$} are linearly-independent, so the octagon basis is strictly {\it over-complete}, but there are only non-trivial relations among integrands in the parity-odd sector.

\subsection{Integration of Manifestly-Finite Octagons}
It is not hard to directly evaluate the general octagon integral $I_8(i,j,k,l)$. Consider for example the case $I_8(3,6,9,12)$ for which all indices are separated by at least 3, \vspace{-0.35cm}
\eq{\figBox{0}{-1.6}{0.5}{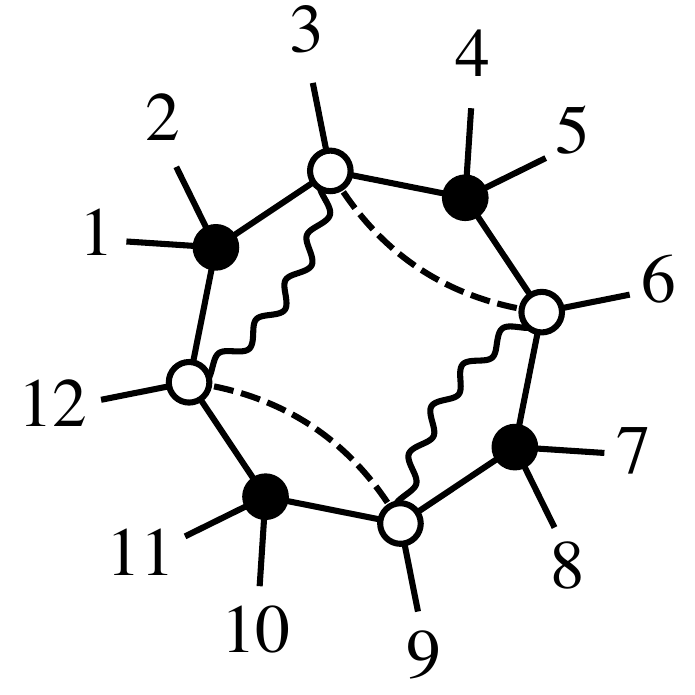}.\vspace{-0.2cm}}
Because of the numerator factors, the only non-vanishing leading singularities of this integral involve cutting at most one of each the pairs of lines $\{(23),(34)\}$, $\{(56),(67)\}$, $\{(89),(9\,10)\}$, and $\{(11\,12),(12\,1)\}$. Therefore, this integral's box-expansion is simply the (manifestly-finite) sum of 16 four-mass box integrals. One disadvantage with this presentation of the integral, however, is that the four-mass box integral logarithmically-diverges when any of its four massive corners becomes massless, and yet we saw above that the general octagon remains manifestly finite upon many such degenerations.

Because of this, we are motivated to replace the four-mass box function with a new function that is free of any divergences over the physical domain of cross ratios. Letting $\Delta_4(u,v)$ denote the familiar four-mass box integral\footnote{If we denote the massive, incoming four-momenta of the box by $K_1,K_2,K_3,$ and $K_4$, and define the canonical Mandelstam variables $s\equiv (K_1+K_2)^2$ and $t\equiv(K_2+K_3)^2$, then we are using $u$ and $v$ to denote the cross ratios $K_1^2K_3^2/(s t),$ and $K_2^2K_4^2/(s t)$, respectively.}---a symmetric function in the two cross-ratios---then let us define the following `modified four-mass' function \eq{\deltatilde(i,j,k,l)\equiv\Delta_4(u_{i,j,k,l},u_{j,k,l,i})-\frac{1}{2}\log(u_{i,j,k,l})\log(u_{j,k,l,i}),}
where \eq{\Delta_4(u,v)\equiv\Li(1-\alpha_+)-\Li(1-\alpha_-)+\frac{1}{2}\log(v)\log(\alpha_+/\alpha_-),}
and \eq{\alpha_{\pm}\equiv\frac{2u}{1+u-v\pm\sqrt{(1-u-v)^2-4 u v}};} here, we have used the four indices $\{i,j,k,l\}$ to signify the (generally time-like separated) spacetime points corresponding to the lines $(i\,i\pl1), (j\,j\pl1), (k\,k\pl1),$ and $(l\,l\pl1)$ in twistor space, which together define the cross-ratios \eq{u_{{\color{cr_red}i},{\color{cr_red}j},{\color{cr_blue}k},{\color{cr_blue}l}}\equiv\frac{\ab{{\color{cr_red}i}\,i\pl1\,{\color{cr_red}j}\,j\pl1}\ab{{\color{cr_blue}k}\,k\pl1\,{\color{cr_blue}l}\,l\pl1}}{\ab{{\color{cr_blue}l}\,l\pl1\,{\color{cr_red}j}\,j\pl1}\ab{{\color{cr_blue}k}\,k\pl1\,{\color{cr_red}i}\,i\pl1}}\quad\mathrm{and}\quad u_{{\color{cr_red}j},{\color{cr_red}k},{\color{cr_blue}l},{\color{cr_blue}i}}\equiv\frac{\ab{{\color{cr_red}j}\,j\pl1\,{\color{cr_red}k}\,k\pl1}\ab{{\color{cr_blue}l}\,l\pl1\,{\color{cr_blue}i}\,i\pl1}}{\ab{{\color{cr_blue}i}\,i\pl1\,{\color{cr_red}k}\,k\pl1}\ab{{\color{cr_blue}l}\,l\pl1\,{\color{cr_red}j}\,j\pl1}}.}

The principle distinction between $\deltatilde(i,j,k,l)$ and the more familiar four-mass box function is that $\deltatilde(i,j,k,l)$ remains finite even when many of the spacetime points become null-separated (or even become identified). In particular, \eq{\hspace{-0.5cm}\lim_{u_{i,j,k,l}\to0}\left(\deltatilde(i,j,k,l)\right)=\Li(1-u_{j,k,l,i})\quad\mathrm{and}\quad\lim_{u_{j,k,l,i}\to0}\left(\deltatilde(i,j,k,l)\right)=\Li(1-u_{i,j,k,l}).}

Of course, if we use $\deltatilde$'s to represent $I_8(3,6,9,12)$, for example, then each four-mass box will contribute a `$\log$-$\log$'-term. It may be worried that this will greatly clutter the final expression, but this turns out to not be the case: taken together, these 16 additional `$\log$-$\log$' terms combine into a single such term.

With this new function, the general octagon integral---together with all its degenerations---becomes extremely simple. Explicitly, the general octagon $I_8(i,j,k,l)$ integral is given by,\vspace{-0.2cm}
\begin{align}\hspace{-1cm}&\nonumber\hspace{-2.75cm}\begin{array}{ll}\raisebox{-1.65cm}{\includegraphics[scale=0.5]{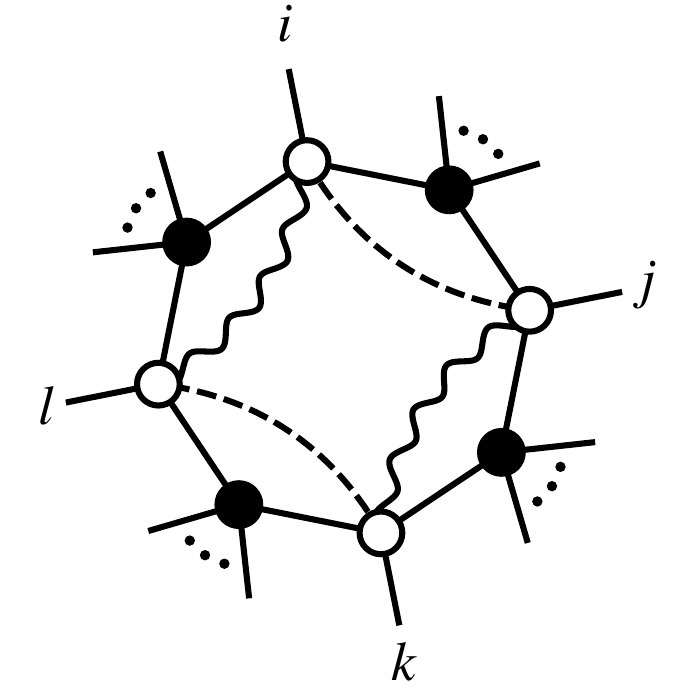}}\!\!\!=&\left\{\!\!\begin{array}{l@{}l@{}l@{}l@{}l}\multicolumn{4}{l}{\phantom{+}\log\left(u_{i\,k-1\,k\,i-1}\right)\log\left(u_{j\,l-1\,l\,j-1}\right)}\\
+\deltatilde(i,j,k,l)&  -\deltatilde(i,j,k,l\mi1)&  -\deltatilde(i,j,k\mi1,l)&  +\deltatilde(i,j,k\mi1,l\mi1)\\-\deltatilde(i,j\mi1,k,l)&  +\deltatilde(i,j\mi1,k,l\mi1)&  +\deltatilde(i,j\mi1,k\mi1,l)&  -\deltatilde(i,j\mi1,k\mi1,l\mi1)\\-\deltatilde(i\mi1,j,k,l)&  +\deltatilde(i\mi1,j,k,l\mi1)&  +\deltatilde(i\mi1,j,k\mi1,l)&  -\deltatilde(i\mi1,j,k\mi1,l\mi1)\\+\deltatilde(i\mi1,j\mi1,k,l)&  -\deltatilde(i\mi1,j\mi1,k,l\mi1)&  -\deltatilde(i\mi1,j\mi1,k\mi1,l)&  +\deltatilde(i\mi1,j\mi1,k\mi1,l\mi1)\end{array}\right.\hspace{-4cm}\end{array}\hspace{-4cm}\end{align}
Although admittedly lengthy, this expression can be considerably compressed in a way which helps illustrate the relative signs appearing in the formula above,
\eq{\hspace{-1.cm}I_8(i,j,k,l)=\log\left(u_{i\,k-1\,k\,i-1}\right)\log\left(u_{j\,l-1\,l\,j-1}\right)+\!\sum_{\sigma_i\in\{1,0\}}(-1)^{(\sigma_1+\sigma_2+\sigma_3+\sigma_4)}\deltatilde(i\mi\sigma_1,j\mi\sigma_2,k\mi\sigma_3,l\mi\sigma_4).\label{general_octagon_integral}}

We can see how the modified four-mass function $\deltatilde(i,j,k,l)$ helps to make all of the octagon's degenerations manifest by looking at a few examples explicitly. For example, consider the 8-point octagon $I_8(2,4,6,8)$; in this case, only 20 of the 34 cross ratios which play a role in the general answer are non-vanishing, converting virtually all the generalized four-mass functions $\deltatilde$'s into $\Li$'s.
\vspace{-0.2cm}{\normalsize\eq{\nonumber\hspace{-1.45cm}\begin{array}{ll}\raisebox{-1.5cm}{\includegraphics[scale=0.5]{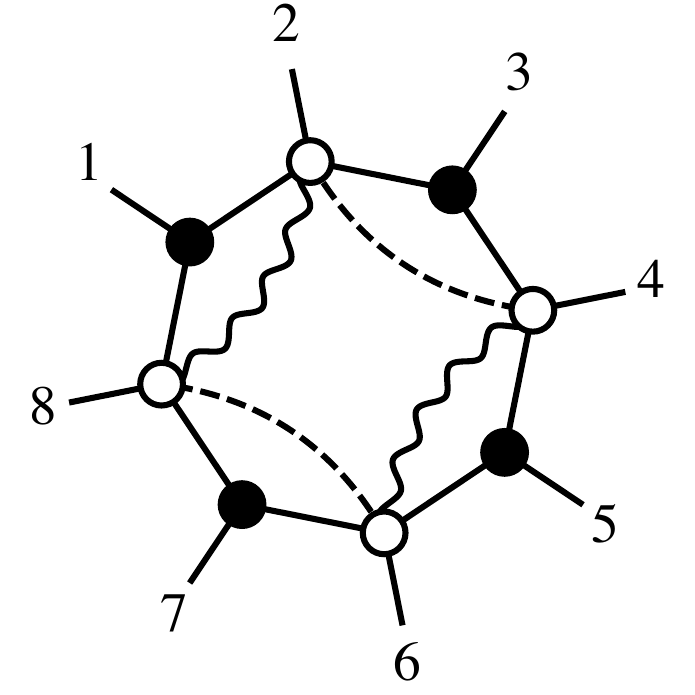}}\!\!\!=&\left\{\!\!\begin{array}{l@{}l@{}l@{}l@{}}\multicolumn{4}{l}{\phantom{+\!}\log\left(u_{{2,5,6,1}}\right) \log\left(u_{{4,7,8,3}}\right)}\\+\;\;\;\deltatilde({2,4,6,8})&  -\Li\left(1-u_{{4,6,7,2}}\right) &  -\Li\left(1-u_{{2,4,5,8}}\right) &  +\Li\left(1-u_{{2,4,5,7}}\right) \\-\Li\left(1-u_{{8,2,3,6}}\right) &  +\Li\left(1-u_{{3,6,7,2}}\right) &  +\Li\left(1-u_{{8,2,3,5}}\right) &  -\Li\left(1-u_{{7,2,3,5}}\right) \\-\Li\left(1-u_{{6,8,1,4}}\right) &  +\Li\left(1-u_{{4,6,7,1}}\right) &  +\Li\left(1-u_{{1,4,5,8}}\right) &  -\Li\left(1-u_{{1,4,5,7}}\right) \\+\Li\left(1-u_{{6,8,1,3}}\right) &  -\Li\left(1-u_{{3,6,7,1}}\right) &  -\Li\left(1-u_{{5,8,1,3}}\right) &  +\;\;\;\deltatilde({1,3,5,7})\end{array}\right.\end{array}}\vspace{-0.3cm}}

Even more simplification occurs for the degenerate `octagons.' Consider for example the general finite heptagon integral, given by $I_8(i,j,k,k+1)$,\vspace{-0.4cm}
\vspace{-0.2cm}\begin{align}\hspace{-1.8cm}&\nonumber\hspace{-2cm}\begin{array}{ll}\raisebox{-1.65cm}{\includegraphics[scale=0.5]{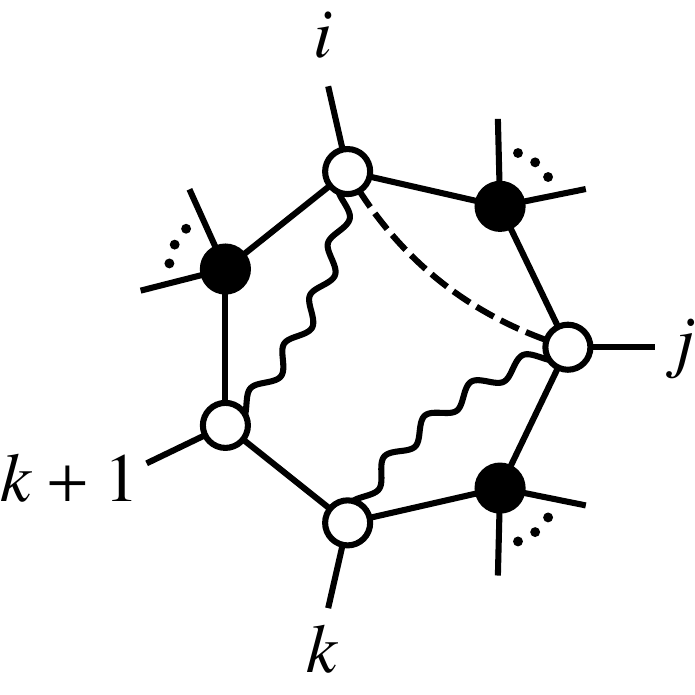}}=&\left\{\!\!\begin{array}{lll@{}l@{}l}\multicolumn{4}{l}{\phantom{+}\log\left(u_{i,k-1,k,i-1}\right) \log\left(u_{j,k,k+1,j-1}\right)}\\
+\Li\left(1-u_{j,k,k+1,i}\right) &  -0&  -\deltatilde(i,j,k\mi1,k\pl1)&  +\Li\left(1-u_{j,k-1,k,i}\right) \\-\Li\left(1-u_{j-1,k,k+1,i}\right) &  +0&  +\deltatilde(i,j\mi1,k\mi1,k\pl1)&  -\Li\left(1-u_{j-1,k-1,k,i}\right) \\-\Li\left(1-u_{j,k,k+1,i-1}\right) &  +0&  +\deltatilde(i\mi1,j,k\mi1,k\pl1)&  -\Li\left(1-u_{j,k-1,k,i-1}\right) \\+\Li\left(1-u_{j-1,k,k+1,i-1}\right) &  -0&  -\deltatilde(i\mi1,j\mi1,k\mi1,k\pl1)&  +\Li\left(1-u_{j-1,k-1,k,i-1}\right) \end{array}\right.\hspace{-4cm}\end{array}\hspace{-4cm}\end{align}~\\[-0.7cm]
Here, because $\deltatilde(i,j,k,k)=\Li(0)=0$, four of the contributions vanish, and eight of the modified four-mass box functions simplify to simple $\Li$'s.

The final class of finite, degenerate octagons are the hexagon integrals---octagons of the form $I_8(i,i+1,k,k+1)$, \vspace{-0.2cm}
\vspace{-0.2cm}\begin{align}\hspace{-1.8cm}&\nonumber\hspace{-2cm}\begin{array}{ll}\raisebox{-1.8cm}{\includegraphics[scale=0.5]{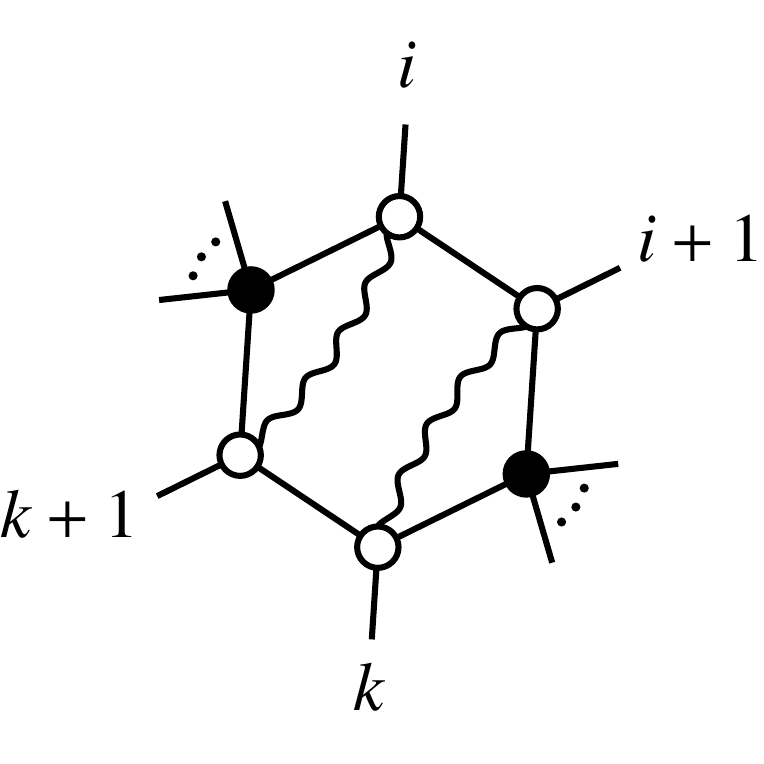}}\!\!\!=&\left\{\!\!\begin{array}{lll@{}l@{}c}\multicolumn{4}{l}{\phantom{+}\log\left(u_{i,k-1,k,i-1}\right) \log\left(u_{i+1,k,k+1,i}\right)}\\
+\Li\left(1-u_{i+1,k,k+1,i}\right) &  -0&  -\Li\left(1-u_{k+1,i,i+1,k-1}\right) &  +\Li\left(1-u_{i+1,k-1,k,i}\right) \\
-0&  +0&  +0&  -0\\-\Li\left(1-u_{i+1,k,k+1,i-1}\right) &  +0&  +\deltatilde(i\mi1,i\pl1,k\mi1,k\pl1)&  -\Li\left(1-u_{i+1,k-1,k,i-1}\right) \\+\Li\left(1-u_{i,k,k+1,i-1}\right) &  -0&  -\Li\left(1-u_{k+1,i-1,i,k-1}\right) &  +\Li\left(1-u_{i,k-1,k,i-1}\right) \end{array}\right.\hspace{-4cm}\end{array}\hspace{-4cm}\end{align}
Just as for the case of the 8-point octagon integral, the general hexagon integral simplifies considerably when potentially-massive corners become massless. As a final illustration, let us see how the general formula for the octagon given above directly yields the result quoted in \mbox{section \ref{finite_integrals_section}} for the 6-point hexagon integral which played such an important role in the 6-point NMHV ratio function:
\begin{align}\hspace{-1.8cm}&\nonumber\hspace{-2cm}\begin{array}{ll}\raisebox{-1.8cm}{\includegraphics[scale=0.5]{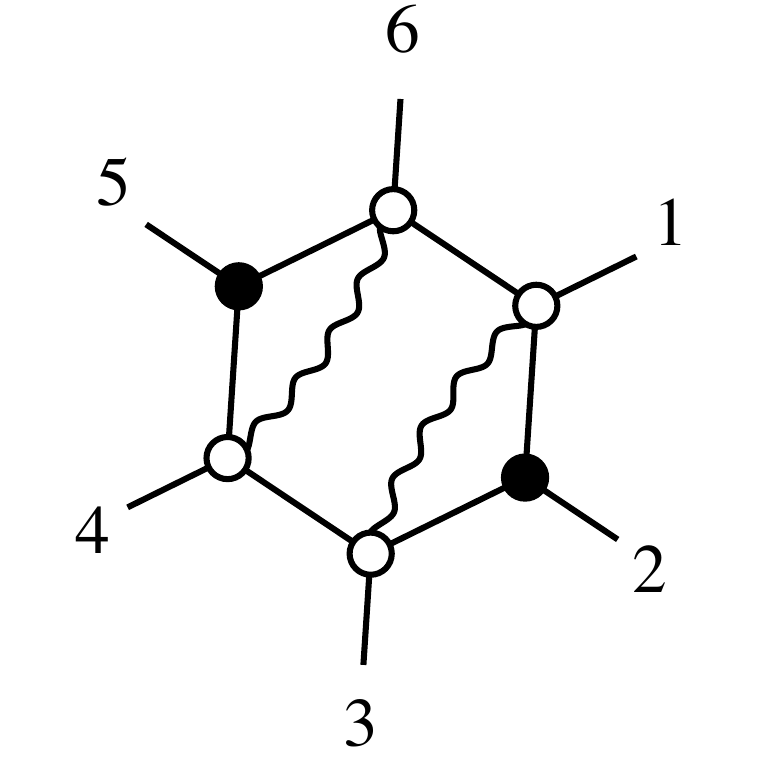}}\!\!\!&\begin{array}{l}=\left\{\!\!\begin{array}{lll@{}l@{}c}\multicolumn{4}{l}{\phantom{+}\log\left(u_{{3,5,6,2}}\right) \log\left(u_{{4,6,1,3}}\right)}\\
+\Li\left(1-u_{{4,6,1,3}}\right) &  -0&  -\Li(1)&  +\Li(1)\\-0&  +0&  +0&  -0\\-\Li(1)&  +0&  +\Li\left(1-u_{{2,4,5,1}}\right) &  -\Li(1)\\+\Li(1)&  -0&  -\Li(1)&  +\Li\left(1-u_{{3,5,6,2}}\right) \end{array}\right.\hspace{-4cm}\\\hline=\phantom{+\,}\Li(1-u_{4,6,1,3})+\Li(1-u_{2,4,5,1})+\Li(1-u_{3,5,6,2})\\\phantom{=}+\log(u_{3,5,6,2})\log(u_{4,6,1,3})-2\Li(1).\end{array}\end{array}\hspace{-4cm}\end{align}


\subsection{Application: the NMHV One-Loop Ratio Function}
As should be clear from the previous subsection, any IR-finite object such as the ratio function will be {\it manifestly finite} when expanded in the basis of octagon integrands. Moreover, since the formula for the completely general octagon integral, \mbox{equation (\ref{general_octagon_integral})}, is free of discontinuities for all the IR-finite degenerations of the octagon, any finite 1-loop integrand expressed in the basis of octagons directly translates into a function that is manifestly dual-conformally invariant.

A very important, manifestly finite function associated with 1-loop scattering amplitudes is the ratio function, \eq{\mathcal{R}_{n,k}^{1\mathrm{-loop}}=\mathcal{A}_{n,k}^{1\mathrm{-loop}}-\mathcal{A}_{n,k}^{\mathrm{tree}}\cdot\mathcal{A}_{n,k=2}^{1\mathrm{-loop}}.}
The most trivial example must be the 5-point 1-loop NMHV ratio function. Expanding into the basis of octagons, the integrand is easily seen to be given by \vspace{-0.2cm}
\eq{\mathcal{R}_{5,3,1}=[1,2,3,4,5]\left(\figBox{0}{-1.15}{0.4}{five_point_one_loop_ratio_1.pdf}-\figBox{0}{-1.15}{0.4}{five_point_one_loop_ratio_2.pdf}\right)+\mathrm{cyclic}.\vspace{-0.2cm}}
Being a parity-odd combination of pentagons, the ratio function is locally free of any divergences at the level of the integrand and is therefore manifestly finite---of course, being parity odd, it also vanishes upon integration.

A less trivial example, and one which we quoted in \mbox{section \ref{finite_integrals_section}}, is the 6-point NMHV 1-loop ratio function. In \mbox{section \ref{finite_integrals_section}} only the parity-even contribution to the ratio function was described; the full integrand is given by, \vspace{-0.7cm}
\begin{align}\nonumber\hspace{-1cm}\mathcal{R}_{6,3,1}=&\hspace{-1.15cm}\phantom{\,+\,}\begin{array}{cc}\begin{array}{c}\phantom{\big(}~\\\figBox{0}{-1.225}{0.5}{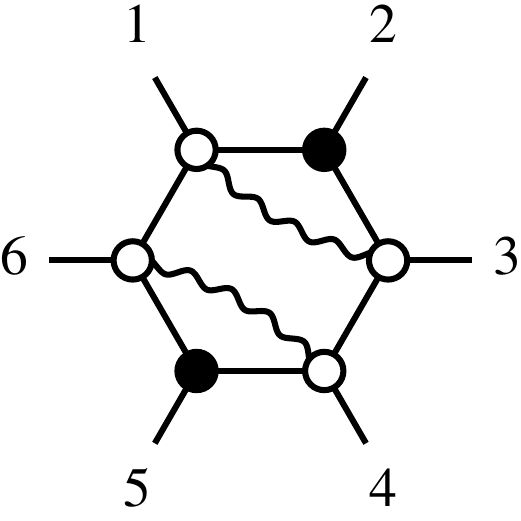}\\\times\frac{1}{2}\big( [1,2,3,4,5]+[1,2,3,5,6]+[1,2,3,6,4]\big)\phantom{\frac{1}{2}\times}\end{array}&\hspace{-0.85cm}+\quad\,\begin{array}{c}\phantom{\big(}~\\\left(\figBox{0}{-1.225}{0.5}{six_point_one_loop_ratio_1.pdf}-\figBox{0}{-1.225}{0.5}{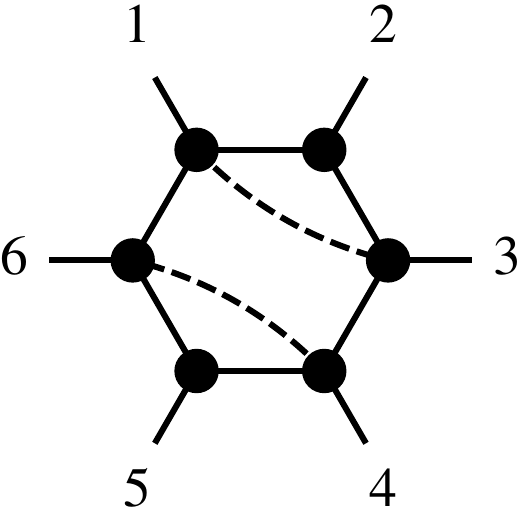}\right)\\\times\frac{1}{6}[1,2,3,4,6]\end{array}\end{array}\\[-0.3cm]
&\begin{array}{cc}+\begin{array}{c}\phantom{\big(}\\\left(\figBox{0}{-1.225}{0.5}{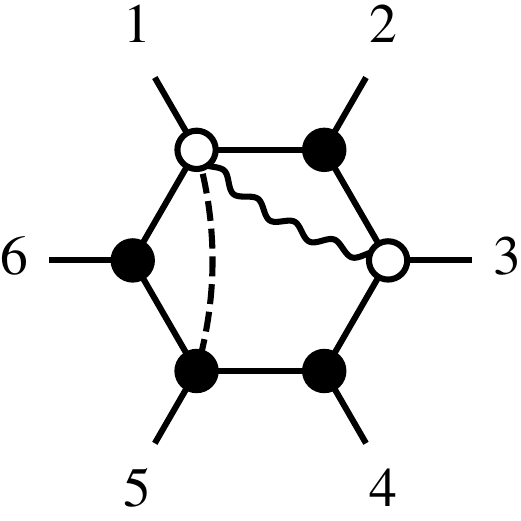}-\figBox{0}{-1.225}{0.5}{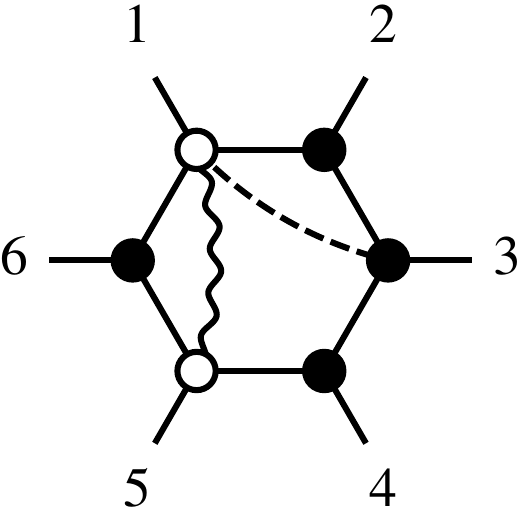}\right)\\\times\frac{1}{6}\big([1,2,3,4,5]-[1,3,4,5,6]\big)\end{array}&\,\,+\quad\,\begin{array}{c}\phantom{\big(}\\\left(\figBox{0}{-1.225}{0.5}{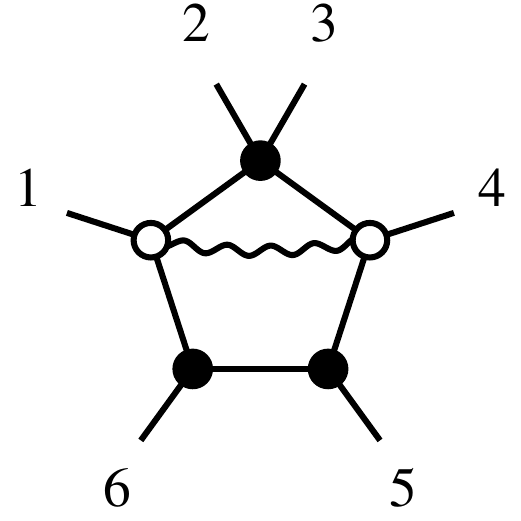}-\figBox{0}{-1.225}{0.5}{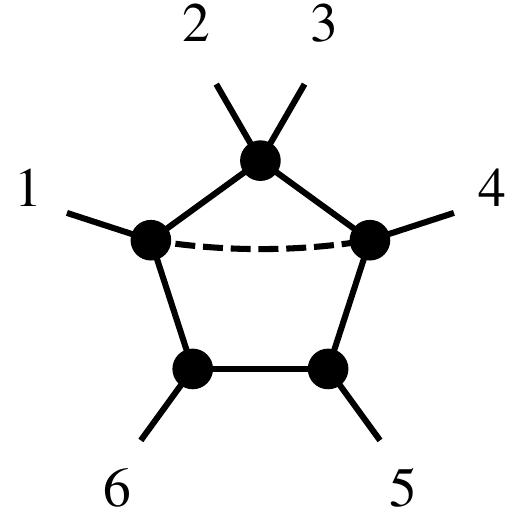}\right)\\\times\frac{1}{6}\big( [1,2,4,5,6]+[1,3,4,5,6]\big)\end{array}\end{array}\nonumber\\&\hspace{6.5cm}+\mathrm{cyclic}.\label{six_point_ratio_integrand}\end{align}
Of course, only the first term in \mbox{equation (\ref{six_point_ratio_integrand})} is non-vanishing when integrated along a parity-invariant contour, reproducing the formula given in \mbox{equation (\ref{six_point_ratio_function})}.

The general formula for the $n$-point NMHV 1-loop ratio function integrand nicely separates into a part which is parity-odd, and another which involves only manifestly finite integrands. In order to best capture the ratio function succinctly, let us introduce one small bit of notation and define \eq{[i,\{i+1,\ldots,j\},\{k,\ldots,l\}]\equiv\sum_{J=\{i+1,i+2\}}^{\{j-1,j\}}\sum_{K=\{k,k+1\}}^{\{l,k\}}[i,J,K];} for example, \eqs{[1,\{2,3,4\},\{5,6,7\}]=&\phantom{\,+\,\,} [1,2,3,5,6]+[1,2,3,6,7]+[1,2,3,7,5]\\&+[1,3,4,5,6]+[1,3,4,6,7]+[1,3,4,7,5].}
Notice that the two-index $J$ ranges over all consecutive pairs between $i+1$ and $j$ inclusively, while the two-index $K$ {\it also} includes a non-consecutively-ordered `wrapping' term. With this notation, it is very easy to write the $n$-point NMHV ratio function integrand:
\eqs{\mathcal{R}_{n,3}^{1\mathrm{-loop}}=&\phantom{\,-\,}\frac{1}{2}\,\,\,\,\,\sum_{i<j<k<l<i}\,\,\,\,[i,\{i+1,\ldots,j\},\{k,\ldots,l\}]I_8(i,j,k,l)\\&-\frac{1}{n}\sum_{i<j<k<l<m<i}[i,j,k,l,m]I_8^{\mathrm{odd}}(i,j,k,l).\label{octagon_NMHV_ratio_integrand}}

Notice that while the first term in \mbox{equation (\ref{octagon_NMHV_ratio_integrand})} appears to include divergent `octagons', only the finite octagons have non-vanishing coefficients. For five-particles, for example, the coefficient of the octagon $I_8(1,3,4,5)$ from \mbox{equation (\ref{octagon_NMHV_ratio_integrand})} would be \mbox{$[1,\{2,3\},\{4,5\}]=[1,2,3,4,5]+[1,2,3,5,4]=0.$}

Combining \mbox{formula (\ref{octagon_NMHV_ratio_integrand})} with the analytic form of the general octagon integral given in \mbox{equation (\ref{general_octagon_integral})} immediately yields a concise, analytic, manifestly dual-conformally invariant, and manifestly-cyclic form of the 1-loop ratio function for any $n$. 

Let us close this section by given another explicit example. The 7-point NMHV 1-loop ratio function is straightforwardly found to be, \eq{\hspace{-4.5cm}\mathcal{R}_{7,3}^{1\mathrm{-loop}}=[1,\{2,3\},\{4,5,6\}]I_8(1,3,4,6)+[1,\{2,3\},\{4,5,6,7\}]I_8(1,3,4,7)+\mathrm{cyclic,}\hspace{-2cm}}
where
\eqs{\hspace{-0.75cm}\begin{array}{l}I_8(1,3,4,6)\equiv\\\phantom{\Big\{}\\\phantom{\Big\{}\end{array}&\begin{array}{llll}\Big\{\!\!\!\phantom{\,+\,}\Li\left(1-u_{1,3,4,6}\right) &+\Li\left(1-u_{2,4,5,1}\right) &+\Li\left(1-u_{4,6,7,2}\right) &+\Li\left(1-u_{7,2,3,5}\right)
\\\phantom{\Big\{\!\!\!}-\Li\left(1-u_{2,4,5,7}\right) &-\Li\left(1-u_{3,6,7,2}\right) &-\Li\left(1-u_{4,6,7,3}\right) &-\Li\left(1-u_{6,1,2,4}\right)\\\multicolumn{2}{l}{\phantom{\Big\{\!\!\!}+\log\left(u_{1,3,4,7}\right) \log\left(u_{3,5,6,2}\right)\Big\};}\end{array}\\
\hspace{-0.75cm}\begin{array}{l}I_8(1,3,4,7)\equiv\\\phantom{\Big\{}\end{array}&\begin{array}{llll}\multicolumn{2}{l}{\Big\{\!\!\phantom{\,+\,}\log\left(u_{1,3,4,7}\right) \log\left(u_{3,6,7,2}\right)-\Li(1)}&-\Li\left(1-u_{1,3,4,6}\right) &-\Li\left(1-u_{4,6,7,2}\right) \\\phantom{\Big\{\!\!\!\,\,}+\Li\left(1-u_{1,3,4,7}\right) &+\Li\left(1-u_{3,6,7,2}\right) &+\Li\left(1-u_{4,6,7,3}\right) &+\Li\left(1-u_{6,1,2,4}\right)\Big\}.\end{array}\nonumber}
(Here, we have not neglected an overall factor of $\frac{1}{2}$: like in the case of the 6-point ratio function---the summand in \mbox{equation (\ref{octagon_NMHV_ratio_integrand})} includes exactly two copies of each term; but this is not generally the case for higher-$n$).

\newpage

\section{Multiloop Amplitudes}\label{multiloop_amplitudes_section}

In this section, we introduce a new strategy for finding local
representations of loop integrands. The idea is closely related to the
leading singularity method, but the philosophy differs in some
important ways. In particular we will {\it not} be guided by
systematically trying to match all the leading singularities of the
integrand. Instead, we will look at a simple subset of
leading singularities defined for generic, large enough number of particles
--- no ``composite" leading singularities will be considered.
We will then find a natural set of pure integrals designed to match
this subset of leading
singularities. We will find that boldly summing over all such objects
miraculously suffices to match the full integrand! In particular,
while the pure integrals are
motivated for a large-enough generic number of external particles,
their degenerations nicely produce all the needed lower-point objects as well.

This method is
heuristic --- we do not yet have a deep understanding for why the
miracles happen. However we have used this strategy successfully to
find stunningly simple expressions for the integrands of all 2- and
3-loop MHV amplitudes as well as all 2-loop NMHV amplitudes, and have
checked that the results are correct by comparing with the form
obtained from the all-loop BCFW recursion.

We will begin by illustrating this strategy by going back to 1-loop
integrands, which will motivate structures for 1-loop
integrands different from the ones we encountered in section 3. For the MHV
integrand, this new form coincides with one of ``polytope
representations" discussed in \cite{polytopePaper}. We will then use this
discussion as a springboard to our treatment of 2- and 3-loop
integrands.

\subsection{New form for the MHV 1-loop Integrand}

Let's begin by going back to the MHV 1-loop integrand, and motivate a
new form for it inspired by straightforwardly matching its leading
singularities, associated with the familiar two-mass-easy colored diagrams
\vspace{-0.4cm}\eqs{\hspace{-0.5cm}\nonumber\raisebox{-2.5cm}{\includegraphics[scale=0.5]
{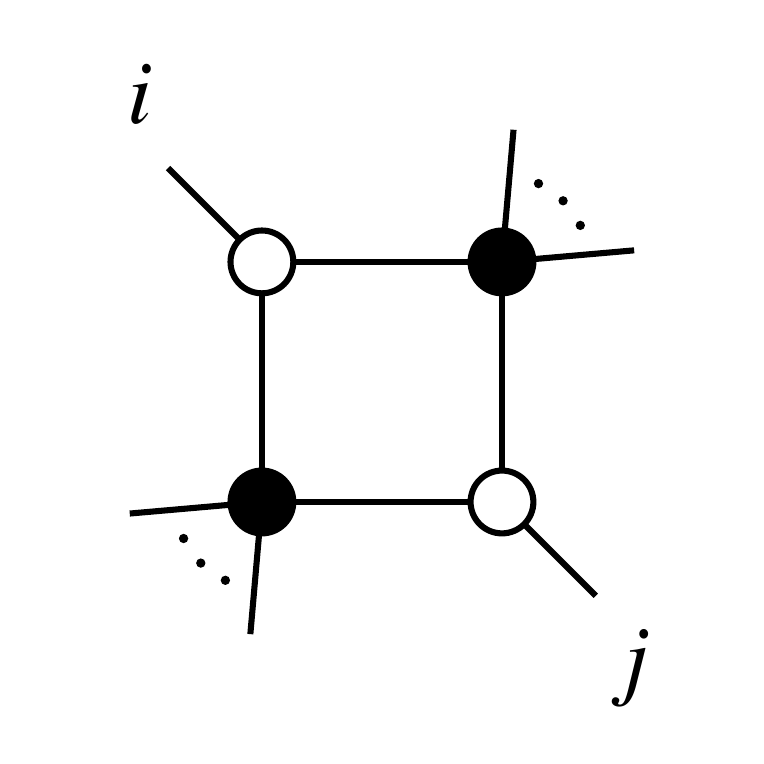}}\raisebox{-2.25cm}{\includegraphics[scale=0.45]
{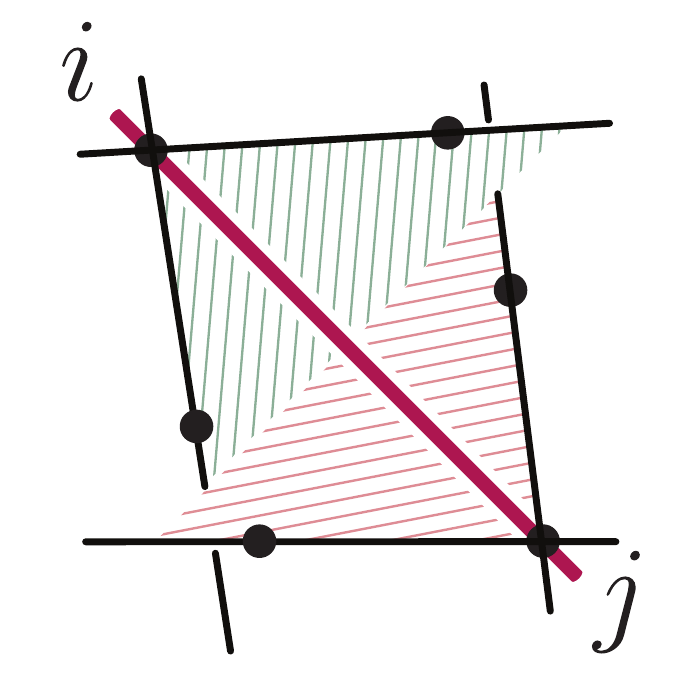}}\raisebox{-2.25cm}{\includegraphics[scale=0.45]
{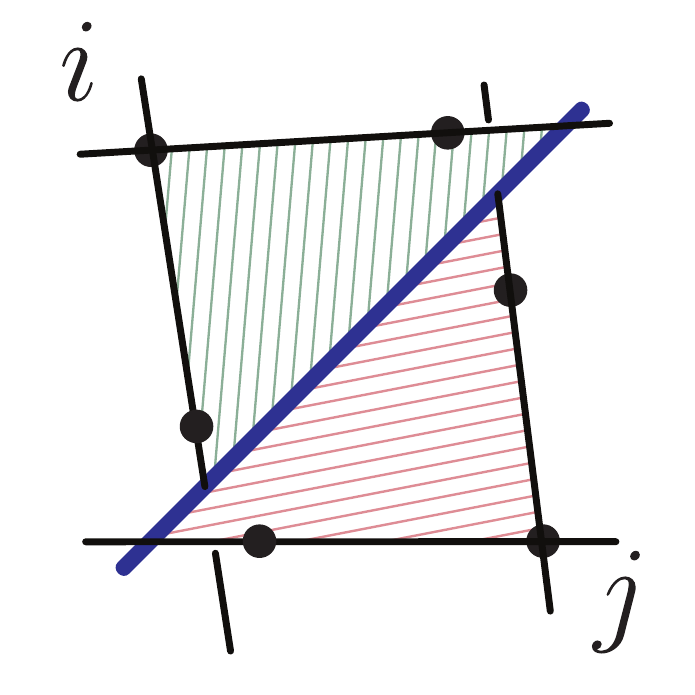}}\raisebox{-2.5cm}{\includegraphics[scale=0.5]
{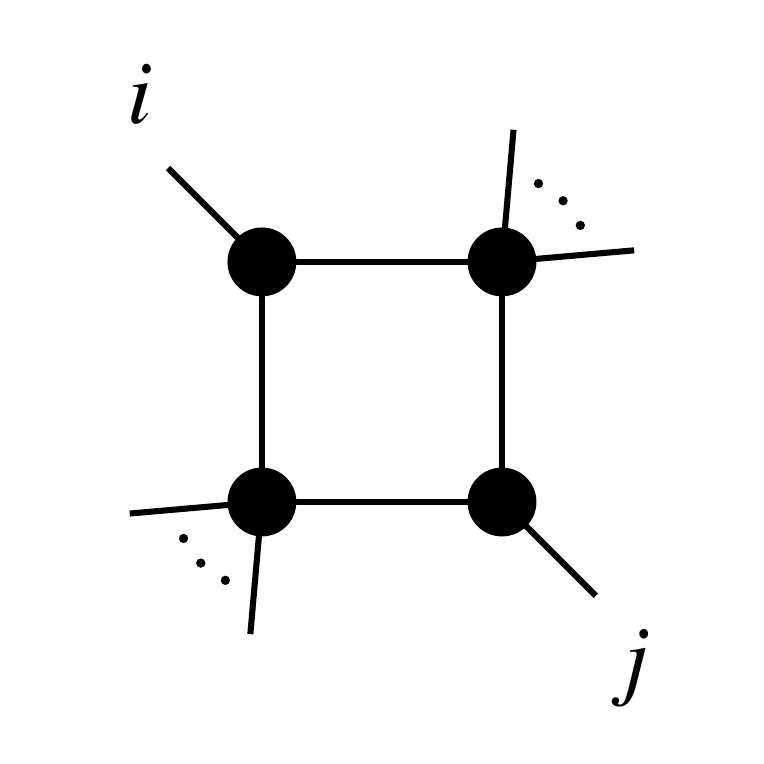}}\vspace{-0.4cm}}
\noindent corresponding to cutting the propagators
\eq{\la AB\,i\mi1\,i\ra\la AB\,i\,i\pl1\ra\la AB\,j\mi1\,j\ra\la
AB\,j\,j\pl1\ra \qquad
\label{cut1}}
The amplitude has unit leading singularity for the first solution of
the Schubert problem $(AB) = (ij)$,
and vanishing leading singularity for the second solution
where$AB=(i\mi1\,i\,i\pl1)\newcap(j\mi1\,j\,j\pl1)$.
We would like to build the integrand out of objects that have exactly
this property. To
beat a dead horse yet again --- it is obvious that the two-mass-easy
box does not do this job because
it is not chiral. The easiest way to do this is to simply insert a
factor in the numerator, $\la AB\,(i\mi1ii\pl1)\newcap(j\mi1jj\pl1)\ra$,
that kills the ``wrong" leading singularity. For correct little-group
weights, we add a factor $\ab{ABX}$ in the denominator,
where $X$ is an
arbitrary bitwistor, and look at an object of the form
\eq{I_{i,j} = \frac{\la
AB\,(i\mi1\,i\,i\pl1)\newcap(j\mi1\,j\,j\pl1)\ra\la X\,i\,j\ra} {\la
AB\,i\mi1\,i\ra\la AB\,i\,i\pl1\ra\la AB\,j\mi1\,j\ra\la
AB\,j\,j\pl1\ra\la AB\,X\ra}}\label{pentaX}
which is just the pentagon already familiar from
section 2, where the local propagator $\la AB\,n\,1\ra$ has been replaced
by $\la ABX\ra$. We denote this graphically as
\vspace{-0.1cm}\eq{\raisebox{-1.5cm}{\includegraphics[scale=0.65]{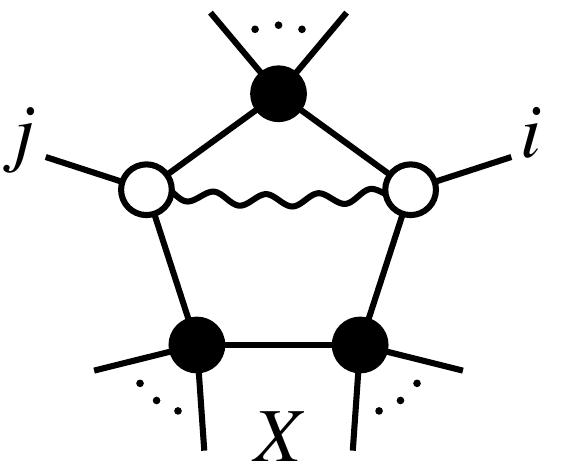}}\vspace{-0.1cm}}
Note that there is in general no significance to the presence of the
legs adjacent to $X$ in this picture. We draw it in this way because
in the special case where $X = (k \, k\!+\!1)$, the legs adjacent to
$X$ are identified with $k$,$k\!+\!1$.

Now consider the Schubert problems associated with cutting four
physical propagators. By construction this object has vanishing
leading singularities on the ``wrong" solution, and can easily be seen
to have unit leading singularity on the ``right" one.  Summing over
all the indices $i<j$---with $|i-j| \geq 2$ corresponding to the
two-mass easy colored graphs---produces an object matching all the
physical leading singularities of the amplitude. Naively this should
give us the integrand, but there is a catch: each term also has
``spurious cuts" where $\ab{ABX}$ is one on the cut propogators.
Indeed, the sum we just described does not match the integrand.

However some wonderful magic happens:
the sum over ${\it all}$ indices $i<j$, including a ``boundary term
"with $j = i \!+\! 1$, which is not included in the sum over colored
graphs,{\it does} reproduce the amplitude! We have
\eq{A_{{\rm MHV}}^{{\rm 1-loop}} = \sum_{i<j} \frac{\la
AB\,(i\mi1ii\pl1)\newcap(j\mi1jj\pl1)\ra\la X\,i\,j\ra} {\la AB\,X\ra
\la AB\,i\mi1\,i\ra\la AB\,i\,i\pl1\ra\la AB\,j\mi1\,j\ra\la
AB\,j\,j\pl1\ra}\label{MHV1loop2}}
or in the form of picture
\eq{\mathcal{A}^{{\rm 1-loop}}_{{\rm
MHV}}=\scalebox{1.05}{{\Large$\displaystyle\sum_{\text{{\footnotesize$i\!<\!j\!<\!i$}}}\,\,\left\{\phantom{\sum_{\text{{\normalsize$i<j$}}}}\hspace{1.5cm}\right\}$}}\hspace{-3.27505cm}\raisebox{-1.05cm}{\includegraphics[scale=0.5]{mhv_one_loop_pentagon.pdf}}\qquad.}
This form is manifestly cyclic but has spurious
$\ab{ABX}$ poles term-by-term. The sum is however independent of $X$.
If we choose $X$ to correspond to one of the external point  $X = (k
\, k \! + \! 1)$, all the poles are manifestly physical but the
formula is not manifestly cyclic invariant.

As mentioned above, this expression follows from a simple ``polytope"
interpretation \cite{polytopePaper}.
The local formula given in \cite{ArkaniHamed:2010kv} is obtained by
choosing $X=kk\pl1$, summing over all $k$ and dividing by $1/n$. The
similar expression in \cite{CaronHuot:2010ek} corresponds to setting
$X=I_{\infty}$ where
$I_{\infty}$ is infinity twistor.

Let us look at the ``boundary term" where $j = i \!+ \!1$ in more
detail--using $\la i\mi 1\,i\,i\pl1\,i\pl2\ra \la AB\,i\,i\pl1\ra =\la
AB\,(i\mi1ii\pl1)\newcap(i\,i\pl1i\pl2)\ra$, we can see that it is just a
(spurious)
box
\eq{\frac{\la i\mi 1\,i\,i\pl1\,i\pl2\ra\la X\,i\,i\pl1\ra} {\la
AB\,X\ra \la AB\,i\mi1\,i\ra\la AB\,i\,i\pl1\ra\la
AB\,i\pl1\,i\pl2\ra} \label{spbox}}
It is instructive to explicitly understand the purpose of this
boundary term in this simple example, since the same phenomenon will
occur in all the rest of our examples in this section. Let us return
to our most naive ansatz, summing only over the pentagons associated
with the colored graphs. Each of the spurious cuts involving
$\ab{ABX}$, such as
\eq{\la ABX\ra\la AB\,i\mi1\,i\ra \la AB\,i\,i\pl1\ra\la
AB\,j\mi1\,j\ra \label{cut2}}
is shared by two pentagons e.g. $I_{i,j-1}$ and
$I_{i,j}$. For generic terms in the sum, these cuts cancel against
each other in pairs.
However, in the limiting
cases when $j=i\pl2$ (or $j=i\mi2$) the
quad-cut is shared by $I_{i,i+2}$ and $I_{i-1,i+1}$ but there is no
cancelation between them because the non-vanishing leading
singularities occur for two different solutions of the Schubert
problems. The spurious
box of (\ref{spbox}) precisely has non-vanishing leading singularities
for these two Schubert problems and completes the cancelation of all
$\ab{ABX}$ poles, ensuring the full sum is independent of $X$.  It is
quite remarkable that the ``new" object needed to
fix the leading singularities and match the amplitude is simply a
degeneration of the pentagon itself.

In our remaining examples, we will not delve into understanding the
details of how
all leading singularities match. We will instead take a class of leading
singularities as a guide for the local integrals to consider, and sum
over all the relevant objects, including boundary terms that do not
directly correspond to any of the leading singularity
pictures that motivated the construction of the objects to begin with.
These formulae are then verified by comparing with the
integrand as computed by BCFW recursion.

Let us finally note a very pretty property of eqn.(\ref{MHV1loop2}): for
generic $X$, all the pentagons
in the double sum are manifestly manifestly IR finite. This ceases to
be true if we make the special choice like $X = (12)$, since the
diagrams with $i=2$ or $j=n$
have an additional massless corner which is not controlled by the
numerator.

\eq{\figBox{0}{-2}{0.7}{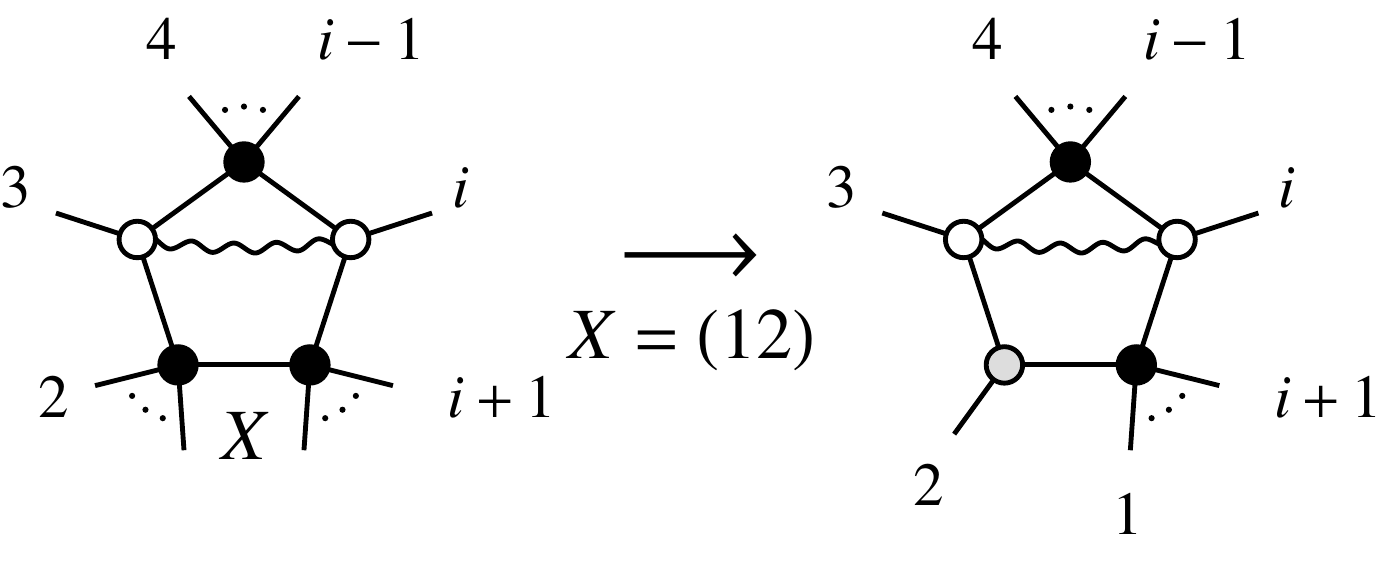}}

\subsection{NMHV 1-loop integrand revisited}

We proceed to use the same strategy to determine a local expression for the
NMHV 1-loop integrand, which will yield a quite different form than we
obtained in section 3. We again
start with the colored graphs for leading singularities. There are two of
them for NMHV amplitudes:
\vspace{-0.2cm}\eq{\figBox{0}{-1.6}{0.45}{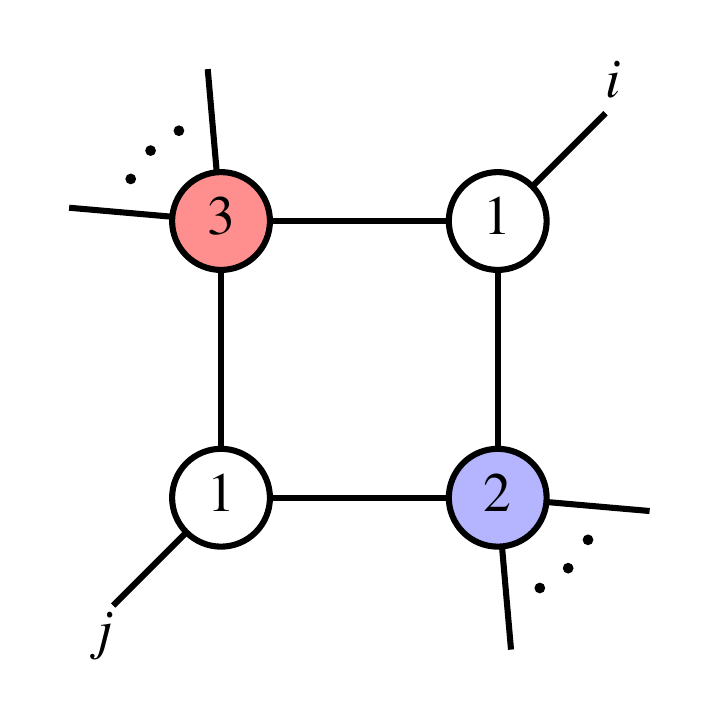}\qquad\mathrm{and}\qquad\figBox{0}{-1.6}{0.45}{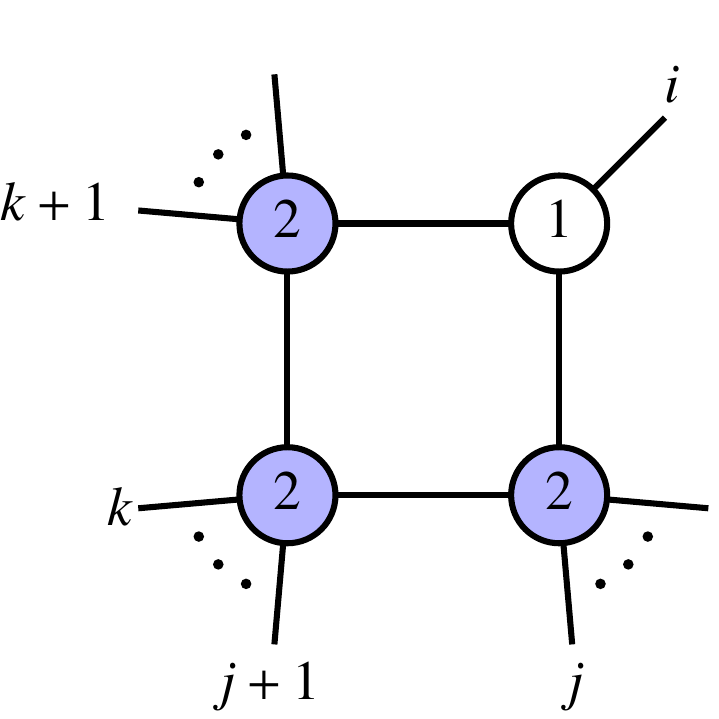}\vspace{-0.2cm}}

Unlike the MHV case where the non-vanishing leading singularities were
``1", here the non-vanishing leading singularities are the
$R$-invariants. The goal is to find objects with non-vanishing support
on the same Schubert problems as the amplitude, and decorate these
with the appropriate $R$-invariants to get a nice ansatz for the
integrand.

The first colored graph correspond to 2-mass easy Schubert problems
and have the same structure as the MHV case. The leading singularity
is just the tree-level amplitude appearing in the
upper-left corner of the figure, $\mathcal{A}^{{\rm tree}}_{{\rm
NMHV}}(j,j\pl1,\ldots
i\mi1,i)$. Thus we expect to have objects in the integrand of the form

\eq{\hspace{-3.5cm}\scalebox{1.00}
{{\Large$\displaystyle\sum_{\text{{\footnotesize$i\!<\!j\!<i$}}}\,\,\left\{\phantom{\sum_{\text{{\normalsize$i<j$}}}}
\hspace{1.6cm}\;\text{{\normalsize$\times\,\mathcal{A}^{{\rm
tree}}_{{\rm NMHV}}(j,j\pl1,\ldots,i\mi1,i)$}}\right\}$}}\hspace{-8.1505cm}\raisebox{-1.05cm}{\includegraphics[scale=0.5]{mhv_one_loop_pentagon.pdf}}\qquad\label{NMHVgraph1}}

Finding an object matching the physical leading singularities of the
second class of colored diagrams is a more interesting exercise. The
cut propogators are
\eq{\la AB\,i\mi1\,i\ra\la ABi\,i\pl1\ra\la ABj\,j\pl1\ra\la
AB\,k\,k\pl1\ra}
The leading singularities vanish for the solution
$AB=(i\mi1\,i\,i\pl1)\newcap(j\,j\pl1)(i\mi1\,i\,i\pl1)\newcap(k\,k\pl1)$,
while for $AB=(i\,j\,j\pl1)\newcap(i\,k\,k\pl1)$ the leading singularity
is $\left[i,j,j\pl1,k,k\pl1 \right]$.

Let us consider objects of the form
\vspace{-0.2cm}\eq{\nonumber I_{i,j,k}\equiv\figBox{0}{-1.04}{0.56}{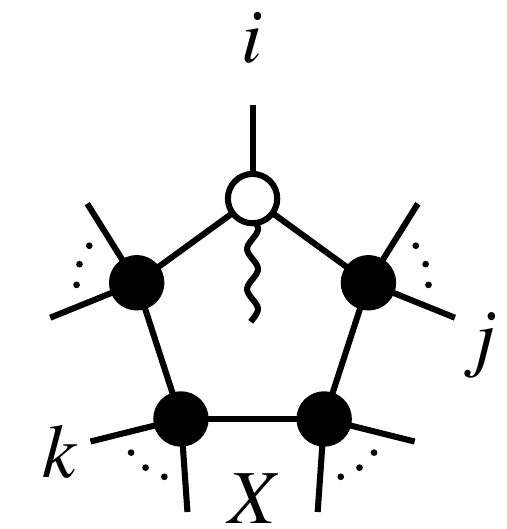}=\int\limits_{AB}\frac{N(i,j,k)}{\ab{AB\,X}\ab{AB\,i\mi1\,i}\ab{AB\,i\,i\pl1}\ab{AB\,j\,j\pl1}\ab{AB\,k\,k\pl1}}\label{NMHVgraph2}}
We are searching for a numerator supported on the same leading singularities
as the amplitude. In addition it should also have unit
leading singularity on all other
spurious quad-cuts. The reason is that the spurious cuts must cancel
in a sum over terms;
since the integrals are multiplied by different $R$-invariants, the
only way this can happen is through residue theorem 6-term identities
between the $R$-invariants.
For instance the spurious quad-cut
\eq{\la ABX\ra \la ABi\,i\pl1\ra\la ABj\,j\pl1\ra\la
ABk\,k\pl1\ra\label{spur1}}
is shared by six different integrals $I_{i;j,k}$, $I_{i\pl1;j,k}$,
$I_{j;i,k}$, $I_{j\pl1;i,k}$, $I_{k;i,j}$ and $I_{k\pl1;i,j}$ that
are multiplied by six different residues. There is a 6-term identity
relating them
\begin{align} \left[i,j,j\pl1,k,k\pl1\right] &+
\left[i\pl1,j,j\pl1,k,k\pl1\right] +
\left[j,i,i\pl1,k,k\pl1\right] \nonumber\\&+
\left[j\pl1,i,i\pl1,k,k\pl1\right] +
\left[k,i,i\pl1,j,j\pl1\right] + \left[k\pl1,i,i\pl1,j,j\pl1\right] =
0 \nonumber
\end{align}
which can only possibly be of help in canceling spurious cuts if the
integrands they multiply have support on the same Schubert problems,
with unit leading singularities.

There is one final guiding principle for determining the structure of
the numerator $N(i,j,k)$. The
topologies occurring in (\ref{NMHVgraph1}) are the same as for the MHV
amplitude, while the second class of integrals is ``purely" NMHV-like.
Since IR-divergences are universal, it would be nice if the IR-divergences could be completely isolated in the MHV-like topology. We should then try to choose the numerator $N(i,j,k)$ to be strictly finite. 
It would be nice if these integrals could be chosen to be manifestly
finite. The only divergence in (\ref{NMHVgraph2}) can come from the
$Z_i$-corner, {\it i.e.}\ the region when
$AB$ crosses point $Z_i$ and lies in the plane $(i\mi1\,i\,i\pl1)$.
In order to control this region the numerator should be of the form
$N = \la AB\,(i\mi1\,i\,i\pl1)\newcap(\dots)\ra$. Combined with the
unit leading singularity constraint, the form of the numerator is
fixed completely:
\eq{\hspace{-0.5cm}N(i,j,k)\equiv\la
AB\,(i\mi1\,i\,i\pl1)\newcap\Sigma_{i,j,k}\ra}
with $\Sigma_{i,j,k}$ a special plane defined according to \eq{\Sigma_{i,j,k}\equiv\frac{1}{2}\big[(j\,j\pl1\,(i\,k\,k\pl1)\newcap
X)-(k\,k\pl1\,(i\,j\,j\pl1)\newcap X)\big]}
This is in fact the only choice we could have made consistent with
little group weights and the desire to treat the $j,k$ indices
symmetrically. We will denote this by,
\vspace{-0.25cm}
{\eqs{\hspace{-1.2cm}\nonumber
\figBox{0}{-1.04}{0.6}{nmhv_one_loop_pentagon.pdf}=\int\limits_{AB}\frac{\la
AB\,(i\mi1\,i\,i\pl1)\newcap\Sigma_{i,j,k}\ra}{\ab{AB\,X}\ab{AB\,i\mi1\,i}\ab{AB\,i\,i\pl1}\ab{AB\,j\,j\pl1}\ab{AB\,k\,k\pl1}}}}

With these objects in hand, we once again brazenly sum over all ranges
of indices, including ``boundary" terms with $j = i \pm 1$ not
directly associated with colored graphs for leading singularities.
The same magic happens as we saw in the MHV case--- this sum agrees with
the 1-loop NMHV amplitude as computed by BCFW recursion, and we find,
\vspace{-0.4cm}\eqs{\hspace{-5cm}\mathcal{A}^{{\rm 1-loop}}_{{\rm
NMHV}}=&\scalebox{1.0}{{\Large$\displaystyle\sum_{\text{{\footnotesize$i\!<\!j\!<\!k\!<i$}}}\,\,\hspace{-0.2cm}\left\{\phantom{\sum_{\text{{\normalsize$i<j$}}}}\hspace{1.25cm}\;\text{{\normalsize$\times\,\left[i,j,j+1,k,k+1\right]$}}\right\}$}}\hspace{-7.215cm}\figBox{0.35}{-0.8}{0.5}{nmhv_one_loop_pentagon.pdf}\qquad\\[-0.1cm]
\hspace{-4cm}&+\!\!\!\scalebox{1.00}{{\Large$\displaystyle\sum_{\text{{\footnotesize$i\!<\!j\!<i$}}}\,\,\left\{\phantom{\sum_{\text{{\normalsize$i<j$}}}}\hspace{1.25cm}\;\text{{\normalsize$\times\,\mathcal{A}^{{\rm
tree}}_{{\rm NMHV}}(j,j\pl1,\ldots,i\mi1,i)$}}\right\}.$}}\hspace{-7.85cm}\raisebox{-1.1cm}{\includegraphics[scale=0.5]{mhv_one_loop_pentagon.pdf}}\label{NMHV1loop}}

Note also that as in the MHV case, the only IR divergent integrals are
in the boundary terms. The (generically) finite integrals for
$I_{i,j,k}$ are given by

\vspace{-0.5cm}
\eqs{\nonumber\hspace{-1cm} I_{i,j,k}=-\Li\left(1-u_{1}\right)
-\Li\left(1-u_{2}\right) +\Li\left(1-u_{3}\right)
+\log\left(u_{4}\right)\log\left(u_{6}\right)+\deltatilde\left(u_{4},v_{4}\right)+\deltatilde\left(u_{5},v_{5}\right)}
where the cross ratios are defined as:
{\footnotesize\eqs{\hspace{-2cm}\nonumber\hspace{-2.5cm}&\hspace{1.9cm}u_{1}\equiv
\frac{\ab{{\color{cr_side1} i\,i\pl1\,}{\color{cr_blue}
j\,j\pl1}}\ab{{\color{cr_red} X}{\color{cr_side2}
i\mi1\,i}}}{\ab{{\color{cr_side1}
i\,i\pl1\,}{\color{cr_red}X}}\ab{{\color{cr_blue}
j\,j\pl1\,}{\color{cr_side2} i\mi1\,i}}},\qquad  u_{2}\equiv
\frac{\ab{{\color{cr_side1}
i\,i\pl1\,}{\color{cr_blue}X}}\ab{{\color{cr_red}
k\,k\pl1\,}{\color{cr_side2} i\mi1\,i}}}{\ab{{\color{cr_side1}
i\,i\pl1\,}{\color{cr_red}
k\,k\pl1}}\ab{{\color{cr_blue}X\,}{\color{cr_side2} i\mi1\,i}}},\qquad
 u_{3}\equiv \frac{\ab{{\color{cr_side1} i\,i\pl1\,}{\color{cr_blue}
j\,j\pl1}}\ab{{\color{cr_red} k\,k\pl1\,}{\color{cr_side2}
i\mi1\,i}}}{\ab{{\color{cr_side1} i\,i\pl1\,}{\color{cr_red}
k\,k\pl1}}\ab{{\color{cr_blue} j\,j\pl1\,}{\color{cr_side2}
i\mi1\,i}}},\qquad  \\
\hspace{-1.75cm}&u_{4}\equiv
\frac{\ab{{\color{cr_side1}X\,}{\color{cr_blue}
k\,k\pl1}}\ab{{\color{cr_red} i\mi1\,i\,}{\color{cr_side2}
j\,j\pl1}}}{\ab{{\color{cr_side1}X\,}{\color{cr_red}
i\mi1\,i}}\ab{{\color{cr_blue} k\,k\pl1\,}{\color{cr_side2}
j\,j\pl1}}},
\;  v_{4}\equiv \frac{\ab{{\color{cr_side1}
j\,j\pl1\,}{\color{cr_blue}X}}\ab{{\color{cr_red}
k\,k\pl1\,}{\color{cr_side2} i\mi1\,i}}}{\ab{{\color{cr_side1}
j\,j\pl1\,}{\color{cr_red}
k\,k\pl1}}\ab{{\color{cr_blue}X\,}{\color{cr_side2} i\mi1\,i}}},
\;  u_{5}\equiv \frac{\ab{{\color{cr_side1}
j\,j\pl1\,}{\color{cr_blue}X}}\ab{{\color{cr_red}
k\,k\pl1\,}{\color{cr_side2} i\,i\pl1}}}{\ab{{\color{cr_side1}
j\,j\pl1\,}{\color{cr_red}
k\,k\pl1}}\ab{{\color{cr_blue}X\,}{\color{cr_side2} i\,i\pl1}}},
\;  v_{5}\equiv \frac{\ab{{\color{cr_side1}X\,}{\color{cr_blue}
k\,k\pl1}}\ab{{\color{cr_red} i\,i\pl1\,}{\color{cr_side2}
j\,j\pl1}}}{\ab{{\color{cr_side1}X\,}{\color{cr_red}
i\,i\pl1}}\ab{{\color{cr_blue} k\,k\pl1\,}{\color{cr_side2}
j\,j\pl1}}}}}

Finally, let us examine the 1-loop NMHV ratio function

\eq{\mathcal{R}^{{\rm 1-loop}}_{{\rm NMHV}} = \mathcal{A}^{{\rm
1-loop}}_{{\rm NMHV}} - \mathcal{A}^{{\rm 1-loop}}_{{\rm MHV}}\cdot
\mathcal{A}^{{\rm tree}}_{{\rm NMHV}}}

Comparing the expressions \ref{MHV1loop2} and \ref{NMHV1loop} we
can see that the ratio function has the same form as NMHV amplitude,
except that
in the first sum we have $\mathcal{A}^{{\rm tree}}_{{\rm
NMHV}}(i,i\pl1,\dots j\mi1,j) -
\mathcal{A}^{{\rm tree}}_{{\rm NMHV}}$ instead of just
$\mathcal{A}^{{\rm tree}}_{{\rm NMHV}}(i,i\pl1,\dots
j\mi1,j)$. The
manifest finiteness is obvious. The only divergent integrals are in
the boundary term $j=i-1$, but their coefficient is given by
$\mathcal{A}^{{\rm tree}}_{{\rm NMHV}}(i,i\pl1,\dots
j\mi1,j)-\mathcal{A}^{{\rm tree}}_{{\rm NMHV}}(1,\ldots,n) =
\mathcal{A}^{{\rm tree}}_{{\rm NMHV}}(i,i\pl1,\dots i\mi2,i\mi1) -
\mathcal{A}^{{\rm tree}}_{{\rm NMHV}}(1,\ldots,n)=0$.
Therefore, the ratio function can be written only using
manifestly finite integrals.

\subsection{The 2-loop MHV amplitude and its logarithm}

Now we turn to the 2-loop case. First we reproduce the MHV amplitude
presented already in \cite{ArkaniHamed:2010kv} and in addition we will write an
expression for the log of the amplitude given in an interesting form
in terms of non-planar diagrams.

We again start with colored graphs,

\eq{\hspace{-0.25cm}\figBox{0}{-1.7}{0.5}{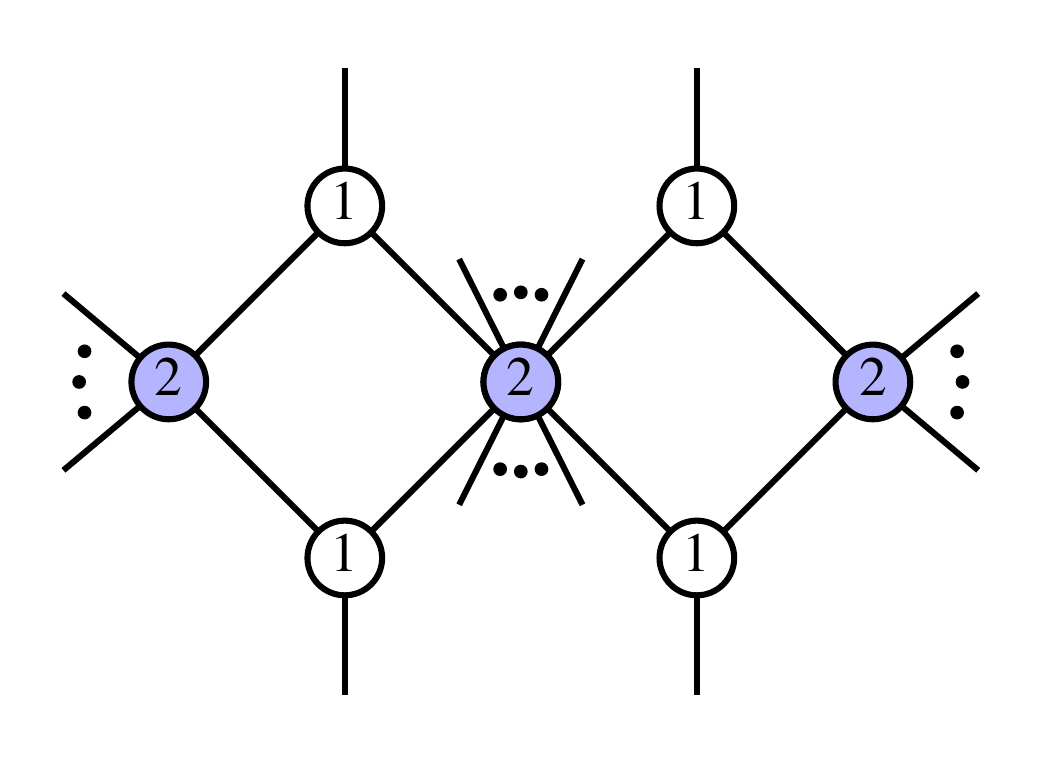}\;\figBox{0}{-1.7}{0.5}{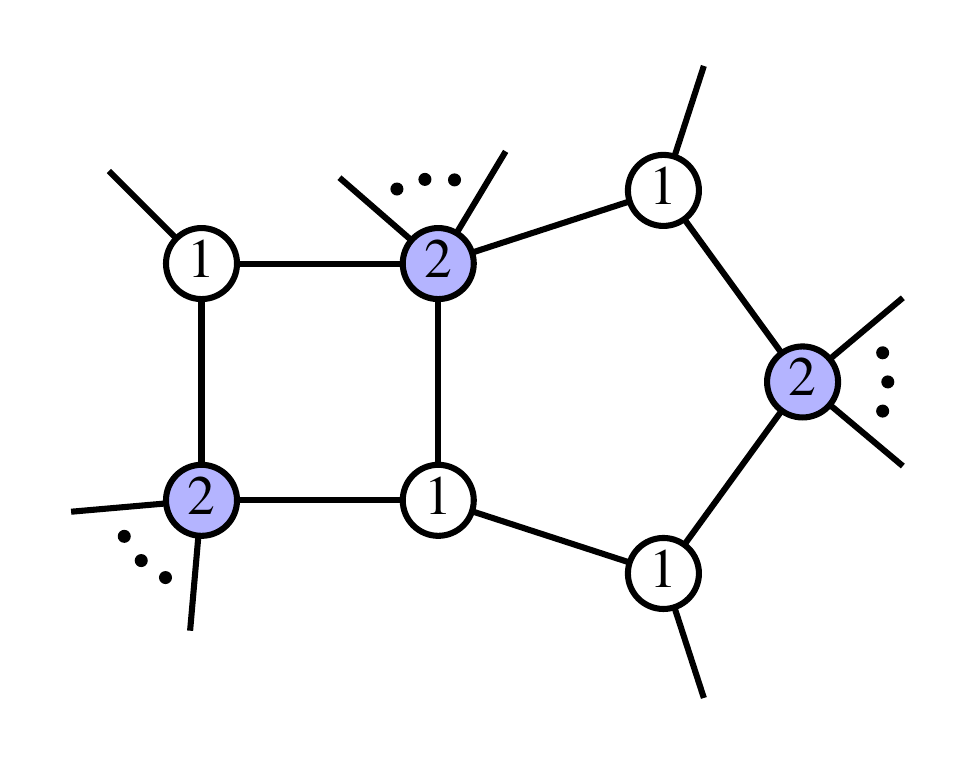}\!\!\figBox{0}{-1.3}{0.5}{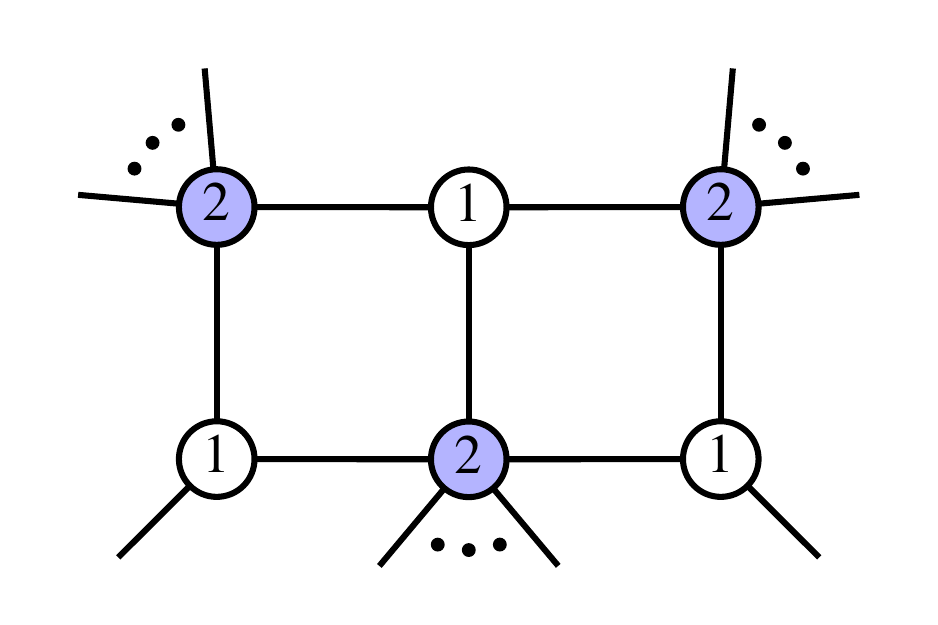}\hspace{-2cm}\label{MHVcol}}

There are more types of graphs in comparison to 1-loop where we had
only boxes. In addition to two glued boxes (also referred to as ``kissing
boxes") we have other topologies---pentaboxes and double-boxes. They
represent cutting the internal $\la ABCD\ra$ propagator once and twice
respectively, the latter case corresponding to ``composite" leading
singularities.

Let us concentrate on the first graph. It looks like a ``squaring" of
the 1-loop cuts with appropriate ranges for indices. And in fact, the
$AB$ part and $CD$ part of the integral are independent, ie. in
order to realize the octa-cut of the first colored graph, we need to
set $AB=ij$ or $AB=(i\mi1\,i\,i\pl1)\newcap(j\mi1\,j\,j\pl1)$ and
$CD=k\ell$ or $CD=(k\mi1\,k\,k\pl1)\newcap(\ell-1\,\ell\,\ell\pl1)$.
Together we have four possible combinations. The amplitude (as we
see from the colored graph) has support just on one of them $(AB)=ij$
and $CD=k\ell$ while for all other it vanishes. It means that
the numerator must vanish whenever
$AB=(i\mi1\,i\,i\pl1)\newcap(j\mi1\,j\,j\pl1)$ or
$CD=(k\mi1\,k\,k\pl1)\newcap(\ell-1\,\ell\,\ell\pl1)$. This
motivates us to start with an integral of the form
\eq{\nonumber\hspace{-1.25cm}\figBox{0}{-1.4}{0.6}{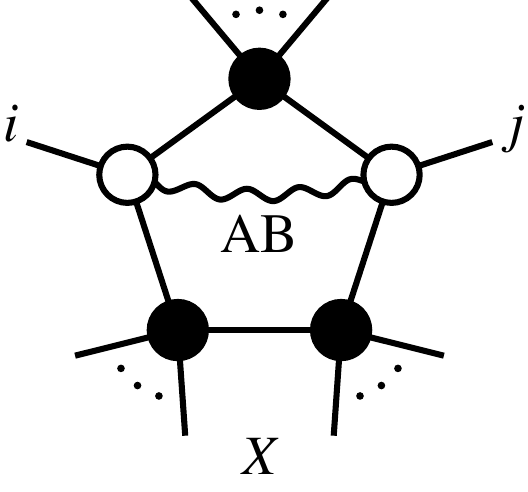}\!\!\times\figBox{-0.25}{-1.4}{0.6}{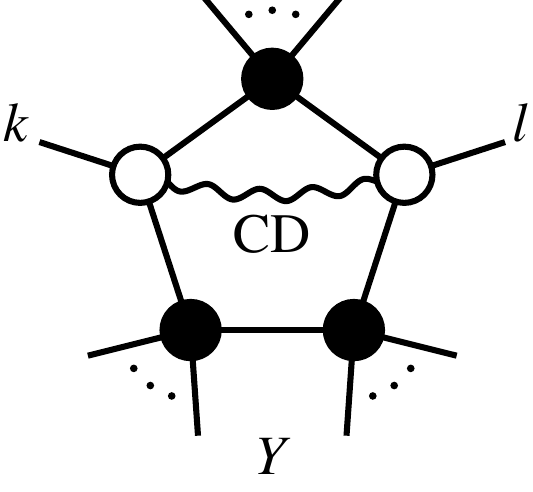}\!\!\!\!\!=\left\{\begin{array}{l}\displaystyle\phantom{\,\times\,}\frac{\ab{AB\,(i\mi1\,i\,i\pl1)\newcap(j\mi1\,j\,j\pl1)}\ab{X\,i\,j}}{\ab{AB\,X}\ab{AB\,i\mi1}\ab{AB\,i\,i\pl1}\ab{AB\,j\mi1\,j}\ab{AB\,j\,j\pl1}}\\\displaystyle\times\frac{\ab{CD\,(k\mi1\,k\,k\pl1)\newcap(l\mi1\,l\,l\pl1)}\ab{Y\,k\,l}}{\ab{CD\,Y}\ab{CD\,k\mi1k}\ab{CD\,k\,k\pl1}\ab{CD\,l\mi1\,l}\ab{CD\,l\,l\pl1}}
\end{array}\right\}}
which has exactly this property. However, there is a better
candidate. Instead of adding $\la ABX\ra$ and $\la CDY\ra$ in the
denominator, we
can add directly the internal propagator $\la ABCD\ra$. That allows
us to write two numerator factors exactly as we need. Therefore, we
consider,
\vspace{-0.2cm}\eq{\nonumber\hspace{-0.6cm}\figBox{0}{-1.82}{0.55}{two_loop_mhv_integrand.pdf}\!\!\!\!\!=\left\{\begin{array}{l}
\displaystyle\frac{\ab{AB\,(i\mi1\,i\,i\pl1)\newcap(j\mi1\,j\,j\pl1)}\ab{i\,j\,k\,l}}{\ab{AB\,i\mi1}\ab{AB\,i\,i\pl1}\ab{AB\,j\mi1\,j}\ab{AB\,j\,j\pl1}\ab{AB\,CD}}\\\displaystyle\times\frac{\ab{CD\,(k\mi1\,k\,k\pl1)\newcap(l\mi1\,l\,l\pl1)}}{\ab{CD\,k\mi1k}\ab{CD\,k\,k\pl1}\ab{CD\,l\mi1\,l}\ab{CD\,l\,l\pl1}}\end{array}\right\}\label{MHVInt}}
Of course, this integral has also many other cuts -- both composite
and non-composite --- that involve the
propagator $\la ABCD\ra$ , and we
have to match other colored graphs in (\ref{MHVcol}) as well. However,
just as in our  1-loop examples, simply summing over all indices with
a planar ordering
reproduces the full amplitude as a cyclic sum over just one integral
topology:
\vspace{-0.5cm}\eq{\mathcal{A}^{{\rm 2-loop}}_{{\rm
MHV}}=\frac{1}{2}\!\!\!\!\scalebox{0.6}{\raisebox{-0.4cm}{\scalebox{3}{$\begin{array}{c}\displaystyle\sum\\[-0.25cm]\scalebox{0.5}{{\normalsize$i\!<\!j\!<\!k\!<\!l\!<\!i$}}\end{array}$}\raisebox{-2.15cm}{\includegraphics[scale=0.8]{two_loop_mhv_integrand.pdf}}}}\vspace{-0.5cm}}
The ``boundary terms" in this case occur for for
$j=i\pl1$ and/or $l=k\pl1$. In these cases the numerator exactly
cancels one of the propagators, leaving us with\footnote{This
simplification was missed in
\cite{ArkaniHamed:2010kv}, and the 2-loop MHV integrand was presented as a
sum over three terms. We would like to thank Johannes Henn for
pointing the simplification out to us.}
\eq{\hspace{-1.85cm}\figBox{0}{-1.7}{0.45}{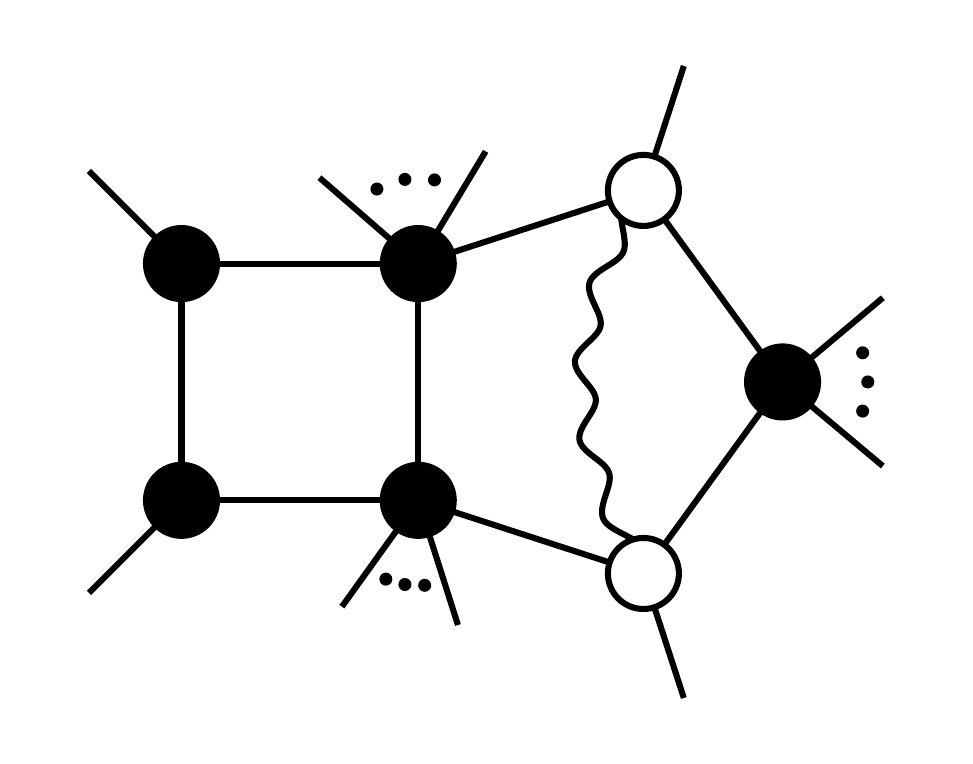}\;\figBox{0}{-1.7}{0.45}{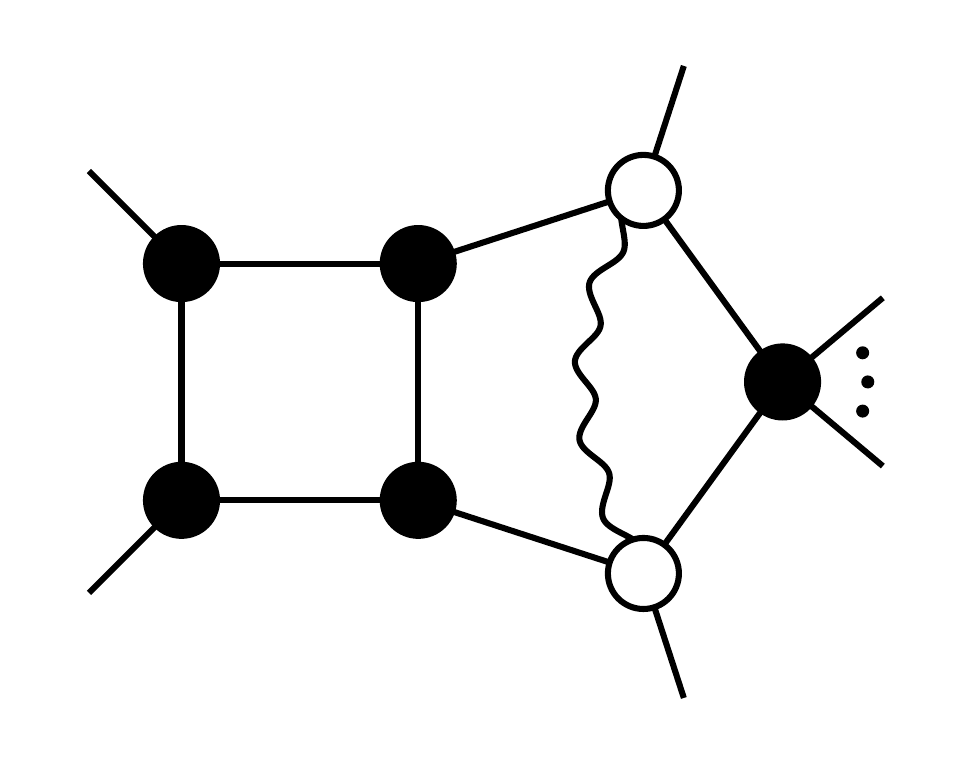}\;\figBox{0}{-1.15}{0.45}{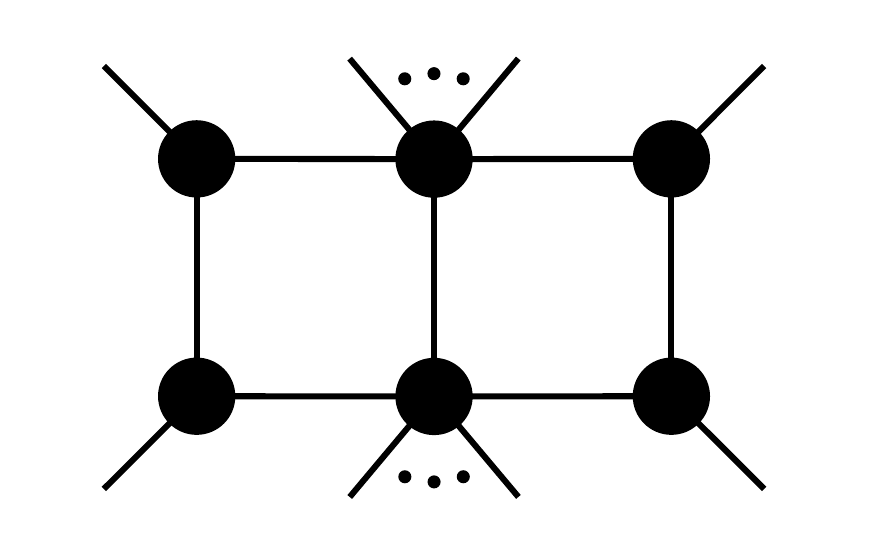}\hspace{-2cm}\label{MHVInt2}}

\subsubsection*{Log of the amplitude}

Finally, we give an interesting new expression for the logarithm of
the amplitude, using a non-planar sum of the same set of objects. At
2-loops, the log of the amplitude is

\eq{[{\rm log}\mathcal{A}]^{{\rm 2-loop}}_{{\rm MHV}} =
\left[\mathcal{A}^{{\rm 2-loop}}_{{\rm MHV}} -
\frac{1}{2}\left(\mathcal{A}^{{\rm 1-loop}}_{{\rm MHV}}\right)^2\right]}

A beautiful expression for the log of the amplitude is made possible by
the existence of a simple relation
between the sum of 1-loop square and 2-loop diagrams:
\eq{\hspace{-1cm}\scalebox{1}{{\huge$\displaystyle\sum_{\text{{\normalsize$i<j$}}}\,\,\hspace{-0.2cm}\left.\phantom{\sum_{\text{{\footnotesize$i<j$}}}}\hspace{3.75cm}\right.$}}\hspace{-5.315cm}\figBox{0}{-1.2}{0.5}{mhv_two_loop_pentagon_pairs_1.pdf}\!\times
\scalebox{1}{{\huge$\displaystyle\sum_{\text{{\normalsize$k<l$}}}\,\,\hspace{-0.2cm}\left.\phantom{\sum_{\text{{\footnotesize$i<j$}}}}\hspace{3.75cm}\right.$}}\hspace{-5.315cm}\figBox{0}{-1.2}{0.5}{mhv_two_loop_pentagon_pairs_2.pdf}\hspace{-0.05cm}=\scalebox{1}{{\huge$\displaystyle\sum_{\substack{\text{{\normalsize$i<j$}}\\\text{{\normalsize$k<l$}}}}\,\,\hspace{-0.2cm}\left.\phantom{\sum_{\substack{\text{{\footnotesize$i<j$}}\\\text{{\footnotesize$k<l$}}}}}\hspace{3.75cm}\right.$}}\hspace{-5.315cm}\figBox{0}{-1.65}{0.5}{two_loop_mhv_integrand.pdf}\hspace{-1cm}}

The left-hand side is just $(\mathcal{A}^{{\rm 1-loop}}_{{\rm
MHV}})^2$ while the right-hand side
contains not only the planar diagrams present
in $\mathcal{A}_{{\rm MHV}}^{{\rm 2-loop}}$ but also non-planar graphs
when for example \mbox{$i<k<j<l$}. In fact,
all planar graphs are equal to $2\mathcal{A}^{{\rm 2-loop}}_{{\rm
MHV}}$ while all non-planar
graphs give us the log of the amplitude in the form

\eq{\hspace{-0.5cm}[\log\mathcal{A}]^{{\rm 2-loop}}_{{\rm
MHV}}=-\hspace{-0.5cm}\scalebox{0.6}{\raisebox{-0.45cm}{\scalebox{3}{$\begin{array}{c}\displaystyle\sum\\[-0.25cm]\scalebox{0.5}{{\footnotesize$i\!<\!k\!<\!j\!<\!l\!<\!i$}}\end{array}$}\raisebox{-2.15cm}{\includegraphics[scale=0.8]{two_loop_mhv_integrand.pdf}}}}\label{pic14}}
The formula found in \cite{Drummond:2010mb} is the 4pt version of this
expression.

Note that naively, all these integrals are IR finite because each
individual 1-loop sub-integral is just
a finite pentagon(which can not shrink to a  box due to the restriction
$j\neq i+1$ and $l\neq k+1$). However, the criteria for finiteness we
described in section 4
applies to planar integrals, while the log contains non-planar terms
which can be IR divergent.

Let us focus on the piece of the integrand of the form
\eq{\frac{\la ABX\ra}{\la ABi\mi1i\ra\la ABii\pl1\ra}\cdot
\frac{1}{\la ABCD\ra}\cdot\frac{\la CDY\ra}{\la CDj\mi1j\ra\la
CDjj\pl1\ra}}
Here $X$ controls the IR divergence of the region where the line $AB$
intersects point $Z_i$ and lies in the plane
$(Z_{i\mi1}Z_iZ_{i\pl1})$, just as $Y$ does for $CD$ sector.
However, if $i=j$ then $AB$ and $CD$ intersect in the point $i$ and the
propagator $\la ABCD\ra$ vanishes. Therefore, finiteness of the
1-loop sub-integrals is not enough. We need an extra condition that
regulates this joint divergence. It is not hard to see that unless
$\la XY\ra=0$, a (mild) IR divergence remains.

As a result, we can find that almost all integrals in (\ref{pic14})
are finite except for the class of diagrams:
\eq{\figBox{0}{-1.45}{0.5}{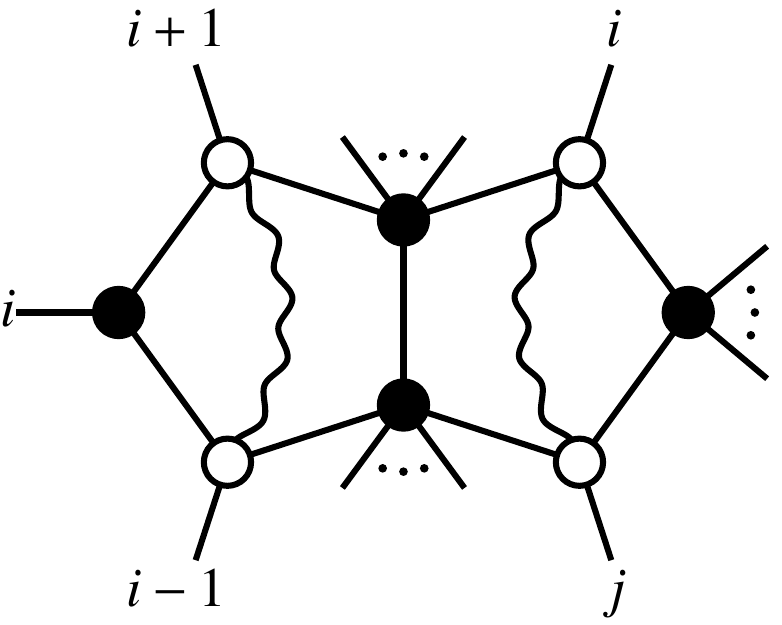}\label{pic15}}

In this case $X=(i\mi2\,i\mi1\,i)\newcap(i\,i\pl1\,i\pl2)$ and
$Y=(i\mi1\,i\,i\pl1)\newcap(j\mi1\,j\,j\pl1)$, so $\la XY\ra \neq 0$. However
the divergence is mild, as observed in the 4-pt result of \cite{Drummond:2010mb}.

\subsection{2-loop NMHV amplitude}

We move on to present the integrand for all 2-loop NMHV amplitudes.
The 6- and 7- point integrands were presented in
\cite{ArkaniHamed:2010kv}, by expanding the BCFW result into a basis
of pure integrals. The parity-even part of the 6- point integrand was
presented using standard (dual) space-time variables in
\cite{Kosower:2010yk}. Here, instead of a brute-force expansion into a
basis of integrals, we follow the same strategy outlined above,
obtaining results vastly simpler than those presented to date, which
also generalize to all $n$.Now 

Let us first start by drawing the colored-graphs that contribute
for general 2-loop NMHV amplitude that do not cut the internal propagator $\ab{AB\,CD}$.
\eq{\nonumber\hspace{-0.6cm}\begin{array}{c@{\hspace{-0.5cm}}c@{\hspace{-0.5cm}}c}
\begin{array}{c}\figBox{0}{-1.825}{0.5}{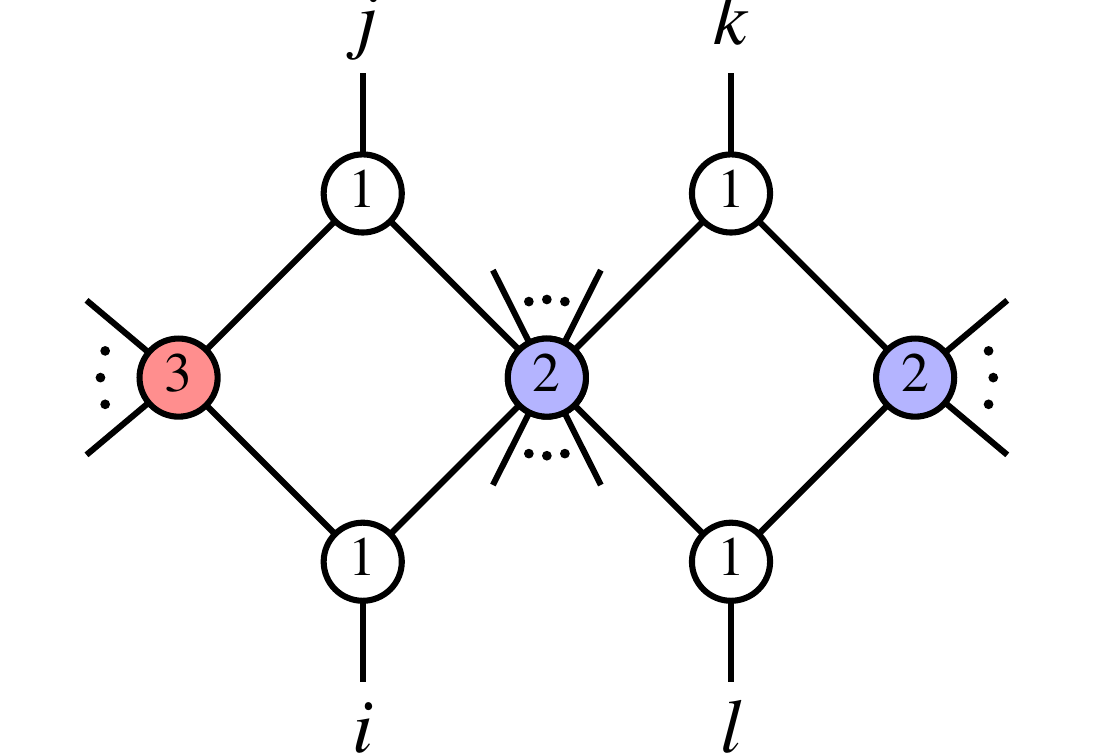}\\\mathcal{A}_{\mathrm{NMHV}}^{\mathrm{tree}}(i,\ldots,j)\end{array}&
\begin{array}{c}\figBox{0}{-1.825}{0.5}{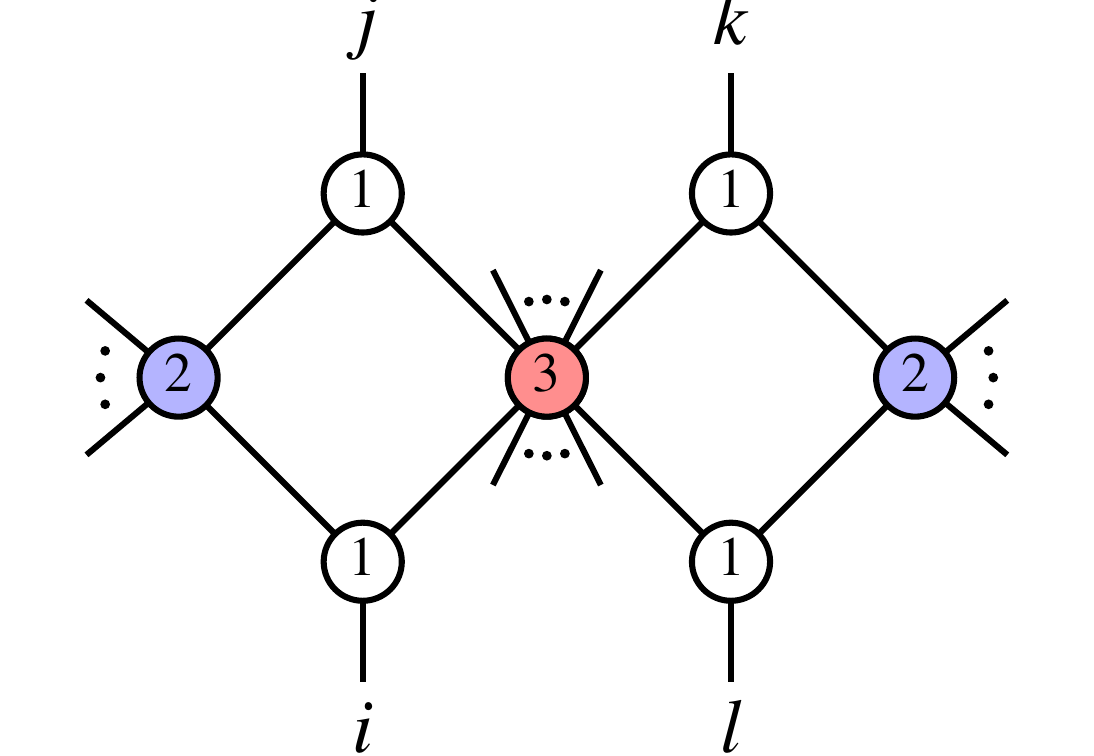}\\\mathcal{A}_{\mathrm{NMHV}}^{\mathrm{tree}}(j,\ldots,k;l,\ldots,i)\end{array}&
\begin{array}{c}\figBox{0}{-1.825}{0.5}{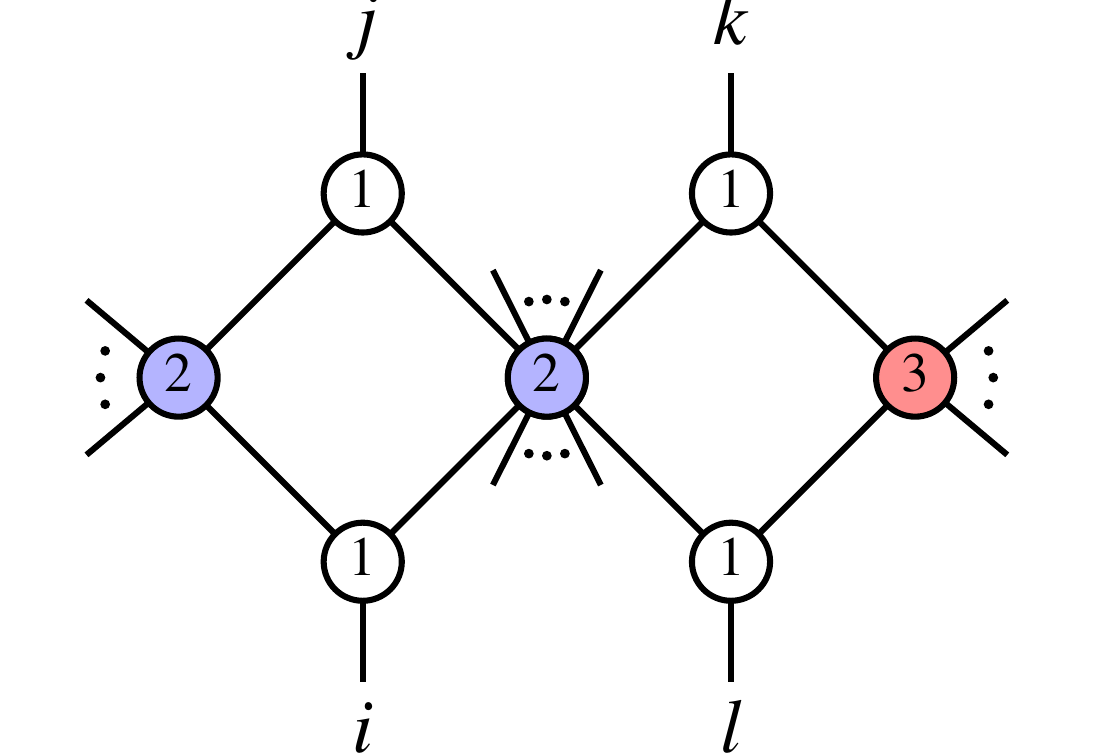}\\\mathcal{A}_{\mathrm{NMHV}}^{\mathrm{tree}}(k,\ldots,l)\end{array}\\[2.05cm]
\begin{array}{c}\figBox{0}{-1.825}{0.5}{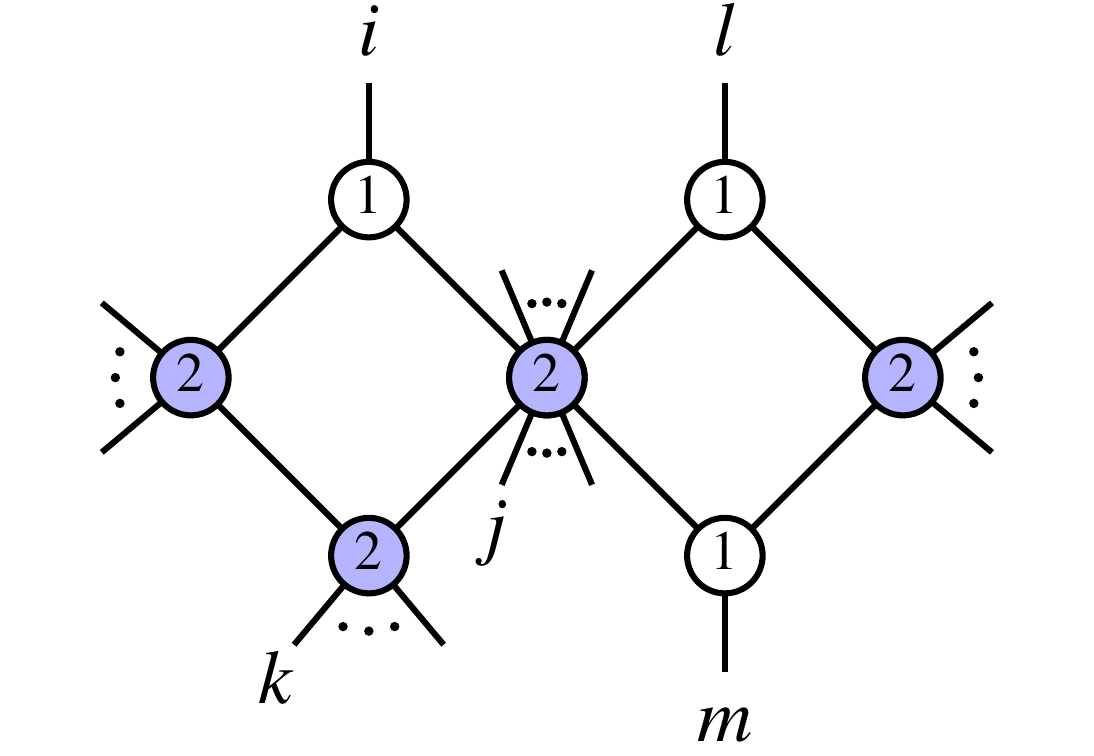}\\\left[i\,j\,j\pl1\,k\,k\pl1\right]\end{array}&
\begin{array}{c}\figBox{0}{-1.825}{0.5}{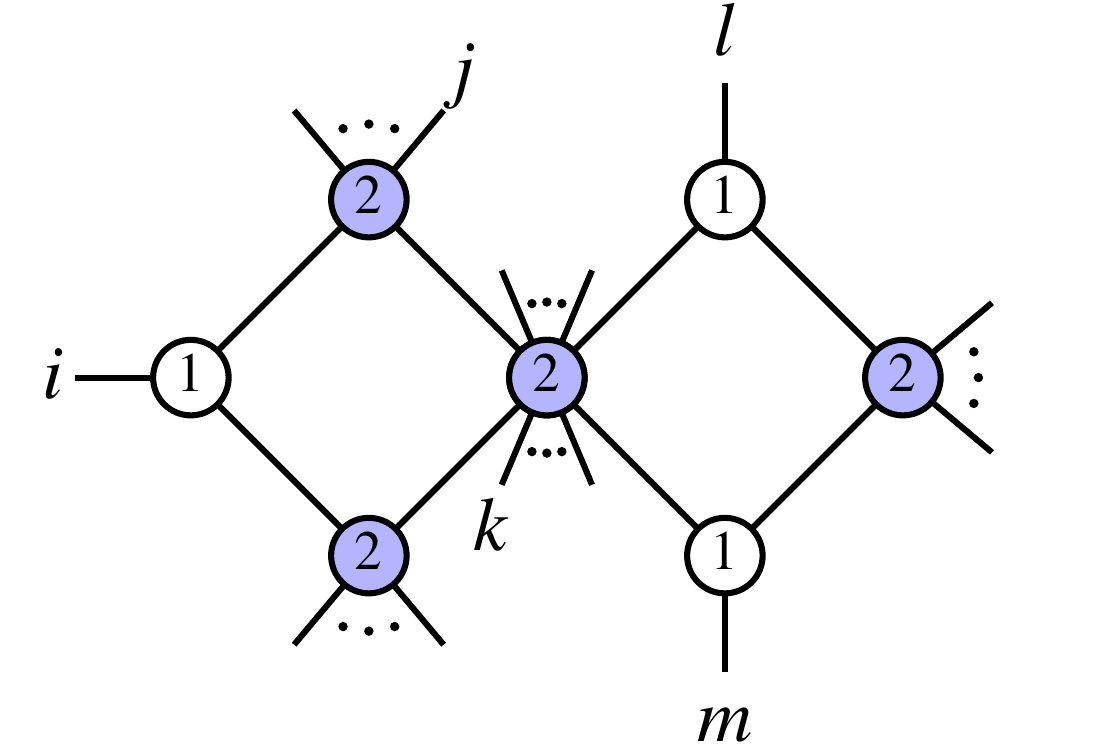}\\\left[i\,j\,j\pl1\,k\,k\pl1\right]\end{array}&
\begin{array}{c}\figBox{0}{-1.825}{0.5}{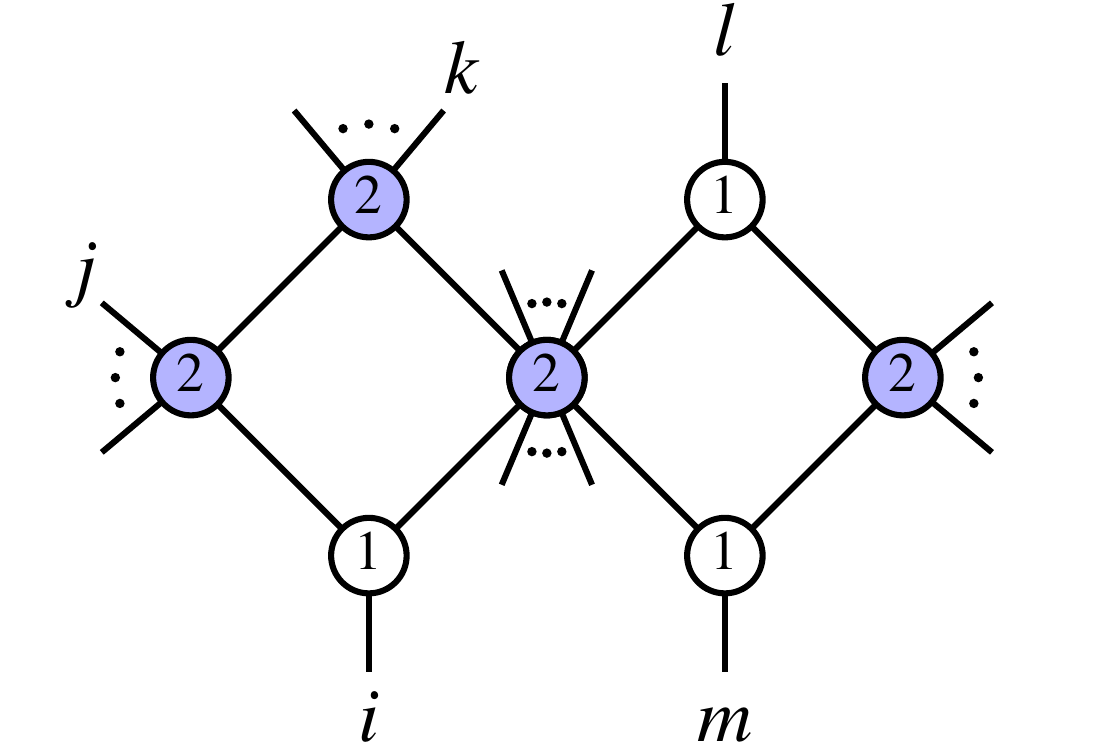}\\\left[i\,j\,j\pl1\,k\,k\pl1\right]\end{array}\end{array}}
Below each colored graph, we have indicated the leading singularity below each. Notice that the coefficient $\mathcal{A}_{{\rm NMHV}}^{\mathrm{tree}}(j,\ldots,k;l,\ldots i)$ is the {\it same} function as an ordinary tree amplitude with particles labelled $(j,\ldots,k;l,\ldots i)$ where $k,l$ and $i,j$ are both treated as if they were adjacently-labelled.

The idea is again to find a set of integrals that each
individually have the same leading singularities as the amplitude on a
given set of octa-cuts.
The first step is to realize that the octa-cuts on the first line of
\ref{NMHV2graphs} respectively looks like the product of NMHV 1-loop
quad-cut $\times$
MHV 1-loop quad-cut and MHV 1-loop quad-cuts $\times$ MHV 1-loop quad-cuts.
Therefore, one might think that the right integrals to start with
look like the product of pentagons that appear in MHV and NMHV
1-loop amplitudes. This strategy worked perfectly in the MHV 2-loop
case, where the amplitude was literary made from double-pentagons
whose origin was in the product of two MHV-like pentagons. So the
natural objects to consider here are the same
double-pentagons as in MHV 2-loop case and also other
double-pentagons that look like NMHV 1-loop $\times$ MHV 1-loop:
\eq{\nonumber\figBox{-1.5}{-2}{0.5}{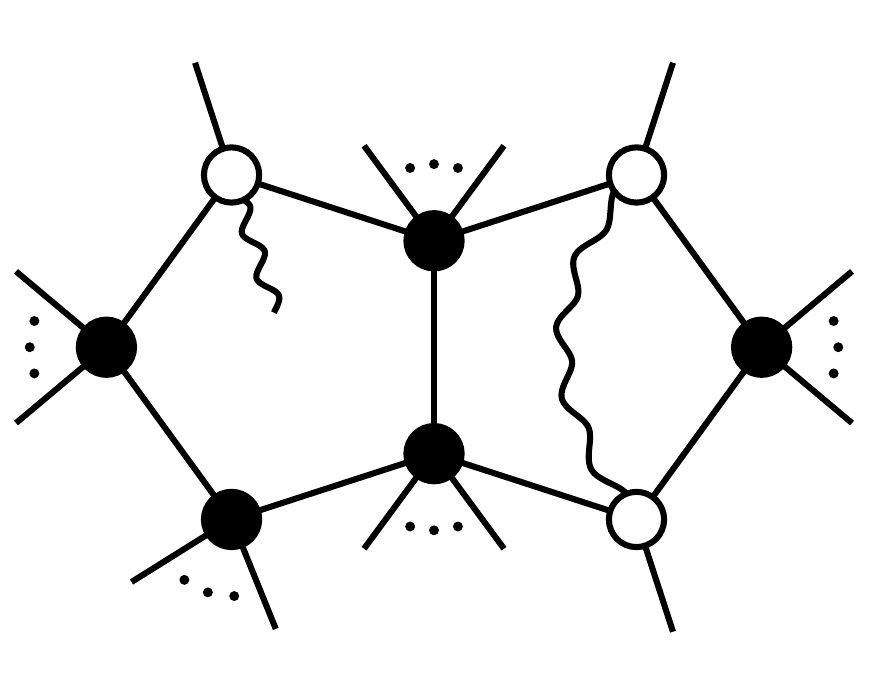}\figBox{0}{-2}{0.5}{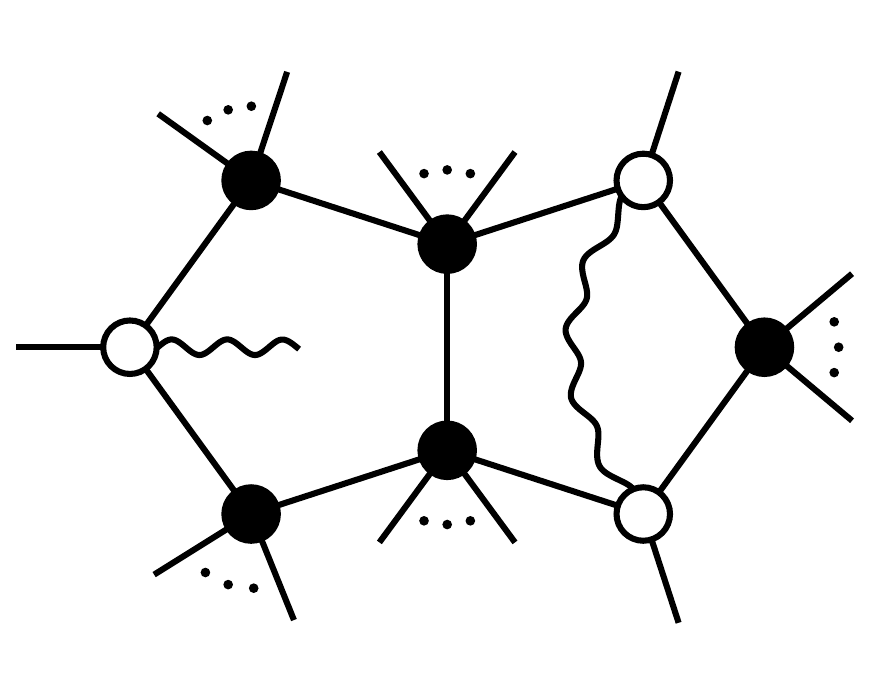}\figBox{0}{-2}{0.5}{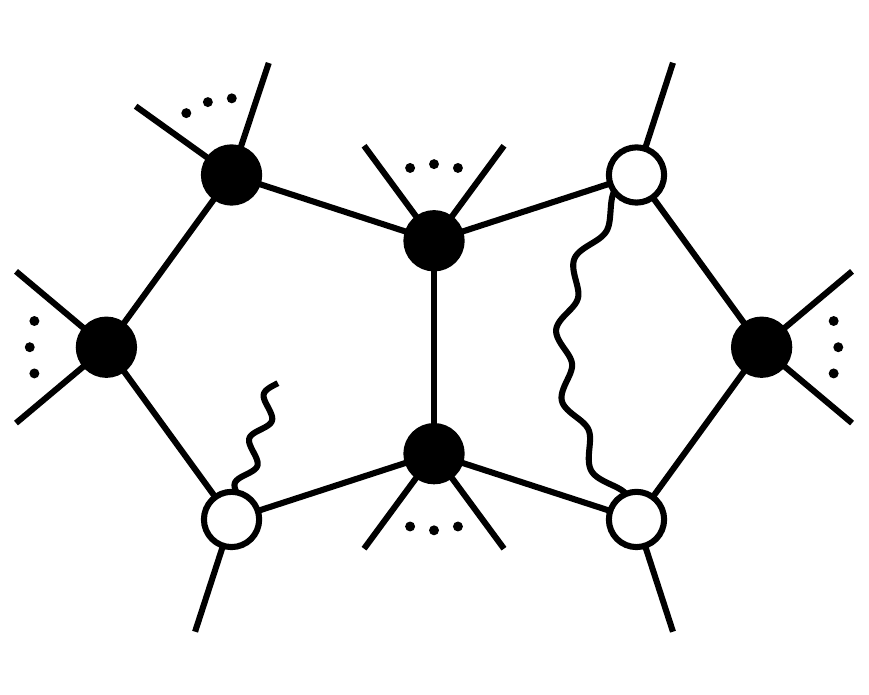}\figBox{0}{-2}{0.5}{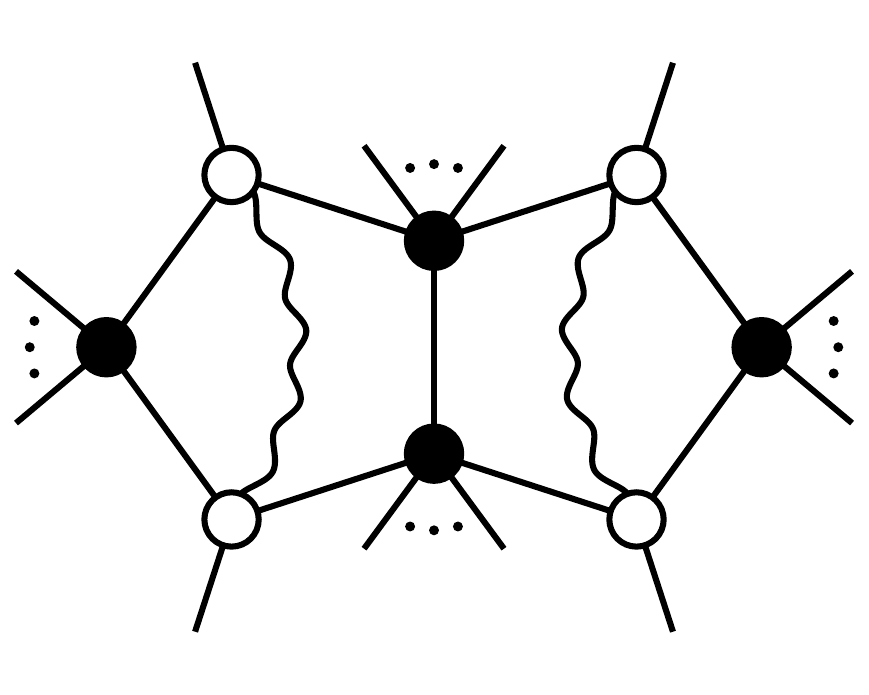}\label{NMHV2graphs}}

The numerators of the first three graphs have the same structure as
the ones that appear in the
NMHV 1-loop integrand.  We provide the complete expressions in \mbox{appendix \ref{nmhv_appendix}}.

Note that first three diagrams are really represented just by single
diagram with permuted indices. For instance, the second one can be obtained
from the first one if we require $k>i$. So, it is non-planar version
of the first graph in the same sense as we saw in the last
subsection in the case of the log of MHV amplitude. We
see that these four graphs are in one-to-one correspondence with the first
four colored graphs in \ref{NMHV2graphs}. If we cut all propagators
except $\la ABCD\ra$ we get not only the same cuts as are in these
colored graphs, but also the support on the correct Schubert problems. These
integrals are definitely the right ones to start with. In order to
get the correct field theory answer we have to multiply them by the
leading singularities of corresponding octa-cuts which are

Now summing over all allowed indices we get,
\vspace{-1.5cm}\eq{\hspace{1.2cm}\!\!\!\displaystyle\underset{\substack{\mathrm{all~allowed}\\i,j,k,l,m}}{\text{{\Huge$\sum$}}}\hspace{-0.2cm}\begin{array}{c}~\\[-0.1cm]\figBox{0}{-1.5}{0.5}{nmhv_two_loop_integrand_2.pdf}\\\times\left[i,j,j+1,k,k+1\right]\phantom{\times}\end{array}\hspace{-0.0cm}+\frac{1}{2}\hspace{-0.2cm}\underset{\substack{i<j<k<l<i}}{\text{{\Huge$\sum$}}}\hspace{-0.6cm}\begin{array}{c}~\\[0.78cm]\figBox{0}{-1.65}{0.5}{nmhv_two_loop_integrand_4.pdf}\\\times\left\{{\footnotesize\begin{array}{c}\mathcal{A}_{\mathrm{NMHV}}^{\mathrm{tree}}(j,\ldots,k;\;l,\ldots,i)\\+\mathcal{A}_{\mathrm{NMHV}}^{\mathrm{tree}}(i,\ldots,j)\\+\mathcal{A}_{\mathrm{NMHV}}^{\mathrm{tree}}(k,\ldots,l)\end{array}}\right\}\phantom{\times}\end{array}\hspace{-0.75cm}}
where the first diagram really represents three as we mentioned
earlier, namely, the complete set of cyclically ordered figures
\eq{\nonumber\hspace{-0.5cm}\figBox{0}{-1.65}{0.5}{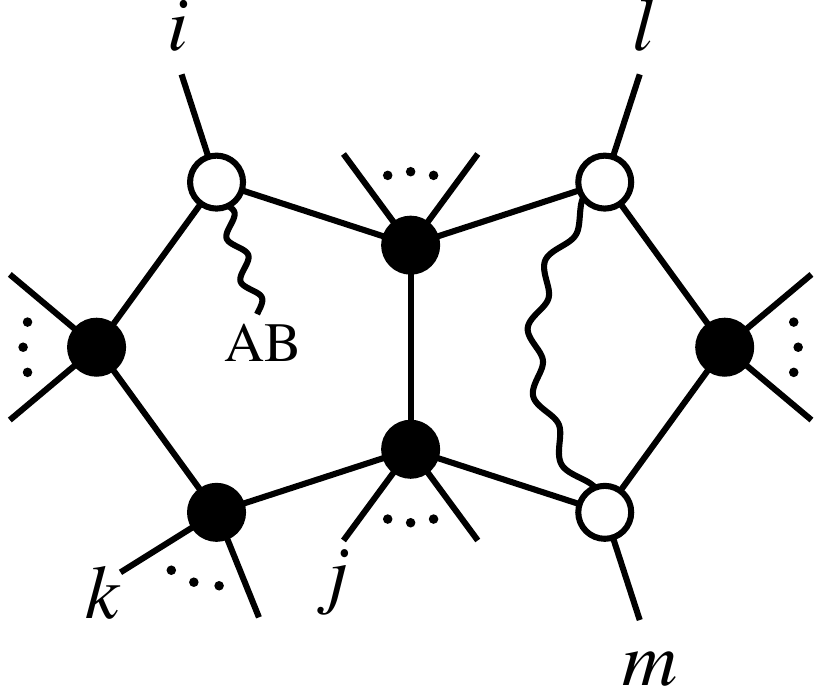}\quad\figBox{0}{-1.65}{0.5}{nmhv_two_loop_integrand_2.pdf}\quad\figBox{0}{-1.65}{0.5}{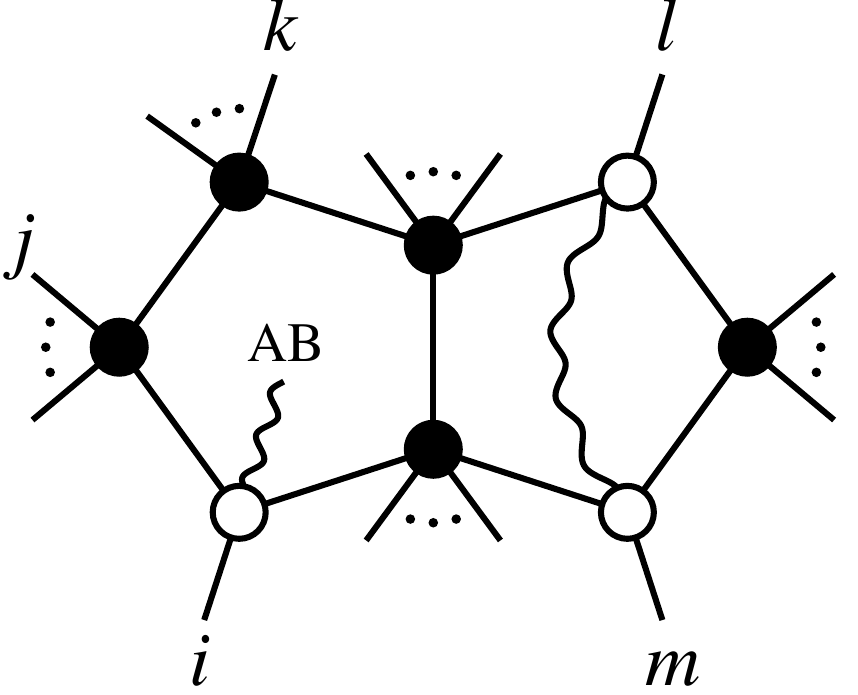}\label{NMHVeq2}}

The rest of the story proceeds in the by now familiar way. Simply
carrying out the sum over the range of indices corresponding to the
colored graphs does not give the right answer, however, a judicious
choice for the range of summation adds the correct ``boundary terms"
to give exactly the right answer, and we finally obtain:
\vspace{-0.1cm}
\vspace{-1.45cm}\eq{\hspace{-0.95cm}\mathcal{A}_{\mathrm{NMHV}}^{2\mathrm{-loop}}=\!\!\!\displaystyle\underset{\substack{i<j<l<m\leq k< i\\i<j<k<l<m\leq i\\i\leq l<m\leq j<k<i}}{\!\!\phantom{\frac{1}{2}}\!\text{{\Huge$\sum$}}\phantom{\!\!\frac{1}{2}\!}}\hspace{-0.2cm}\begin{array}{c}~\\[-0.1cm]\figBox{0}{-1.5}{0.5}{nmhv_two_loop_integrand_2.pdf}\\\times\left[i,j,j+1,k,k+1\right]\phantom{\times}\end{array}\hspace{0.3cm}\hspace{-0.3cm}+\,\,\frac{1}{2}\underset{\substack{i<j<k<l<i}}{\!\!\phantom{\frac{1}{2}}\!\text{{\Huge$\sum$}}\phantom{\frac{1}{2}\!\!\!}}\hspace{-0.6cm}\begin{array}{c}~\\[0.78cm]\figBox{0}{-1.65}{0.5}{nmhv_two_loop_integrand_4.pdf}\\\times\left\{{\footnotesize\begin{array}{c}\mathcal{A}_{\mathrm{NMHV}}^{\mathrm{tree}}(j,\ldots,k;\;l,\ldots,i)\\+\mathcal{A}_{\mathrm{NMHV}}^{\mathrm{tree}}(i,\ldots,j)\\+\mathcal{A}_{\mathrm{NMHV}}^{\mathrm{tree}}(k,\ldots,l)\end{array}}\right\}\phantom{\times}\end{array}\hspace{-0.75cm}\label{NMHV2final}}

These two terms represent the general 2-loop NMHV amplitude for any
number of external particles. The explicit forms of the integrals in
term of momentum-twistors are presented in \mbox{appendix \ref{nmhv_appendix}}.

\subsection{3-loop MHV amplitudes}

Finally, we present the integrand for all 3-loop
MHV amplitudes. These amplitudes were studied in the past, the 4pt
formula for the integrand was given in \cite{Bern:2005iz} and the 5pt in \cite{Spradlin:2008uu}. The
4pt and 5pt amplitudes were also determined using BCFW recursion and
translated into pure momentm-twistor integrals in
\cite{ArkaniHamed:2010kv}. However once again our new strategy
will both yield vastly simpler expressions for these integrands and
also generalize to all $n$.

We begin as always by drawing the colored graphs that contribute to
general 3-loop
amplitude. While there are a large number of them, our experience with
the 2-loop NMHV calculation
tells us that for the purpose of ``translating" the graphs into the
integrals, one
needs to focus on the colored graphs without internal propagators.
There are just two of these:
\vspace{-0.7cm}\eq{\figBox{0}{-1.8}{0.35}{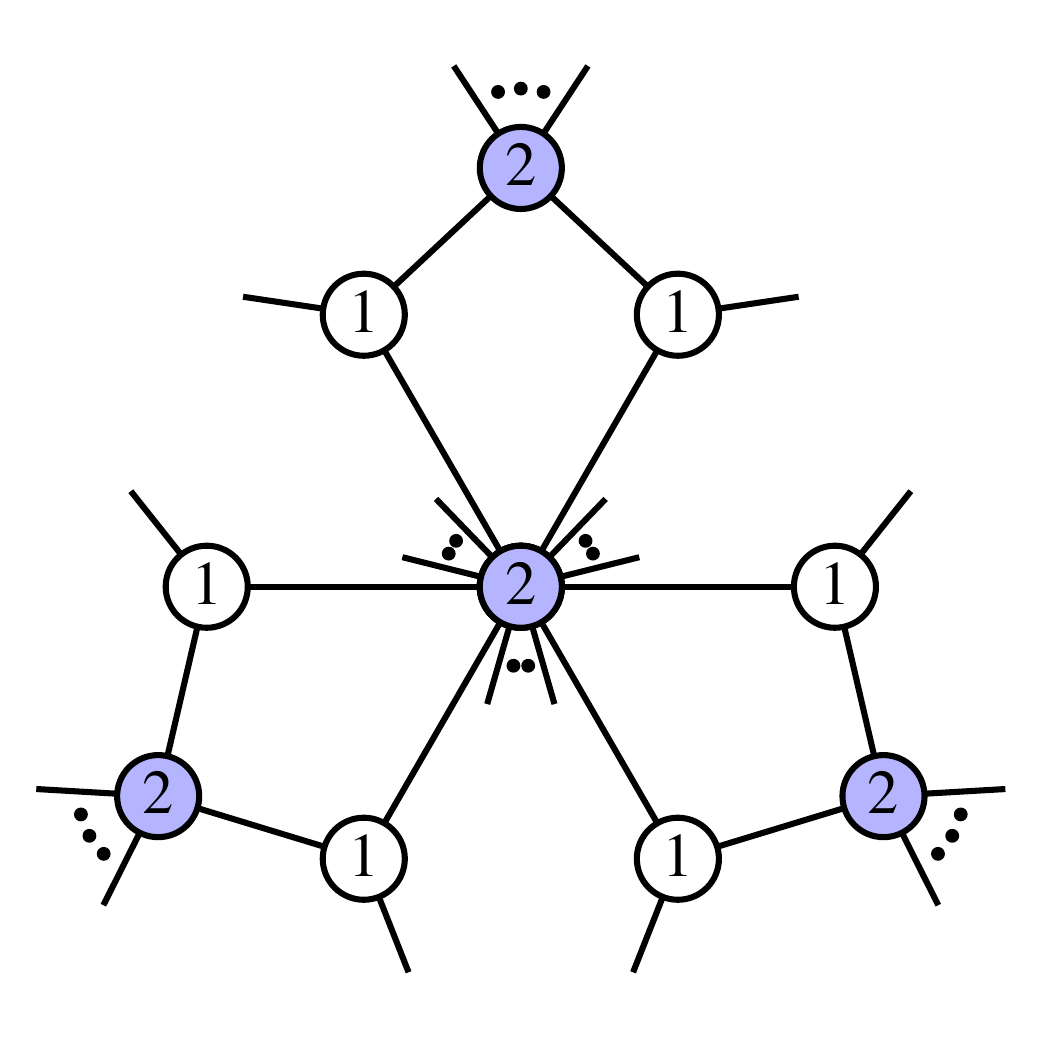}\figBox{0}{-1.65}{0.5}{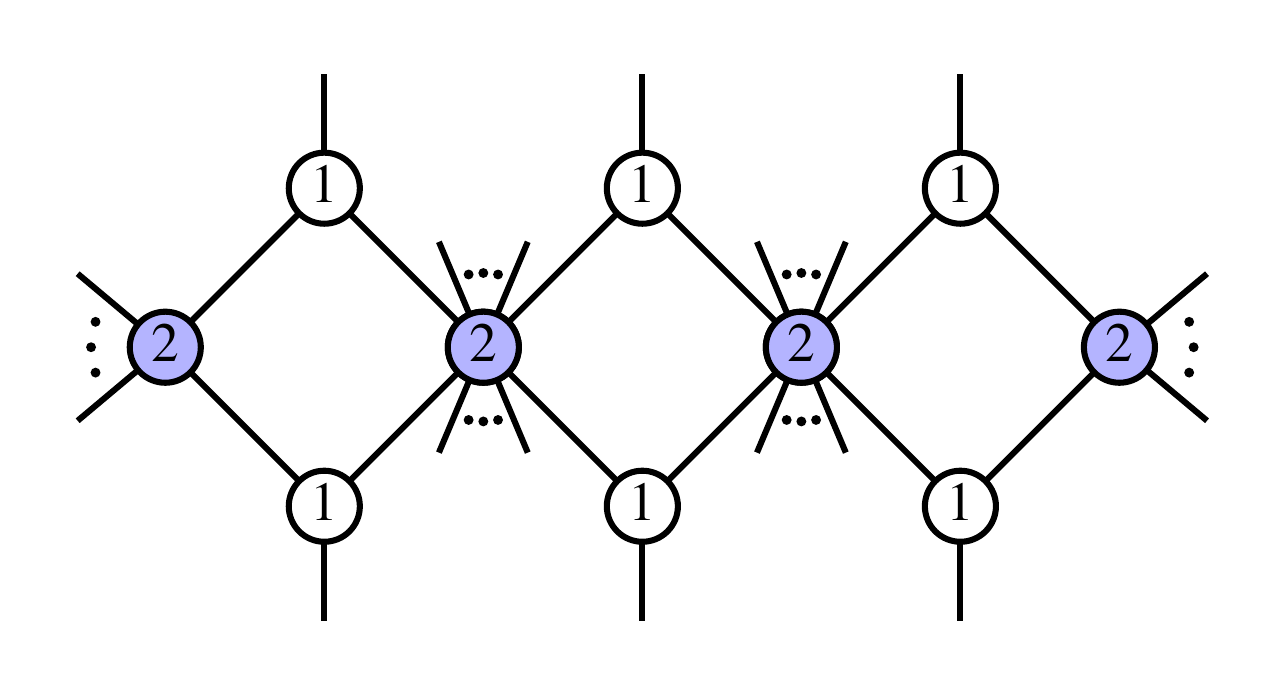}\label{Dodekacuts}\vspace{-0.7cm}}
The colored graphs suggest that the correct 3-loop integral must
correspond to ``gluing" together three 1-loop MHV integrals.
But these can not be just pentagons because of number of
internal propagators, we would also need hexagons. Fortunately, in the
``polytope picture" of \cite{polytopePaper}, the most
natural form of MHV amplitude {\it is} written using hexagons. We
leave the detailed exploration of this gluing procedure to future
work. It suffices to say that we can indeed find objects which have
support on the correct leading dodecacuts (\ref{Dodekacuts}). Having
identified them, the magic happens again: to get the full 3-loop
amplitude, we need only to identify the correct ranges for the
summations involved.
As a result, we can write the general 3-loop MHV amplitude for any number of
external particles as a sum of two structures,
\vspace{-0.4cm}\eq{\nonumber\hspace{-1cm}\mathcal{A}_{\mathrm{MHV}}^{3\mathrm{-loop}}=\!\!\!\!\displaystyle\underset{\substack{i_1\leq i_2<j_1\leq\\\leq j_2<k_1\leq k_2<i_1}}{\frac{1}{3}\!\!\text{{\Huge$\sum$}}\phantom{\frac{1}{2}\!\!}}\!\!\!\figBox{0}{-2.15}{0.55}{mhv_three_loop_integrand_1.pdf}+\!\!\displaystyle\underset{\substack{i_1\leq j_1< k_1<\\< k_2\leq j_2< i_2<i_1}}{\frac{1}{2}\!\text{{\Huge$\sum$}}\phantom{\frac{1}{4}\!\!}}\!\!\!\!\figBox{0}{-1.8}{0.55}{mhv_three_loop_integrand_5.pdf}\vspace{-0.cm}}
The explicit formulas for these graphs with all numerator factors
are given in the appendix \ref{3-loop-appendix}.\\~\\

\section*{Acknowledgements}
Its our please to acknowledge stimulating discussions with Simon Caron-Huot, James Drummond, Johannes Henn, Andrew Hodges, Lionel Mason, and Dave Skinner. N.A.-H. is supported by the DOE under grant DE-FG02-
91ER40654, F.C. was supported in part by the NSERC of Canada, MEDT of Ontario and by The
Ambrose Monell Foundation. J.T. is supported by the U.S. Department of State through a Fulbright Science and Technology Award.

\newpage
\appendix

\section{Residue Computations in Momentum-Twistor Space}\label{residue_appendix}

In section 3, we gave a heuristic argument for the form of the Jacobian in the computation of the residue of a pentagon integral. The actual computation is essentially trivial but it might serve as yet one more way to get used to momentum twistors. This is why we carry it out in detail in this appendix.

Recall that the non-vanishing residue of the pentagon integral for a contour which `encircles' the isolated pole $(AB)=(24)$ is computed using
\be
\oint\limits_{|(AB)-(24)|=\epsilon}\!\!\!\!\!\!\!\!\frac{\langle AB\,13\rangle \langle 12\,45\rangle\langle 23\,45\rangle}{\langle AB\,12\rangle\langle AB\,23\rangle\langle AB\,34\rangle\langle AB\,45\rangle\langle AB\,51\rangle},
\ee

As with all multidimensional residues, the entire computation amounts to Jacobians. Let us choose to expand $Z_A$ and $Z_B$ using the twistors $\{Z_5,Z_1,Z_2,Z_4\}$ as a basis; this parameterization introduces a Jacobian $\underset{AB\to(5\,1\,2\,4)}{J}=\ab{5\,1\,2\,4}^{-2}$. Exploiting the $GL(2)$-redundancy of the integrand, may therefore parameterize $Z_A$ and $Z_B$ according to \eqs{Z_A\equiv& \alpha_1 Z_5+\alpha_2Z_1+Z_2;\\Z_B\equiv& \beta_1 Z_5+\beta_2 Z_1+Z_4;}
Of course, the contour being evaluated corresponds to the choice of maps $f_i$ given by  $\vec{f}\equiv\{\ab{AB\,12},\ab{AB\,23},\ab{AB\,34},\ab{AB\,45}\}$; using these coordinates for $Z_A,Z_B$, the contour will be evaluated around the pole at the origin: $\alpha_i=\beta_i=0$.

With this, the integral in question has become fully gauge-fixed and concrete:
\eq{\oint\limits_{\substack{|\alpha_i|=|\beta_i|=\epsilon}}\!\!\!\!\!\!\!\!d^2\alpha_id^2\beta_i\,\,\frac{\langle AB\,13\rangle \langle 12\,45\rangle^3\langle 23\,45\rangle}{\langle AB\,12\rangle\langle AB\,23\rangle\langle AB\,34\rangle\langle AB\,45\rangle\langle AB\,51\rangle}.}
Because the contour encircles the origin, the Jacobian appearing the definition of a multidimensional residue will be evaluated at the origin. This means that for our purposes, we need only compute the maps $\vec{f}$ to linear-order in $(\alpha_i,\beta_i)$ to compute the residue.

Doing this in complete detail, we see that
\eq{\nonumber\hspace{-0.0cm}\begin{array}{lclllll}
f_1=&\ab{AB\,12}&=\alpha_1\ab{5\,4\,1\,2}&&&&+\ldots\\
f_2=&\ab{AB\,23}&=\alpha_1\ab{5\,4\,1\,3}&+\alpha_2\ab{1\,4\,2\,3}&&&+\ldots\\
f_3=&\ab{AB\,34}&=&&\beta_1\ab{2\,5\,3\,4}&+\beta_2\ab{2\,1\,3\,4}&+\ldots\\
f_4=&\ab{AB\,45}&=&&&\phantom{+}\beta_2\ab{2\,1\,4\,5}&+\ldots\\
\end{array}}
where `$\ldots$' stands for terms quadratic in $\alpha_i,\beta_i$. From this, it is trivial to read-off the Jacobian: \eq{\left.\phantom{\Big(}J\right|_{(AB)=(24)}=\ab{5\,4\,1\,2}\ab{1\,4\,2\,3}\ab{2\,5\,3\,4}\ab{2\,1\,4\,5}=\ab{1\,2\,4\,5}^2\ab{1\,2\,3\,4}\ab{2\,3\,4\,5};} combining this with the rest of the integrand---e.g.\ $\ab{AB\,13}/\ab{AB\,51}$ evaluated on $(AB)=(24)$---we find that
\eq{\nonumber\oint\limits_{|(AB)-(24)|=\epsilon}\!\!\!\!\!\!\!\!\frac{\langle AB\,13\rangle \langle 12\,45\rangle\langle 23\,45\rangle}{\langle AB\,12\rangle\langle AB\,23\rangle\langle AB\,34\rangle\langle AB\,45\rangle\langle AB\,51\rangle}=-\frac{\ab{2\,4\,1\,3}\ab{1\,2\,4\,5}^3\ab{2\,3\,4\,5}}{\ab{1\,2\,4\,5}^3\ab{1\,2\,3\,4}\ab{2\,3\,4\,5}}=1.}

\section{Explicit form of the 2-loop NMHV Amplitude}\label{nmhv_appendix}

In this appendix, we will provide all the details that go into the formula for the $n$-point 2-loop NMHV amplitude, which can be graphically represented as follows: \vspace{-1.25cm}\eq{\hspace{-0.75cm}\mathcal{A}_{\mathrm{NMHV}}^{2\mathrm{-loop}}=\!\!\!\displaystyle\underset{\substack{i<j<l<m\leq k< i\\i<j<k<l<m\leq i\\i\leq l<m\leq j<k<i}}{\!\!\phantom{\frac{1}{2}}\!\text{{\Huge$\sum$}}\phantom{\!\!\frac{1}{2}\!}}\hspace{-0.2cm}\begin{array}{c}~\\[-0.1cm]\figBox{0}{-1.5}{0.5}{nmhv_two_loop_integrand_2.pdf}\\\times\left[i,j,j+1,k,k+1\right]\phantom{\times}\end{array}\hspace{0.3cm}\hspace{-0.3cm}+\frac{1}{2}\,\,\underset{\substack{i<j<k<l<i}}{\!\!\phantom{\frac{1}{2}}\!\text{{\Huge$\sum$}}\phantom{\frac{1}{2}\!\!\!}}\hspace{-0.6cm}\begin{array}{c}~\\[0.78cm]\figBox{0}{-1.65}{0.5}{nmhv_two_loop_integrand_4.pdf}\\\times\left\{{\footnotesize\begin{array}{c}\mathcal{A}_{\mathrm{NMHV}}^{\mathrm{tree}}(j,\ldots,k;\;l,\ldots,i)\\+\mathcal{A}_{\mathrm{NMHV}}^{\mathrm{tree}}(i,\ldots,j)\\+\mathcal{A}_{\mathrm{NMHV}}^{\mathrm{tree}}(k,\ldots,l)\end{array}}\right\}\phantom{\times}\end{array}\hspace{-0.75cm}}
Of these two terms, only the first requires any comment, because the second summand involves only the familiar double-pentagons which generate the MHV two-loop amplitude's integrand,
\eq{\hspace{-1.9cm}\begin{array}{c}\figBox{0}{-1.65}{0.5}{nmhv_two_loop_integrand_4.pdf}\\\displaystyle\Leftrightarrow\hspace{-0.35cm}\int\limits_{(AB,CD)}\hspace{-0.5cm}\frac{\ab{AB\,(i\mi1\,i\,i\pl1)\newcap(j\mi1\,j\,j\pl1)}\ab{CD\,(k\mi1\,k\,k\pl1)\newcap(l\mi1\,l\,l\pl1)}\ab{i\,j\,k\,l}}{\ab{AB\,i\mi1\,i}\ab{AB\,i\,i\pl1}\ab{AB\,j\mi1\,j}\ab{AB\,j\,j\pl1}\ab{AB\,CD}\ab{CD\,k\mi1\,k}\ab{CD\,k\,k\pl1}\ab{CD\,l\mi1\,l}\ab{CD\,l\,l\pl1}}\phantom{\Leftrightarrow}\end{array}\hspace{-4cm}}

As indicated by the ranges of the summation, the first sum actually represents a sum over three distinct cyclic orderings of the labels $(i,j,k,l,m)$, corresponding to each of the following cyclically-ordered integrands, 

\vspace{0.4cm}\mbox{\hspace{-2.5cm}\begin{tabular}{lccc}
Integrand:&\figBox{0}{-1.65}{0.5}{nmhv_two_loop_integrand_2.pdf}&\figBox{0}{-1.65}{0.5}{nmhv_two_loop_integrand_1.pdf}&\figBox{0}{-1.65}{0.5}{nmhv_two_loop_integrand_3.pdf}\\
Range:&$i<j<l<m\leq k< i$&$i\leq l<m\leq j<k<i$&$i<j<k<l<m\leq i$\\
$\begin{array}{l}{\rm Boundary}\\{\rm terms:}\end{array}$&$\left\{\begin{array}{ll}\mathbf{A}&i\pl1=j\\\mathbf{B}&i\mi1=k\pl1\end{array}\right\}$&$\left\{\begin{array}{ll}\mathbf{A}&i=l\\\mathbf{B}&i\mi1=k\pl1\end{array}\right\}$&$\left\{\begin{array}{ll}\mathbf{A}&i\pl1=j\\\mathbf{B}&i=m\end{array}\right\}$
\end{tabular}\vspace{0.2cm}}~\\~\\

\noindent For each range of indices, there are {\it boundary-terms} for which the general integrand's numerator must change slightly; these have been indicated in the table above. Given the ranges and boundaries indicated above, the numerators for these contributions to the 2-loop NMHV amplitude are given by,
\eq{\begin{array}{lc}\text{term}&\mathrm{numerator}\\\hline\text{non-boundary}&\ab{AB\,(i\mi1\,i\,i\pl1)\newcap(\Sigma_{i,j,k})}\\\mathbf{A}\,\mathrm{boundary}&\ab{AB\,i\pl1(i\mi1\,i) \newcap(\Sigma_{i,j,k})}\\\mathbf{B}\,\mathrm{boundary}&\ab{AB\,i\mi1(i\,i\pl1) \newcap(\Sigma_{i,j,k})}\\
\mathbf{A\&B}\,\mathrm{boundary}&\ab{AB\,i\pl1\,i\mi1}\ab{i\,\Sigma_{i,j,k}}
\end{array}}
where in all these cases the special plane $\Sigma_{i,j,k}$ is given by the same object encountered at one-loop, but with the arbitrary bitwistor $X$ replaced by $(lm)$, \eq{\Sigma_{i,j,k}\equiv\frac{1}{2}\Big[(j\,j\pl1)\Big((i\,k\,k\pl1)\newcap(lm)\Big)-(k\,k\pl1)\Big((i\,j\,j\pl1)\newcap(lm)\Big)\Big].}

\newpage
\section{Explicit form of the 3-Loop MHV Amplitude}\label{3-loop-appendix}
In this appendix, we present the explicit form of the $n$-point 3-loop MHV amplitude, which we represent graphically graphically represented as follows: 
\vspace{-0.2cm}\eq{\nonumber\hspace{-1cm}\mathcal{A}_{\mathrm{MHV}}^{3\mathrm{-loop}}=\!\!\!\!\displaystyle\underset{\substack{i_1\leq i_2<j_1\leq\\\leq j_2<k_1\leq k_2<i_1}}{\frac{1}{3}\!\!\text{{\Huge$\sum$}}\phantom{\frac{1}{2}\!\!}}\!\!\!\figBox{0}{-2.15}{0.55}{mhv_three_loop_integrand_1.pdf}+\!\!\displaystyle\underset{\substack{i_1\leq j_1< k_1<\\< k_2\leq j_2< i_2<i_1}}{\frac{1}{2}\!\text{{\Huge$\sum$}}\phantom{\frac{1}{4}\!\!}}\!\!\!\!\figBox{0}{-1.8}{0.55}{mhv_three_loop_integrand_5.pdf}\vspace{-0.2cm}}

As described in the body of the paper, the `boundary terms' of the summands above require some comment. We will discuss the two topologies separately, starting with with the first summand in the equation above. Because when any two of the indices become identified in the first graph the wavy-line numerators become ill-defined, special consideration must be made for each of the degenerations allowed in the range of the summand---that is, all the cases where two or more of the indices are identified. Separating each type of such degeneration that is allowed in the first summand,
\vspace{-0.2cm}\eq{\hspace{-1.5cm}\nonumber\left.\hspace{-0.0cm}\displaystyle\underset{\substack{i_1\leq i_2<j_1\leq\\\leq j_2<k_1\leq k_2<i_1}}{\frac{1}{3}\text{{\Huge$\sum$}}\phantom{\frac{1}{4}\!\!}}\!\!\!\figBox{0}{-2.15}{0.55}{mhv_three_loop_integrand_1.pdf}\right.=\left\{
\begin{array}{l@{\hspace{-0.2cm}}clc}
1\times\frac{1}{3}&\underset{{\substack{i_1<i_2<j_1<\\<j_2<k_1<k_2<i_1}}}{\text{{\Large$\sum$}}}&\mathcal{I}_{1}^A(i_1,i_2,j_1,j_2,k_1,k_2)&\left(\begin{array}{c}\text{{\small all~indices}}\\\text{{\small distinct}}\end{array}\right)\\
3\times\!\!\frac{1}{3}&\underset{{\substack{i_1<i_2<j_1<\\<j_2<k<i_1}}}{\text{{\Large$\sum$}}}&\mathcal{I}_{2}^A(i_1,i_2,j_1,j_2,k)&{\small\left(k_1=k_2\equiv k\right)}\\
3\times\!\!\frac{1}{3}&\underset{{\substack{i_1<i_2<j<k<i_1}}}{\text{{\Large$\sum$}}}&\mathcal{I}_{3}^A(i_1,i_2,j,k)&{\small\left(\begin{array}{c}k_1=k_2\equiv k\\j_1=j_2\equiv j\end{array}\right)}\\
1\times\frac{1}{3}&\underset{{\substack{i<j<k<i}}}{\text{{\Large$\sum$}}}&\mathcal{I}_{4}^A(i,j,k)&{\small\left(\begin{array}{c}k_1=k_2\equiv k\\j_1=j_2\equiv j\\i_1=i_2\equiv i\end{array}\right)}\end{array}\right.}
Here, the overall factor of `$\tfrac{1}{3}$' reflects the $\mathbb{Z}_3$-symmetry of the loop integrand (recall that {\it every term in the sum is understood to be fully-symmetrized with respect to the $3!$ permutations of the loop-variable labels}); although every term in the summand has the same factor of $\tfrac{1}{3}$, the boundary terms for which {\it e.g.}\ $k_1=k_2$ in the sum are equivalent to those where $j_1=j_2$ or $i_1=i_2$, allowing us to represent all three degenerations with a single integrand---$\mathcal{I}_2^A$ in this case, and similarly for $\mathcal{I}_3^A$. 

Let us now carefully define the contributions to this class of graph each in turn. First, we have the generic integrand:

\noindent\hspace{-0.95cm}$\bullet\;\;$$\underset{\mathrm{for~}i_1<i_2<j_1<j_2<k_1<k_2<i_1}{\mathcal{I}_1^A(i_1,i_2,j_1,j_2,k_1,k_2)}\Longleftrightarrow\figBox{0}{-2.15}{0.55}{mhv_three_loop_integrand_1.pdf}\;\;\begin{array}{c}\text{Numerator}\\\mathrm{Tr}\left[(i_1\,|AB|\,i_2)(j_1\,|CD|\,j_2)(k_1\,|EF|\,k_2)\right]\\~\end{array}$\\
\indent Here, we have left implicit the twelve propagators shown in the figure by solid lines, and the three `wavy-line' numerators $\ab{AB\,(i_1\mi1\,i_1\,i_1\pl1)\newcap(i_2\mi1\,i_2\,i_2\pl1)}$ etc. Observe that we have introduced a new notation for remaining tensor components of the numerator for this integrand. Letting `$\bullet$' denote an arbitrary bitwistor, we may define a `trace' over a pair of such auxiliary bitwistors: \mbox{$\mathrm{Tr}\left[(a\,b\,\bullet)(\bullet\,c\,d)\right]\equiv\ab{a\,b\,c\,d}$}; that is, the trace is nothing but the completely-antisymmetric contraction of bitwistors which are dual to a pair of auxiliary bitwistors, which are indicated by `$\bullet$' in the corresponding formula.\footnote{The idea of `tracing' over auxiliary bitwistors turns out to be a very powerful generalization of the four-bracket. Indeed, all the four-brackets in this paper could be translated directly into traces, and often with considerable simplification.}

\indent It may be helpful to illustrate the meaning of this numerator using the familiar notation of Wick contraction; in this notation, the tensor numerator of $\mathcal{I}_1^A(i_1,i_2,j_1,j_2,k_1,k_2)$ corresponds to: $\scalebox{0.2}{$\newcap$}$
\eqs{\nonumber\hspace{-1.5cm}\mathrm{Tr}\left[(i_1\,|AB|\,i_2)(j_1\,|CD|\,j_2)(k_1\,|EF|\,k_2)\right]\equiv\Wwick{11}{\langle AB\,(i_1\, \bullet)\cap(i_2\,<1\bullet)\rangle\langle CD\,(j_1\,>1\bullet)\cap(j_2\,<2\bullet)\rangle \langle EF\,(k_1\,>2\bullet)\cap(k_2\,\bullet)\rangle}{1}{\langle AB\,(i_1\,<+\bullet)\cap(i_2\,\bullet)\rangle\langle CD\,(j_1\,\bullet)\cap(j_2\,\bullet)\rangle \langle EF\,(k_1\,\bullet)\cap(k_2\,>+\bullet)\rangle};} alternatively, the numerator can be written in any one of the following equivalent forms (the equality of which offering further justification for calling this a `trace'):
\eqs{\nonumber\hspace{-1.5cm}&\hspace{-2cm}\mathrm{Tr}\left[(i_1\,|AB|\,i_2)(j_1\,|CD|\,j_2)(k_1\,|EF|\,k_2)\right]\\&\equiv\ab{i_2\,j_1\Big[\Big(j_2\,k_1\,\big(\left(k_2\,i_1 A\right)\newcap(FE)\big)\Big)\newcap(DC)\Big]B}-(A\leftrightarrow B);\\
&=\ab{j_2\,k_1\Big[\Big(k_2\,i_1\,\big(\left(i_2\,j_1 C\right)\newcap(BA)\big)\Big)\newcap(FE)\Big]D}-(C\leftrightarrow D);\\
&=\ab{k_2\,i_1\Big[\Big(i_2\,j_1\,\big(\left(j_2\,k_1 E\right)\newcap(DC)\big)\Big)\newcap(BA)\Big]F}-(E\leftrightarrow F).}

As we will see presently, this numerator will change only very slightly for the boundary terms in the summand. Always leaving the propagators and wavy-line implicit from the the corresponding figures, the remaining integrands are defined according to the following:

\noindent\mbox{\hspace{-0.75cm}\begin{tabular}{@{$\bullet\;$}l@{$\;\Longleftrightarrow$}c@{}c}
$\underset{\mathrm{for~}i_1<i_2<j_1<j_2<k<i_1}{\mathcal{I}_2^A(i_1,i_2,j_1,j_2,k)}$&\figBox{0}{-2.35}{0.55}{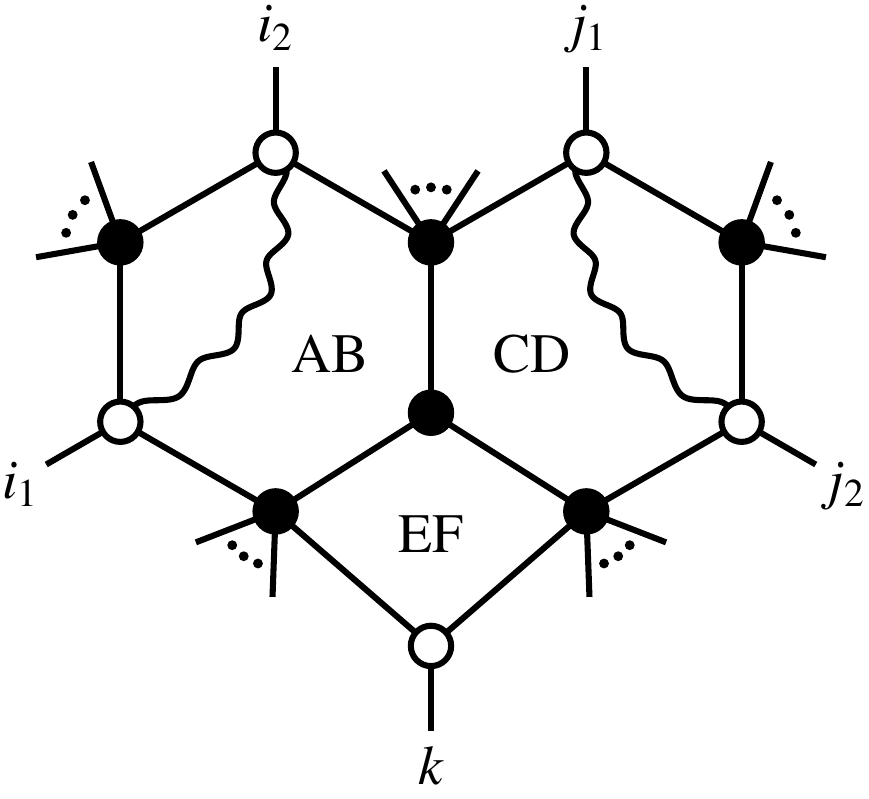}&$\begin{array}{c}\text{Numerator}\\\mathrm{Tr}\left[(i_1\,|AB|\,i_2)(j_1\,|CD|\,j_2)(k\,|k\mi1\,k\pl1|\,k)\right]\\~\end{array}$\\
$\underset{\mathrm{for~}i_1<i_2<j<k<i_1}{\mathcal{I}_3^A(i_1,i_2,j,k)}$&\figBox{0}{-1.85}{0.55}{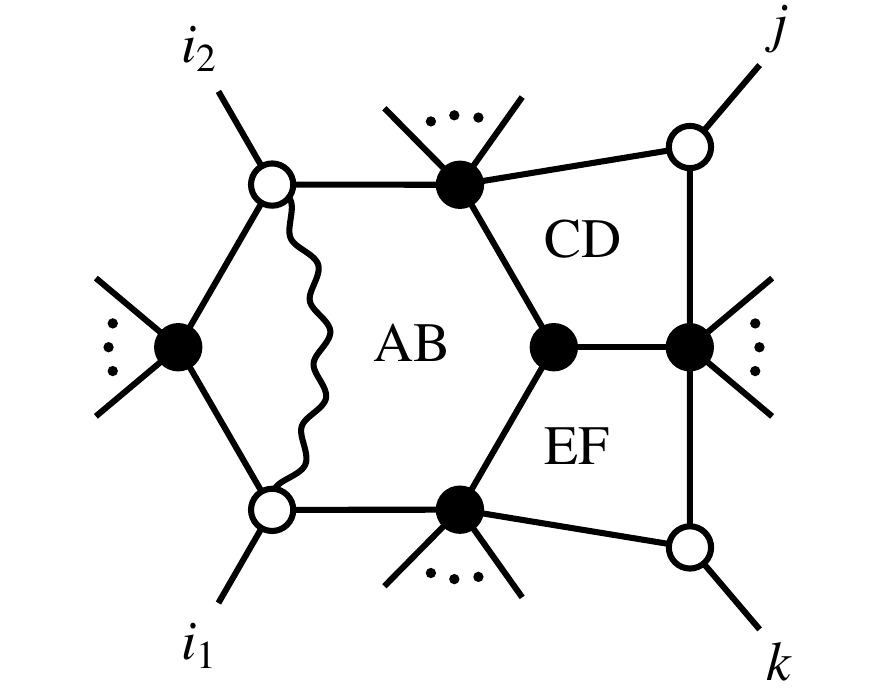}&$\begin{array}{c}\text{Numerator}\\\mathrm{Tr}\left[(i_1\,|AB|\,i_2)(j\,|j\mi1\,j\pl1|\,j)(k\,|k\mi1\,k\pl1|\,k)\right]\\~\end{array}$\\
$\underset{\mathrm{for~}i<j<k<i}{\mathcal{I}_4^A(i,j,k)}$&\figBox{0}{-2.125}{0.55}{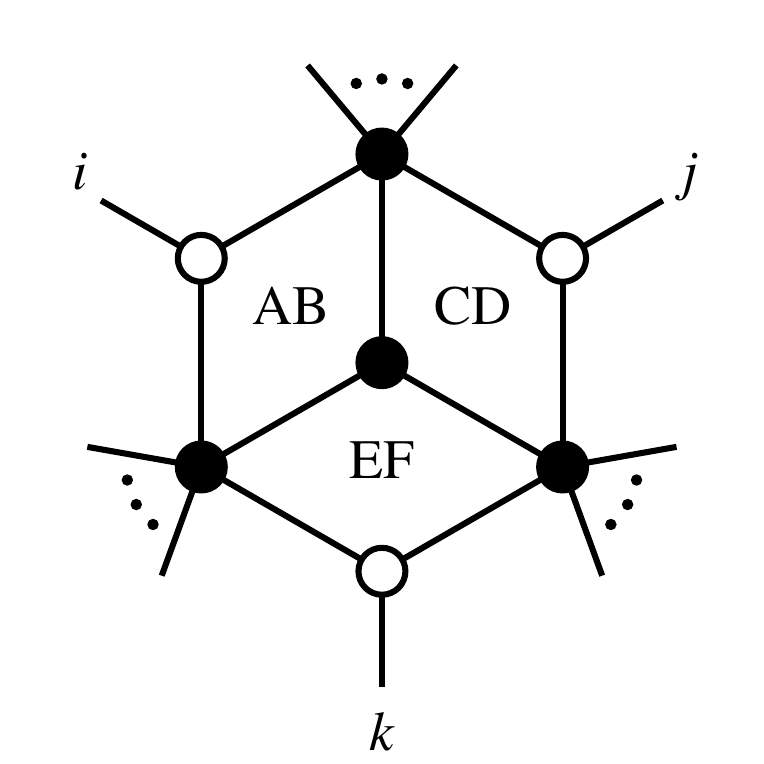}&$\begin{array}{c}\text{Numerator}\\\mathrm{Tr}\left[(i\,|i\mi1\,i\pl1|\,i)(j\,|j\mi1\,j\pl1|\,j)(k\,|k\mi1\,k\pl1|\,k)\right]\\~\end{array}$
\end{tabular}}

For the second topology, the boundary terms in the summand lead to just three separate contributions that must be specifically addressed. 
\vspace{-0.3cm}\eq{\hspace{-1.25cm}\nonumber\left.\hspace{-0.0cm}\displaystyle\underset{\substack{i_1\leq j_1< k_1<\\< k_2\leq j_2< i_2<i_1}}{\frac{1}{2}\!\text{{\Huge$\sum$}}\phantom{\frac{1}{4}\!\!}}\!\!\figBox{0}{-1.8}{0.55}{mhv_three_loop_integrand_5.pdf}\right.=\!\!\left\{
\begin{array}{l@{\hspace{-0.25cm}}c@{\hspace{-0.05cm}}lc}
1\times\frac{1}{2}&\underset{{\substack{i_1<j_1<k_1<\\< k_2< j_2< i_2<i_1}}}{\text{{\Large$\sum$}}}&\mathcal{I}_{1}^B(i_1,j_1,k_1,k_2,j_2,i_2)&\left(\begin{array}{c}\text{{\small all~indices}}\\\text{{\small distinct}}\end{array}\right)\\[0.75cm]
2\times\!\!\frac{1}{2}&\underset{{\substack{i_1<j_1<k_1<\\<k_2<i_2<i_1}}}{\text{{\Large$\sum$}}}&\mathcal{I}_{2}^B(i_1,j_1,k_1,k_2,i_2)&{\small\left(k_2=j_2\equiv k_2\right)}\\[0.6cm]
1\times\!\!\frac{1}{2}&\underset{{\substack{i_1<k_1<\\<k_2<i_2<i_1}}}{\text{{\Large$\sum$}}}&\mathcal{I}_{3}^B(i_1,k_1,k_2,i_2)&{\small\left(\begin{array}{c}i_1=j_1\equiv i_1\\k_2=j_2\equiv k_2\end{array}\right)}\end{array}\right.}
As above, the overall factor of `$\tfrac{1}{2}$' reflects the $\mathbb{Z}_2$-symmetry of the integrand (we remind the reader that each term in the summand is to be fully-symmetrized with respect to the $3!$ permutations of the loop variables). As before, we have exploited the symmetry of the integrand to identify various boundary terms: the degenerations $i_1=j_1$ and $k_2=j_2$, being equivalent in the cyclic sum, they can be combined into the single summand $\mathcal{I}_2^B$---which explains its relative factor of $2$. 

With this, we can directly present the three classes of integrands of the second topology which contribute to the 3-loop MHV amplitude:

\noindent\mbox{\hspace{-0.95cm}\begin{tabular}{@{$\bullet\;$}l@{$\;\Longleftrightarrow\;$}c@{$\;$}c}
$\underset{\mathrm{for~}i_1<j_1<k_1<k_2<j_2<i_2<i_1}{\mathcal{I}_1^B(i_1,j_1,k_1,k_2,j_2,i_2)}$&\figBox{0}{-1.825}{0.55}{mhv_three_loop_integrand_5.pdf}&$\begin{array}{c}\text{Numerator}\\\ab{AB\,(i_2\,i_1\,j_2)\newcap(j_1\mi1\,j_1\,j_1\pl1)}\\\times\ab{AB\,(j_2\mi1\,j_2\,j_2\pl1)\newcap(j_1\,k_1\,k_2)}\phantom{\times}\\~\end{array}$\\[1.95cm]
$\underset{\mathrm{for~}i_1<j_1<k_1<k_2<i_2<i_1}{\mathcal{I}_2^B(i_1,j_1,k_1,k_2,i_2)}$&\figBox{0}{-1.725}{0.55}{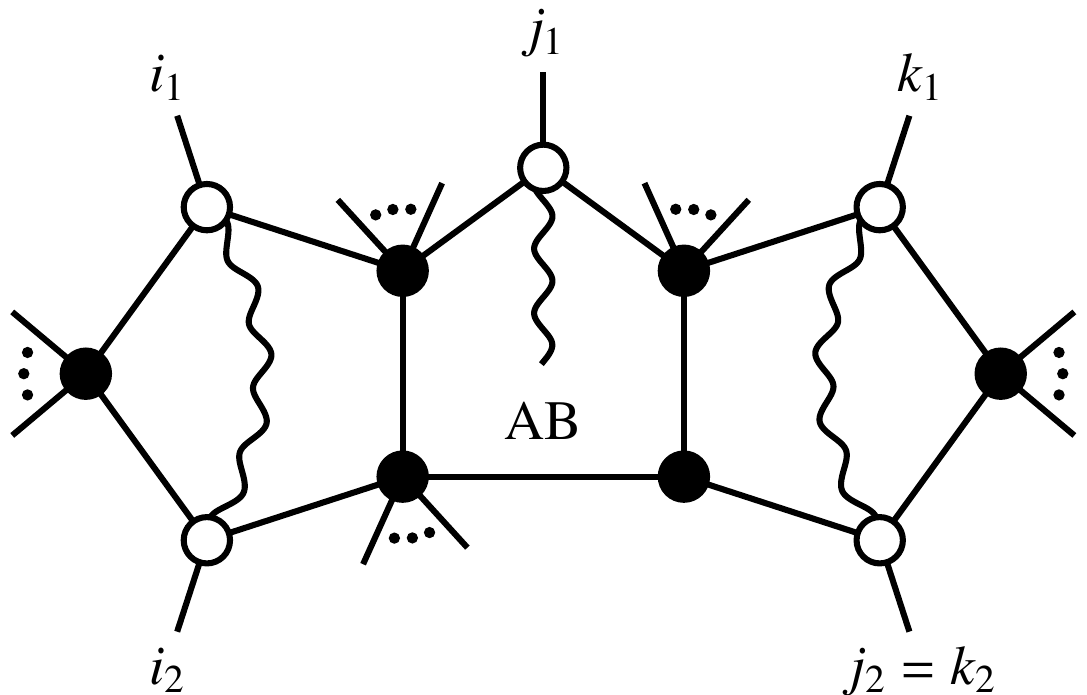}&$\begin{array}{c}\text{Numerator}\\\ab{AB\,(i_2\,i_1\,k_2)\newcap(j_1\mi1\,j_1\,j_1\pl1)}\\\times\ab{k_2\pl1\,j_1\,k_1\,k_2}\phantom{\times}\\~\end{array}$\\[1.95cm]
$\underset{\mathrm{for~}i_1<k_1<k_2<i_2<i_1}{\mathcal{I}_3^B(i_1,k_1,k_2,i_2)}$&\figBox{0}{-1.825}{0.55}{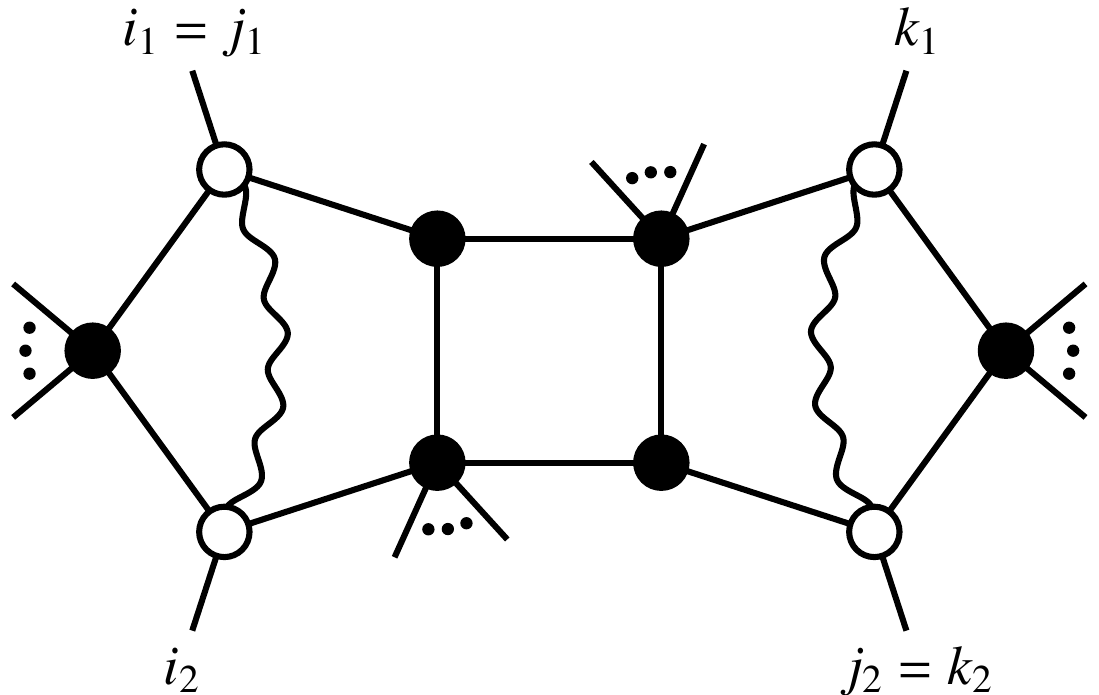}&$\begin{array}{c}\text{Numerator}\\\ab{k_2\,i_2\,i_1\,i_1\pl1}\\\times\ab{k_2\pl1\,i_1\,k_1\,k_2}\phantom{\times}\\~\end{array}$
\end{tabular}}

\newpage

\begin{thebibliography}{10}

\bibitem{ArkaniHamed:2010kv}
N.~Arkani-Hamed, J.~L. Bourjaily, F.~Cachazo, S.~Caron-Huot, and J.~Trnka,
  ``{The All-Loop Integrand For Scattering Amplitudes in Planar $\mathcal{N}=4$
  SYM},'' 2010, arXiv:1008.2958~[hep-th].

\bibitem{Britto:2004ap}
R.~Britto, F.~Cachazo, and B.~Feng, ``{New Recursion Relations for Tree
  Amplitudes of Gluons},'' {\em Nucl. Phys.}, vol.~B715, pp.~499--522, 2005,
  hep-th/0412308.

\bibitem{Britto:2005fq}
R.~Britto, F.~Cachazo, B.~Feng, and E.~Witten, ``{Direct Proof of Tree-Level
  Recursion Relation in Yang- Mills Theory},'' {\em Phys. Rev. Lett.}, vol.~94,
  p.~181602, 2005, hep-th/0501052.

\bibitem{ArkaniHamed:2009dn}
N.~Arkani-Hamed, F.~Cachazo, C.~Cheung, and J.~Kaplan, ``{A Duality for The
  S-Matrix},'' 2009, arXiv:0907.5418~[hep-th].

\bibitem{CaronHuot:2010ek}
S.~Caron-Huot, ``{Notes on the Scattering Amplitude / Wilson Loop Duality},''
  2010, arXiv:1010.1167~[hep-th].

\bibitem{Mason:2010yk}
L.~Mason and D.~Skinner, ``{The Complete Planar S-Matrix of $\mathcal{N}=4$ SYM
  as a Wilson Loop in Twistor Space},'' {\em JHEP}, vol.~12, p.~018, 2010,
  arXiv:1009.2225~[hep-th].

\bibitem{Alday:2008yw}
L.~F. Alday and R.~Roiban, ``{Scattering Amplitudes, Wilson Loops and the
  String/Gauge Theory Correspondence},'' {\em Phys.Rept.}, vol.~468,
  pp.~153--211, 2008, arXiv:0807.1889~[hep-th].
\newblock * Temporary entry *.

\bibitem{Kosower:2010yk}
D.~A. Kosower, R.~Roiban, and C.~Vergu, ``{The Six-Point NMHV Amplitude in
  Maximally Supersymmetric Yang-Mills Theory},'' 2010,
  arXiv:1009.1376~[hep-th].

\bibitem{polytopePaper}
N.~Arkani-Hamed, J.~L. Bourjaily, F.~Cachazo, A.~P. Hodges, and
  J.~Trnka. ``{A Note on Polytopes for Scattering Amplitudes},'' 2010.

\bibitem{Alday:2010vh}
L.~F. Alday, J.~Maldacena, A.~Sever, and P.~Vieira, ``{Y-System for Scattering
  Amplitudes},'' 2010, arXiv:1002.2459~[hep-th].

\bibitem{Alday:2010ku}
L.~F. Alday, D.~Gaiotto, J.~Maldacena, A.~Sever, and P.~Vieira, ``{An Operator
  Product Expansion for Polygonal Null Wilson Loops},'' 2010,
  arXiv:1006.2788~[hep-th].

\bibitem{Gaiotto:2010fk}
D.~Gaiotto, J.~Maldacena, A.~Sever, and P.~Vieira, ``{Bootstrapping Null
  Polygon Wilson Loops},'' 2010, arXiv:1010.5009~[hep-th].
\newblock * Temporary entry *.

\bibitem{Hodges:2009hk}
A.~Hodges, ``{Eliminating Spurious Poles from Gauge-Theoretic Amplitudes},''
  2009, arXiv:0905.1473~[hep-th].

\bibitem{Drummond:2008vq}
J.~M. Drummond, J.~Henn, G.~P. Korchemsky, and E.~Sokatchev, ``{Dual
  Superconformal Symmetry of Scattering Amplitudes in $\mathcal{N}=4$
  Super-Yang-Mills Theory},'' 2008, arXiv:0807.1095~[hep-th].

\bibitem{Berends:1981rb}
F.~A. Berends, R.~Kleiss, P.~De~Causmaecker, R.~Gastmans, and T.~T. Wu,
  ``{Single Bremsstrahlung Processes in Gauge Theories},'' {\em Phys. Lett.},
  vol.~B103, p.~124, 1981.

\bibitem{DeCausmaecker:1981bg}
P.~De~Causmaecker, R.~Gastmans, W.~Troost, and T.~T. Wu, ``{Multiple
  Bremsstrahlung in Gauge Theories at High-Energies. 1. General Formalism for
  Quantum Electrodynamics},'' {\em Nucl. Phys.}, vol.~B206, p.~53, 1982.

\bibitem{Kleiss:1985yh}
R.~Kleiss and W.~J. Stirling, ``{Spinor Techniques for Calculating
  $p\,\bar{p}\to W^\pm Z^0\,+$ Jets},'' {\em Nucl. Phys.}, vol.~B262,
  pp.~235--262, 1985.

\bibitem{Gunion:1985vca}
J.~F. Gunion and Z.~Kunszt, ``{Improved Analytic Techniques for Tree Graph
  Calculations and the $G\, g\, q\, \bar{q}$ Lepton anti-Lepton Subprocess},''
  {\em Phys. Lett.}, vol.~B161, p.~333, 1985.

\bibitem{Xu:1986xb}
Z.~Xu, D.-H. Zhang, and L.~Chang, ``{Helicity Amplitudes for Multiple
  Bremsstrahlung in Massless Nonabelian Gauge Theories},'' {\em Nucl. Phys.},
  vol.~B291, p.~392, 1987.

\bibitem{Drummond:2006rz}
J.~M. Drummond, J.~Henn, V.~A. Smirnov, and E.~Sokatchev, ``{Magic Identities
  for Conformal Four-Point Integrals},'' {\em JHEP}, vol.~01, p.~064, 2007,
  hep-th/0607160.

\bibitem{Penrose:1967wn}
R.~Penrose, ``{Twistor Algebra},'' {\em J. Math. Phys.}, vol.~8, p.~345, 1967.

\bibitem{Hodges:2010kq}
A.~Hodges, ``{The box integrals in momentum-twistor geometry},'' 2010,
  arXiv:1004.3323~[hep-th].

\bibitem{Dixon:1996wi}
L.~J. Dixon, ``{Calculating Scattering Amplitudes Efficiently},'' 1996,
  hep-ph/9601359.

\bibitem{Mason:2010pg}
L.~Mason and D.~Skinner, ``{Amplitudes at Weak Coupling as Polytopes in
  $AdS_5$},'' 2010, arXiv:1004.3498~[hep-th].

\bibitem{vanNeerven:1983vr}
W.~L. van Neerven and J.~A.~M. Vermaseren, ``{Large Loop Integrals},'' {\em
  Phys. Lett.}, vol.~B137, p.~241, 1984.

\bibitem{Bern:1994zx}
Z.~Bern, L.~J. Dixon, D.~C. Dunbar, and D.~A. Kosower, ``{One-Loop $n$-Point
  Gauge Theory Amplitudes, Unitarity and Collinear Limits},'' {\em Nucl.
  Phys.}, vol.~B425, pp.~217--260, 1994, hep-ph/9403226.

\bibitem{Drummond:2010mb}
J.~M. Drummond and J.~M. Henn, ``{Simple Loop Integrals and Amplitudes in
  $\mathcal{N}=4$ SYM},'' 2010, arXiv:1008.2965~[hep-th].

\bibitem{Alday:2009zm}
L.~F. Alday, J.~M. Henn, J.~Plefka, and T.~Schuster, ``{Scattering into the
  Fifth Dimension of $\mathcal{N}=4$ Super Yang- Mills},'' {\em JHEP}, vol.~01,
  p.~077, 2010, arXiv:0908.0684~[hep-th].

\bibitem{ELOP}
R.~J. Eden, P.~V. Landshoff, D.~I. Olive, and J.~C. Polkinghorne, {\em The
  Analytic S-Matrix}.
\newblock Cambridge University Press, 1966.

\bibitem{Britto:2004nc}
R.~Britto, F.~Cachazo, and B.~Feng, ``{Generalized Unitarity and One-Loop
  Amplitudes in $\mathcal{N} = 4$ Super-Yang-Mills},'' {\em Nucl. Phys.},
  vol.~B725, pp.~275--305, 2005, hep-th/0412103.

\bibitem{ArkaniHamed:2008gz}
N.~Arkani-Hamed, F.~Cachazo, and J.~Kaplan, ``{What is the Simplest Quantum
  Field Theory?},'' 2008, arXiv:0808.1446~[hep-th].

\bibitem{Cachazo:2008vp}
F.~Cachazo, ``{Sharpening The Leading Singularity},'' 2008,
  arXiv:0803.1988~[hep-th].

\bibitem{Griffiths:1978a}
P.~Griffiths and J.~Harris, {\em Principles of Algebraic Geometry}.
\newblock New York: Wiley, 1978.

\bibitem{Buchbinder:2005wp}
E.~I. Buchbinder and F.~Cachazo, ``{Two-Loop Amplitudes of Gluons and Octa-Cuts
  in $\mathcal{N} = 4$ Super Yang-Mills},'' {\em JHEP}, vol.~11, p.~036, 2005,
  hep-th/0506126.

\bibitem{ArkaniHamed:2009si}
N.~Arkani-Hamed, F.~Cachazo, C.~Cheung, and J.~Kaplan, ``{The S-Matrix in
  Twistor Space},'' 2009, arXiv:0903.2110~[hep-th].

\bibitem{ArkaniHamed:2009vw}
N.~Arkani-Hamed, F.~Cachazo, and C.~Cheung, ``{The Grassmannian Origin of Dual
  Superconformal Invariance},'' 2009, arXiv:0909.0483~[hep-th].

\bibitem{Mason:2009qx}
L.~Mason and D.~Skinner, ``{Dual Superconformal Invariance, Momentum Twistors
  and Grassmannians},'' {\em JHEP}, vol.~11, p.~045, 2009,
  arXiv:0909.0250~[hep-th].

\bibitem{Drummond:2010uq}
J.~M. Drummond and L.~Ferro, ``{The Yangian {O}rigin of the Grassmannian
  {I}ntegral},'' 2010, arXiv:1002.4622~[hep-th].

\bibitem{Schubert:1879}
H.~Schubert, {\em {Kalk\"{u}l der Abz\"{a}hlenden Geometrie}}.
\newblock Verlag von B. G. Teubner, 1879.

\bibitem{schubert:Bio}
W.~Burau and B.~Renschuch, {\em {Erg\"{a}nzungen zur Biographie von Hermann
  Schubert}}.
\newblock Mitt. Math. Ges. Hamb. 13, pp. 63-65, ISSN 0340-4358, 1993.

\bibitem{Drummond:2008bq}
J.~M. Drummond, J.~Henn, G.~P. Korchemsky, and E.~Sokatchev, ``{Generalized
  Unitarity for $\mathcal{N}=4$ Super-Amplitudes},'' 2008,
  arXiv:0808.0491~[hep-th].

\bibitem{Brandhuber:2009kh}
A.~Brandhuber, P.~Heslop, and G.~Travaglini, ``{Proof of the Dual Conformal
  Anomaly of One-Loop Amplitudes in $\mathcal{N}=4$ SYM},'' {\em JHEP},
  vol.~10, p.~063, 2009, arXiv:0906.3552~[hep-th].

\bibitem{Elvang:2009ya}
H.~Elvang, D.~Z. Freedman, and M.~Kiermaier, ``{Dual Conformal Symmetry of
  1-Loop NMHV Amplitudes in $\mathcal{N}=4$ SYM Theory},'' 2009,
  arXiv:0905.4379~[hep-th].

\bibitem{Bern:2005iz}
Z.~Bern, L.~J. Dixon, and V.~A. Smirnov, ``{Iteration of Planar Amplitudes in
  Maximally Supersymmetric Yang-Mills Theory at Three Loops and Beyond},'' {\em
  Phys. Rev.}, vol.~D72, p.~085001, 2005, hep-th/0505205.

\bibitem{Spradlin:2008uu}
M.~Spradlin, A.~Volovich, and C.~Wen, ``{Three-Loop Leading Singularities and
  BDS Ansatz for Five Particles},'' {\em Phys. Rev.}, vol.~D78, p.~085025,
  2008, arXiv:0808.1054~[hep-th].

\end{thebibliography}

\end{document}